\documentclass[reprint,prd,nofootinbib,superscriptaddress]{revtex4}
\usepackage{amssymb,wrapfig,amsmath,hyperref,graphicx,epsfig,bm,color,verbatim,slashed,mathtools,xparse,adjustbox,dcolumn,float,relsize,tikz,cancel}
\usepackage{caption,booktabs,array}
\usepackage{subcaption,multirow}
\usepackage{float}
\usepackage[compat=1.1.0]{tikz-feynman}
\usepackage{subcaption}
\usepackage[english]{babel}
\usepackage[utf8]{inputenc}
\usepackage[mathletters]{ucs}

\newcolumntype{C}{>{$}c<{$}}
\AtBeginDocument{
\heavyrulewidth=.08em
\lightrulewidth=.05em
\cmidrulewidth=.03em
\belowrulesep=.65ex
\belowbottomsep=0pt
\aboverulesep=.4ex
\abovetopsep=0pt
\cmidrulesep=\doublerulesep
\cmidrulekern=.5em
\defaultaddspace=.5em
}

\pdfoutput=1
\newcommand{\func}{\operatorname}
\DeclareUnicodeCharacter{2212}{-}
\input{tcilatex}
\begin{document}

\title{Can the muon anomalous magnetic moment and the $B$ anomalies be simultaneously explained in a minimally extended $Z^{\prime}$ model?}
\author{H. Lee}
\email{Huchan.Lee@ncbj.gov.pl}
\affiliation{National Centre for Nuclear Research,\\
Pasteura 7, 02-093 Warsaw, Poland}

\author{A. E. C\'{a}rcamo Hern\'{a}ndez}
\email{antonio.carcamo@usm.cl}
\affiliation{Departamento de F\'{\i}sica, Universidad T\'{e}cnica Federico Santa Mar\'{\i}a,\\
 Casilla 110-V, Valpara\'{\i}so, Chile }
\affiliation{{Centro Cient\'{\i}fico-Tecnol\'ogico de Valpara\'{\i}so, Casilla 110-V,
Valpara\'{\i}so, Chile}}
\affiliation{{Millennium Institute for Subatomic Physics at High-Energy Frontier
(SAPHIR), Fern\'andez Concha 700, Santiago, Chile}}

\date{\today }

\begin{abstract}
We analyze the implications of an extended 2HDM theory in quark masses and mixings, electroweak precision tests, charged lepton flavor violating (CLFV) decays, neutrino trident production, B meson, $g-2$ muon and W anomalies. In the extended 2HDM theory considered in this work, the SM gauge symmetry is enlarged by the local $U(1)^{\prime}$ symmetry, whereas its scalar and fermion sectors are augmented by the inclusion of a gauge singlet scalar and vectorlike fermions, respectively. Within the framework of this model, we determine the allowed ranges of the $Z^{\prime}$ gauge boson mass consistent with the constraints arising from neutrino trident production, $R_{K^{*}}$ anomaly, $B_{s}$ meson oscillation. We conclude the $R_{K^{*}}$ and muon $g-2$ anomalies cannot be simultaneously explained by the same new physics under all the constraints. Besides that, we found that the new physics sources, the $Z^{\prime}$ and non SM scalars cannot be connected via the scalar potential. Furthermore, we have found that the model under consideration can successfully comply with the constraints arising from oblique $T,S,U$ parameters as well as with the $W$ mass  and $g-2$ muon anomalies. In particular we found that the muon $g-2$ anomaly, whose dominant contribution arises from the one loop level scalar exchange, 
 can be successfully accommodated within the $2\sigma$ experimentally allowed range.
\end{abstract}

\maketitle

\section{INTRODUCTION} \label{sec:I}
The Standard Model (SM) has explained many observables with a great consistency, so being established as a cornerstone to understand our nature. In spite of its great successes, there are a few critical limitations in the SM and one of them is muon and electron anomalies. Until the new result for muon and electron $g-2$ were released in 2020, explaining the two anomalies simultaneously had been regarded as a potential new physics signal due to their relative different overall sign and magnitude. The old experimental constraints for the muon (the Brookhaven E821 experiment at BNL~\cite{Muong-2:2006rrc}) and electron $g-2$ (the Berkeley 2018 experiment~\cite{Parker:2018vye}) at $1\sigma$ read:
\begingroup
\begin{equation}
\begin{split}
\Delta a_{\mu} &= a_{\mu}^{\func{exp}} - a_{\mu}^{\func{SM}} = \left( 26.1 \pm 8.0 \right) \times 10^{-10},
\\
\Delta a_{e} &= a_{e}^{\func{exp}} - a_{e}^{\func{SM}} = \left( -0.88 \pm 0.36 \right) \times 10^{-12}.
\label{eqn:old_muon_electrong2}
\end{split}
\end{equation}
\endgroup
The old muon $g-2$ experimental constraint reports $3.3\sigma$, whereas the old electron $g-2$ reports $-2.4\sigma$ deviation from the SM prediction. One of our works~\cite{Hernandez:2021tii} tried to explain both anomalies in a unified way, based on the old experimental constraints, and showed that the SM $W$ exchange at one-loop level is not possible, however the CP-even and -odd scalar exchange at one-loop level  
allows to successfully 
explain both anomalies within the $1\sigma$ experimentally allowed range. 
This situation has changed significantly 
when the new experimental constraints for the muon 
 anomalous magnetic moment 
 was released in 2020. The renewed muon $g-2$ constraint is given by FNAL~\cite{Muong-2:2021ojo} 
\begingroup
\begin{equation}
\begin{split}
\Delta a_{\mu} &= a_{\mu}^{\func{exp}} - a_{\mu}^{\func{SM}} = \left( 25.1 \pm 5.9 \right) \times 10^{-10},
\label{eqn:new_muong2}
\end{split}
\end{equation}
\endgroup
which reports $4.2\sigma$ deviation from the SM prediction~\cite{Aoyama:2020ynm,Aoyama:2012wk,Aoyama:2019ryr,Czarnecki:2002nt,Gnendiger:2013pva,Davier:2017zfy,Keshavarzi:2018mgv,Colangelo:2018mtw,Hoferichter:2019mqg,Davier:2019can,Keshavarzi:2019abf,Kurz:2014wya,Melnikov:2003xd,Masjuan:2017tvw,Colangelo:2017fiz,Hoferichter:2018kwz,Gerardin:2019vio,Bijnens:2019ghy,Colangelo:2019uex,Blum:2019ugy,Colangelo:2014qya}. Around the same time, the renewed electron $g-2$ constraint is given by the LKB 2020 experiment~\cite{Morel:2020dww}
\begingroup
\begin{equation}
\begin{split}
\Delta a_{e} &= a_{e}^{\func{exp}} - a_{e}^{\func{SM}} = \left( 0.48 \pm 0.30 \right) \times 10^{-12},
\label{eqn:new_electrong2}
\end{split}
\end{equation}
\endgroup
which reports $1.6\sigma$ SM deviation. Comparing the old constraints for the muon and electron $g-2$ with the renewed results, one can know that the muon $g-2$ gets more important whereas the electron $g-2$ gets less important due to the reduced room for physics beyond the SM. Especially, the muon $g-2$ with $4.2\sigma$ SM deviation can be studied in the context of $Z^{\prime}$ models~\cite{Altmannshofer:2016brv,CarcamoHernandez:2019ydc,Belanger:2015nma,CarcamoHernandez:2019xkb,Allanach:2015gkd,Raby:2017igl,Kawamura:2019rth,Kawamura:2019hxp,Kawamura:2021ygg} or of leptoquarks~\cite{Cheung:2001ip,ColuccioLeskow:2016dox,Crivellin:2020tsz} or of an augmented scalar sector with/without vectorlike fermions~\cite{Arnan:2019uhr,Crivellin:2018qmi,Crivellin:2021rbq,Hernandez:2021tii}. Instead of considering the electron $g-2$ for new physics, the $B$ anomalies start  getting more attention and interest by the particle physics community due to its sizeable SM deviation corresponding to $3-4\sigma$ at most. The $B$ anomalies we consider in this work are the $R_{K}$ anomaly which has been studied by BABAR~\cite{BaBar:2012mrf}, Belle~\cite{BELLE:2019xld}, LHC~\cite{LHCb:2014vgu,LHCb:2019hip}, and LHCb~\cite{LHCb:2021trn} and $R_{K^{*}}$ by LHCb~\cite{LHCb:2017avl}. Specially, the $R_{K}$ anomaly which has been studied by LHCb~\cite{LHCb:2021trn} reports a recent measurement in the dilepton mass-squared range $1.1\func{GeV}^{2} < q^{2} < 6.0\func{GeV}^{2}$
\begingroup
\begin{equation}
R_{K}^{\left[ 1.1, 6.0 \right]} = \frac{\func{BR}\left( B \rightarrow K \mu^{+} \mu^{-} \right)}{\func{BR}\left( B \rightarrow K e^{+} e^{-} \right)} = 0.846_{-0.041}^{+0.044},
\end{equation}
\endgroup
which corresponds for $2.4-2.5\sigma$ SM deviation and appears to indicate 
the breaking of SM lepton universality. The breaking of SM lepton universality can also be seen by the $R_{K^{*}}$ anomaly, which has been explored by LHCb~\cite{LHCb:2017avl}, with increased SM deviation of $3.1\sigma$
\begingroup
\begin{equation}
R_{K^{*}}^{\left[ 1.1, 6.0 \right]} = \frac{\func{BR}\left( B \rightarrow K^{*} \mu^{+} \mu^{-} \right)}{\func{BR}\left( B \rightarrow K^{*} e^{+} e^{-} \right)} = 0.690_{-0.120}^{+0.160}.
\end{equation}
\endgroup
The $B$ anomaly can be accessed at tree level via the virtual exchange of a neutral and massive $Z^{\prime}$ gauge boson~\cite{Crivellin:2015mga,Crivellin:2015lwa,Chiang:2017hlj,King:2017anf,King:2018fcg,Falkowski:2018dsl}, or via the leptoquark exchange ~\cite{Becirevic:2017jtw,deMedeirosVarzielas:2018bcy,DeMedeirosVarzielas:2019nob,King:2021jeo} for the purpose of enhancing its theoretical sensitivity so that it can be seen by close future experiments. The sizeable anomalies, muon $g-2$ and $B$ anomalies, are an important clue in the search for physics beyond the SM and simultaneously explaining both of them in a well-motivated BSM theory can be an intuitive study for new physics. These attempts to consider both anomalies in an unified way can be performed  
via $Z^{\prime}$ gauge boson~\cite{Navarro:2021sfb} or leptoquark.
\\~\\
We start from this question: is it possible to explain the muon $g-2$ and $B$ anomalies in an extended $Z^{\prime}$ model while fullfilling all the well-known SM constraints? Answering to this question is not easy as there can exist very diverse $Z^{\prime}$ models and one of them, like the one of Ref.~\cite{Navarro:2021sfb} which has been a good motivation for this work and a guideline, tells that it is possible to explain both anomalies in a simplified fermionic $Z^{\prime}$ model with a fewer number of parameters. However, the simplified $Z^{\prime}$ model assumes no direct mixing between the SM fermions, so not considering CLFV constraints such as $\tau \rightarrow \mu \gamma$ and $\tau \rightarrow 3\mu$, which appears in our work as one of the most significant constraints, therefore drawing an opposite conclusion that the muon $g-2$ and $R_{K^{*}}$ anomalies cannot be explained by the same new physics as there is no overlapped $Z^{\prime}$ mass range in the model under consideration. 
\\~\\
In this work, what we want to achieve is to confirm whether the muon anomalous magnetic moment $g-2$ and $B$ anomalies can be explained in a unified way, taking into account 
many SM constraints as possible, in an extended $Z^{\prime}$ model. The extended $Z^{\prime}$ model features both a fermionic $Z^{\prime}$ model~\cite{CarcamoHernandez:2019ydc,Belanger:2015nma,CarcamoHernandez:2019xkb,Allanach:2015gkd,Raby:2017igl,Kawamura:2019rth,Kawamura:2019hxp,Kawamura:2021ygg,Navarro:2021sfb} and an extend 2HDM model~\cite{Hernandez:2021tii,CarcamoHernandez:2021yev}. To be more specific, in the extended 2HDM theory considered in this work, the SM gauge symmetry is enhanced by the inclusion of the $U(1)^{\prime}$ local symmetry. Its scalar sector is the same as the scalar sector of the extended 2HDM theory of~\cite{Hernandez:2021tii}, thus allowing to simultaneously accommodate both muon and electron $g-2$ anomalies via the non-SM scalar exchange as in~\cite{Hernandez:2021tii}. However we consider a $U(1)^{\prime}$ local symmetry instead of the global $U(1)^{\prime}$ of the model of~\cite{Hernandez:2021tii}. It is worth mentioning that in the model considered in~\cite{Hernandez:2021tii}, the CKM mixing matrix can be explained to a good approximation~\cite{CarcamoHernandez:2021yev}, predicting the mass range of vectorlike charged leptons as well as vectorlike quarks by charged lepton flavor violation (CLFV) decays and flavor changing neutral currents (FCNCs). Given that in the extended 2HDM theory considered in this work, a local $U(1)^{\prime}$ symmetry is considered, its gauge sector includes a neutral massive fermionic $Z^{\prime}$ gauge boson. 
Therefore, the BSM model under consideration has two new physics sources, which are $Z^{\prime}$ gauge boson and the non-SM scalars (CP-even and -odd scalars), and then we try to answer whether the muon anomaly and $B$ anomalies can be simultaneously explained in the BSM model.
\\~\\
For the investigation, the SM constraints that will be analyzed in this work are: 
\begin{itemize}
\item The SM lepton sector : Muon $g-2$, Neutrino trident process, CLFV constraint $\tau \rightarrow \mu \gamma$ and $\tau \rightarrow 3\mu$
\item The SM quark sector : $R_{K(K^{*})}$ anomaly, $B_{s}$ meson oscillation, Collider constraint, The CKM mixing matrix
\item Electroweak precision data corresponding to the oblique $T$, $S$ and $U$ parameters as well as the SM $W$ mass anomaly and the tree level vacuum stability of the scalar potential.
\end{itemize}
The SM fermion sector in this BSM model is extended by a complete vectorlike family and one vectorlike family contains two fermionic seesaw mediators that will trigger an seesaw mechanism to produce the masses of the second and third generation of the SM fermions~\cite{Hernandez:2021tii,CarcamoHernandez:2021yev}. And that is why the SM constraints shown above consist of the second and third generation of the SM fermions. Adding one more vectorlike family can give rise to two extra fermionic mediators, allowing the whole generation of the SM fermions to acquire masses, however this approach makes our analysis much more complicated since we want to diagonalize the mass matrices in a very concise, accurate, and complete way without any assumptions to increase predictability of the model. For this reason, we restrict our attention to the second and third generation of SM fermions. On top of that, it is worth mentioning that this BSM model is a simple and economical BSM model as it can explain well the SM mixing parameters as well as the SM fermion masses, thus allowing a comprehensive and complete analysis of the anomalies. First of all, we try to determine the highest order of the $Z^{\prime}$ coupling constant in the well-motivated BSM model using the charged lepton sector observables, which are the muon $g-2$, the branching ratio of the CLFV $\tau \rightarrow \mu \gamma$ and $\tau \rightarrow 3\mu$ decays, and then to find the highest order can reach up to unity. Based on the determined order of $g_{X}$, $M_{Z^{\prime}}$ ranges are derived from the neutrino trident production, $R_{K^{*}}$ anomaly, and $B_{s}$ meson oscillation and we find there is no overlapped region which can satisfy all constraints, thus considering only one obtained from $R_{K^{*}}$ anomaly and regarding one from $B_{s}$ meson oscillation to be explained by some other new physics sources at higher energies. At this stage, we separated $Z^{\prime}$ mass range derived from each fit of $R_{K^{*}}$ anomaly by two cases where the first is one that the experimental CMS and upper-limit of $\tau \rightarrow \mu \gamma$ are considered together, and the second is one considering only the experimental upper-limit of $\tau \rightarrow \mu \gamma$ decay constraint and we call the second ``theoretically interesting $Z^{\prime}$ mass range. The reason why we distinguish two cases is it is evident that the $R_{K^{*}}$ anomaly and muon $g-2$ can not be explained simultaneously by the same new physics under all the SM constraints as there is no overlapped $Z^{\prime}$ mass range (as reflected in the first case), whereas the $Z^{\prime}$ for the $R_{K^{*}}$ anomaly and non-SM scalars for the muon $g-2$ can be connected via the scalar potential under consideration in this work while fitting the oblique parameters as well as the $W$ mass anomaly in the ``theoretically interesting $Z^{\prime}$ mass range". Finally, we explore the electroweak precision observables as well as muon $g-2$ anomaly mediated by one loop level scalar exchange by fitting all the relevant scalar- and gauge-mediated observables and confirm that they can be explained at their $2\sigma$ constraints at most, thus concluding that the muon $g-2$ and $R_{K^{*}}$ anomalies can not be explained by the same new physics, however their new physics sources $Z^{\prime}$ and non-SM scalars (CP-even and -odd scalars) are closely interconnected via the scalar potential as well as the electroweak precision observables in the ``theoretically interesting $Z^{\prime}$ mass range".
\\~\\
The rest of this paper is organized as follows. In section~\ref{sec:II}, we introduce the particle spectrum and the relevant Lagrangians. In section~\ref{sec:III}, we construct the mass matrices for the charged lepton and quark sectors and we perform their diagonalization. In section~\ref{sec:IV}, we determine the $Z^{\prime}$ couplings in the interaction basis and we derive the $Z^{\prime}$ interactions in the mass basis. In section~\ref{sec:V}, the lepton sector phenomenology is analyzed by the massive neutral $Z^{\prime}$ gauge boson, determining the upper bound of the $Z^{\prime}$ coupling constant $g_{X}$ considering the $Z^{\prime}$ mass as a free parameter. In section~\ref{sec:VI}, the quark sector phenomenology is explored with the obtained $Z^{\prime}$ coupling constant and the $Z^{\prime}$ mass ranges are derived from the quark sector observables like the $R_{K^{*}}$ anomaly and the $B_{s}$ meson oscillation parameters as well as from the neutrino trident production. In section~\ref{sec:VII}, we analyze the electroweak precision data and fit all the relevant scalar- and gauge-mediated observables including muon $g-2$ and $W$ mass anomalies to their experimental bounds and then finally we conclude that the muon $g-2$ and $R_{K^{*}}$ anomaly can not be explained simultaneously by the same new physics, however they can be successfully accommodated within their $2\sigma$ experimentally allowed range in the ``theoretically interesting $Z^{\prime}$ mass range". Finally, we state our conclusions in section~\ref{sec:VIII}. We relegate all the derived theoretical predictions for the vacuum polarization amplitudes $\Pi$s contributing to the oblique parameters $T,S,U$ to Appendices~\ref{app:A} to \ref{app:E}.
\section{AN EXTENDED $Z^{\prime}$ MODEL WITH A FOURTH vectorlike FAMILY} \label{sec:II}
The SM, which has survived against many experiments, is an important theory which has successfully described the electromagnetic, weak and strong interactions. However the SM is not a complete answer for our nature due to the fact that it does not provide a mechanism for explaining the tiny neutrino masses, the strong SM fermion mass hierarchy, the quark and lepton mixings as well as the flavor anomalies and the $g-2$ anomalies. In order to address these issues,   we consider an extension of the SM motivated to provide a mechanism that explain the SM fermion mass hierarchy as well as the other issues previously mentioned. The well-measured SM mass parameters shows a very strong hierarchy between the SM neutrino mass and the top quark mass. Especially, the tiny SM neutrino masses are regarded to be explained by the see-saw mechanism mediated by heavy right-handed neutrinos, rather than by the Yukawa interaction. As in the SM neutrinos, it can be thought that the SM charged fermions masses can also be generated by a seesaw mechanism mediated by heavy charged vectorlike fermions. Therefore, all the SM fermion masses in the BSM theory considered in this work can be explained by the same dynamical mechanism. This approach has several advantages: 1) The
 effective Yukawa couplings are proportional to a product of two other dimensionless couplings at each vertex, so a moderate hierarchy in those couplings can yield a quadratically larger hierarchy in the effective couplings, which together with the mass of the heavy vectorlike fermions allow to successfully explain the strong SM fermion mass hierarchy. 2) Given that SM charged fermion masses will scale with the inverse of heavy vectorlike fermion masses, one can obtain a range for the mass scale of heavy vectorlike fermions. 3) The vectorlike charged leptons which mediate the seesaw mechanism that yields the SM charged lepton masses are also crucial for radiatively generating the muon and electron anomalous magnetic moments, whose magnitudes are not explained by Standard Model. For the reasons mentioned, this economical BSM theory is quite interesting and well-motivated. The particle content with their assigments under the $SU(3)_C\times SU(2)_L\times U(1)_Y\times U(1)^\prime$ gauge symmetry of the BSM theory considered in this work are shown in Table~\ref{tab:BSM_model}:
\begin{table}[H]
\resizebox{\textwidth}{!}{
\centering\renewcommand{\arraystretch}{1.3} 
\begin{tabular}{*{21}{c}}
\toprule
\toprule
Field & $Q_{iL}$ & $u_{iR}$ & $d_{iR}$ & $L_{iL}$ & $e_{iR}$ & $Q_{kL}$ & $%
u_{kR}$ & $d_{kR}$ & $L_{kL}$ & $e_{kR} $ & $\nu_{kR}$ & $\widetilde{Q}_{kR}$
& $\widetilde{u}_{kL}$ & $\widetilde{d}_{kL}$ & $\widetilde{L}_{kR}$ & $%
\widetilde{e}_{kL}$ & $\widetilde{\nu}_{kR}$ & $\phi$ & $H_u$ & $H_d$ \\ 
\midrule
$SU(3)_C$ & $\mathbf{3}$ & $\mathbf{3}$ & $\mathbf{3}$ & $\mathbf{1}$
& $\mathbf{1}$ & $\mathbf{3}$ & $\mathbf{3}$ & $\mathbf{3}$ & $\mathbf{1}$ & 
$\mathbf{1}$ & $\mathbf{1}$ & $\mathbf{3}$ & $\mathbf{3}$ & $\mathbf{3}$ & $%
\mathbf{1}$ & $\mathbf{1}$ & $\mathbf{1}$ & $\mathbf{1}$ & $\mathbf{1}$ & $%
\mathbf{1}$ \\ 
$SU(2)_L$ & $\mathbf{2}$ & $\mathbf{1}$ & $\mathbf{1}$ & $\mathbf{2}$
& $\mathbf{1}$ & $\mathbf{2}$ & $\mathbf{1}$ & $\mathbf{1}$ & $\mathbf{2}$ & 
$\mathbf{1}$ & $\mathbf{1}$ & $\mathbf{2}$ & $\mathbf{1}$ & $\mathbf{1}$ & $%
\mathbf{2}$ & $\mathbf{1}$ & $\mathbf{1}$ & $\mathbf{1}$ & $\mathbf{2}$ & $%
\mathbf{2}$ \\ 
$U(1)_Y$ & $\frac{1}{6}$ & $\frac{2}{3}$ & $-\frac{1}{3}$ & $-\frac{1%
}{2}$ & $1$ & $\frac{1}{6}$ & $\frac{2}{3}$ & $-\frac{1}{3}$ & $-\frac{1}{2}$
& $-1$ & $0$ & $\frac{1}{6}$ & $\frac{2}{3}$ & $-\frac{1}{3}$ & $-\frac{1}{2}
$ & $-1$ & $0$ & $0$ & $\frac{1}{2}$ & $-\frac{1}{2}$ \\ 
$U(1)^\prime$ & $0$ & $0$ & $0$ & $0$ & $0$ & $1$ & $-1$ & $-1$ & $1$
& $-1$ & $-1$ & $1$ & $-1$ & $-1$ & $1$ & $-1$ & $-1$ & $1$ & $-1$ & $-1$ \\ 
\bottomrule
\bottomrule
\end{tabular}}%
\caption{An extended $Z^{\prime}$ model with one complete vectorlike family, two SM-like Higgses, and one singlet flavon. This BSM theory is exactly the same as the ones in Refs.~\cite{Hernandez:2021tii,CarcamoHernandez:2021yev}, however in this work the SM gauge symmetry is extended by the local $U(1)^{\prime}$ symmetry instead of global, therefore this BSM theory features the massive neutral $Z^{\prime}$ gauge boson. This BSM theory is a combination of an extended 2HDM model and a $Z^{\prime}$ model. After spontaneous symmetry breaking (SSB) takes place, this BSM theory gives rise to the effective SM Yukawa interactions at the electroweak energy scale.}
\label{tab:BSM_model}
\end{table}
With the particle content shown in table \ref{tab:BSM_model}, the renormalizable Yukawa interactions of this model at high energy scale yield the following effective SM Yukawa interactions at the electroweak energy scale: 
\begingroup
\begin{equation}
\mathcal{L}_{\func{eff}}^{\func{Yukawa}} =y_{ik}^\psi (M_{\psi^\prime}^{-1})_{kl} {x_{lj}^{\psi^\prime}
\left\langle \phi \right\rangle} \overline{%
\psi}_{iL} \widetilde{H} \psi_{jR} + {x_{ik}^{\psi} \left\langle \phi
\right\rangle}(M_{\psi}^{-1} )_{kl}  y_{lj}^\psi  \overline{\psi}_{iL} \widetilde{H}
\psi_{jR} + \func{h.c.}  \label{eqn:the_effective_Yukawa_Lagrangian}
\end{equation}
\endgroup
where the indices $i,j=1,2,3$ and $k,l=4$ and $\psi,\psi^\prime = Q, u, d, L, e$ and $M$ means heavy vectorlike mass. The corresponding Feynman diagrams are shown in Figure~\ref{fig:mass_insertion_diagrams}.
\begingroup
\begin{figure}[H]
\centering
\begin{subfigure}{0.48\textwidth}
	\includegraphics[width=1.0\textwidth]{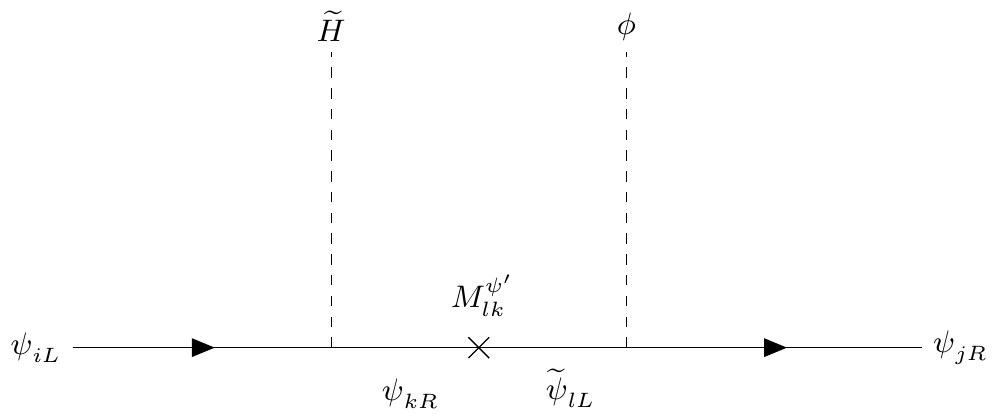}
\end{subfigure} \hspace{0.1cm} 
\begin{subfigure}{0.48\textwidth}
	\includegraphics[width=1.0\textwidth]{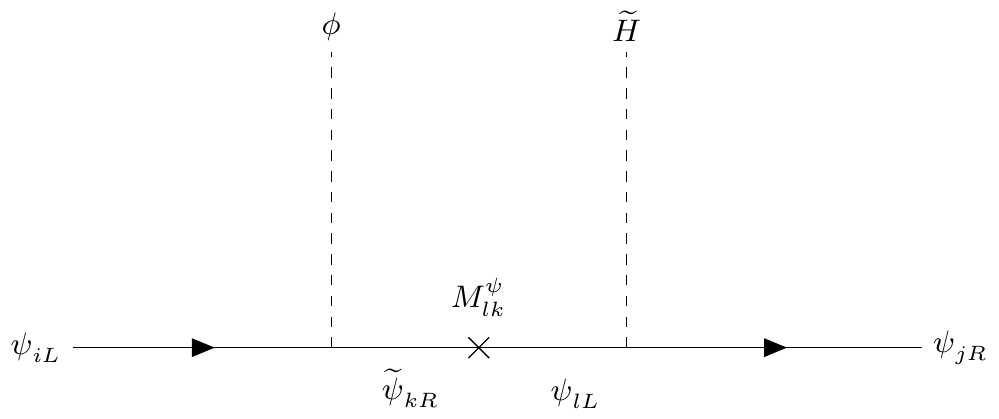}
\end{subfigure}
\caption{Diagrams in this model which lead to the effective Yukawa
interactions, where $\protect\psi,\protect\psi^\prime = Q,u,d,L,e$(neutrinos
will be treated separately) $i,j=1,2,3$, $k,l=4$, $M_{lk}$ is vectorlike
mass and $\widetilde{H} = i\protect\sigma_2 H^*, H = H_{u,d}$}
\label{fig:mass_insertion_diagrams}
\end{figure}
\endgroup
Notice that the left and right diagrams of Figure~\ref{fig:mass_insertion_diagrams} are mediated by vectorlike singlet and vectorlike doublet fermions, respectively. Therefore, a complete vectorlike family can give rise to two fermionic seesaw operators and this feature leads to massless first generation of SM fermions, whereas the second and third families of SM fermions do acquire their masses. Adding one more vectorlike family will imply the inclusion of two extra fermionic seesaw operators thus allowing all SM fermions to acquire their masses. However we only consider one complete vectorlike family for the purpose of keeping this BSM theory economical and the phenomenological analysis as simple as possible.
\subsection{Effective Yukawa interactions for the SM fermions} \label{sec:II_1}
The renormalizable Lagrangian for the quark sector is given by:
\begingroup
\begin{equation}
\begin{split}
\mathcal{L}_{q}^{\func{Yukawa+Mass}} &= y_{ik}^{u} \overline{Q}_{iL} 
\widetilde{H}_u u_{kR} + x_{ki}^{u} \phi \overline{\widetilde{u}}_{kL}
u_{iR} + x_{ik}^Q \phi \overline{Q}_{iL} \widetilde{Q}_{kR} + y_{ki}^u 
\overline{Q}_{kL} \widetilde{H}_u u_{iR} \\
&+ y_{ik}^{d} \overline{Q}_{iL} \widetilde{H}_d d_{kR} + x_{ki}^{d} \phi 
\overline{\widetilde{d}}_{kL} d_{iR} + y_{ki}^d \overline{Q}_{kL} \widetilde{%
H}_d d_{iR} \\
&+ M_{kl}^{u} \overline{\widetilde{u}}_{lL} u_{kR} + M_{kl}^{d} \overline{%
\widetilde{d}}_{lL} d_{kR} + M_{kl}^Q \overline{Q}_{kL} \widetilde{Q}_{lR} + 
\func{h.c.}  \label{eqn:general_Quark_Yukawa__Mass_Lagrangian}
\end{split}%
\end{equation}
\endgroup
where $i,j=1,2,3$ and $k,l = 4,5$ and $\widetilde{H} = i\sigma^{2} H^{*}$. After the singlet flavon $\phi$ develops its vacuum expectation value (vev) and the vectorlike fermions are integrated out, the renormalizable quark Yukawa terms give rise to the effective SM quark Yukawa interaction. The corresponding effective SM Yukawa interactions are given in Figure~\ref{fig:diagrams_quark_mass_insertion}:
\begingroup
\begin{figure}[H]
\centering
\begin{subfigure}{0.48\textwidth}
	\includegraphics[width=1.0\textwidth]{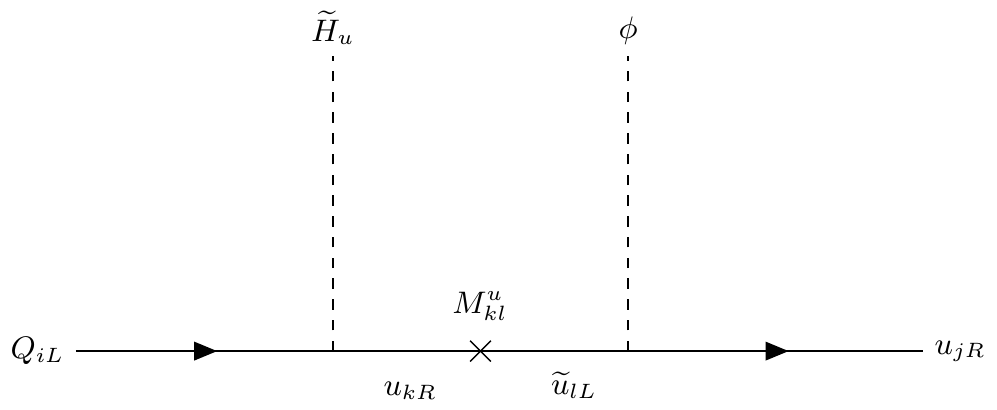}
\end{subfigure} \hspace{0.1cm} 
\begin{subfigure}{0.48\textwidth}
	\includegraphics[width=1.0\textwidth]{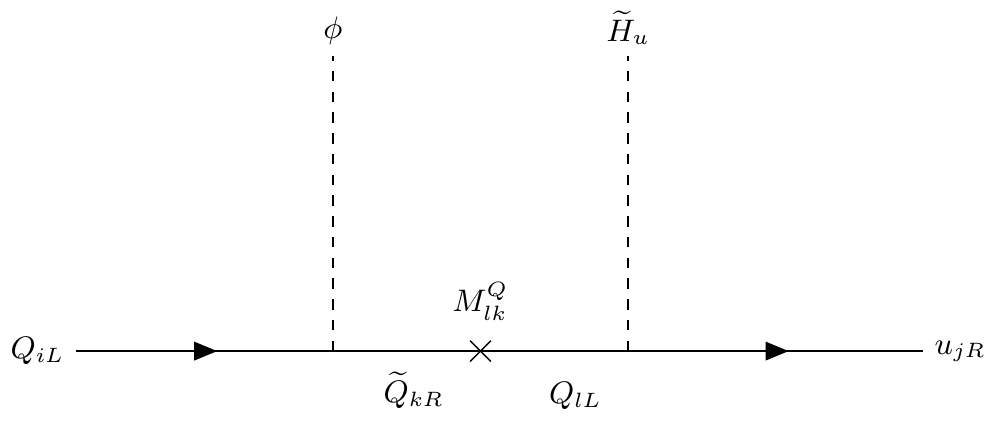}
\end{subfigure}
\begin{subfigure}{0.48\textwidth}
	\includegraphics[width=1.0\textwidth]{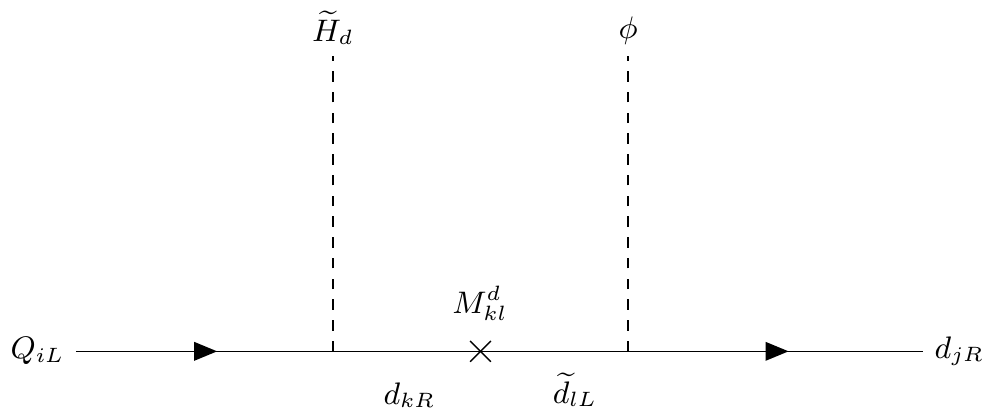}
\end{subfigure} \hspace{0.1cm} 
\begin{subfigure}{0.48\textwidth}
	\includegraphics[width=1.0\textwidth]{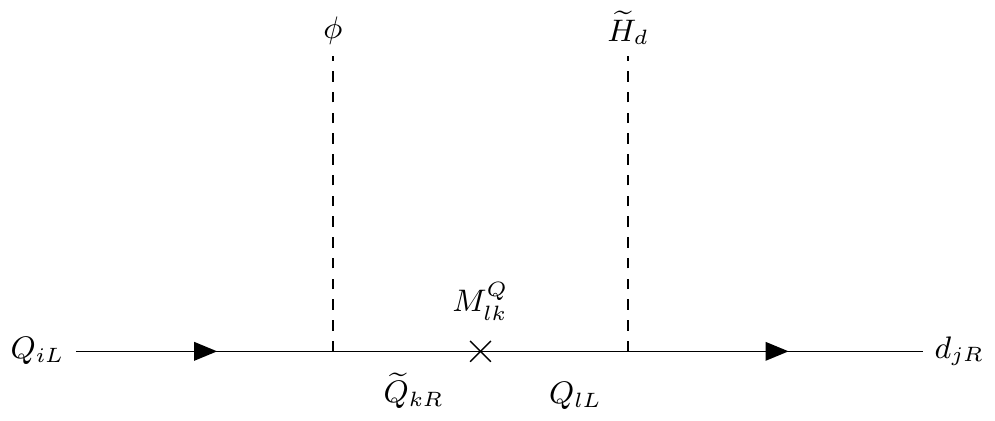}
\end{subfigure}
\caption{Diagrams in this model which lead to the effective Yukawa
interactions for the up quark sector(two above diagrams) and the down quark
sector(two below diagrams) in mass insertion formalism, where $i,j=1,2,3$
and $k,l=4$ and $M_{lk}$ is vectorlike mass.}
\label{fig:diagrams_quark_mass_insertion}
\end{figure}
\endgroup
As in the quark sector, the renormalizable Yukawa terms for the charged lepton sector can be written as follows:
\begingroup
\begin{equation}
\begin{split}
\mathcal{L}_{e}^{\func{Yukawa+Mass}} &= y_{ik}^{e} \overline{L}_{iL} 
\widetilde{H}_{d} e_{kR} + x_{ki}^{e} \phi \overline{\widetilde{e}}_{kL}
e_{iR} + x_{ik}^L \phi \overline{L}_{iL} \widetilde{L}_{kR} + y_{ki}^e 
\overline{L}_{kL} \widetilde{H}_{d} e_{iR} 
\\
&+ M_{kl}^{e} \overline{\widetilde{%
e}}_{lL} e_{kR} + M_{kl}^L \overline{L}_{kL} \widetilde{L}_{lR} + \func{h.c.},
\end{split}
\label{eqn:general_charged_lepton_Yukawa_Mass_Lagrangian}
\end{equation}
\endgroup
which give rise to the effective SM charged lepton Yukawa interactions in Figure~\ref{fig:diagrams_charged_leptons_mass_insertion}:
\begingroup
\begin{figure}[H]
\centering
\begin{subfigure}{0.48\textwidth}
	\includegraphics[width=1.0\textwidth]{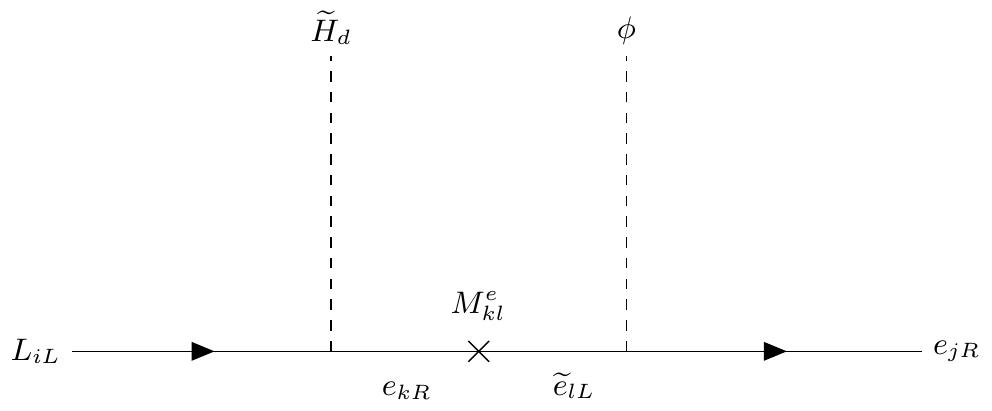}
\end{subfigure} \hspace{0.1cm} 
\begin{subfigure}{0.48\textwidth}
	\includegraphics[width=1.0\textwidth]{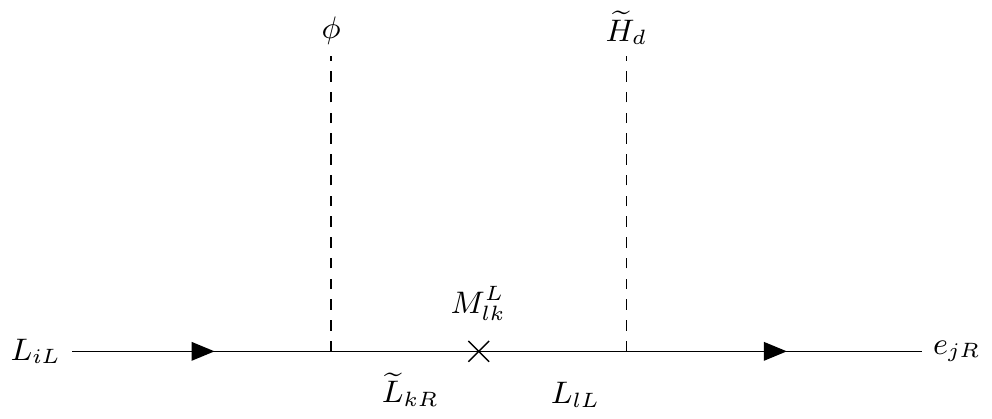}
\end{subfigure}
\caption{Diagrams in this model which lead to the effective Yukawa
interactions for the charged lepton sector in mass insertion formalism,
where $i,j=1,2,3$ and $k,l=4$ and $M_{lk}$ is vectorlike mass}
\label{fig:diagrams_charged_leptons_mass_insertion}
\end{figure}
\endgroup
As for the SM neutrino sector, it requires a different approach compared to the SM charged fermions since we assume that the SM left-handed neutrinos are Majorana particles and they have their masses via expansion through the vectorlike neutrinos. The renormalizable neutrino Yukawa terms are given by: 
\begingroup
\begin{equation}
\begin{split}
\mathcal{L}_{\nu}^{\func{Yukawa+Mass}} = y_{ik}^{\nu} \overline{L}_{iL} 
\widetilde{H}_u \nu_{kR} + x_{ik}^L \overline{L}_{iL} H_d \overline{%
\widetilde{\nu}}_{kR} + M_{kl}^{M} \overline{\widetilde{\nu}}_{lR} \nu_{kR}
+ \func{h.c.}
\end{split}
\label{eqn:general_neutrinos_Yukawa_Mass_Lagrangian}
\end{equation}
\endgroup
After the heavy vectorlike neutrinos are integrated out, the renormalizable Lagrangian induces the Weinberg-like operator which generates the ``type 1b seesaw mechanism" in Figure~\ref{fig:diagrams_neutrinos_mass_insertion} to differentiate it from the original Weinberg operator corresponding to the type 1a seesaw mechanism which does not work in the model under consideration since the SM-like Higgses are charged under the local $U(1)^{\prime}$ symmetry.
\begingroup
\begin{figure}[H]
\centering
\begin{subfigure}{0.48\textwidth}
\includegraphics[width=1.0\textwidth]{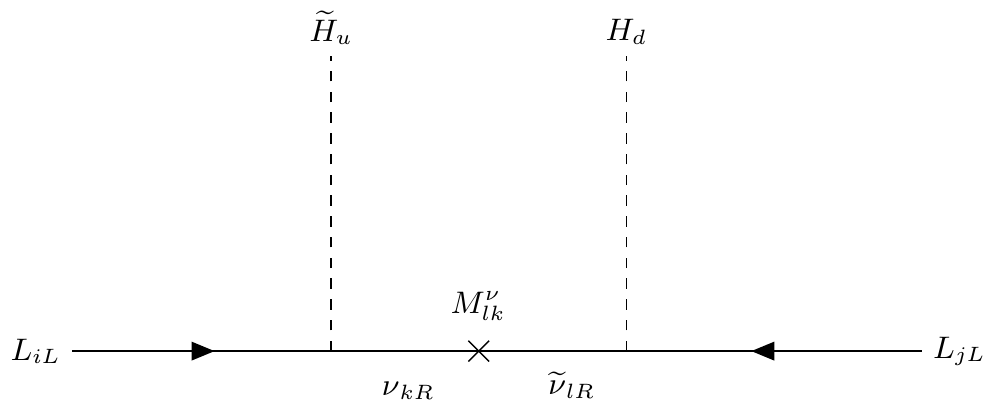}
\end{subfigure}
\caption{Type Ib seesaw mechanism~\cite{Hernandez:2021tii,Hernandez-Garcia:2019uof} which leads to the effective Yukawa
interactions for the Majorana neutrinos in mass insertion formalism,
where $i,j=1,2,3$ and $k,l=4,5$ and $M_{lk}$ is vectorlike mass.}
\label{fig:diagrams_neutrinos_mass_insertion}
\end{figure}
\endgroup
There are three important features we can read off from the vectorlike mass in Figure~\ref{fig:diagrams_neutrinos_mass_insertion}. The first one is the nature of the vectorlike mass consisting of two different vectorlike neutrinos, which is differentiated by the Majorana mass consisting of the same neutrinos. The second one is the vectorlike mass in Figure~\ref{fig:diagrams_neutrinos_mass_insertion} which also violates the lepton number conservation as the Majorana mass does and this feature can be easily confirmed by checking the lepton number in the renormalizable neutrino Yukawa interaction of Equation~\ref{eqn:general_neutrinos_Yukawa_Mass_Lagrangian}. The third one is that the tiny active neutrino masses are explained from the type 1b seesaw mechanism where each interaction vertex involves two SM Higgs vevs $v_{u}, v_{d}$ ($v_{u}^{2} + v_{d}^{2} = \left( 246 \func{GeV} \right)^{2}$) and allowing two different vertex values, therefore it is possible to significantly lower the order of predicted right-handed neutrinos from $10^{14}\func{GeV}$ up to $\func{TeV}$ scale. 
\\~\\
We call the mass insertion formalism introduced from Figure~\ref{fig:diagrams_quark_mass_insertion} to \ref{fig:diagrams_neutrinos_mass_insertion} ``low energy scale seesaw mechanism" and this mechanism can explain both the SM fermion masses and their mixings as well. On top of that, we also showed that the SM $Z$ gague boson can cause the FCNCs in both quark and lepton sectors with one complete vectorlike family and how the unitarity of the first row of the CKM matrix can be relaxed in an analytical way in one of our works~\cite{CarcamoHernandez:2021yev}. Based on what we have discussed, we construct the mass matrices in both quark and charged lepton sectors and then diagonalize them in the next section.
\section{EFFECTIVE YUKAWA MATRICES USING A MIXING FORMALISM}
\label{sec:III}
The mass matrices for the SM fermions and their diagonalization are addressed in this section and they are well described in section~\ref{sec:III} of Ref.~\cite{CarcamoHernandez:2021yev}, so we focus on important features of the mass matrices instead of explaining every details about them. The general mass matrix for fermions in the flavor basis is given by:
\begingroup
\begin{equation}
M^{\psi }=\left( 
\begin{array}{c|ccccc}
& \psi _{1R} & \psi _{2R} & \psi _{3R} & \psi _{4R} & \widetilde{\psi }_{4R} \\[0.5ex] \hline
\overline{\psi }_{1L} & 0 & 0 & 0 & y_{14}^{\psi }\langle \widetilde{H}%
^{0}\rangle & x_{14}^{\psi }\langle \phi \rangle \\[1ex]
\overline{\psi }_{2L} & 0 & 0 & 0 & y_{24}^{\psi }\langle \widetilde{H}%
^{0}\rangle & x_{24}^{\psi }\langle \phi \rangle \\[1ex] 
\overline{\psi }_{3L} & 0 & 0 & 0 & y_{34}^{\psi }\langle \widetilde{H}%
^{0}\rangle & x_{34}^{\psi }\langle \phi \rangle \\[1ex]
\overline{\psi }_{4L} & y_{41}^{\psi }\langle \widetilde{H}%
^{0}\rangle & y_{42}^{\psi }\langle \widetilde{H}^{0}\rangle
& y_{43}^{\psi }\langle \widetilde{H}^{0}\rangle & 0 & 
M_{44}^{\psi } \\[1ex]
\overline{\widetilde{\psi }}_{4L} & x_{41}^{\psi ^{\prime }}\langle
\phi \rangle & x_{42}^{\psi ^{\prime }}\langle \phi \rangle
& x_{43}^{\psi ^{\prime }}\langle \phi \rangle & M_{44}^{\psi
^{\prime }} & 0 \\ 
\end{array}%
\right) ,  
\label{eqn:general_55_mass_matrix}
\end{equation}
\endgroup
where $\psi,\psi^{\prime} = Q,u,d,L,e$ and the zeros appearing in the upper-left $3 \times 3$ block means the SM fermions remain massless without extension of the SM, which is implemented by assigning $U(1)^{\prime}$ charges to the SM-like Higgses $H_{u,d}$, and lastly the mass matrix involves three different mass scales $\langle H^{0} \rangle, \langle \phi \rangle$ and $M$ so it can naturally accommodate the strong SM hierarchy in this BSM model. In order to dynamically implement the SM hierarchy, it requires to rotate the mass matrix of Equation~\ref{eqn:general_55_mass_matrix} maximally and this feature will be discussed across the next subsections.
\subsection{Diagonalizing the charged lepton mass matrix} \label{sec:III_1}
After all scalars develop its vevs ($v_{d} = \langle H_{d}^{0} \rangle$ and $v_{\phi} = \langle \phi \rangle$) and then the mass matrix for the charged lepton is fully rotated, it can be written as follows:
\begingroup
\begin{equation}
M^{e }=\left( 
\begin{array}{c|ccccc}
& e _{1R} & e _{2R} & e _{3R} & e _{4R} & \widetilde{L }_{4R} \\[0.5ex] \hline
\overline{L }_{1L} & 0 & 0 & 0 & 0 & 0 \\[1ex]
\overline{L }_{2L} & 0 & 0 & 0 & y_{24}^{e } v_{d} & 0 \\[1ex] 
\overline{L }_{3L} & 0 & 0 & 0 & y_{34}^{e } v_{d} & x_{34}^{L } v_{\phi} \\[1ex]
\overline{L }_{4L} & 0 & 0 & y_{43}^{e } v_{d} & 0 & M_{44}^{L } \\[1ex]
\overline{\widetilde{e }}_{4L} & 0 & x_{42}^{e} v_{\phi}
& x_{43}^{e} v_{\phi} & M_{44}^{e} & 0 \\ 
\end{array}%
\right),
\label{eqn:cl_1}
\end{equation}
\endgroup
where it is worth mentioning that the fully rotated mass matrix can provide a dynamical explanation for the SM hierarchy and that with one complete vectorlike family can give rise to two fermionic seesaw mediators, which means that two of the SM generations of fermions, i.e., the second and the third one can acquire masses. Adding one extra vectorlike family can allow for the whole generations of SM fermions to be massive, as done in one of our works~\cite{Hernandez:2021tii}, however this approach requires some assumptions to simplify the resulting mass matrices  since they include many fields. For the purpose of avoiding any assumptions while diagonalizing the mass matrices for the SM fermions and also keeping those as economical as possible, we constrain our attention to the massive SM second and third generations with one complete vectorlike family. It is convenient to rewrite the Yukawa terms by simple mass parameters and then switch the column of $e_{4R}$ and $\widetilde{L}_{4R}$ in order for the heavy vectorlike masses to locate at the diagonal position as follows:
\begingroup
\begin{equation}
M^{e }
=
\left( 
\begin{array}{c|ccccc}
& e _{1R} & e _{2R} & e _{3R} & e _{4R} & \widetilde{L }_{4R} \\[0.5ex] \hline
\overline{L }_{1L} & 0 & 0 & 0 & 0 & 0 \\[1ex]
\overline{L }_{2L} & 0 & 0 & 0 & m_{24} & 0 \\[1ex] 
\overline{L }_{3L} & 0 & 0 & 0 & m_{34} & m_{35} \\[1ex]
\overline{L }_{4L} & 0 & 0 & m_{43} & 0 & M_{45}^{L } \\[1ex]
\overline{\widetilde{e }}_{4L} & 0 & m_{52} & m_{53} & M_{54}^{e} & 0 \\ 
\end{array}%
\right)
=
\left( 
\begin{array}{c|ccccc}
& e _{1R} & e _{2R} & e _{3R} & \widetilde{L }_{4R} & e _{4R} \\[0.5ex] \hline
\overline{L }_{1L} & 0 & 0 & 0 & 0 & 0 \\[1ex]
\overline{L }_{2L} & 0 & 0 & 0 & 0 & m_{24} \\[1ex] 
\overline{L }_{3L} & 0 & 0 & 0 & m_{35} & m_{34} \\[1ex]
\overline{L }_{4L} & 0 & 0 & m_{43} & M_{45}^{L } & 0 \\[1ex]
\overline{\widetilde{e }}_{4L} & 0 & m_{52} & m_{53} & 0 & M_{54}^{e} \\ 
\end{array}%
\right), 
\label{eqn:cl_2}
\end{equation}
\endgroup
where the left-handed fields follow the order of $12345$ whereas the right-handed fields follow that of $12354$ (the tilde particles have the index $5$). The mass matrix of Equation~\ref{eqn:cl_2} is diagonalized by two methods which are the numerical singular value decomposition (SVD) and the step-by-step analytical diagonalization and it is confirmed that subtracting from a result from one method to that from the other method is quite ignorable~\cite{CarcamoHernandez:2021yev}. In this work, we mainly make use of the numerical singular value decomposition method to focus further on the relevant phenomenology. The rearranged mass matrix for the charged lepton sector is diagonalized by:
\begingroup
\begin{equation}
M_{\func{diag}}^{e} = \func{diag}\left( 0, m_{\mu}, m_{\tau}, M_{E_4}, M_{\widetilde{E}_4} \right) = V^{L} M^{e} (V^{e})^{\dagger},
\end{equation}
\endgroup
where $V^{L}(V^{e})$ is the left-handed(right-handed) mixing matrix for the charged lepton fields and they connect the flavor basis and the physical basis as follows:
\begingroup
\begin{equation}
\begin{split}
\begin{pmatrix}
e_{L} \\[0.5ex]
\mu_{L} \\[0.5ex]
\tau_{L} \\[0.5ex]
E_{4L} \\[0.5ex]
\widetilde{E}_{4L}
\end{pmatrix}
=
V^L  
\begin{pmatrix}
e_{1L} \\[0.5ex]
e_{2L} \\[0.5ex]
e_{3L} \\[0.5ex]
e_{4L} \\[0.5ex]
\widetilde{e}_{4L}
\end{pmatrix},
\qquad
\begin{pmatrix}
e_{R} \\[0.5ex]
\mu_{R} \\[0.5ex]
\tau_{R} \\[0.5ex]
\widetilde{E}_{4R} \\[0.5ex]
E_{4R}
\end{pmatrix}
=
V^e
\begin{pmatrix}
e_{1R} \\[0.5ex]
e_{2R} \\[0.5ex]
e_{3R} \\[0.5ex]
\widetilde{e}_{4R} \\[0.5ex]
e_{4R}
\end{pmatrix},
\label{eqn:cl_mixing}
\end{split}
\end{equation}
\endgroup

\subsection{Diagonalizing the up-type quark mass matrix} \label{sec:III_2}
The mass matrix for the up-type quark sector can be approached as in the charged lepton sector. The rearranged form is given by:
\begingroup
\begin{equation}
M^{u }
=
\left( 
\begin{array}{c|ccccc}
& u _{1R} & u _{2R} & u _{3R} & u _{4R} & \widetilde{Q }_{4R} \\[0.5ex] \hline
\overline{Q }_{1L} & 0 & 0 & 0 & 0 & 0 \\[1ex]
\overline{Q }_{2L} & 0 & 0 & 0 & y_{24}^{u } v_{u} & 0 \\[1ex] 
\overline{Q }_{3L} & 0 & 0 & 0 & y_{34}^{u } v_{u} & x_{34}^{Q } v_{\phi} \\[1ex]
\overline{Q }_{4L} & 0 & 0 & y_{43}^{u } v_{u} & 0 & M_{44}^{Q } \\[1ex]
\overline{\widetilde{u }}_{4L} & 0 & x_{42}^{u} v_{\phi}
& x_{43}^{u} v_{\phi} & M_{44}^{u} & 0 \\ 
\end{array}%
\right)
=
\left( 
\begin{array}{c|ccccc}
& u _{1R} & u _{2R} & u _{3R} & \widetilde{Q }_{4R} & u _{4R} \\[0.5ex] \hline
\overline{Q }_{1L} & 0 & 0 & 0 & 0 & 0 \\[1ex]
\overline{Q }_{2L} & 0 & 0 & 0 & 0 & m_{24}^{u} \\[1ex] 
\overline{Q }_{3L} & 0 & 0 & 0 & m_{35}^{u} & m_{34}^{u} \\[1ex]
\overline{Q }_{4L} & 0 & 0 & m_{43}^{u} & M_{44}^{Q } & 0 \\[1ex]
\overline{\widetilde{u }}_{4L} & 0 & m_{52}^{u}
& m_{53}^{u} & 0 & M_{44}^{u} \\ 
\end{array}%
\right)
, \label{eqn:uq_1}
\end{equation}
\endgroup
Here, the whole form is exactly consistent with the one for the charged lepton sector except for a few substitutions such as $y^{e} \rightarrow y^{u}, v_{d} \rightarrow v_{u}, x^{L} \rightarrow x^{Q}$, and $x^{e} \rightarrow x^{u}$. Therefore, we can simply reuse the derived result for the charged lepton sector. The up-type quark mass matrix is diagonalized by:
\begingroup
\begin{equation}
M_{\func{diag}}^{u} = \func{diag}\left( 0, m_{c}, m_{t}, M_{U_4}, M_{\widetilde{U}_4} \right) = V_{L}^{u} M^{u} (V_{R}^{u})^{\dagger},
\end{equation}
\endgroup
where the mixing matrices $V_{L,R}^{u}$ are defined as follows:
\begingroup
\begin{equation}
\begin{split}
\begin{pmatrix}
u_{L} \\[0.5ex]
c_{L} \\[0.5ex]
t_{L} \\[0.5ex]
U_{4L} \\[0.5ex]
\widetilde{U}_{4L}
\end{pmatrix}
=
V_{L}^{u}  
\begin{pmatrix}
u_{1L} \\[0.5ex]
u_{2L} \\[0.5ex]
u_{3L} \\[0.5ex]
u_{4L} \\[0.5ex]
\widetilde{u}_{4L}
\end{pmatrix}
, \qquad
\begin{pmatrix}
u_{R} \\[0.5ex]
c_{R} \\[0.5ex]
t_{R} \\[0.5ex]
\widetilde{U}_{4R} \\[0.5ex]
U_{4R}
\end{pmatrix}
=
V_{R}^{u}
\begin{pmatrix}
u_{1R} \\[0.5ex]
u_{2R} \\[0.5ex]
u_{3R} \\[0.5ex]
\widetilde{u}_{4R} \\[0.5ex]
u_{4R}
\end{pmatrix}.
\label{eqn:up_mixing}
\end{split}
\end{equation}
\endgroup
\subsection{Diagonalizing the down-type quark mass matrix} \label{sec:III_3}
The down-type quark mass matrix requires some attention as its form is somewhat different, compared to that for the charged lepton or up-type quark sector. For comparison, it is convenient to look at the up- and down-type quark mass matrices together
\begingroup
\begin{equation}
M^{u }=\left( 
\begin{array}{c|ccccc}
& u _{1R} & u _{2R} & u _{3R} & u _{4R} & \widetilde{Q }_{4R} \\[0.5ex] \hline
\overline{Q }_{1L} & 0 & 0 & 0 & 0 & 0 \\[1ex]
\overline{Q }_{2L} & 0 & 0 & 0 & y_{24}^{u } v_{u} & 0 \\[1ex] 
\overline{Q }_{3L} & 0 & 0 & 0 & y_{34}^{u } v_{u} & x_{34}^{Q } v_{\phi} \\[1ex]
\overline{Q }_{4L} & 0 & 0 & y_{43}^{u } v_{u} & 0 & M_{44}^{Q } \\[1ex]
\overline{\widetilde{u }}_{4L} & 0 & x_{42}^{u} v_{\phi}
& x_{43}^{u} v_{\phi} & M_{44}^{u} & 0 \\ 
\end{array}%
\right), 
\quad
M^{d }=\left( 
\begin{array}{c|ccccc}
& d _{1R} & d _{2R} & d _{3R} & d _{4R} & \widetilde{Q }_{4R} \\[0.5ex] \hline
\overline{Q }_{1L} & 0 & 0 & 0 & y_{14}^{d } v_{d} & 0 \\[1ex]
\overline{Q }_{2L} & 0 & 0 & 0 & y_{24}^{d } v_{d} & 0 \\[1ex] 
\overline{Q }_{3L} & 0 & 0 & 0 & y_{34}^{d } v_{d} & x_{34}^{Q } v_{\phi} \\[1ex]
\overline{Q }_{4L} & 0 & 0 & y_{43}^{d } v_{d} & 0 & M_{44}^{Q } \\[1ex]
\overline{\widetilde{d }}_{4L} & 0 & x_{42}^{d} v_{\phi}
& x_{43}^{d} v_{\phi} & M_{44}^{d} & 0 \\ 
\end{array}%
\right).
\label{eqn:diff_up_down}
\end{equation}
\endgroup
The first important feature is the fifth column of the up-type mass matrix is exactly consistent with that of the down-type mass matrix~\cite{King:2018fcg}. What this implements is the left-handed vectorlike quark doublets contribute to the right-handed vectorlike doublets equally, so the vectorlike quark doublets get to have very degenerate masses, and this feature plays a crucial role in constraining violation of the custodial symmetry as we will see in the oblique parameter section. The second feature is the presence of the Yukawa term $y_{14}^{d} v_{d}$. For the up-quark sector, we can rotate $Q_{1L}$ and $Q_{2L}$ fields further to vanish the up-type Yukawa term $y_{14}^{u} v_{u}$, however this rotation simply remixes the down-type Yukawa terms $y_{14}^{d} v_{d}$ and $y_{24}^{d} v_{d}$ and therefore both the down-type Yukawa terms survive. This rotation between $Q_{1L}$ and $Q_{2L}$ fields can be done first in the down-type mass matrix, remaining the unvanished up-type Yukawa term, however this different approach does not change our analyses since we found that the analyses which will be explored are basis-independent as confirmed in the SM $Z$ physics in our previous work~\cite{CarcamoHernandez:2021yev}. Therefore, we consider it is phenomenologically reasonable to consider the unvanished Yukawa term in the down-type mass matrix in order to explain the CKM mixing matrix (or equally, the CKM mixing matrix is mainly based on the down-type quark mixings and this is one of our findings in the previous work~\cite{CarcamoHernandez:2021yev}). The last feature to discuss is the down-type left-handed quark fields can access to all the SM mixings even thought the first generation of the down-type quarks ($d$ quark) remains massless in the model under consideration. It can be easily confirmed by looking at a partially diagonalized mass matrix for the up- and down-type quark sector. After integrating out the heavy degrees of freedom in both the mass matrices, the mass matrices have the forms as follows:
\begingroup
\begin{equation}
M^{u \prime}=\left( 
\begin{array}{c|ccccc}
& u _{1R} & u _{2R}^{\prime} & u _{3R}^{\prime} & \widetilde{Q }_{4R}^{\prime} & u _{4R}^{\prime} \\[0.5ex] \hline
\overline{Q }_{1L} & 0 & 0 & 0 & 0 & 0 \\[1ex]
\overline{Q }_{2L}^{\prime} & 0 & m_{22}^{u \prime} & m_{23}^{u \prime} & 0 & 0 \\[1ex] 
\overline{Q }_{3L}^{\prime} & 0 & m_{32}^{u \prime} & m_{33}^{u \prime} & 0 & 0 \\[1ex]
\overline{Q }_{4L}^{\prime} & 0 & 0 & 0 & M_{44}^{Q } & 0 \\[1ex]
\overline{\widetilde{u }}_{4L}^{\prime} & 0 & 0
& 0 & 0 & M_{44}^{u} \\ 
\end{array}%
\right), 
\quad
M^{d \prime}=\left( 
\begin{array}{c|ccccc}
& d _{1R} & d _{2R}^{\prime} & d _{3R}^{\prime} & \widetilde{Q }_{4R}^{\prime} & d _{4R}^{\prime} \\[0.5ex] \hline
\overline{Q }_{1L}^{\prime} & 0 & m_{12}^{d \prime} & m_{13}^{d \prime} & 0 & 0 \\[1ex]
\overline{Q }_{2L}^{\prime} & 0 & m_{22}^{d \prime} & m_{23}^{d \prime} & 0 & 0 \\[1ex] 
\overline{Q }_{3L}^{\prime} & 0 & m_{32}^{d \prime} & m_{33}^{d \prime} & 0 & 0 \\[1ex]
\overline{Q }_{4L}^{\prime} & 0 & 0 & 0 & M_{44}^{Q } & 0 \\[1ex]
\overline{\widetilde{d }}_{4L}^{\prime} & 0 & 0
& 0 & 0 & M_{44}^{d} \\ 
\end{array}%
\right),
\label{eqn:partially_rotated_Mu_Md}
\end{equation}
\endgroup
where the primed field means the field is rotated. The partially diagonalized down-type mass matrix has two more Yukawa terms $m_{12,13}^{d\prime}$ compared to the up-type mass matrix and this term allows left-handed mixing with the first generation of the down-type quarks even though the first generation remains massless (note that right-handed mixing in the down-type mass matrix is only 23 mixing as in the other mass matrices). Then the down-type mass matrix is diagonalized by:
\begingroup
\begin{equation}
M_{\func{diag}}^{d} = \func{diag}\left( 0, m_{s}, m_{b}, M_{D_4}, M_{\widetilde{D}_4} \right) = V_{L}^{d} M^{d} (V_{R}^{d})^{\dagger},
\end{equation}
\endgroup
where the mixing matrices $V_{L,R}^{d}$ are defined by:
\begingroup
\begin{equation}
\begin{split}
\begin{pmatrix}
d_{L} \\
s_{L} \\
b_{L} \\
D_{4L} \\
\widetilde{D}_{4L}
\end{pmatrix}
=
V_{L}^{d}  
\begin{pmatrix}
d_{1L} \\
d_{2L} \\
d_{3L} \\
d_{4L} \\
\widetilde{d}_{4L}
\end{pmatrix}
, \qquad
\begin{pmatrix}
d_{R} \\
s_{R} \\
b_{R} \\
D_{4R} \\
\widetilde{D}_{4R}
\end{pmatrix}
=
V_{R}^{d}
\begin{pmatrix}
d_{1R} \\
d_{2R} \\
d_{3R} \\
\widetilde{d}_{4R} \\
d_{4R}
\end{pmatrix}.
\label{eqn:down_mixing}
\end{split}
\end{equation}
\endgroup

\section{THE $Z^{\prime}$ GAUGE BOSON INTERACTIONS WITH THE VECTORLIKE FAMILY}
\label{sec:IV}
As the model under consideration features the extended gauge symmetry by the $U(1)^{\prime}$ local symmetry, this BSM model involves a neutral massive $Z^{\prime}$ gauge boson. Considering a new physics source $Z^{\prime}$ for the anomalies is interesting, since it can make a direct connection to the SM observables and anomalies. 

\subsection{FCNC mediated by the $Z^{\prime}$ gauge boson with the fourth vectorlike fermions} \label{sec:IV_1}
The $Z^{\prime}$ interactions with the fourth vectorlike fermions in the flavor basis are given by:
\begin{equation}
\mathcal{L}_{\func{flavor}}^{Z^{\prime}} = g_{X} Z_{\mu}^{\prime} \left( \overline{Q}_{iL} D_{Q} \gamma^{\mu} Q_{iL} + \overline{u}_{iR} D_{u} \gamma^{\mu} u_{iR} + \overline{d}_{iR} D_{d} \gamma^{\mu} d_{iR} + \overline{L}_{iL} D_{L} \gamma^{\mu} L_{iL} + \overline{e}_{iR} D_{e} \gamma^{\mu} e_{iR} + \overline{\nu}_{iR} D_{\nu} \gamma^{\mu} \nu_{iR} \right)
\end{equation}
where $i=1,2,3,4,5$ and the $U(1)^{\prime}$ charge matrices, $D$s, are defined in the model under consideration as follows:
\begin{equation}
\begin{split}
D_{Q} = \func{diag}\left(0,0,0,1,-1\right), \quad D_{u} &= \func{diag}\left(0,0,0,1,-1\right), \quad \quad D_{d} = \func{diag}\left(0,0,0,1,-1\right), \\
D_{L} = \func{diag}\left(0,0,0,1,-1\right), \quad D_{e} &= \func{diag}\left(0,0,0,1,-1\right), \quad \quad D_{\nu} = \func{diag}\left(0,0,0,-1,-1\right),
\label{eqn:Ds}
\end{split}
\end{equation}
where it is worth mentioning that the left-handed fields $Q_{iL}, L_{iL}$ follow the order of $12345$ and the right-handed fields $u_{iR}, d_{iR}, e_{iR}, \nu_{iR}$ follow the order of $12354$ as defined in the mass matrix diagonalization of section~\ref{sec:III} and the orders determine each $D$ of Equation~\ref{eqn:Ds}. This BSM model follows the fermionic $Z^{\prime}$ model~\cite{CarcamoHernandez:2019ydc}, which means that the SM fermions do not interact with the $Z^{\prime}$ gauge boson in the flavor basis however it starts to have interactions with the SM fermions via mixings in the physical basis. The $Z^{\prime}$ interactions in the physical basis are given by:
\begingroup
\begin{equation}
\mathcal{L}_{\func{physical}}^{Z^{\prime}} = g_{X} Z_{\mu}^{\prime} \left( \overline{Q}_{L} D_{Q}^{\prime} \gamma^{\mu} Q_{L} + \overline{u}_{R} D_{uR}^{\prime} \gamma^{\mu} u_{R} + \overline{d}_{R} D_{dR}^{\prime} \gamma^{\mu} d_{R} + \overline{L}_{L} D_{L}^{\prime} \gamma^{\mu} L_{L} + \overline{e}_{R} D_{eR}^{\prime} \gamma^{\mu} e_{R} + \overline{\nu}_{R} D_{\nu}^{\prime} \gamma^{\mu} \nu_{R} \right),
\end{equation}
\endgroup
where the primed $D$s are determined as follows ($D_{Q}^{\prime}$ is separated to $D_{uL}^{\prime}$ and $D_{dL}^{\prime}$ after SSB):
\begingroup
\begin{equation}
\begin{split}
D_{uL}^{\prime} &= V_{L}^{u} D_{Q} V_{L}^{u\dagger}, \quad D_{uR}^{\prime} = V_{R}^{u} D_{u} V_{R}^{u\dagger}, \\
D_{dL}^{\prime} &= V_{L}^{d} D_{Q} V_{L}^{d\dagger}, \quad D_{dR}^{\prime} = V_{R}^{d} D_{d} V_{R}^{d\dagger}, \\
D_{e L}^{\prime} &= V_{L}^{e} D_{L} V_{L}^{e\dagger}, \quad D_{e R}^{\prime} = V_{R}^{e} D_{e} V_{R}^{e\dagger}, \\
D_{\nu}^{\prime} &= V^{\nu} D_{\nu} V^{\nu \dagger}, \\
\label{eqn:Dprimes}
\end{split}
\end{equation}
\endgroup
The derived $D^{\prime}$s of Equation~\ref{eqn:Dprimes} induce the flavor changing neutral currents (FCNCs) so this feature allows to access tree level contributions to the $\tau \rightarrow \mu \gamma$ decay or $R_{K}$ anomaly, etc. and those will be discussed in detail in the next sections.

\subsection{FCNC mediated by the $Z^{\prime}$ gauge boson in the neutrino sector with the fourth vectorlike neutrinos} \label{sec:IV_2}

The neutrino sector in the model under consideration requires another approach compared to the charged fermion sectors. We build the mass matrices for the charged fermion sectors explicitly since our analyses on the observables which will be explored are sensitive to mixing parameters coming from their mixing matrices. As for the neutrino sector, what is required is the $Z^{\prime}$ coupling to $\nu_{\mu}$ pair for the neutrino trident constraint. Therefore, we do not need to have a full mass matrix for the neutrino sector instead we consider only the coupling constant of $Z^{\prime} \nu_{\mu} \nu_{\mu}$ interaction. As covered in section~\ref{sec:III}, the neutrino sector in this BSM model is explained by the type $1$b seesaw mechanism and the analysis was done in one of our works~\cite{Hernandez:2021tii}. The $D_{\nu}^{\prime}$ is given by ($\alpha = \begin{pmatrix}
-1 & 0 \\
0 & -1 
\end{pmatrix}$):
\begingroup
\begin{equation}
\begin{split}
D_{\nu}^{\prime} &= V D_{\nu} V^{T} \\
&= V \begin{pmatrix}
0_{3 \times 3} & 0_{3 \times 2} \\
0_{2 \times 3} & \alpha
\end{pmatrix} V^{T} \\
&= \begin{pmatrix}
U_{\func{PMNS}} & 0 \\
0 & I
\end{pmatrix}
\begin{pmatrix}
I - \frac{\Theta \Theta^{\dagger}}{2} & \Theta \\
-\Theta^{\dagger} & I - \frac{\Theta \Theta^{\dagger}}{2}
\end{pmatrix}
\begin{pmatrix}
0_{3 \times 3} & 0_{3 \times 2} \\
0_{2 \times 3} & \alpha
\end{pmatrix}
\begin{pmatrix}
I - \frac{\Theta \Theta^{\dagger}}{2} & \Theta \\
-\Theta^{\dagger} & I - \frac{\Theta \Theta^{\dagger}}{2}
\end{pmatrix}^{T}
\begin{pmatrix}
U_{\func{PMNS}} & 0 \\
0 & I
\end{pmatrix}^{T} \\
&= \begin{pmatrix}
U_{\func{PMNS}} & 0 \\
0 & I
\end{pmatrix}
\begin{pmatrix}
I - \eta_{ij} & \Theta_{ik} \\
-\Theta_{ki}^{\dagger} & I - \eta_{kl}
\end{pmatrix}
\begin{pmatrix}
0_{3 \times 3} & 0_{3 \times 2} \\
0_{2 \times 3} & \alpha
\end{pmatrix}
\begin{pmatrix}
I - \eta_{ij} & \Theta_{ik} \\
-\Theta_{ki}^{\dagger} & I - \eta_{kl}
\end{pmatrix}^{T}
\begin{pmatrix}
U_{\func{PMNS}} & 0 \\
0 & I
\end{pmatrix}^{T} \\
&= \begin{pmatrix}
U_{\func{PMNS}} & 0 \\
0 & I
\end{pmatrix}
\begin{pmatrix}
\Theta_{ik} \alpha \Theta_{ki}^{T} & \Theta_{ik} \alpha \left( I - \eta_{lk}^{T} \right) \\
\left( I - \eta_{kl} \right) \alpha \Theta_{ki}^{T} & \left( I - \eta_{kl} \right) \alpha \left( I - \eta_{lk}^{T} \right)
\end{pmatrix}
\begin{pmatrix}
U_{\func{PMNS}} & 0 \\
0 & I
\end{pmatrix}^{T},
\end{split}
\end{equation}
\endgroup
where the indices $i,j=1,2,3$ and $k,l=4,5$ and $U_{\func{PMNS}}$ is the well-known Pontecorvo-Maki-Nakagawa-Sakata matrix and is parametrized as follows ($c_{ij} = \cos\theta_{ij}, s_{ij} = \sin\theta_{ij}$ and $\delta_{\func{CP}}$ : the Dirac phase, $\alpha^{\prime}, \alpha$ : the Majorana phases)~\cite{Hernandez-Garcia:2019uof,Chau:1984fp}:
\begingroup
\begin{equation}
U_{\func{PMNS}} = 
\begin{pmatrix}
1 & 0 & 0 \\
0 & c_{23} & s_{23} \\
0 & -s_{23} & c_{23} 
\end{pmatrix}
\begin{pmatrix}
c_{13} & 0 & s_{13} e^{-i\delta_{\func{CP}}} \\
0 & 1 & 0 \\
-s_{13} e^{i\delta_{\func{CP}}} & 0 & c_{13}
\end{pmatrix}
\begin{pmatrix}
c_{12} & s_{12} & 0 \\
-s_{12} & c_{12} & 0 \\
0 & 0 & 1
\end{pmatrix}
\begin{pmatrix}
e^{-i\alpha^{\prime}/2} & 0 & 0 \\
0 & e^{-i\alpha/2}_{12} & 0 \\
0 & 0 & 1
\end{pmatrix},
\end{equation}
\endgroup
and $\Theta$ and the deviation of unitarity $\eta$ coming from the dimension six operators are defined as follows~\cite{Blennow:2011vn,Fernandez-Martinez:2007iaa,Broncano:2002rw}:
\begingroup
\begin{equation}
\Theta \simeq m_{D}^{\dagger} M_{N}^{-1}, \quad \eta = \frac{\Theta \Theta^{\dagger}}{2} = \frac{1}{2} \frac{m_{D}^{\dagger} m_{D}}{M_{N}^{2}}.
\end{equation}
\endgroup
We do not need to care about all the elements of $D_{\nu}^{\prime}$ instead focus on the upper-left $3 \times 3$ block. Going back to the derivation, it is given by:
\begingroup
\begin{equation}
\begin{split}
D_{\nu}^{\prime} &= U_{\func{PMNS}} \Theta_{ik} \alpha \Theta_{ki}^{T} U_{\func{PMNS}}^{T}, \quad \text{for the upper-left $3 \times 3$ block} \\
&= U_{\func{PMNS}} 
\begin{pmatrix}
\Theta_{14} & \Theta_{15} \\
\Theta_{24} & \Theta_{25} \\
\Theta_{34} & \Theta_{35} \\
\end{pmatrix}
\begin{pmatrix}
-1 & 0 \\
0 & -1
\end{pmatrix}
\begin{pmatrix}
\Theta_{41}^{T} & \Theta_{42}^{T} & \Theta_{43}^{T} \\
\Theta_{51}^{T} & \Theta_{52}^{T} & \Theta_{53}^{T}
\end{pmatrix}
U_{\func{PMNS}}^{T} \\
&= -U_{\func{PMNS}}
\begin{pmatrix}
\Theta_{14} \Theta_{41}^{T} + \Theta_{15} \Theta_{51}^{T} & \Theta_{14} \Theta_{42}^{T} + \Theta_{15} \Theta_{52}^{T} & \Theta_{14} \Theta_{43}^{T} + \Theta_{15} \Theta_{53}^{T} \\
\Theta_{24} \Theta_{41}^{T} + \Theta_{25} \Theta_{51}^{T} & \Theta_{24} \Theta_{42}^{T} + \Theta_{25} \Theta_{52}^{T} & \Theta_{24} \Theta_{43}^{T} + \Theta_{25} \Theta_{53}^{T} \\
\Theta_{34} \Theta_{41}^{T} + \Theta_{35} \Theta_{51}^{T} & \Theta_{34} \Theta_{42}^{T} + \Theta_{35} \Theta_{52}^{T} & \Theta_{34} \Theta_{43}^{T} + \Theta_{35} \Theta_{53}^{T} \\
\end{pmatrix}
U_{\func{PMNS}}^{T} \\
&\simeq -U_{\func{PMNS}}
\begin{pmatrix}
\Theta_{14} \Theta_{41}^{T} & \Theta_{14} \Theta_{42}^{T} & \Theta_{14} \Theta_{43}^{T} \\
\Theta_{24} \Theta_{41}^{T} & \Theta_{24} \Theta_{42}^{T} & \Theta_{24} \Theta_{43}^{T} \\
\Theta_{34} \Theta_{41}^{T} & \Theta_{34} \Theta_{42}^{T} & \Theta_{34} \Theta_{43}^{T} \\
\end{pmatrix}
U_{\func{PMNS}}^{T} \\
&= -2U_{\func{PMNS}}
\begin{pmatrix}
\eta_{11} & \eta_{12} & \eta_{13} \\
\eta_{21} & \eta_{22} & \eta_{23} \\
\eta_{31} & \eta_{32} & \eta_{33} \\
\end{pmatrix}
U_{\func{PMNS}}^{T},
\end{split}
\end{equation}
\endgroup
where at the fourth equality the latter terms including index $5$ can be safely ignored due to the relative smallness of $v_{d}$ compared to $v_{u}$ and the suppression factor $\epsilon$~\cite{Hernandez:2021tii}. The experimental upper limits for the deviation of unitarity $\eta_{ij}$ at $1\sigma$ are given by~\cite{Fernandez-Martinez:2016lgt}:
\begingroup
\begin{table}[H]
\centering
\resizebox{0.3\textwidth}{!}{
\centering\renewcommand{\arraystretch}{1.3} 
\begin{tabular}{*{3}{c}}
\toprule
\toprule
  & NH & IH \\ 
\midrule
$\eta_{11}$ & $4.2 \times 10^{-4}$ & $4.8 \times 10^{-4}$ \\
$\eta_{22}$ & $2.9 \times 10^{-7}$ & $2.4 \times 10^{-7}$ \\
$\eta_{33}$ & $9.2 \times 10^{-4}$ & $6.8 \times 10^{-4}$ \\
$\eta_{12}$ & $8.4 \times 10^{-6}$ & $8.4 \times 10^{-6}$ \\
$\eta_{13}$ & $6.5 \times 10^{-4}$ & $6.5 \times 10^{-4}$ \\
$\eta_{23}$ & $2.5 \times 10^{-5}$ & $1.3 \times 10^{-5}$ \\
\bottomrule
\bottomrule
\end{tabular}}%
\caption{Experimental upper limits for the deviation of unitarity $\eta$. The off-diagonal elements of $\eta_{ij}$ where $i \neq j$ are calculated by the Schwarz inequality $\eta_{ij} \lesssim \sqrt{\eta_{ii} \eta_{jj}}$}
\end{table}
\endgroup
Taking the experimental neutrino mixing angles from NuFIT 5.0~\cite{Esteban:2020cvm}, they read off (NH):
\begingroup
\begin{equation}
\begin{split}
\sin^2\theta_{12} &= 0.304, \quad \sin^2\theta_{23} = 0.570, \quad \sin^2\theta_{13} = 0.02221
\\[1ex]
U_{\func{PMNS}} &= U_{23} U_{13} U_{12}
\\[1ex]
U_{\func{PMNS}} &= \left(
\begin{array}{ccc}
 0.82495 & -0.545205 & -0.14903 \\
 0.267684 & 0.609102 & -0.746552 \\
 0.497798 & 0.575975 & 0.648421 \\
\end{array}
\right)
\label{eqn:neutrino_mixing}
\end{split}
\end{equation}
\endgroup 
From Equation~\ref{eqn:neutrino_mixing}, we can calculate the $Z^{\prime}$ coupling constant to muon neutrino pair when $g_{X}$ is assumed to be order of unity:
\begingroup
\begin{equation}
g_{\nu_{\mu} \nu_{\mu}} \simeq -5.263 \times 10^{-4}.
\label{eqn:gnumunumu_gX1}
\end{equation}
\endgroup

\section{LEPTON SECTOR PHENOMENOLOGY} \label{sec:V}

In this section, we analyze the SM observables in the charged lepton sector with the $Z^{\prime}$ coupling constants defined in section \ref{sec:IV}. There are many $Z^{\prime}$ models in the particle physics community, however most of them do not lead to a definite prediction as the $Z^{\prime}$ coupling constants and its mass are not determined experimentally. In spite of this drawback, the $Z^{\prime}$ has gotten a lot of attention since it can be a new possible FCNC source and thus might be able to solve the well-known anomalies such as $R_{K(K^{*})}$. We studied the FCNC interactions mediated by the SM $Z$ gauge boson with one complete vectorlike family~\cite{CarcamoHernandez:2021yev} in the same model frame except that the $U(1)^{\prime}$ local symmetry is replaced by the global $U(1)^{\prime}$ symmetry and what we found there is the SM $Z$ FCNC interactions generally give suppressed sensitivities for the diverse CLFV decays ($\tau \rightarrow \mu \gamma, \tau \rightarrow 3\mu$, and $Z \rightarrow \mu \tau$), compared to their experimental bounds. In this time, we assume that the CLFV decays are mainly mediated by the $Z^{\prime}$ gauge boson and the first task to do is to determine the highest order of the $Z^{\prime}$ coupling constant $g_{X}$ and the coupling constant can be constrained by investigating both the muon $g-2$ and the CLFV decays.

\subsection{Muon anomalous magnetic moment $g-2$}

The first observable we consider in the charged lepton sector is the muon anomalous magnetic moment $g-2$, regarded as one of the most important new physics signal with $4.2\sigma$ deviation~\cite{Muong-2:2021ojo}. The muon $g-2$ in the model under consideration gets the leading order contributions from the one loop diagrams in Figure~\ref{fig:muong2_Zp}:
\begingroup
\begin{figure}[H]
\centering
\begin{subfigure}{0.49\textwidth}
	\scalebox{0.9}{
	\includegraphics[keepaspectratio,width=1.0\textwidth]{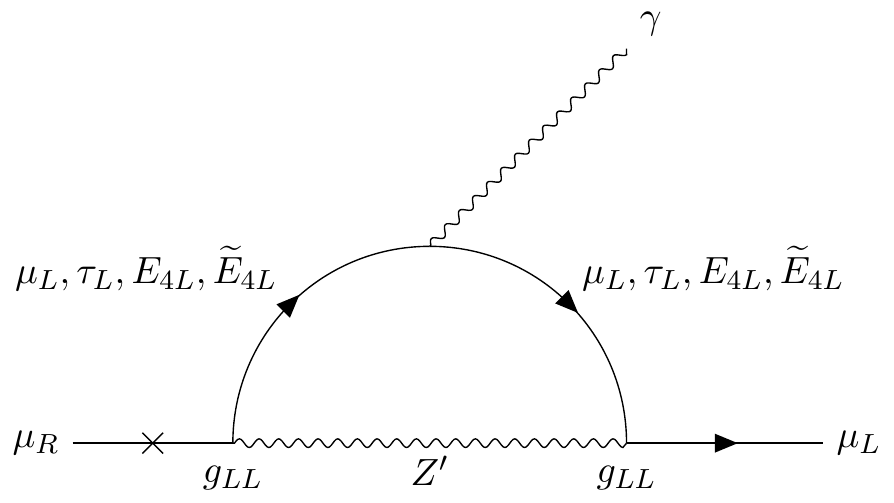} 
	}
\end{subfigure} 
\begin{subfigure}{0.49\textwidth}
	\scalebox{0.9}{
	\includegraphics[keepaspectratio,width=1.0\textwidth]{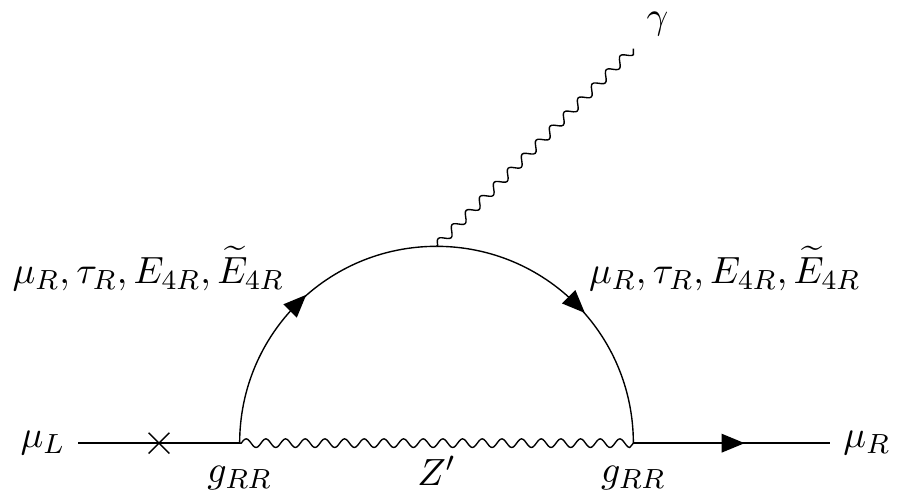}
	} 
\end{subfigure} \par
\begin{subfigure}{0.49\textwidth}
	\scalebox{0.9}{
	\includegraphics[keepaspectratio,width=1.0\textwidth]{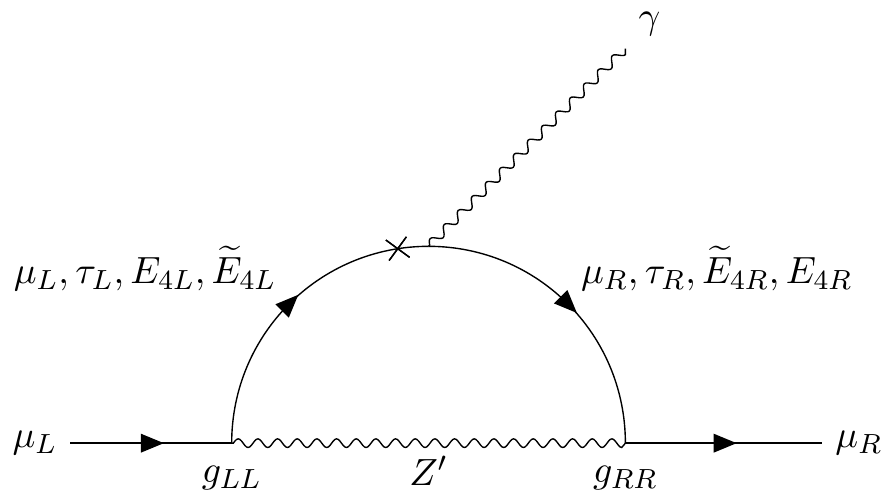}
	}
\end{subfigure}
\caption{Feynman diagrams contributing to the muon anomalous magnetic moment $g-2$ at one loop level. The cross notation in each diagram means the helicity flip process.}
\label{fig:muong2_Zp}
\end{figure}
\endgroup
The corresponding analytic expression for the muon $g-2$ reads off~\cite{Raby:2017igl,CarcamoHernandez:2019ydc}:
\begingroup
\begin{equation}
	\begin{split}
	\Delta a_{\mu}^{Z^{\prime}} = -\frac{m_{\mu}^2}{8\pi^2 M_{Z^{\prime}}^2} \bigg[& \big( \vert g_{\mu\mu}^{L} \vert^2 + \vert g_{\mu\mu}^{R} \vert^2 \big) F(x_{\mu}) 
		+ \big( \vert g_{\mu\tau}^{L} \vert^2 + \vert g_{\mu\tau}^{R} \vert^2 \big) F(x_{\tau}) \\
		&+ \big( \vert g_{\mu E_{4}}^{L} \vert^2 + \vert g_{\mu E_{4}}^{R} \vert^2 \big) F(x_{E_4}) 
		+ \big( \vert g_{\mu \widetilde{E}_{4}}^{L} \vert^2 + \vert g_{\mu \widetilde{E}_{4}}^{R} \vert^2 \big) F(x_{\widetilde{E}_{4}}) \\
		&+ \operatorname{Re}\big( g_{\mu\mu}^{L} g_{\mu\mu}^{R*} \big) \frac{m_{\mu}}{m_{\mu}} G(x_{\mu}) 
		+ \operatorname{Re}\big( g_{\mu\tau}^{L} g_{\mu\tau}^{R*} \big) \frac{m_{\tau}}{m_{\mu}} G(x_{\tau}) \\
		&+ \operatorname{Re}\big( g_{\mu E_{4}}^{L} g_{\mu \widetilde{E}_{4}}^{R*} \big) \frac{M_{E_4}}{m_{\mu}} G(x_{E_{4}})
		+  \operatorname{Re}\big( g_{\mu \widetilde{E}_{4}}^{L} g_{\mu E_{4}}^{R*} \big) \frac{M_{\widetilde{E}_4}}{m_{\mu}} G(x_{\widetilde{E}_{4}}) \bigg],
	\label{eqn:muon_g2_Zp}
	\end{split}
\end{equation}
\endgroup
where the coupling constant $g_{ij}^{{L}(R)}$ means the left-handed (right-handed) $Z^{\prime}$ coupling constant to the particle $i$ and $j$, and $F$ and $G$ are the loop functions defined as follows ($m_{\func{loop}}$ means mass of the particle running in the loop):
\begingroup
\begin{equation}
\begin{split}
F(x) &= \frac{5x^4 - 14x^3 + 39x^2 - 38x - 18x^2 \func{ln}x + 8}{12(1-x)^4} \\[1ex]
G(x) &= \frac{x^3 + 3x - 6x \func{ln}x - 4}{2(1-x)^3}, \qquad x = \frac{m_{\func{loop}}^2}{M_{Z}^{\prime 2}}.
\end{split}
\end{equation}
\endgroup
The most dominant contributions in the analytic expression of Equation~\ref{eqn:muon_g2_Zp} come from the terms involving the enhancement factors $M_{E_{4}}/m_{\mu}$ or $M_{\widetilde{E}_{4}}/m_{\mu}$. However, these terms can not keep increasing as the vectorlike masses get heavier since the relevant coupling constants get weaker at the same time, so these terms yield a balanced contribution to the muon $g-2$ prediction and this approach is also applied to the CLFV $\tau \rightarrow \mu \gamma$ decay as we discuss in the next subsection. The most recent experimental bound for the muon $g-2$ at $1\sigma$ is given by FNAL~\cite{Muong-2:2021ojo}:
\begingroup
\begin{equation}
\begin{split}
\Delta a_{\mu} &= a_{\mu}^{\func{exp}} - a_{\mu}^{\func{SM}} = \left( 25.1 \pm 5.9 \right) \times 10^{-10},
\end{split}
\end{equation}
\endgroup

\subsection{CLFV $\tau \rightarrow \mu \gamma$ decay}

As this BSM model features the massive second and third generation of the SM, the most stringent constraint in the charged lepton sector comes from the CLFV $\tau \rightarrow \mu \gamma$ decay. The leading order contributions to the CLFV $\tau \rightarrow \mu \gamma$ decay mainly arise at one loop level as given in Figure~\ref{fig:taumugamma}:
\begingroup
\begin{figure}[H]
\centering
\begin{subfigure}{0.49\textwidth}
	\scalebox{0.9}{
	\includegraphics[keepaspectratio,width=1.0\textwidth]{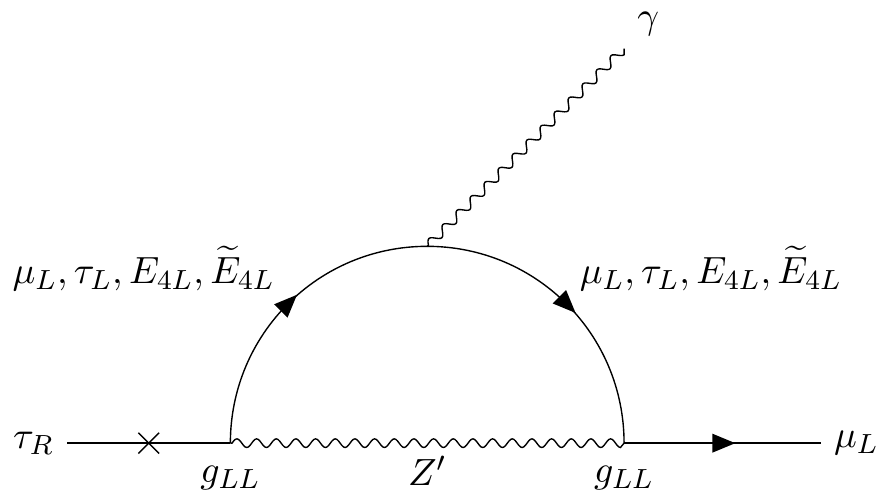} 
	}
\end{subfigure} 
\begin{subfigure}{0.49\textwidth}
	\scalebox{0.9}{
	\includegraphics[keepaspectratio,width=1.0\textwidth]{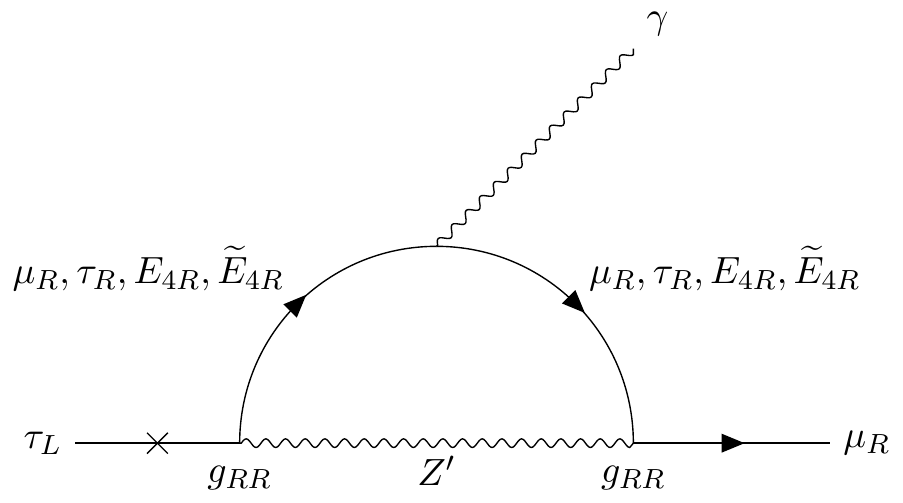}
	} 
\end{subfigure} \par
\begin{subfigure}{0.49\textwidth}
	\scalebox{0.9}{
	\includegraphics[keepaspectratio,width=1.0\textwidth]{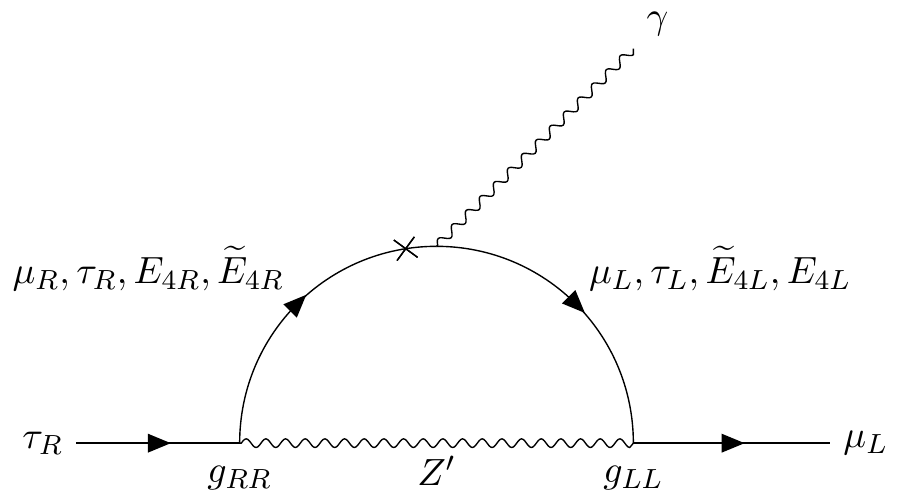}
	}
\end{subfigure} 
\begin{subfigure}{0.49\textwidth}
	\scalebox{0.9}{
	\includegraphics[keepaspectratio,width=1.0\textwidth]{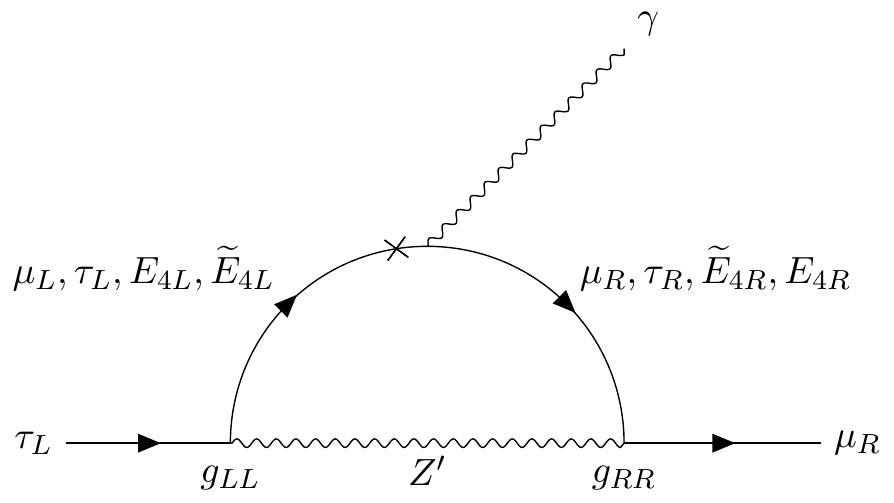}
	}
\end{subfigure} \par
\begin{subfigure}{0.49\textwidth}
	\scalebox{0.9}{
	\includegraphics[keepaspectratio,width=1.0\textwidth]{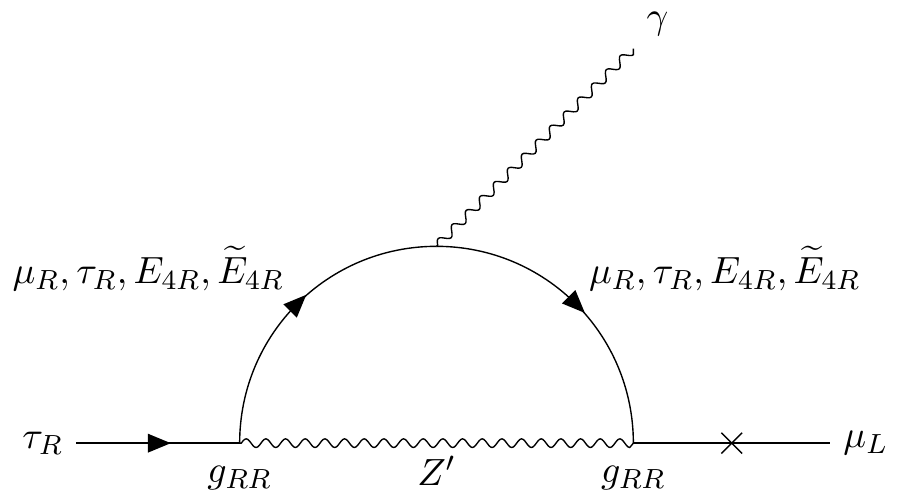}
	}
\end{subfigure} 
\begin{subfigure}{0.49\textwidth}
	\scalebox{0.9}{
	\includegraphics[keepaspectratio,width=1.0\textwidth]{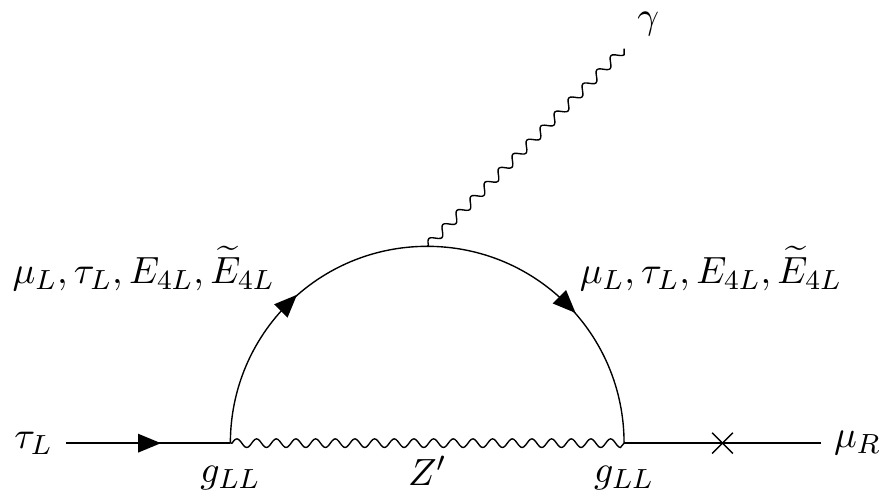}
	} 
\end{subfigure}
\caption{Feynman diagrams contributing to the CLFV $\tau \rightarrow \mu \gamma$ decay at one loop level. The cross notation in each diagram means the helicity flip process.}
\label{fig:taumugamma}
\end{figure}
\endgroup
The corresponding analytic expression for the branching ratio of the $\tau \rightarrow \mu \gamma$ decay reads~\cite{CarcamoHernandez:2019ydc,Raby:2017igl,Lindner:2016bgg,Lavoura:2003xp,Chiang:2011cv}:
\begin{equation}
\begin{split}
\func{BR} \left( \tau \rightarrow \mu \gamma \right) &= \frac{\alpha_{\func{em}}}{1024\pi^4} \frac{m_{\tau}^5}{M_Z^{\prime 4} \Gamma_{\tau}} \Big( \lvert g_{\tau \mu}^L g_{\mu \mu}^L F(x_\mu) + g_{\tau \tau}^L g_{\tau \mu}^L F(x_\tau) + g_{\tau E_4}^L g_{E_4 \mu}^L F(x_{E_4}) + g_{\tau \widetilde{E}_4}^L g_{\widetilde{E}_4 \mu}^L F(x_{\widetilde{E}_4}) \\
&+ \frac{m_{\mu}}{m_{\tau}} g_{\tau \mu}^L g_{\mu \mu}^L F(x_\mu) + \frac{m_{\mu}}{m_{\tau}} g_{\tau \tau}^L g_{\tau \mu}^L F(x_\tau) + \frac{m_{\mu}}{m_{\tau}} g_{\tau E_4}^L g_{E_4 \mu}^L F(x_{E_4}) + \frac{m_{\mu}}{m_{\tau}} g_{\tau \widetilde{E}_4}^L g_{\widetilde{E}_4 \mu}^L F(x_{\widetilde{E}_4}) \\
&+ \frac{m_{\mu}}{m_{\tau}} g_{\tau \mu}^L g_{\mu \mu}^R G(x_{\mu}) + \frac{m_{\tau}}{m_{\tau}} g_{\tau \tau}^L g_{\tau \mu}^R G(x_{\mu}) + \frac{M_{E_4}}{m_{\tau}} g_{\tau E_4}^L g_{\widetilde{E}_4 \mu}^R G(x_{E_4}) + \frac{M_{\widetilde{E}_4}}{m_{\tau}} g_{\tau \widetilde{E}_4}^L g_{E_4 \mu}^R G(x_{\widetilde{E}_4}) \rvert^2 \\
&+ \lvert g_{\tau \mu}^R g_{\mu \mu}^R F(x_\mu) + g_{\tau \tau}^R g_{\tau \mu}^R F(x_\tau) + g_{\tau E_4}^R g_{E_4 \mu}^R F(x_{E_4}) + g_{\tau \widetilde{E}_4}^R g_{\widetilde{E}_4 \mu}^R F(x_{\widetilde{E}_4}) \\
&+ \frac{m_{\mu}}{m_{\tau}} g_{\tau \mu}^R g_{\mu \mu}^R F(x_\mu) + \frac{m_{\mu}}{m_{\tau}} g_{\tau \tau}^R g_{\tau \mu}^R F(x_\tau) + \frac{m_{\mu}}{m_{\tau}} g_{\tau E_4}^R g_{E_4 \mu}^R F(x_{E_4}) + \frac{m_{\mu}}{m_{\tau}} g_{\tau \widetilde{E}_4}^R g_{\widetilde{E}_4 \mu}^R F(x_{\widetilde{E}_4}) \\
&+ \frac{m_{\mu}}{m_{\tau}} g_{\tau \mu}^R g_{\mu \mu}^L G(x_{\mu}) + \frac{m_{\tau}}{m_{\tau}} g_{\tau \tau}^R g_{\tau \mu}^L G(x_{\mu}) + \frac{M_{\widetilde{E}_4}}{m_{\tau}} g_{\tau E_4}^R g_{\widetilde{E}_4 \mu}^L G(x_{\widetilde{E}_4}) + \frac{M_{E_4}}{m_{\tau}} g_{\tau \widetilde{E}_4}^R g_{E_4 \mu}^L G(x_{E_4}) \rvert^2 \Big),
\label{eqn:analytic_tmgdecay}
\end{split}
\end{equation}
where $\alpha_{\func{em}}$ is the fine structure constant, $\Gamma_{\tau}$ is the total decay width of the tau lepton ($\Gamma_{\tau} = 5\Gamma\left( \tau_{L}^{-} \rightarrow \nu_{\tau} e_{L}^{-} \overline{\nu}_{e} \right)) = 2.0 \times 10^{-12}\func{GeV}$). The most recent experimental upper-limit for the CLFV $\tau \rightarrow \mu \gamma$ decay is given by~\cite{Crivellin:2020ebi,MEG:2016leq,BaBar:2009hkt}:
\begingroup
\begin{equation}
\func{BR} \left( \tau \rightarrow \mu \gamma \right)_{\func{exp}} = 4.4 \times 10^{-8}.
\label{eqn:exp_taumugamma}
\end{equation}
\endgroup
\subsection{CLFV $\tau \rightarrow \mu \mu \mu$ decay}
The next observable we consider is the CLFV $\tau \rightarrow \mu\mu\mu$ decay. The CLFV $\tau \rightarrow 3\mu$ decay mediated by the neutral massive $Z^{\prime}$ gauge boson can have the leading order contributions at tree level as given in Figure~\ref{fig:tau3mu_Zp}:
\begingroup
\begin{figure}[H]
\centering
\begin{subfigure}{0.49\textwidth}
	\scalebox{0.9}{
	\includegraphics[keepaspectratio,width=1.0\textwidth]{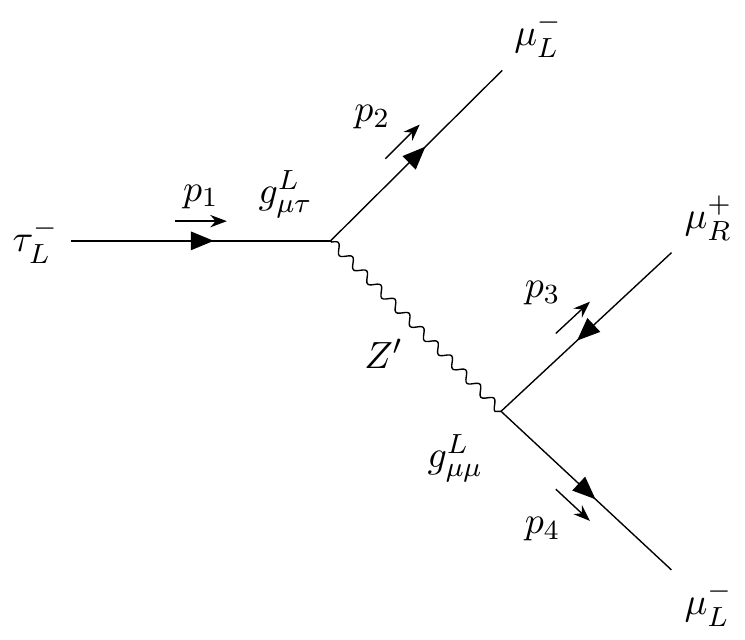} 
	}
\end{subfigure} 
\begin{subfigure}{0.49\textwidth}
	\scalebox{0.9}{
	\includegraphics[keepaspectratio,width=1.0\textwidth]{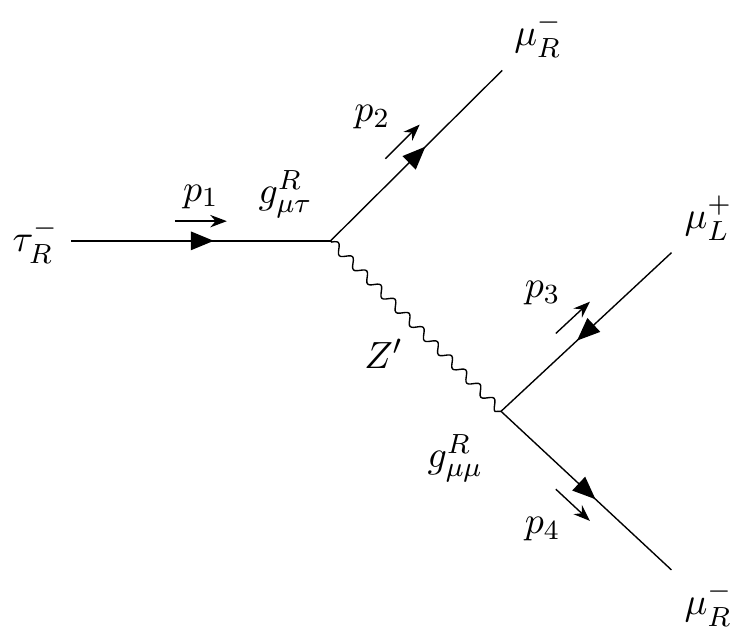}
	} 
\end{subfigure} \par
\begin{subfigure}{0.49\textwidth}
	\scalebox{0.9}{
	\includegraphics[keepaspectratio,width=1.0\textwidth]{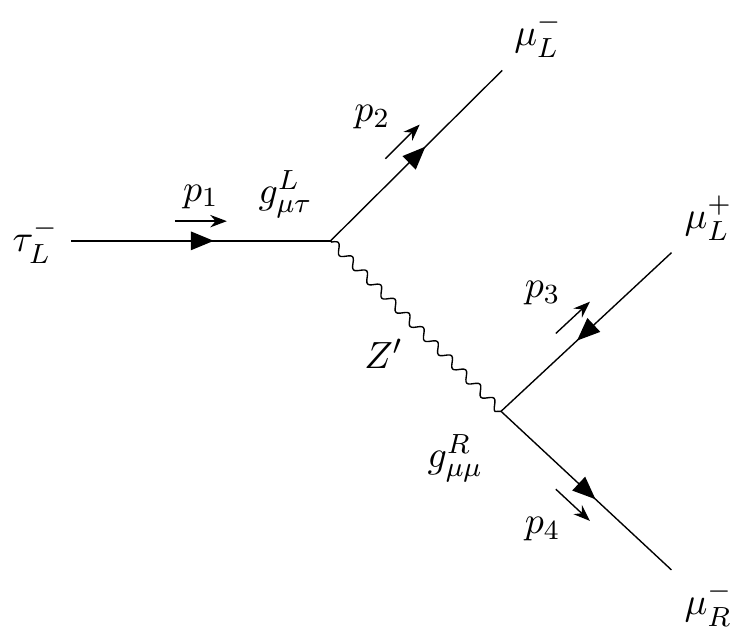}
	}
\end{subfigure} 
\begin{subfigure}{0.49\textwidth}
	\scalebox{0.9}{
	\includegraphics[keepaspectratio,width=1.0\textwidth]{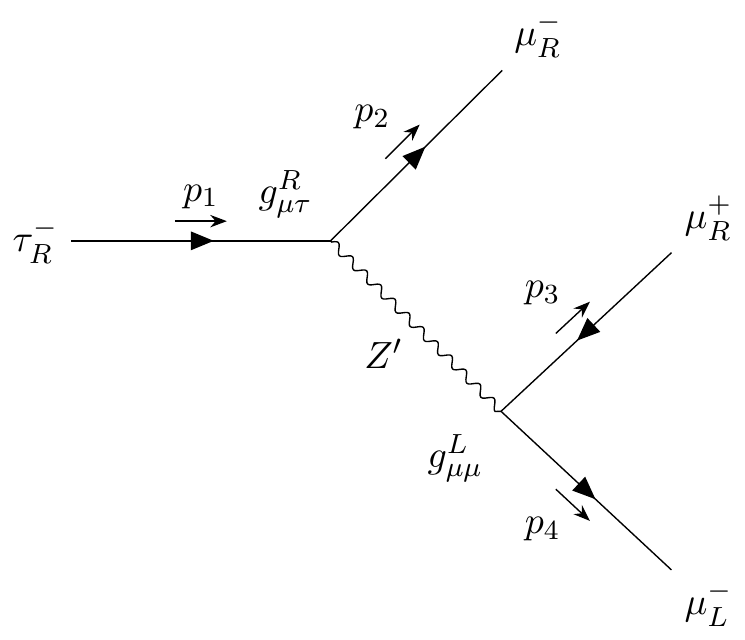}
	}
\end{subfigure}
\caption{Feynman diagrams contributing to the CLFV $\tau \rightarrow 3\mu$ decay at tree level.}
\label{fig:tau3mu_Zp}
\end{figure}
\endgroup
We derived the analytic expression for the branching ratio of the CLFV $\tau \rightarrow 3\mu$ decay mediated by the SM $Z$ gauge boson in one of our works~\cite{CarcamoHernandez:2021yev}, so we can simply make use of the derived result after replacing the SM $Z$ gauge boson in the prediction by the neutral massive $Z^{\prime}$ gauge boson as follows:
\begingroup
\begin{equation}
\func{BR}\left( \tau \rightarrow 3 \mu \right) = \frac{1}{64 \pi^3 m_{\tau} \Gamma_{\tau}} \int_{m_{\mu}^2/m_{\tau}}^{\frac{1}{2}m_{\tau}} \int_{\frac{1}{2}m_{\tau}+\frac{m_{\mu}^2}{m_{\tau}}-E_2}^{\frac{1}{2}m_{\tau}} \left( \frac{1}{2}\sum_{\func{spin}} \lvert \mathcal{M} \left( g_1, g_2, E_2, E_4 \right) \rvert^2 \right) dE_4 dE_2,
\label{eqn:BRtau3mu_Zp}
\end{equation}
\endgroup
where the amplitude squared is defined by~\cite{Hernandez:2021tii} ($g_1 = g_{\mu \tau}^{L2} g_{\mu \mu}^{L2} + g_{\mu \tau}^{R2} g_{\mu \mu}^{R2}, g_2 = g_{\mu \tau}^{L2} g_{\mu \mu}^{R2} + g_{\mu \tau}^{R2} g_{\mu \mu}^{L2}$)
\begingroup
\begin{equation}
\begin{split}
\frac{1}{2}\sum_{\func{spin}} \lvert \mathcal{M} \left( g_1, g_2, E_2, E_4 \right) \rvert^2 &= \frac{4}{M_{Z^{\prime}}^{4}} \Big[ g_1 \left( m_{\tau}^2 - m_{\tau} (E_2 + E_4) \right) \left( -m_{\tau}^2 - m_{\mu}^2 + 2m_{\tau} (E_2 + E_4) \right) \\
&+ g_2 \left( m_{\tau} E_4 \right) \left( m_{\tau}^2 - m_{\mu}^2 - 2m_{\tau} E_4 \right) \Big]
\end{split}
\end{equation}
\endgroup
and $E_{2}$ and $E_{4}$ are the energies defined in the rest frame as follows:
\begingroup
\begin{equation}
\begin{split}
p_1 &=  ( m_{\tau}, \vec{0} ) \\
p_2 &= \left( E_2, \vec{p}_2 \right) \\
p_3 &= \left( m_{\tau} - E_2 - E_4, -\vec{p}_2 -\vec{p}_4  \right) \\
p_4 &= \left( E_4, \vec{p}_4 \right) \\
\end{split}
\end{equation}
\endgroup
and the energy $E_{4}$ runs from $\frac{1}{2} m_{\tau} + \frac{m_{\mu}^{2}}{m_{\tau}} - E_{2}$ to $\frac{1}{2} m_{\tau}$ and the other energy $E_{2}$ runs from $\frac{m_{\mu}^{2}}{m_{\tau}}$ to $\frac{1}{2} m_{\tau}$ as seen from the integration range of Equation~\ref{eqn:BRtau3mu_Zp}. The experimental upper-limit of the branching ratio of the CLFV $\tau \rightarrow 3\mu$ decay is given by~\cite{Hayasaka:2010np}:
\begingroup
\begin{equation}
\func{BR}\left( \tau \rightarrow 3\mu \right)_{\func{exp}} = 2.1 \times 10^{-8}
\end{equation}
\endgroup
\subsection{Neutrino trident production}
One more constraint for the $Z^{\prime}$ gauge boson in the lepton sector arise from the neutrino trident production. The relevant diagram for the neutrino trident production reads in Figure~\ref{fig:neutrino_trident_production}:
\begingroup
\begin{figure}[H]
\centering
\includegraphics[keepaspectratio,width=0.8\textwidth]{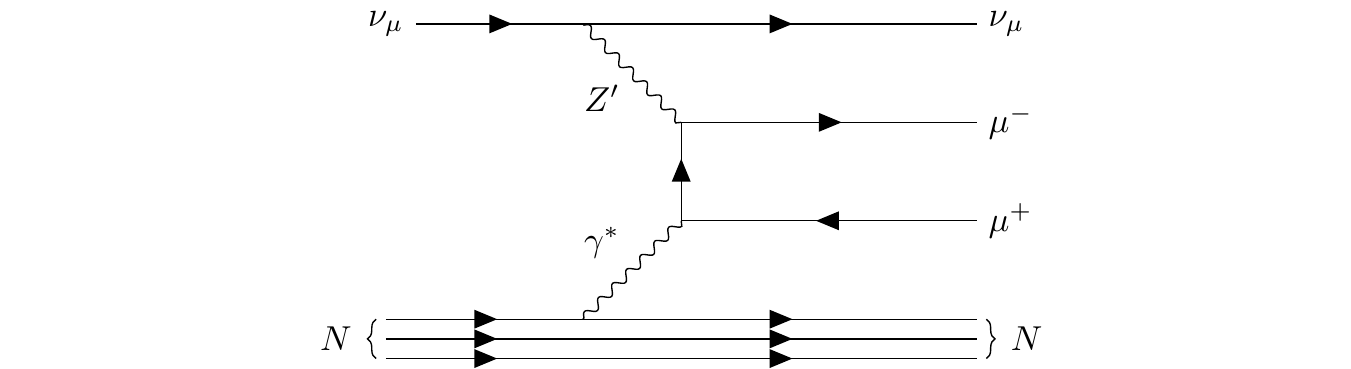}
\caption{Feynman diagram for the neutrino trident production.}
\label{fig:neutrino_trident_production}
\end{figure}
\endgroup
The constraint for the neutrino trident production is given by the effective four fermion interaction by exchanging the neutral massive $Z^{\prime}$ gauge boson as follows~\cite{CarcamoHernandez:2019ydc,CHARM-II:1990dvf,CCFR:1991lpl,Altmannshofer:2014pba}:
\begingroup
\begin{equation}
\mathcal{L}_{\nu \func{Tri}}^{Z^{\prime},\func{eff}} = -\frac{g_{\nu_{\mu} \nu_{\mu}}^{L} g_{\mu \mu}^{L}}{M_{Z^{\prime}}^2} \left( \overline{\mu}_{L} \gamma_{\mu} \mu_{L} \right) \left( \overline{\nu}_{\mu L} \gamma^{\mu} \nu_{\mu L} \right) - \frac{g_{\nu_{\mu} \nu_{\mu}}^{L} g_{\mu \mu}^{R}}{M_{Z^{\prime}}^2} \left( \overline{\mu}_{R} \gamma_{\mu} \mu_{R} \right) \left( \overline{\nu}_{\mu L} \gamma^{\mu} \nu_{\mu L} \right)
\end{equation}
\endgroup
The experimental bound for the neutrino trident production at $2\sigma$ reads~\cite{Falkowski:2017pss}:
\begingroup
\begin{equation}
\begin{split}
-\frac{1}{\left( 390 \func{GeV} \right)^{2}} &< \frac{g_{\nu_{\mu} \nu_{\mu}}^{L} g_{\mu\mu}^{L} + g_{\nu_{\mu} \nu_{\mu}}^{L} g_{\mu\mu}^{R}}{M_{Z^{\prime}}^{2}} < \frac{1}{\left( 370 \func{GeV} \right)^{2}}
\\
-\frac{1}{\left( 390 \func{GeV} \right)^{2}} &< \frac{g_{\nu_{\mu} \nu_{\mu}}^{L} \left( g_{\mu\mu}^{L} + g_{\mu\mu}^{R} \right)}{M_{Z^{\prime}}^{2}} < \frac{1}{\left( 370 \func{GeV} \right)^{2}}
\label{eqn:const_neutrino_trident_production}
\end{split}
\end{equation}
\endgroup
It is possible to simplify Equation~\ref{eqn:const_neutrino_trident_production} further by using the calculated coupling constant $g_{\nu_{\mu} \nu_{\mu}}$ of Equation~\ref{eqn:gnumunumu_gX1}:
\begingroup
\begin{equation}
-\frac{1}{\left( 390 \func{GeV} \right)^{2}} < \frac{-5.263 \times 10^{-4} \left( g_{\mu\mu}^{L} + g_{\mu\mu}^{R} \right)}{M_{Z^{\prime}}^{2}} < \frac{1}{\left( 370 \func{GeV} \right)^{2}},
\label{eqn:neutrino_trident_const_final}
\end{equation}
\endgroup
where it is worth mentioning that the coupling constant $g_{\nu_{\mu} \nu_{\mu}}$ is given when we assume the $Z^{\prime}$ coupling constant $g_{X}$ is the order of unity. As we will see soon in the numerical analysis for the charged lepton sector, the coupling constant $g_{X}$ can reach its highest order up to unity (or equally $g_{X} = 1$). Therefore, we can determine the $Z^{\prime}$ mass range from the neutrino trident production constraint once a numerical range of summing over the left-handed and right-handed muon coupling constant is given by the numerical analysis and this will be discussed in the next subsection.

\subsection{Numerical analysis in the charged lepton sector}
In this subsection, we carry out numerical scan in the charged lepton sector. What we want to achieve via this numerical scan is to determine the highest order of the $Z^{\prime}$ coupling constant $g_{X}$ and this process is quite important since the determined order of the coupling constant $g_{X}$ affects the other analyses which will be carried out in the quark sector as well as in the electroweak precision observables. The possible range of the $Z^{\prime}$ mass can be constrained in the quark sector via $R_{K^*}$ anomaly as well as $B_{s}$ meson oscillation and we leave constraining the $Z^{\prime}$ mass until we carry out the quark sector investigation and regard the $Z^{\prime}$ mass as a free parameter in this subsection.
\subsubsection{Free parameter setup}
The free parameter setup given in Table~\ref{tab:parameter_setup_chargedleptonsector} follows the exactly same form defined in one of our works~\cite{CarcamoHernandez:2021yev} except for the $Z^{\prime}$ mass:
\begingroup
\begin{table}[H]
\centering
\resizebox{0.45\textwidth}{!}{
\centering\renewcommand{\arraystretch}{1.3} 
\begin{tabular}{*{2}{c}}
\toprule
\toprule
\textbf{Mass parameter} & \textbf{Scanned Region($\func{GeV}$)} \\
\midrule
$y_{24}^{e} v_{d} = m_{24}$ & $\pm [1,10]$ \\[1ex]
$y_{34}^{e} v_{d} = m_{34}$ & $\pm [1,10]$ \\[1ex]
$y_{43}^{e} v_{d} = m_{43}$ & $\pm [1,10]$ \\[1ex]
$x_{34}^{L} v_{\phi} = m_{35}$ & $\pm [50,200]$ \\[1ex]
$x_{42}^{e} v_{\phi} = m_{52}$ & $\pm [50,200]$ \\[1ex]
$x_{43}^{e} v_{\phi} = m_{53}$ & $\pm [50,200]$ \\[1ex]
$M_{45}^{L}$ & $\pm [150,2000]$ \\[1ex]
$M_{54}^{e}$ & $\pm [150,2000]$ \\[1ex]
\midrule
$M_{Z^{\prime}}$ & $[75, 3000]$ \\[1ex]
\bottomrule
\bottomrule
\end{tabular}}%
\caption{Parameter setup to determine the highest order of the $Z^{\prime}$ coupling constant $g_{X}$ by the muon $g-2$ and the CLFV $\tau \rightarrow \mu \tau$ and $\tau \rightarrow 3\mu$ decays.}
\label{tab:parameter_setup_chargedleptonsector}
\end{table}
\endgroup
The most relevant features about the parameter setup of Table~\ref{tab:parameter_setup_chargedleptonsector} are summarized as follows:
\begingroup
\begin{enumerate}
\item As the model under consideration features an extended 2HDM, the up-type Higgs vev $v_{u}$, very close to $246\func{GeV}$, and the down-type Higgs vev $v_{d}$, assumed to run from $1$ to $10\func{GeV}$, satisfy the relation $v_{u}^{2} + v_{d}^{2} = \left( 246\func{GeV} \right)^{2}$. The down-type Higgs $H_{d}$ interacts with the down-type quarks as well as the charged leptons and thus the mass parameters involving the down-type Higgs vev $v_{d}$ are determined under the assumption.
\item The singlet flavon vev $v_{\phi}$ and vectorlike masses $M_{45}^{L}$ and $M_{54}^{e}$ and lastly the $Z^{\prime}$ mass $M_{Z^{\prime}}$ are free parameters and mass parameters including the flavon vev $v_{\phi}$ are determined to be consistent with the observed charged lepton hierarchy. Here a main role of the free parameter $M_{Z^{\prime}}$ is to see whether it can be constrained by the muon $g-2$ and CLFV $\tau \rightarrow \mu \gamma$ and $\tau \rightarrow 3\mu$ decays. As for the lower bound of the $Z^{\prime}$ mass in Table~\ref{tab:parameter_setup_chargedleptonsector}, it is set to avoid the low mass constraint of $Z^{\prime}$ mass $5 \lesssim M_{Z^{\prime}} \lesssim 70\func{GeV}$ resulting from the decay process  $pp \rightarrow Z \rightarrow 4\mu$~\cite{Navarro:2021sfb,Falkowski:2018dsl,Altmannshofer:2014pba,Altmannshofer:2014cfa,Altmannshofer:2016jzy}.
\item What we constrain in this numerical scan is the muon and tau mass plus the $23$ mixing angle. The theoretically predicted muon and tau mass should be put between $\left[ 1 \pm 0.1 \right] \times m_{\mu,\tau}^{\func{exp}}$. On top of that, we also constrain the $23$ mixing angle and determine the upper-limit for the mixing angle to be $0.2$ since the sizeable off-diagonal components of the PMNS mixing matrix mainly arise from the neutrino sector.
\end{enumerate}
\endgroup
\subsubsection{Numerical scan result for the charged lepton sector}
The numerical scan result for the muon $g-2$ and for the CLFV $\tau \rightarrow \mu \gamma, \tau \rightarrow 3\mu$ decays when the $Z^{\prime}$ coupling constant $g_{X} = 1$ is given in Figure~\ref{fig:numerical_scan_chargedlepton1}:
\begingroup
\begin{figure}[H]
\centering
\begin{subfigure}{0.49\textwidth}
	\scalebox{0.90}{
	\includegraphics[keepaspectratio,width=1.0\textwidth]{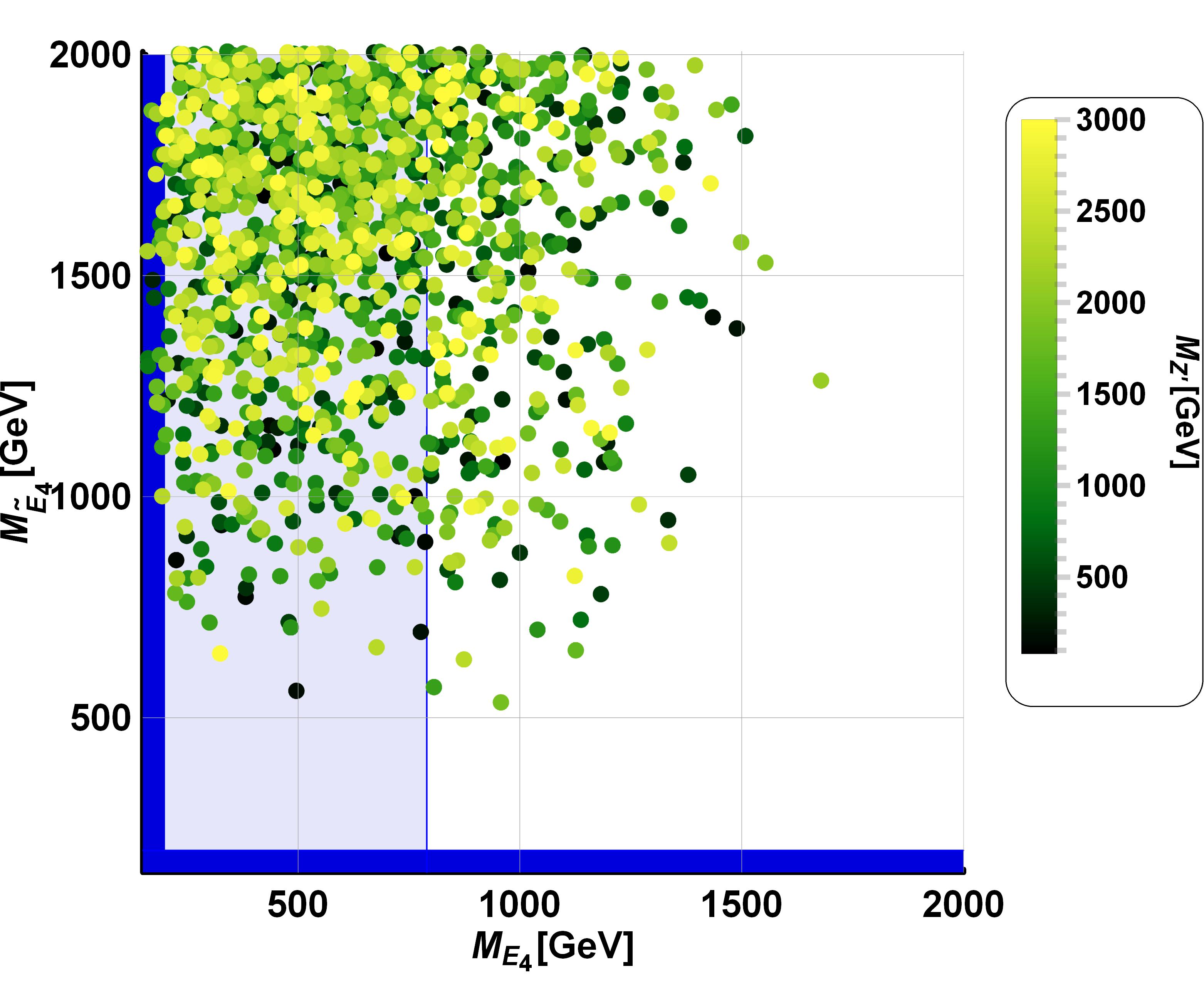} 
	}
\end{subfigure} 
\begin{subfigure}{0.49\textwidth}
	\scalebox{0.90}{
	\includegraphics[keepaspectratio,width=1.0\textwidth]{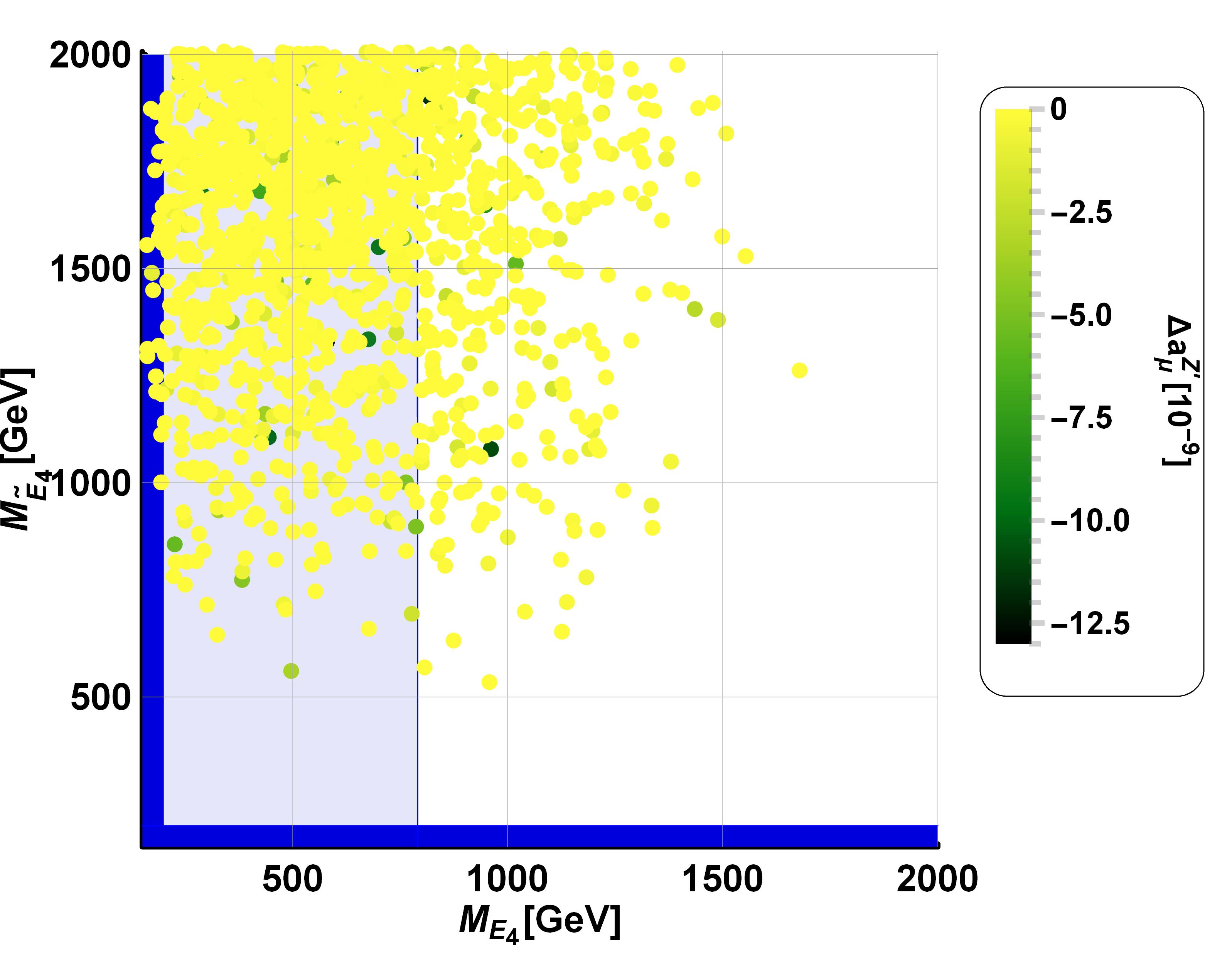}
	} 
\end{subfigure} \par
\begin{subfigure}{0.49\textwidth}
	\scalebox{0.90}{
	\includegraphics[keepaspectratio,width=1.0\textwidth]{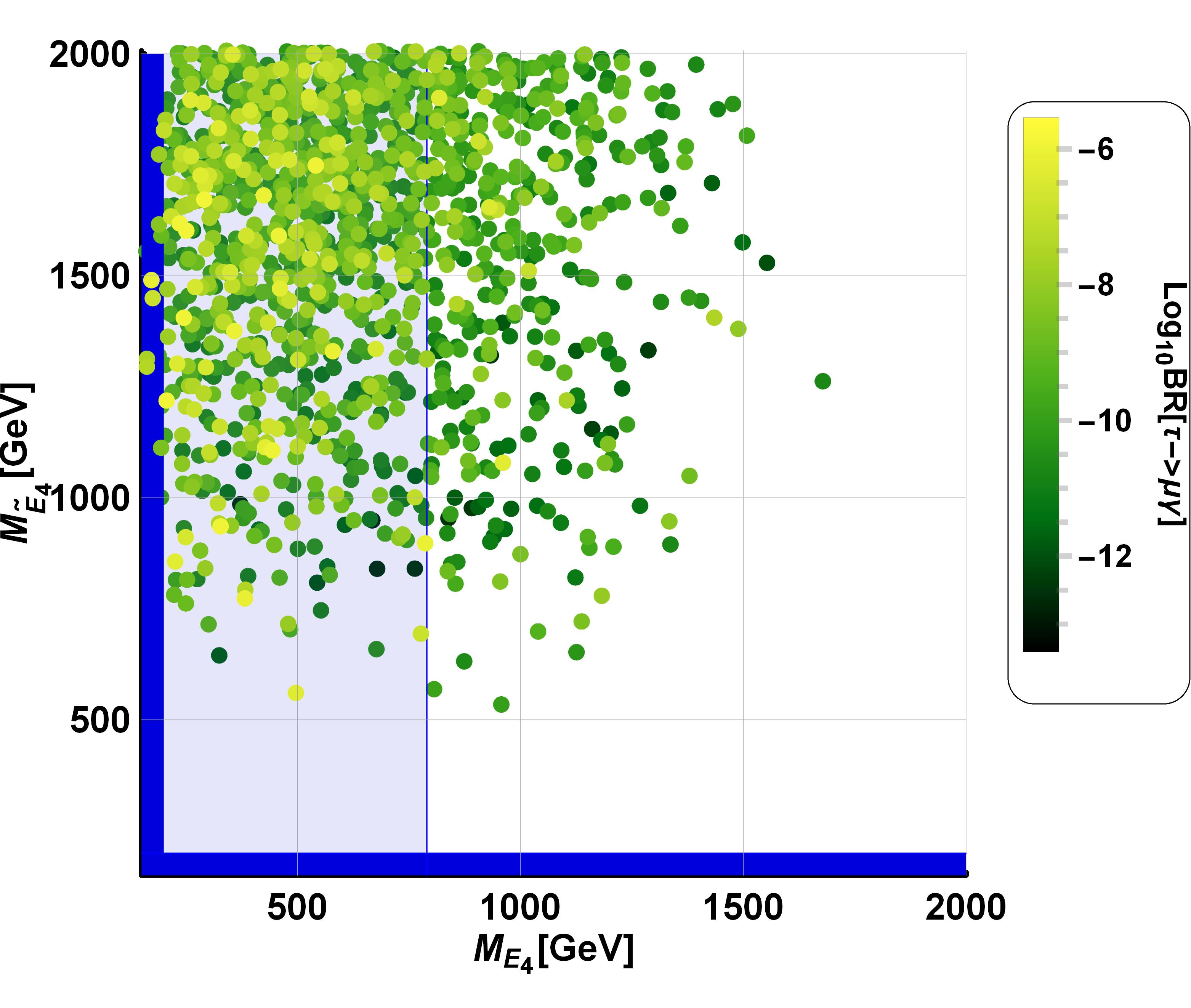}
	}
\end{subfigure} 
\begin{subfigure}{0.49\textwidth}
	\scalebox{0.90}{
	\includegraphics[keepaspectratio,width=1.0\textwidth]{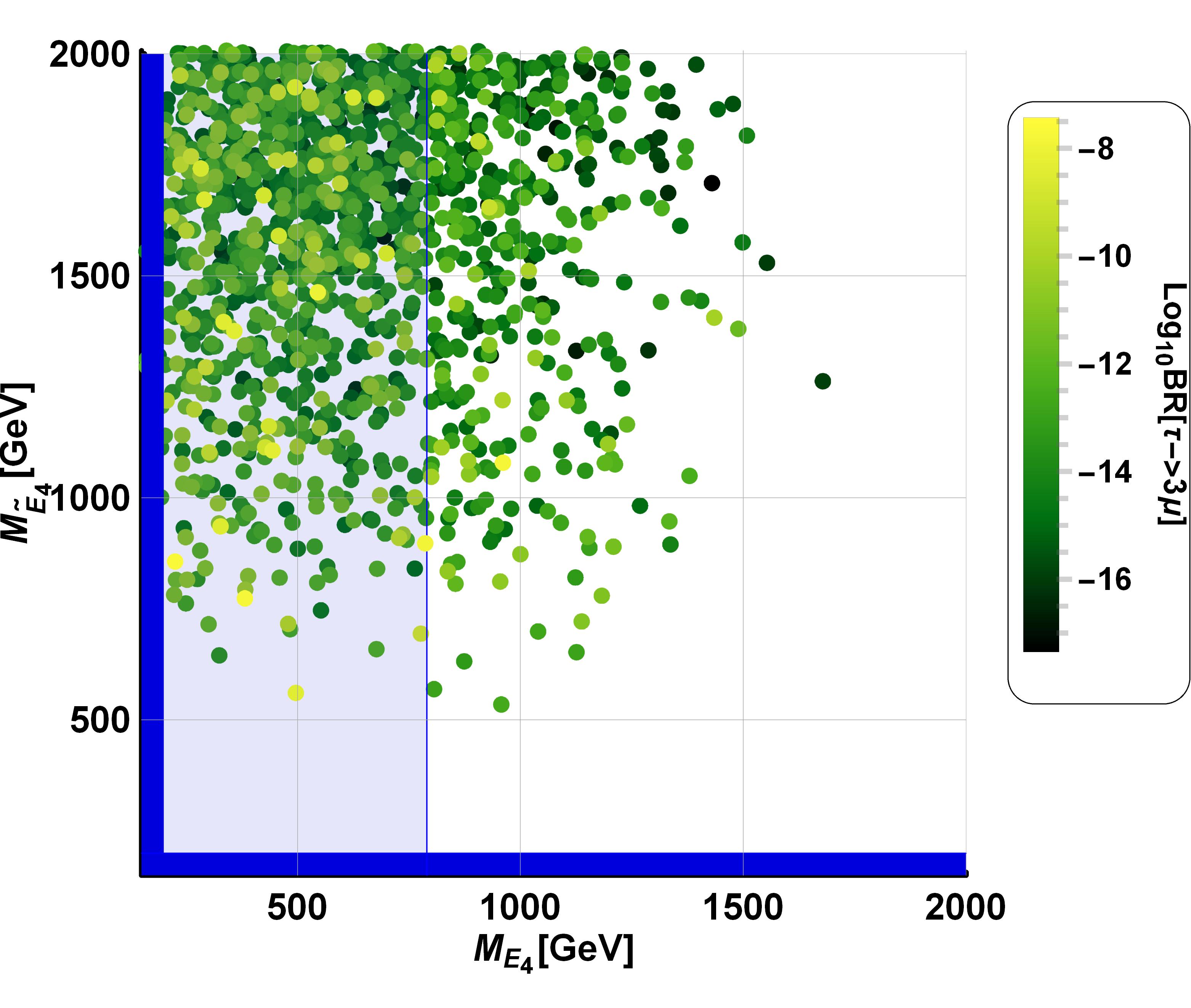}
	}
\end{subfigure}
\caption{Numerical scan result for $Z^{\prime}$ mass, free parameter, and for the muon $g-2$ as well as the CLFV $\tau \rightarrow \mu \gamma, \tau \rightarrow 3\mu$ decays when the $Z^{\prime}$ coupling constant $g_{X}$ is fixed by unity. $M_{E_{4}}$ appearing in $x$ axis is mass of the vectorlike doublet charged lepton $E_{4}$ and $M_{\widetilde{E}_{4}}$ is mass of the vectorlike singlet charged lepton $\widetilde{E}_{4}$. The darker blue region in each plot indicates that the vectorlike doublet or singlet charged lepton mass should be excluded up to $200\func{GeV}$~\cite{Xu:2018pnq,Guedes:2021oqx}. The brighter blue region in each plot means that the vectorlike doublet mass $M_{E_{4}}$ should be excluded up to $790\func{GeV}$ by the CMS experimental constraint~\cite{Bhattiprolu:2019vdu,CMS:2019hsm}.}
\label{fig:numerical_scan_chargedlepton1}
\end{figure}
\endgroup
The first feature we discuss in Figure~\ref{fig:numerical_scan_chargedlepton1} is the $Z^{\prime}$ mass. Here, the scanned $Z^{\prime}$ mass range appears to be the same as the initial setup $\left[ 75, 3000\func{GeV} \right]$, which means that the charged lepton observables (muon $g-2$, the CLFV $\tau \rightarrow \mu \gamma, \tau \rightarrow 3\mu$ decays) can not constrain the free parameter $M_{Z^{\prime}}$ at all. However, constraining the $Z^{\prime}$ mass can be done by constraints arising in the quark sector. The second is the scanned muon $g-2$. Interestingly, the scanned muon $g-2$ in this BSM reports only negative contributions with order of mostly $10^{-11}$, which implies that the $Z^{\prime}$ physics is definitely not an answer for the muon anomaly since it only deteriorates our muon prediction and this feature is one of big differences, compared to \cite{Navarro:2021sfb}, as we can not have positive contributions to the muon $g-2$ regardless of how large the $Z^{\prime}$ coupling constant to right-handed muon pair is. We will try to explain the muon $g-2$ via scalar exchange and this will be discussed in the next section. Next we discuss our predictions for the CLFV $\tau \rightarrow \mu \gamma$ decay. Considering its logarithmic experimental bound, $\func{Log}_{10}\func{BR}\left( \tau \rightarrow \mu \gamma \right) \approx -7.4$~\cite{Crivellin:2020ebi,MEG:2016leq,BaBar:2009hkt}, our predictions for the $\tau \rightarrow \mu \gamma$ decay are not significantly constrained by its experimental constraint, implying that the $Z^{\prime}$ coupling constant $g_{X}$ can be order of unity at most. The last CLFV observable we discuss is the $\tau \rightarrow 3\mu$ decay and its situation is similar to the CLFV $\tau \rightarrow \mu \gamma$ decay in the sense that it can not constrain $Z^{\prime}$ gauge boson at all. Excluding the benchmark points, which exceed the experimental upper-limit of the $\tau \rightarrow \mu \gamma$ decay, the result is given in Figure~\ref{fig:numerical_scan_chargedlepton2}:
\begingroup
\begin{figure}[H]
\centering
\begin{subfigure}{0.49\textwidth}
	\scalebox{0.90}{
	\includegraphics[keepaspectratio,width=1.0\textwidth]{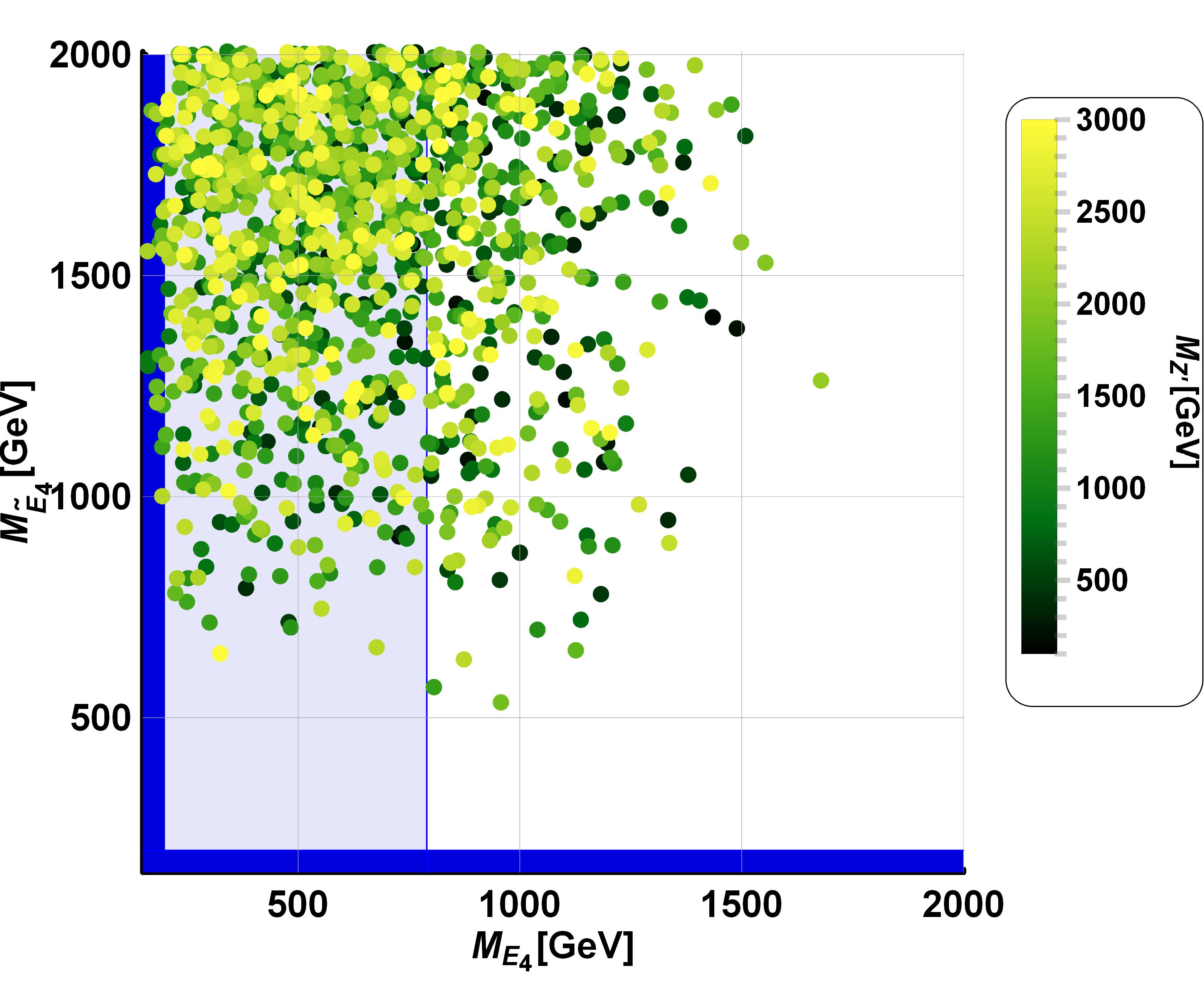} 
	}
\end{subfigure} 
\begin{subfigure}{0.49\textwidth}
	\scalebox{0.90}{
	\includegraphics[keepaspectratio,width=1.0\textwidth]{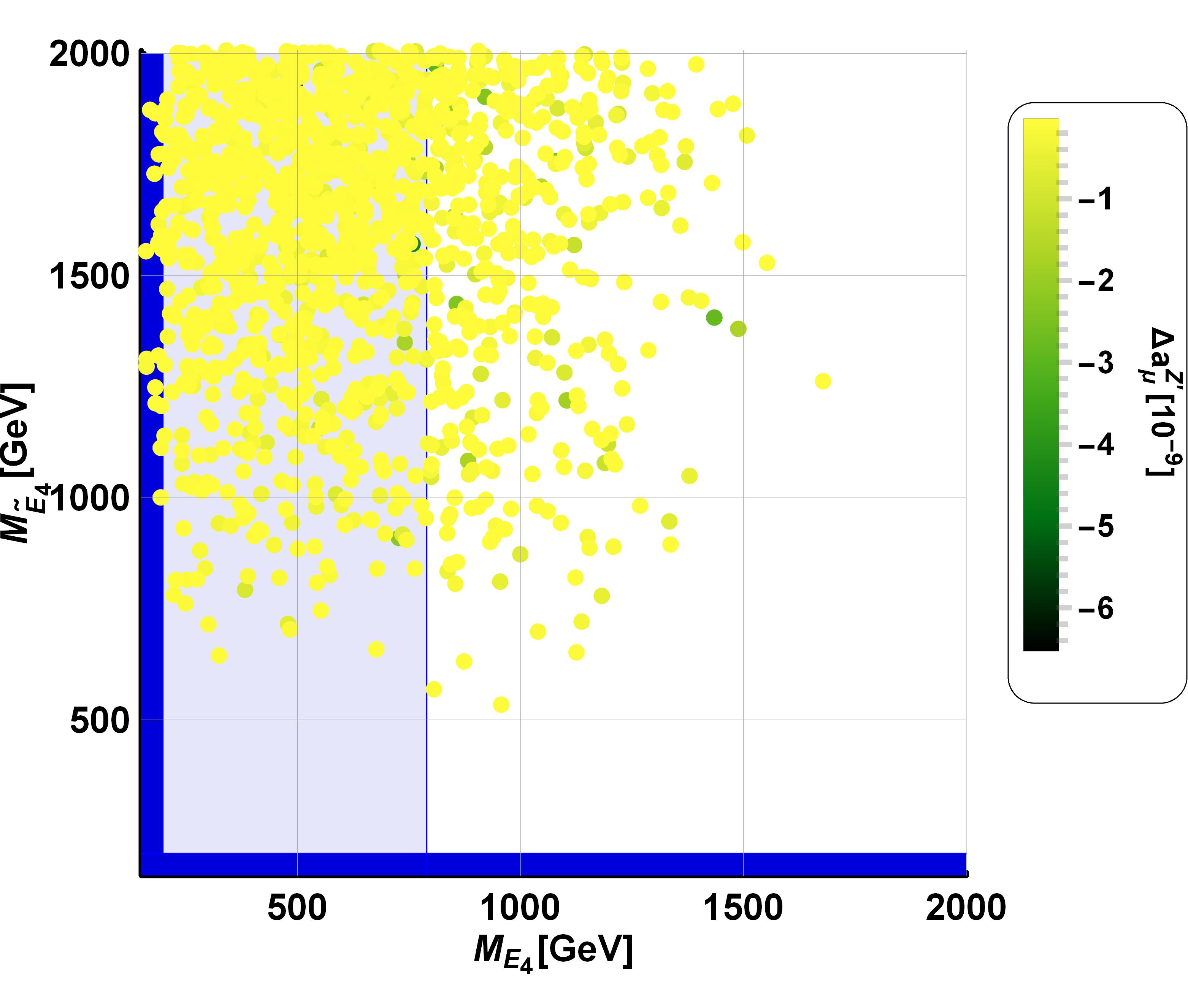}
	} 
\end{subfigure} \par
\begin{subfigure}{0.49\textwidth}
	\scalebox{0.90}{
	\includegraphics[keepaspectratio,width=1.0\textwidth]{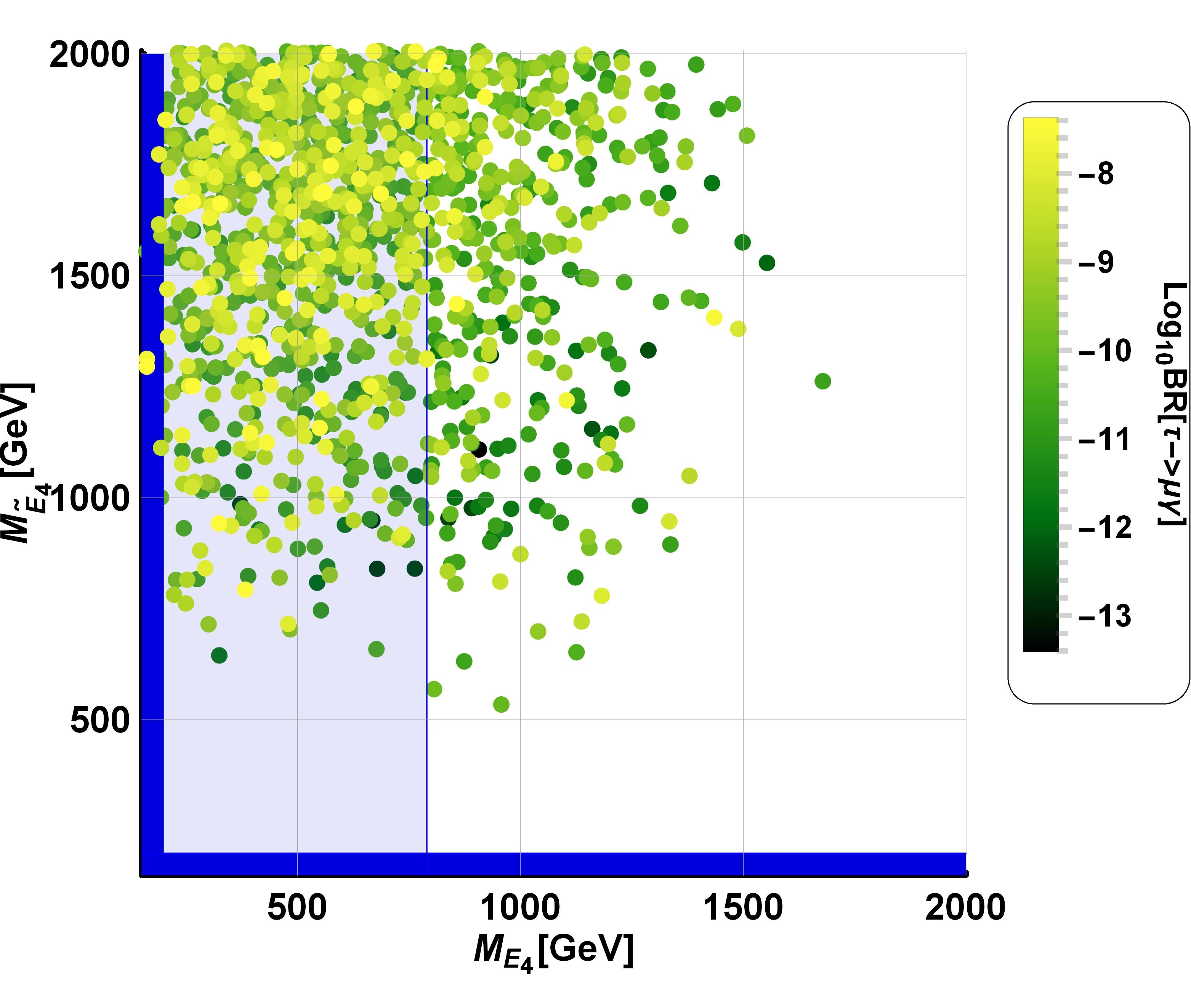}
	}
\end{subfigure} 
\begin{subfigure}{0.49\textwidth}
	\scalebox{0.90}{
	\includegraphics[keepaspectratio,width=1.0\textwidth]{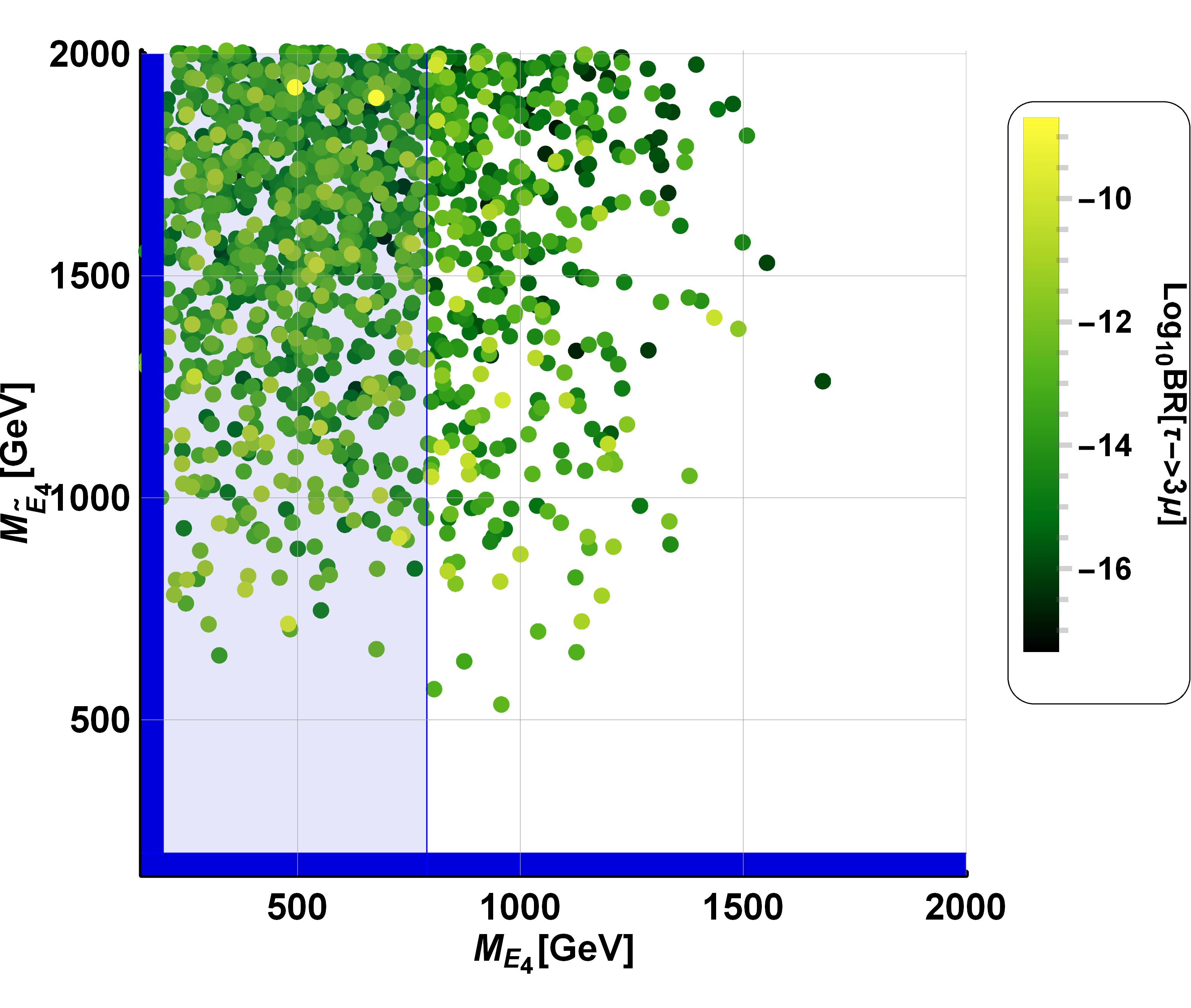}
	}
\end{subfigure}
\caption{Numerical scan result for the charged lepton observables after excluding the benchmark points which exceed the upper-limit of the $\tau \rightarrow \mu \gamma$ decay. After excluding the benchmark points which exceed the experimental upper-limit of $\tau \rightarrow \mu \gamma$ decay, most of other benchmark points survive and this implies that the highest order of coupling constant $g_{X} = 1$ is not ruled out by the charged lepton observables.}
\label{fig:numerical_scan_chargedlepton2}
\end{figure}
\endgroup
From the survived benchmark points of Figure~\ref{fig:numerical_scan_chargedlepton2}, we can derive a numerical range of the $Z^{\prime}$ coupling constant to the left- or right-handed muon pair $g_{\mu\mu}^{L,R}$ as follows:
\begin{table}[H]
\centering
\resizebox{0.6\textwidth}{!}{
\centering\renewcommand{\arraystretch}{1.3} 
\begin{tabular}{*{2}{c}}
\toprule
\toprule
Coupling constant & Scanned range \\ 
\midrule
$g_{\mu\mu}^{L}$ & $4.888 \times 10^{-6} < g_{\mu\mu}^{L} < 2.113 \times 10^{-3}$ \\
$g_{\mu\mu}^{R}$ & $-1.195 \times 10^{-2} < g_{\mu\mu}^{R} < -3.756 \times 10^{-4}$ \\
$g_{\mu\mu}^{L} + g_{\mu\mu}^{R}$ & $-1.175 \times 10^{-2} < g_{\mu\mu}^{L} + g_{\mu\mu}^{R} < 1.179 \times 10^{-3}$ \\
\bottomrule
\bottomrule
\end{tabular}}%
\caption{Scanned range of the $Z^{\prime}$ coupling constant to left-handed (or right-handed) muon pair when the $Z^{\prime}$ coupling constant $g_{X}$ is set to be order of unity, required to derive mass range of $Z^{\prime}$ gauge boson from the neutrino trident constraint ($g_{\mu\mu}^{L} + g_{\mu\mu}^{R}$) as well as the $R_{K^{*}}$ anomaly ($g_{\mu\mu}^{L}$ or $g_{\mu\mu}^{R}$).}
\label{tab:gmumuLR}
\end{table}
where it should be clarified that the upper-limit of the left-handed muon pair coupling constant $g_{L}^{\mu\mu}$ is given by ruling out the benchmark points which exceed the CMS and the $\tau \rightarrow \mu \gamma$ experimental constraints. Without taking the CMS constraint into account, the left-handed coupling constant $g_{L}^{\mu\mu}$ can reach up to a very similar magnitude of the lower-limit of the right-handed coupling constant, $g_{L}^{\mu\mu} = 1.339 \times 10^{-2}$, and we can connect the physical $Z^{\prime}$ gauge boson and the CP-odd scalar arising in this BSM model via the scalar potential in this case as we will see in the electroweak numerical study. However, the upper-limit of the left-handed coupling constant must be suppressed from $1.339 \times 10^{-2}$ to $2.113 \times 10^{-3}$ by the CMS experimental constraint and what this implements is connecting the $Z^{\prime}$ mass to the CP-odd scalar becomes much more challenging as $Z^{\prime}$ mass range derived from the $R_{K^{*}}$ anomaly gets lower further.
\subsubsection{$Z^{\prime}$ mass range derived from the neutrino trident constraint}
As we have derived a numerical range of summing over the $Z^{\prime}$ coupling constants to muon pair in Table~\ref{tab:gmumuLR}, it can evaluate $Z^{\prime}$ mass range derived from the neutrino trident constraint. Reminding the neutrino trident constraint of Equation~\ref{eqn:neutrino_trident_const_final}, it reads:
\begingroup
\begin{equation}
-\frac{1}{\left( 390 \func{GeV} \right)^{2}} < \frac{-5.263 \times 10^{-4} \left( g_{\mu\mu}^{L} + g_{\mu\mu}^{R} \right)}{M_{Z^{\prime}}^{2}} < \frac{1}{\left( 370 \func{GeV} \right)^{2}}.
\end{equation}
\endgroup
Using the derived result of $g_{\mu\mu}^{L} + g_{\mu\mu}^{R}$, it is possible to determine $Z^{\prime}$ mass range as given in Table~\ref{tab:MZp_range}:
\begin{table}[H]
\centering
\resizebox{0.5\textwidth}{!}{
\centering\renewcommand{\arraystretch}{1.3} 
\begin{tabular}{*{2}{c}}
\toprule
\toprule
Coupling constant & $M_{Z^{\prime}}$ range \\ 
\midrule
$g_{\mu\mu}^{L} + g_{\mu\mu}^{R} = -1.175 \times 10^{-2}$ & $M_{Z^{\prime}} > 0.920\func{GeV}$ \\
$g_{\mu\mu}^{L} + g_{\mu\mu}^{R} = 1.179 \times 10^{-3}$ & $M_{Z^{\prime}} > 0.307\func{GeV}$ \\
\bottomrule
\bottomrule
\end{tabular}}%
\caption{$M_{Z^{\prime}}$ mass range derived from the neutrino trident production constraint with the $Z^{\prime}$ coupling constants to muon pair summed over.}
\label{tab:MZp_range}
\end{table}
Considering both results for the derived $Z^{\prime}$ mass range in Table~\ref{tab:MZp_range}, it concludes
\begingroup
\begin{equation}
M_{Z^{\prime}} > 0.920\func{GeV}.
\label{eqn:MZp_nutri}
\end{equation}
\endgroup
\section{QUARK SECTOR PHENOMENOLOGY} \label{sec:VI}
Now that we have defined the preferred order of the $Z^{\prime}$ coupling constant $g_{X}$, the next task is to constrain mass range of the $Z^{\prime}$ and it can be done by considering constraints arising in the quark sector, which are the $R_{K^{*}}$ anomaly, $B_{s}$ meson oscillation, collider, and CKM mixing matrix.
\subsection{$R_{K^{*}}$ anomaly}
The first observable we consider is the $R_{K^{*}}$ anomaly. The $R_{K^{*}}$ anomaly is also regarded as a potential new physics signal in the quark sector due to its sizeable SM deviation corresponding to $3.1\sigma$. The $R_{K^{*}}$ anomaly and the $B_{s} \rightarrow \overline{\mu} \mu$ decay, connected to the $R_{K^{*}}$ anomaly by the crossing symmetry, can be explained by the effective four fermion operator consisting of left- and right-handed muon pair and left-handed $b$ and $s$ quarks after integrating out the degree of freedom of $Z^{\prime}$ gauge boson as follows~\cite{Navarro:2021sfb}:
\begingroup
\begin{equation}
\mathcal{L}_{R_{K^{*}}}^{\func{eff}} = G_{bs\mu}^{L} \left( \overline{s}_{L} \gamma^{\mu} b_{L} \right) \left( \overline{\mu}_{L} \gamma_{\mu} \mu_{L} \right) + G_{bs\mu}^{R} \left( \overline{s}_{L} \gamma^{\mu} b_{L} \right) \left( \overline{\mu}_{R} \gamma_{\mu} \mu_{R} \right) + \func{H.c.},
\label{eqn:Lagrangian_RK}
\end{equation}
\endgroup
where the coupling constants $G_{bs\mu}^{L,R}$ are known as the Wilson coefficients and are defined by~\cite{Navarro:2021sfb}:
\begingroup
\begin{equation}
\begin{split}
G_{bs\mu}^{L} &= -\frac{g_{bs}^{L} g_{\mu\mu}^{L}}{M_{Z^{\prime}}^{2}}, \\
G_{bs\mu}^{R} &= -\frac{g_{bs}^{L} g_{\mu\mu}^{R}}{M_{Z^{\prime}}^{2}}.
\end{split}
\end{equation}
\endgroup
The Feynman diagrams contributing to the $R_{K^{*}}$ anomaly and the $B_{s} \rightarrow \overline{\mu} \mu$ decay read in Figure~\ref{fig:RKanomaly_Bsmumu}:
\begingroup
\begin{figure}[H]
\centering
\begin{subfigure}{0.49\textwidth}
	\scalebox{0.9}{
	\includegraphics[keepaspectratio,width=1.0\textwidth]{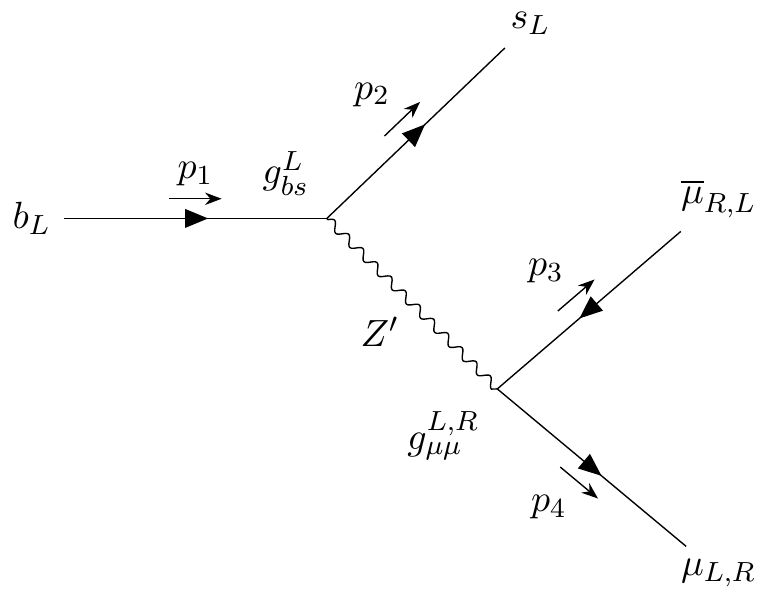} 
	}
\end{subfigure} 
\begin{subfigure}{0.49\textwidth}
	\scalebox{0.9}{
	\includegraphics[keepaspectratio,width=1.0\textwidth]{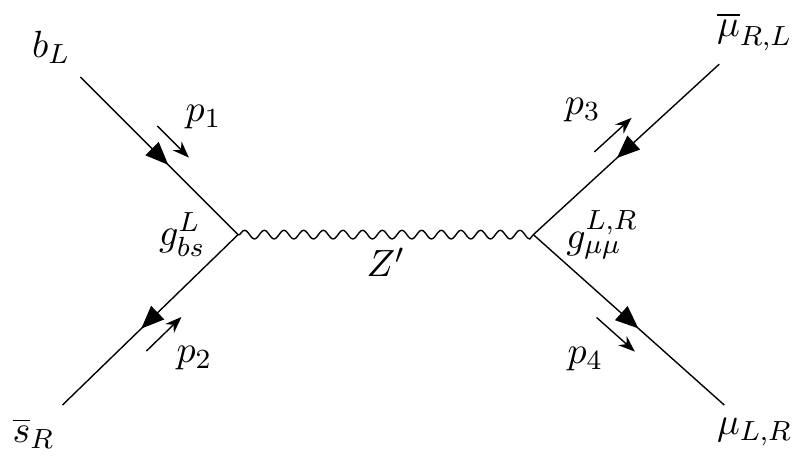}
	} 
\end{subfigure}
\caption{Feynman diagrams contributing to the $R_{K^{*}}$ anomaly (left) and the $B_{s} \rightarrow \overline{\mu}\mu$ decay (right).}
\label{fig:RKanomaly_Bsmumu}
\end{figure}
\endgroup
The effective interactions in Equation~\ref{eqn:Lagrangian_RK} can be rewritten in terms of the vector and axial-vector effective operators~\cite{Navarro:2021sfb,Geng:2021nhg}:
\begingroup
\begin{equation}
\begin{split}
\mathcal{H}_{\func{eff}} = N \left[ \delta C_{9} \left( \overline{s}_{L} \gamma^{\mu} b_{L} \right) \left( \overline{\mu} \gamma_{\mu} \mu \right) + \delta C_{10} \left( \overline{s}_{L} \gamma^{\mu} b_{L} \right) \left( \overline{\mu} \gamma_{\mu} \gamma^{5} \mu \right) \right] + \func{H.c.},
\end{split}
\end{equation}
\endgroup
where $N$ is the normalization factor~\cite{Navarro:2021sfb}:
\begingroup
\begin{equation}
N = -\frac{G_{F}}{\sqrt{2}} V_{tb} V_{ts}^{*} \frac{e^{2}}{16\pi^{2}}.
\end{equation}
\endgroup
The coupling constants $\delta C_{9,10}$ can be rearranged in terms of the Wilson coefficients $G_{bs\mu}^{L,R}$:
\begingroup
\begin{equation}
\begin{split}
G_{bs\mu}^{L} &= N \left( \delta C_{9} - \delta C_{10} \right)
\\
G_{bs\mu}^{R} &= N \left( \delta C_{9} + \delta C_{10} \right)
\end{split}
\end{equation}
\endgroup
As the BSM model under consideration can gain access to the coupling constants $g_{bs}^{L}$ and $g_{\mu\mu}^{L,R}$ in person, it requires a numerical range of the Wilson coefficients $G_{bs\mu}^{L,R}$ and they can be determined by using the numerical results of $\delta C_{9,10}$ which has been fitted to explain the $R_{K^{*}}$ anomaly at $1\sigma$. The coefficients $\delta C_{9,10}$ has been fitted by two cases: one of which is the ``theoretically clean fit", meaning that this fitted result can explain the $R_{K^{*}}$ anomaly and the $B_{s} \rightarrow \overline{\mu}\mu$ decay in a unified way without theoretical uncertainties, and the other is the ``global fit" which includes more substantial theoretical uncertainties when compared to the $B_{s} \rightarrow \overline{\mu}\mu$ data as well as the ratios of lepton universality violation and the two fits are displayed in Table~\ref{tab:theoreically_clean_fit} and \ref{tab:global_fit}:
\begingroup
\begin{table}[H]
\centering
\resizebox{\textwidth}{!}{
\centering\renewcommand{\arraystretch}{1.5} 
\begin{tabular*}{500pt}{@{\extracolsep{\fill}}*{3}{c}}
\toprule
\toprule
 & \textbf{Best fit} & \textbf{at $1\sigma$} \\[1ex]
\midrule
$( \delta C_{9}, \delta C_{10} )$ & $( -0.11, 0.59 )$ & $\left( -0.41 < \delta C_{9} < 0.17 \right), \left( 0.38 < \delta C_{10} < 0.81 \right)$ \\
$( G_{bs\mu}^{L}, G_{bs\mu}^{R} )$ & $\left( \frac{1}{\left( 42.50\func{TeV} \right)^{2}}, -\frac{1}{\left( 51.3\func{TeV} \right)^{2}} \right)$ & $\left( \frac{1}{\left( 44.44\func{TeV} \right)^{2}} < G_{bs\mu}^{L} < \frac{1}{\left( 40\func{TeV} \right)^{2}} \right), \left( -\frac{1}{\left( 35.9\func{TeV} \right)^{2}} < G_{bs\mu}^{R} < \frac{1}{\left( 205\func{TeV} \right)^{2}} \right)$ \\[1ex]
\bottomrule
\bottomrule
\end{tabular*}}
\caption{``Theoretically clean fit" for the $R_{K^{*}}$ anomaly and the $B_{s} \rightarrow \overline{\mu}\mu$ decay~\cite{Navarro:2021sfb,Geng:2021nhg}.} 
\label{tab:theoreically_clean_fit}
\end{table}
\endgroup
\begingroup
\begin{table}[H]
\centering
\resizebox{\textwidth}{!}{
\centering\renewcommand{\arraystretch}{1.5} 
\begin{tabular*}{500pt}{@{\extracolsep{\fill}}*{3}{c}}
\toprule
\toprule
 & \textbf{Best fit} & \textbf{at $1\sigma$} \\[1ex]
\midrule
$( \delta C_{9}, \delta C_{10} )$ & $( -0.56, 0.30 )$ & $\left( -0.79 < \delta C_{9} < -0.31 \right), \left( 0.15 < \delta C_{10} < 0.49 \right)$ \\
$( G_{bs\mu}^{L}, G_{bs\mu}^{R} )$ & $\left( \frac{1}{\left( 38.34\func{TeV} \right)^{2}}, -\frac{1}{\left( 69.73\func{TeV} \right)^{2}} \right)$ & $\left( \frac{1}{\left( 39.75\func{TeV} \right)^{2}} < G_{bs\mu}^{L} < \frac{1}{\left( 36.67\func{TeV} \right)^{2}} \right), \left( -\frac{1}{\left( 83.8\func{TeV} \right)^{2}} < G_{bs\mu}^{R} < \frac{1}{\left( 44.44\func{TeV} \right)^{2}} \right)$ \\[1ex]
\bottomrule
\bottomrule
\end{tabular*}}
\caption{``Global fit" for the $R_{K^{*}}$ anomaly and the $B_{s} \rightarrow \overline{\mu}\mu$ decay~\cite{Navarro:2021sfb,Geng:2021nhg}.} 
\label{tab:global_fit}
\end{table}
\endgroup
What we can confirm via the two fits is the different behavior between $G_{bs\mu}^{L}$ and $G_{bs\mu}^{R}$: the left-handed Wilson coefficients $G_{bs\mu}^{L}$ at $1\sigma$ in both fits report the order of $\left( 40\func{TeV} \right)^{-2}$ on average even though the ``theoretically clean fit" is slightly tighter than the ``global fit", whereas the right-handed Wilson coefficient $G_{bs\mu}^{R}$ in the ``theoretically clean fit" prefers negative contributions more while that in the ``global fit" prefers positive contributions further. The inconsistent behavior of the right-handed Wilson coefficient $G_{bs\mu}^{R}$ in both fits might be hint at that the $R_{K^{*}}$ anomaly is explained by the only left-handed Wilson coefficient $G_{bs\mu}^{L}$ and interactions consisting of the only left-handed fermionic fields~\cite{King:2017anf,King:2018fcg,Falkowski:2018dsl}. We confirm this feature by applying the scanned range of $Z^{\prime}$ coupling constant to left- or right-handed muon pair as well as left-handed $b$ and $s$ quarks to the Wilson coefficients $G_{bs\mu}^{L,R}$ and then determine a possible mass range of $Z^{\prime}$. Considering the constraints coming from the ``theoretically clean fit", they read
\begingroup
\begin{equation}
\begin{split}
\frac{1}{\left( 44.44\func{TeV} \right)^{2}} &< G_{bs\mu}^{L} = -\frac{g_{bs}^{L} g_{\mu\mu}^{L}}{M_{Z^{\prime}}^{2}} < \frac{1}{\left( 40\func{TeV} \right)^{2}}, \\
-\frac{1}{\left( 35.9\func{TeV} \right)^{2}} &< G_{bs\mu}^{R} = -\frac{g_{bs}^{L} g_{\mu\mu}^{R}}{M_{Z^{\prime}}^{2}} < \frac{1}{\left( 205\func{TeV} \right)^{2}}, \\
\label{eqn:RK_const_theoretical_clean_fit}
\end{split}
\end{equation}
\endgroup
and the constraints arising from the ``global fit" are given by:
\begingroup
\begin{equation}
\begin{split}
\frac{1}{\left( 39.75\func{TeV} \right)^{2}} &< G_{bs\mu}^{L} = -\frac{g_{bs}^{L} g_{\mu\mu}^{L}}{M_{Z^{\prime}}^{2}} < \frac{1}{\left( 36.67\func{TeV} \right)^{2}}, \\
-\frac{1}{\left( 83.8\func{TeV} \right)^{2}} &< G_{bs\mu}^{R} = -\frac{g_{bs}^{L} g_{\mu\mu}^{R}}{M_{Z^{\prime}}^{2}} < \frac{1}{\left( 44.44\func{TeV} \right)^{2}}. \\
\label{eqn:RK_const_global_fit}
\end{split}
\end{equation}
\endgroup
\subsection{$B_{s}-\overline{B}_{s}$ meson mixing oscillation}
Next we consider the $B_{s}-\overline{B}_{s}$ meson mixing oscillation and this observable gives rise to the most strict constraint in the quark sector. As in the $R_{K^{*}}$ anomaly, the effective four fermion interactions after integrating out the $Z^{\prime}$ gauge boson take the form~\cite{Navarro:2021sfb}:
\begingroup
\begin{equation}
\begin{split}
\mathcal{L}_{B_{s}-\overline{B}_{s}}^{\func{eff}} = -\frac{G_{bs}}{2} \left( \overline{s} \gamma_{\mu} b \right)^{2} + \func{H.c.},
\label{eqn:Lagrangian_BsBbars}
\end{split}
\end{equation}
\endgroup
where the coupling constant $G_{bs}$ reads:
\begingroup
\begin{equation}
G_{bs} = \frac{g_{bs}^{L2} + g_{bs}^{L} g_{bs}^{R*} + g_{bs}^{R2}}{M_{Z^{\prime}}^{2}},
\end{equation}
\endgroup
and it requires to clarify that we count all polarized bases, meaning that not just left-handed $b$ and $s$ quarks but also right-handed $b$ and $s$ quarks are considered together unlike~\cite{Navarro:2021sfb}. The effective $B_{s}-\overline{B}_{s}$ meson mixing operators in Equation~\ref{eqn:Lagrangian_BsBbars} yield the following diagrams at tree level in Figure~\ref{fig:BsBbars_Zp}:
\begingroup
\begin{figure}[H]
\centering
\begin{subfigure}{0.49\textwidth}
	\scalebox{0.9}{
	\includegraphics[keepaspectratio,width=1.0\textwidth]{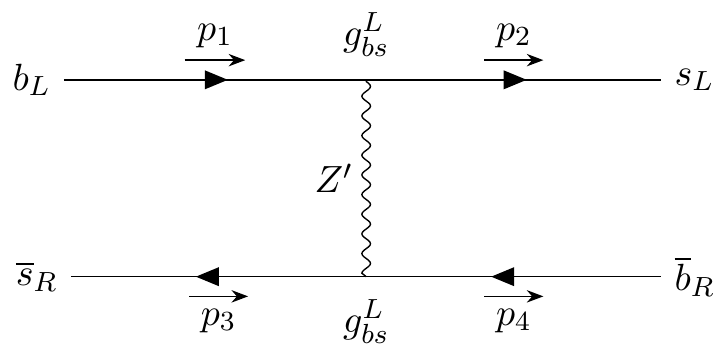} 
	}
\end{subfigure} 
\begin{subfigure}{0.49\textwidth}
	\scalebox{0.9}{
	\includegraphics[keepaspectratio,width=1.0\textwidth]{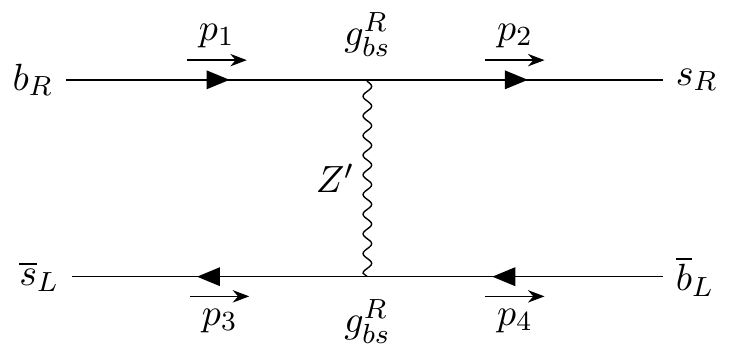}
	} 
\end{subfigure}
\begin{subfigure}{0.49\textwidth}
	\scalebox{0.9}{
	\includegraphics[keepaspectratio,width=1.0\textwidth]{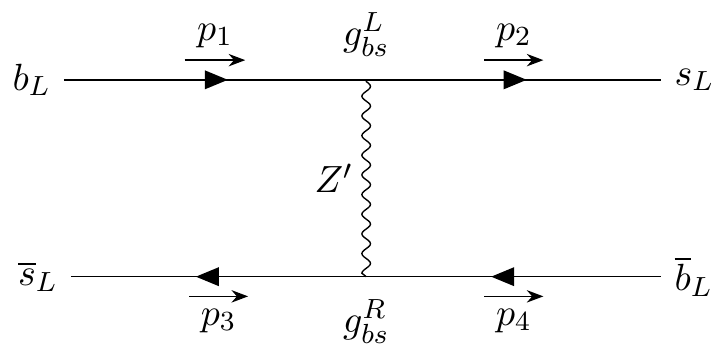} 
	}
\end{subfigure} 
\begin{subfigure}{0.49\textwidth}
	\scalebox{0.9}{
	\includegraphics[keepaspectratio,width=1.0\textwidth]{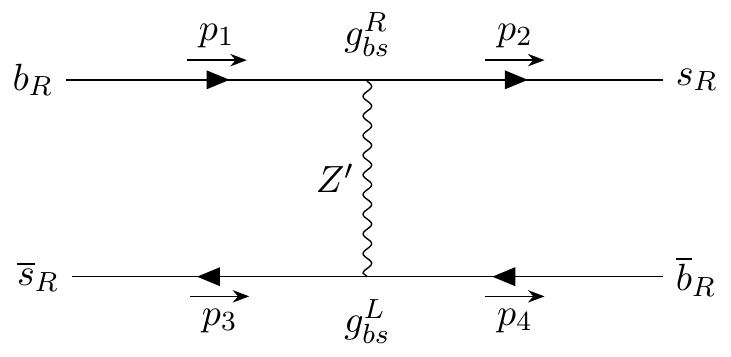}
	} 
\end{subfigure}
\caption{Feynman diagrams contributing to the $B_{s} - \overline{B}_{s}$ meson mixing oscillation in the polarized basis.}
\label{fig:BsBbars_Zp}
\end{figure}
\endgroup
The experimental mass difference for the $B_{s} - \overline{B}_{s}$ meson mixing oscillation is given by two theoretical constraints, which are achieved by either lattice simulations~\cite{DiLuzio:2019jyq} or lattice simulations plus sum rule results~\cite{DiLuzio:2019jyq}
\begingroup
\begin{equation}
\begin{split}
\Delta M_{s}^{\func{FLAG}^{\prime}19} &= \left( 1.13_{-0.07}^{+0.07} \right) \Delta M_{s}^{\func{exp}},
\\
\Delta M_{s}^{\func{Average}^{\prime}19} &= \left( 1.04_{-0.09}^{+0.04} \right) \Delta M_{s}^{\func{exp}},
\label{eqn:theoretical_const_BsBbars}
\end{split}
\end{equation}
\endgroup
The above one of Equation~\ref{eqn:theoretical_const_BsBbars} is analyzed by the lattice simulations and gives the strong bound~\cite{Navarro:2021sfb}
\begingroup
\begin{equation}
G_{bs} < \frac{1}{\left( 330\func{TeV} \right)^{2}},
\label{eqn:Bs_const_theor_1}
\end{equation}
\endgroup
and the below one is analyzed by both the lattice simulations and sum rule results and gives the less constraining bound~\cite{Navarro:2021sfb}
\begingroup
\begin{equation}
G_{bs} < \frac{1}{\left( 220\func{TeV} \right)^{2}},
\label{eqn:Bs_const_theor_2}
\end{equation}
\endgroup
where it is worth mentioning that the less constraining theoretical constraint is more consistent with the experimental bound. Instead of taking the theoretical constraints into account, it is also possible to evaluate the experimental bound in person. The experimental and SM bound for the $B_{s}$ meson oscillation reads~\cite{Dowdall:2019bea}:
\begingroup
\begin{equation}
\begin{split}
\Delta M_{s}^{\func{exp}} &= 17.757(21) \func{ps}^{-1} \\
\Delta M_{s}^{\func{SM}} &= 17.59 
\begin{pmatrix}
+0.33 \\
-1.22
\end{pmatrix} \left( 0.78 \right) \func{ps}^{-1}.
\end{split}
\end{equation}
\endgroup
Then, the new physics effect for the $B_{s}$ meson oscillation can be evaluated by subtracting the SM bound from the experimental bound as follows:
\begingroup
\begin{equation}
\begin{gathered}
\Delta M_{s}^{\func{NP}} = \Delta M_{s}^{\func{exp}} - \Delta M_{s}^{\func{SM}} = 0.167
\begin{pmatrix}
+1.22 \\
-0.33
\end{pmatrix} \left( -0.78 \right) \func{ps}^{-1}, \\
-1.07237 \times 10^{-13} \func{GeV} < \Delta M_{s}^{\func{NP}} < 9.125 \times 10^{-13} \func{GeV}
\label{eqn:Bs_const_exp}
\end{gathered}
\end{equation}
\endgroup
We will constrain $Z^{\prime}$ mass range from all the three constraints once scanned numerical ranges of the coupling constants $g_{bs}^{L,R}$ are obtained in the quark sector simulation. The final task in this subsection is to rewrite the new physics contribution in terms of the physical input parameters~\cite{Dowdall:2019bea,Branco:2021vhs,CarcamoHernandez:2020pnh}:
\begingroup
\begin{equation}
f_{B_{s}} = 0.225 \func{GeV}, \quad B_{B_{s}} = 1.26, \quad \eta_{B_{s}} = 0.55, \quad m_{B_{s}} = 5.3663 \func{GeV}.
\end{equation}
\endgroup
Then, the new physics contribution to the $B_{s}$ meson oscillation reads in terms of the physical input parameters:
\begingroup
\begin{equation}
\begin{split}
\Delta M_{s}^{\func{NP}} &= G_{bs} f_{B_{s}}^{2} B_{B_{s}} \eta_{B_{s}} m_{B_{s}} 
\\
&= \frac{g_{bs}^{L2} + g_{bs}^{L} g_{bs}^{R*} + g_{bs}^{R2}}{M_{Z^{\prime}}^{2}} \times \left( 0.225 \func{GeV} \right)^{2} \times 1.26 \times 0.55 \times \left( 5.3663 \func{GeV} \right)
\\
&\simeq \frac{g_{bs}^{L2} + g_{bs}^{L} g_{bs}^{R*} + g_{bs}^{R2}}{M_{Z^{\prime}}^{2}} \times 0.188 \func{GeV}^{3}
\end{split}
\end{equation}
\endgroup
\subsection{Collider experimental constraints}
As the model under consideration features the massive second and third generation of the SM fermions, the electron collider constraints arising from the LEP experiments can not be applied to this BSM model. The most relevant collider constrains arise from LHC experiments. Regarding light $Z^{\prime}$ masses, it can be constrained by the decay process $pp \rightarrow Z \rightarrow 4\mu$ and give rise to the constraint:
\begingroup
\begin{equation}
5\func{GeV} < M_{Z^{\prime}} < 70\func{GeV},
\end{equation}
\endgroup
where it is worth mentioning that we already reflected this condition when we carried out the numerical scan for the charged lepton sector, thus providing the condition on $Z^{\prime}$ mass which is $M_{Z^{\prime}} > 75\func{GeV}$. The $Z^{\prime}$ mass can be further constrained by LHC dimuon resonance measurements~\cite{Abdullah:2017oqj,Alonso:2017uky}, which are $pp \rightarrow Z^{\prime} \rightarrow \overline{\mu}\mu$. Following the argument given in~\cite{Navarro:2021sfb}, the main portal for generating $Z^{\prime}$ is $\overline{b}b$ channel and this feature can be applied to our BSM model. Using the narrow width approximation, the cross section for the decay process $pp \rightarrow Z^{\prime} \rightarrow \overline{\mu}\mu$ is given for any $g_{bb}$~\cite{Navarro:2021sfb}:
\begingroup
\begin{equation}
\begin{split}
\sigma \left( pp \rightarrow Z^{\prime} \rightarrow \overline{\mu}\mu \right) &\approx \sigma \left( pp \rightarrow Z^{\prime} \right) \func{BR} \left( Z^{\prime} \rightarrow \overline{\mu}\mu \right),
\\
\func{BR} \left( Z^{\prime} \rightarrow \overline{\mu}\mu \right) &= \frac{\Gamma_{Z^{\prime} \rightarrow \overline{\mu}\mu}}{\Gamma_{Z^{\prime} \rightarrow \overline{\mu}\mu}+\Gamma_{Z^{\prime} \rightarrow \overline{\nu}_{\mu}\nu_{\mu}}+\Gamma_{Z^{\prime} \rightarrow \overline{b}b}+\Gamma_{Z^{\prime} \rightarrow \overline{s}b}+\Gamma_{Z^{\prime} \rightarrow \overline{t}t}},
\end{split}
\end{equation}
\endgroup
where~\cite{Navarro:2021sfb}
\begingroup
\begin{equation}
\begin{split}
\Gamma \left( Z^{\prime} \rightarrow \overline{\mu}\mu \right) &= \frac{M_{Z^{\prime}}}{24\pi} \left[ (g_{\mu\mu}^{L})^{2} + (g_{\mu\mu}^{R})^{2} \right],
\\
\Gamma \left( Z^{\prime} \rightarrow \overline{\nu}_{\mu}\nu_{\mu} \right) &= \frac{M_{Z^{\prime}}}{24\pi} \left[ (g_{\nu_{\mu}\nu_{\mu}}^{L})^{2} \right],
\\
\Gamma \left( Z^{\prime} \rightarrow \overline{b}b \right) &= \frac{M_{Z^{\prime}}}{8\pi} \left[ (g_{bb}^{L})^{2} + (g_{bb}^{R})^{2} \right],
\\
\Gamma \left( Z^{\prime} \rightarrow \overline{s}b \right) &= \frac{M_{Z^{\prime}}}{8\pi} \left[ (g_{sb}^{L})^{2} + (g_{sb}^{R})^{2} \right],
\\
\Gamma \left( Z^{\prime} \rightarrow \overline{t}t \right) &= \frac{M_{Z^{\prime}}}{8\pi} \left[ (g_{bb}^{L})^{2} + (g_{bb}^{R})^{2} \right] \left( 1 - \frac{m_{t}^{2}}{M_{Z^{\prime}}^{2}} \right) \sqrt{1 - \frac{4m_{t}^{2}}{M_{Z^{\prime}}^{2}}},
\end{split}
\end{equation}
\endgroup
where the $Z^{\prime}$ coupling constant $g_{bs}$ can be approximated to $g_{bb} V_{ts}$ and the CKM mixing element $V_{tb} \approx 0.04$ play a role of suppressing the $Z^{\prime}\overline{s}b$ interactions, therefore the main decay channel is $Z^{\prime}\overline{b}b$. Given that the dimuon resonance data from ATLAS~\cite{ATLAS:2017fih} gives the experimental bound for $Z^{\prime}$ mass, the whole experimental bound reads~\cite{Navarro:2021sfb}:
\begingroup
\begin{equation}
150 \func{GeV} < M_{Z^{\prime}} < 5\func{TeV}.
\end{equation}
\endgroup
\subsection{CKM mixing matrix}
The CKM mixing matrix plays a quite crucial role in constraining the quark sector in the model under consideration. As the investigation for the CKM mixing matrix was done in one of our works~\cite{CarcamoHernandez:2021yev} in detail, we follow the same logic covered there. The renormalizable Lagrangian for the CKM mixing matrix with the extended fermion spectrum by the vectorlike family reads~\cite{CarcamoHernandez:2021yev}:
\begin{equation}
\begin{split}
\mathcal{L}_{\func{BSM}}^{W} &= 
\frac{g}{\sqrt{2}}
\begin{pmatrix}
\overline{u}_{1L} & \overline{u}_{2L} & \overline{u}_{3L} & \overline{u}_{4L} & \overline{\widetilde{u}}_{4L}
\end{pmatrix}
\gamma_{\mu}
\begin{pmatrix}
1 & 0 & 0 & 0 & 0 \\
0 & 1 & 0 & 0 & 0 \\
0 & 0 & 1 & 0 & 0 \\
0 & 0 & 0 & 1 & 0 \\
0 & 0 & 0 & 0 & 0 
\end{pmatrix}
\begin{pmatrix}
d_{1L} \\
d_{2L} \\
d_{3L} \\
d_{4L} \\
\widetilde{d}_{4L}
\end{pmatrix}
W^{\mu +},
\\
&= 
\frac{g}{\sqrt{2}}
\begin{pmatrix}
\overline{u}_{1L} & \overline{u}_{2L} & \overline{u}_{3L} & \overline{u}_{4L} & \overline{\widetilde{u}}_{4L}
\end{pmatrix}
V_{L}^{u \dagger} V_{L}^{u}
\gamma_{\mu}
\begin{pmatrix}
1 & 0 & 0 & 0 & 0 \\
0 & 1 & 0 & 0 & 0 \\
0 & 0 & 1 & 0 & 0 \\
0 & 0 & 0 & 1 & 0 \\
0 & 0 & 0 & 0 & 0 
\end{pmatrix}
V_{L}^{d \dagger} V_{L}^{d}
\begin{pmatrix}
d_{1L} \\
d_{2L} \\
d_{3L} \\
d_{4L} \\
\widetilde{d}_{4L}
\end{pmatrix}
W^{\mu +},
\\
&= 
\frac{g}{\sqrt{2}}
\begin{pmatrix}
\overline{u}_{L} & \overline{c}_{L} & \overline{t}_{L} & \overline{U}_{4L} & \overline{\widetilde{U}}_{4L}
\end{pmatrix}
\gamma_{\mu}
V_{L}^{u}
\begin{pmatrix}
1 & 0 & 0 & 0 & 0 \\
0 & 1 & 0 & 0 & 0 \\
0 & 0 & 1 & 0 & 0 \\
0 & 0 & 0 & 1 & 0 \\
0 & 0 & 0 & 0 & 0 
\end{pmatrix}
V_{L}^{d \dagger}
\begin{pmatrix}
d_{L} \\
s_{L} \\
b_{L} \\
D_{4L} \\
\widetilde{D}_{4L}
\end{pmatrix}
W^{\mu +},
\end{split}
\label{eqn:CKM_analytic}
\end{equation}
where the zero in the matrix between the mixing matrix $V_{L}^{u}$ and $V_{L}^{d}$ arises from the fact that the vectorlike singlet quarks $\widetilde{U}_{4}$ and $\widetilde{D}_{4}$ do not interact with the SM $W$ gauge boson, therefore featuring non-unitarity for the CKM mixing matrix in this BSM model. The non-unitarity $\Delta$ reads~\cite{Branco:2021vhs,Belfatto:2021jhf} as follows:
\begingroup
\begin{equation}
\Delta = 1 - V_{ud}^{2} - V_{us}^{2} - V_{ub}^{2}
\end{equation}
\endgroup
What we found via the numerical scan in the previous work~\cite{CarcamoHernandez:2021yev} is the non-unitarity $\Delta$ is quite small ($\Delta = 0.00035$~\cite{CarcamoHernandez:2021yev}) compared to its experimental upper-limit ($\Delta_{\func{exp}} = 0.04$~\cite{Belfatto:2019swo}). The most fitted CKM prediction $(\chi_{\func{CKM}}^{2} = 956.828)$ corresponding to the non-unitarity $0.00035$ has the form of~\cite{CarcamoHernandez:2021yev}:
\begingroup
\begin{equation}
V_{\func{CKM}}^{\func{pred}} 
=
\begin{pmatrix}
0.97409 & 0.22602 & 0.00799 & -6.38471 \times 10^{-6} & -0.00036 \\[0.5ex]
0.22615 & -0.97372 & -0.02697 & 3.80102 \times 10^{-5} & 0.00147 \\[0.5ex]
0.00166 & 0.02815 & -0.99880 & -0.00766 & 0.00874 \\[0.5ex]
0.00003 & 0.00019 & -0.00916 & 0.99919 & -0.01773 \\[0.5ex]
-0.00057 & 0.00112 & 0.03812 & 0.03539 & -0.00096 
\end{pmatrix}.
\label{eqn:CKM_bestfit}
\end{equation} 
\endgroup
Restricting our attention up to the upper-left $3 \times 3$ block of Equation~\ref{eqn:CKM_bestfit}, the partial block can be compared to the experimental CKM mixing matrix featuring non-unitarity~\cite{Branco:2021vhs,ParticleDataGroup:2020ssz}:
\begingroup
\setlength\arraycolsep{5pt}
\begin{equation}
\lvert K_{\func{CKM}} \rvert
=
\begin{pmatrix}
0.97370 \pm 0.00014 & 0.22450 \pm 0.00080 & 0.00382 \pm 0.00024 \\[1.5ex]
0.22100 \pm 0.00400 & 0.98700 \pm 0.01100 & 0.04100 \pm 0.00140 \\[1.5ex]
0.00800 \pm 0.00030 & 0.03880 \pm 0.00110 & 1.01300 \pm 0.03000 
\end{pmatrix},
\label{eqn:CKM_exp_wx_uni}
\end{equation}
\endgroup
and the $31,32,13,23$ element of Equation~\ref{eqn:CKM_bestfit} can not be fitted to their experimental bounds at $3\sigma$ by a small difference. However, considering this CKM prediction is an effective approach to the CKM mixing matrix since we only have 23 mixing angle in the up-quark sector, the prediction is a good approximation. For the purpose of making the numerical analysis for the quark sector which will be carried out soon economical, we reuse the conditions imposed in the previous work~\cite{CarcamoHernandez:2021yev} that we collect benchmark points which satisfy $\chi_{\func{CKM}}^{2} < 980$ and the predicted $m_{c,s,b}$ and $m_{t}$ should be put between $\left[ 1 \pm 0.1 \right] \times m_{c,s,b}^{\func{exp}}$ and $\left[ 1 \pm 0.01 \right] \times m_{t}^{\func{exp}}$ respectively.
\subsection{Numerical analysis in the quark sector}
A main goal of this numerical scan for the quark sector is to constrain the $Z^{\prime}$ coupling constant to $b$ and $s$ quarks, in order to determine a numerical $Z^{\prime}$ mass range from the $R_{K^{*}}$ anomaly (Equation~\ref{eqn:RK_const_theoretical_clean_fit} and \ref{eqn:RK_const_global_fit}) and from the $B_{s}$ meson oscillation experimental constraints (Equation~\ref{eqn:Bs_const_theor_1}, \ref{eqn:Bs_const_theor_2} and \ref{eqn:Bs_const_exp}), and then we try to find an overlapped region among the numerical $Z^{\prime}$ mass ranges. For the task, we reuse the best fitted benchmark point ($\chi_{\func{CKM}}^{2} = 956.828$) given in our previous work~\cite{CarcamoHernandez:2021yev} instead of setting numerical scan up from scratch since this BSM model can also gain access to the CKM mixing matrix in the same way as done in~\cite{CarcamoHernandez:2021yev}
\begingroup
\begin{equation}
\begin{gathered}
M^{u} = 
\begin{pmatrix}
0 & 0 & 0 & 0 & 0 \\[0.5ex]
0 & 0 & 0 & 0 & 14.474 \\[0.5ex]
0 & 0 & 0 & 1206.340 & 277.563 \\[0.5ex]
0 & 0 & 273.503 & -1775.200 & 0 \\[0.5ex]
0 & 550.990 & 434.462 & 0 & -5624.050 
\end{pmatrix}
\quad
M^{d} = 
\begin{pmatrix}
0 & 0 & 0 & 0 & -0.938 \\[0.5ex]
0 & 0 & 0 & 0 & -4.041 \\[0.5ex]
0 & 0 & 0 & 1206.340 & -27.427 \\[0.5ex]
0 & 0 & -5.636 & -1775.200 & 0 \\[0.5ex]
0 & 72.915 & -75.760 & 0 & 2623.620
\end{pmatrix}
\\[2ex]
M_{\func{diag}}^{u} = 
\begin{pmatrix}
0 & 0 & 0 & 0 & 0 \\[0.5ex]
0 & 1.255 & 0 & 0 & 0 \\[0.5ex]
0 & 0 & 171.303 & 0 & 0 \\[0.5ex]
0 & 0 & 0 & 2155.890 & 0 \\[0.5ex]
0 & 0 & 0 & 0 & 5674.840 
\end{pmatrix}
\quad
M_{\func{diag}}^{d} = 
\begin{pmatrix}
0 & 0 & 0 & 0 & 0 \\[0.5ex]
0 & 0.094 & 0 & 0 & 0 \\[0.5ex]
0 & 0 & 3.875 & 0 & 0 \\[0.5ex]
0 & 0 & 0 & 2146.190 & 0 \\[0.5ex]
0 & 0 & 0 & 0 & 2625.960 
\end{pmatrix},
\label{eqn:MuMd_diagMuMd}
\end{gathered}
\end{equation}
\endgroup
which gives rise to the best fitted CKM prediction of Equation~\ref{eqn:CKM_bestfit}. We vary mass parameters in the best fitted benchmark point of Equation~\ref{eqn:MuMd_diagMuMd} by a factor of $\left[ 1 \pm \kappa \right]$ where $\kappa = 0.1$ in order to find the mixing matrices $V_{L,R}^{u}$ and $V_{L,R}^{d}$, required to derive the $Z^{\prime}$ coupling constant to $b$ and $s$ quarks, and then collect the benchmark points which satisfy $\chi_{\func{CKM}}^{2} < 980$ as done in \cite{CarcamoHernandez:2021yev}. The varying process is shown in Table~\ref{tab:parameter_region_initial_q_second}:
\begingroup
\begin{table}[H]
\centering
\resizebox{0.45\textwidth}{!}{
\renewcommand{\arraystretch}{1.3} 
\begin{tabular}{cc}
\toprule
\toprule
\textbf{Mass parameter} & \textbf{Scanned Region($\func{GeV}$)} \\ 
\midrule
$y_{24}^{u} v_{u} = m_{24}^{u}$ & $\left[ 1 \pm \kappa \right] \times m_{24r}^{u}$ \\[1.0ex]
$y_{34}^{u} v_{u} = m_{34}^{u}$ & $\left[ 1 \pm \kappa \right] \times m_{34r}^{u}$ \\[1.0ex]
$y_{43}^{u} v_{u} = m_{43}^{u}$ & $\left[ 1 \pm \kappa \right] \times m_{43r}^{u}$ \\[1.0ex]
$x_{34}^{Q} v_{\phi} = m_{35}^{Q}$ & $m_{35}^{Q}$ \\[1.0ex]
$x_{42}^{u} v_{\phi} = m_{52}^{u}$ & $\left[ 1 \pm \kappa \right] \times m_{42r}^{u}$ \\[1.0ex]
$x_{43}^{u} v_{\phi} = m_{53}^{u}$ & $\left[ 1 \pm \kappa \right] \times m_{43r}^{u}$ \\[1.0ex]
$M_{45}^{Q}$ & $M_{45}^{Q}$ \\[1.0ex]
$M_{54}^{u}$ & $\left[ 1 \pm \kappa \right] \times M_{54r}^{u}$ \\[1.0ex]
\midrule
$y_{14}^{d} v_{d} = m_{14}^{d}$ & $\left[ 1 \pm \kappa \right] \times m_{14r}^{d}$ \\[1.0ex]
$y_{24}^{d} v_{d} = m_{24}^{d}$ & $\left[ 1 \pm \kappa \right] \times m_{24r}^{d}$ \\[1.0ex]
$y_{34}^{d} v_{d} = m_{34}^{d}$ & $\left[ 1 \pm \kappa \right] \times m_{34r}^{d}$ \\[1.0ex]
$y_{43}^{d} v_{d} = m_{43}^{d}$ & $\left[ 1 \pm \kappa \right] \times m_{43r}^{d}$ \\[1.0ex]
$x_{34}^{Q} v_{\phi} = m_{35}^{Q}$ & $\left[ 1 \pm \kappa \right] \times m_{35r}^{Q}$ \\[1.0ex]
$x_{42}^{d} v_{\phi} = m_{52}^{d}$ & $\left[ 1 \pm \kappa \right] \times m_{42r}^{d}$ \\[1.0ex]
$x_{43}^{d} v_{\phi} = m_{53}^{d}$ & $\left[ 1 \pm \kappa \right] \times m_{43r}^{d}$ \\[1.0ex]
$M_{45}^{Q}$ & $\left[ 1 \pm \kappa \right] \times M_{45r}^{Q}$ \\[1.0ex]
$M_{54}^{d}$ & $\left[ 1 \pm \kappa \right] \times M_{54r}^{d}$ \\[1.0ex]
\midrule
$\kappa$ & $0.1$ \\
\bottomrule
\bottomrule 
\end{tabular}}
\caption{Parameter setup, based on the best fitted benchmark point of Equation~\ref{eqn:MuMd_diagMuMd}, to find the mixing matrices $V_{L,R}^{u}$ and $V_{L,R}^{d}$ required to derive the $Z^{\prime}$ coupling constant to $b$ and $s$ quarks. The imposed conditions to fit the mass parameters to the SM are the predicted $c,s,b$ and $t$ quark masses are put between $\left[ 1 \pm 0.1 \right] \times m_{c,s,b}^{\func{exp}}$ and between $\left[ 1 \pm 0.01 \right] \times m_{t}^{\func{exp}}$ and the CKM mixing matrix~\cite{CarcamoHernandez:2021yev}.} 
\label{tab:parameter_region_initial_q_second}
\end{table}
\endgroup
\subsubsection{Numerical scan result and ranges of the $Z^{\prime}$ coupling constant to $b$ and $s$ quarks}
The scanned vectorlike quark masses are shown in Figure~\ref{fig:scanned_vectorlike_quarks}:
\begingroup
\begin{figure}[H]
\centering
\begin{subfigure}{0.49\textwidth}
	\scalebox{0.9}{
	\includegraphics[keepaspectratio,width=1.0\textwidth]{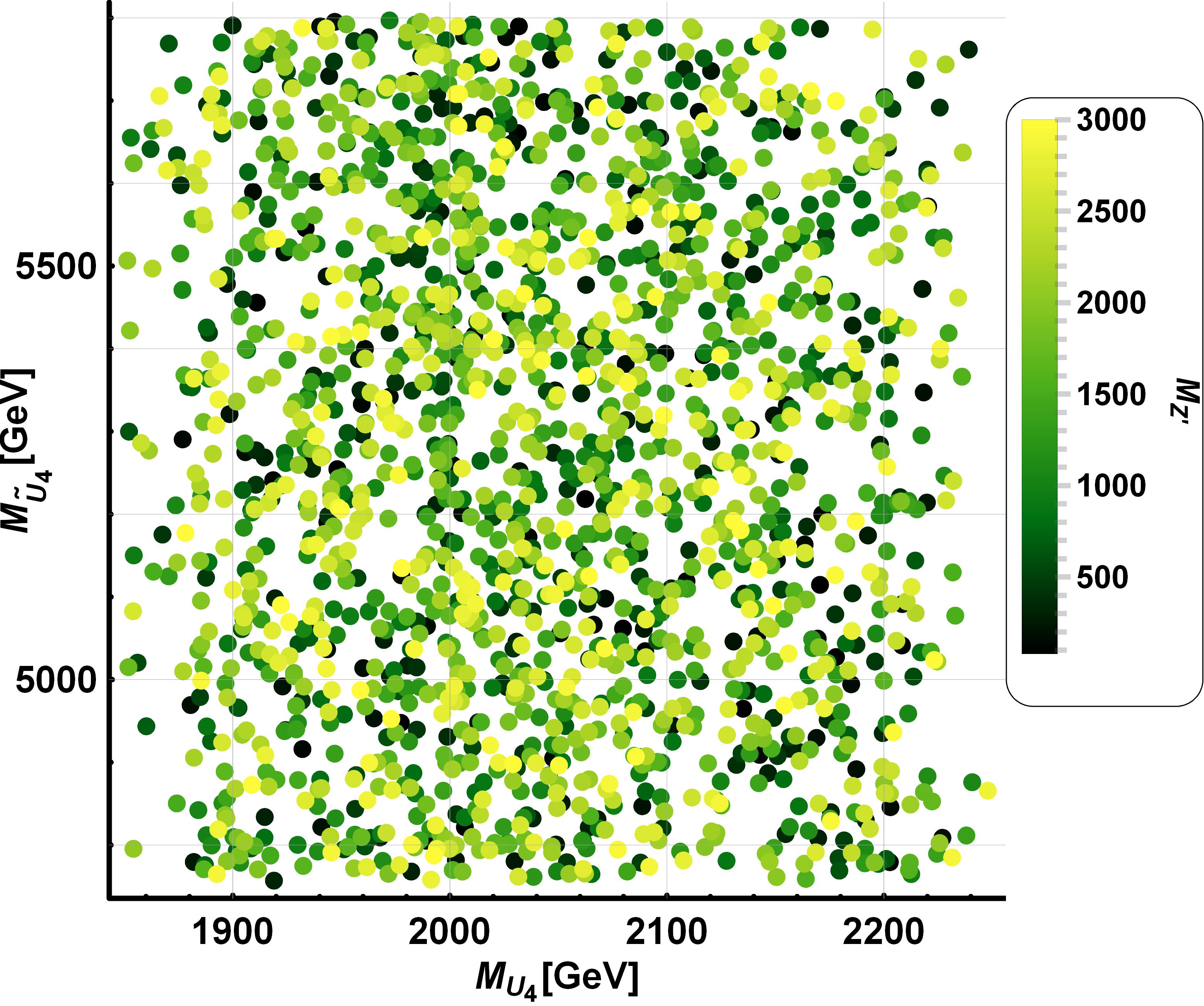}
	}
\end{subfigure} 
\begin{subfigure}{0.49\textwidth}
	\scalebox{0.9}{
	\includegraphics[keepaspectratio,width=1.0\textwidth]{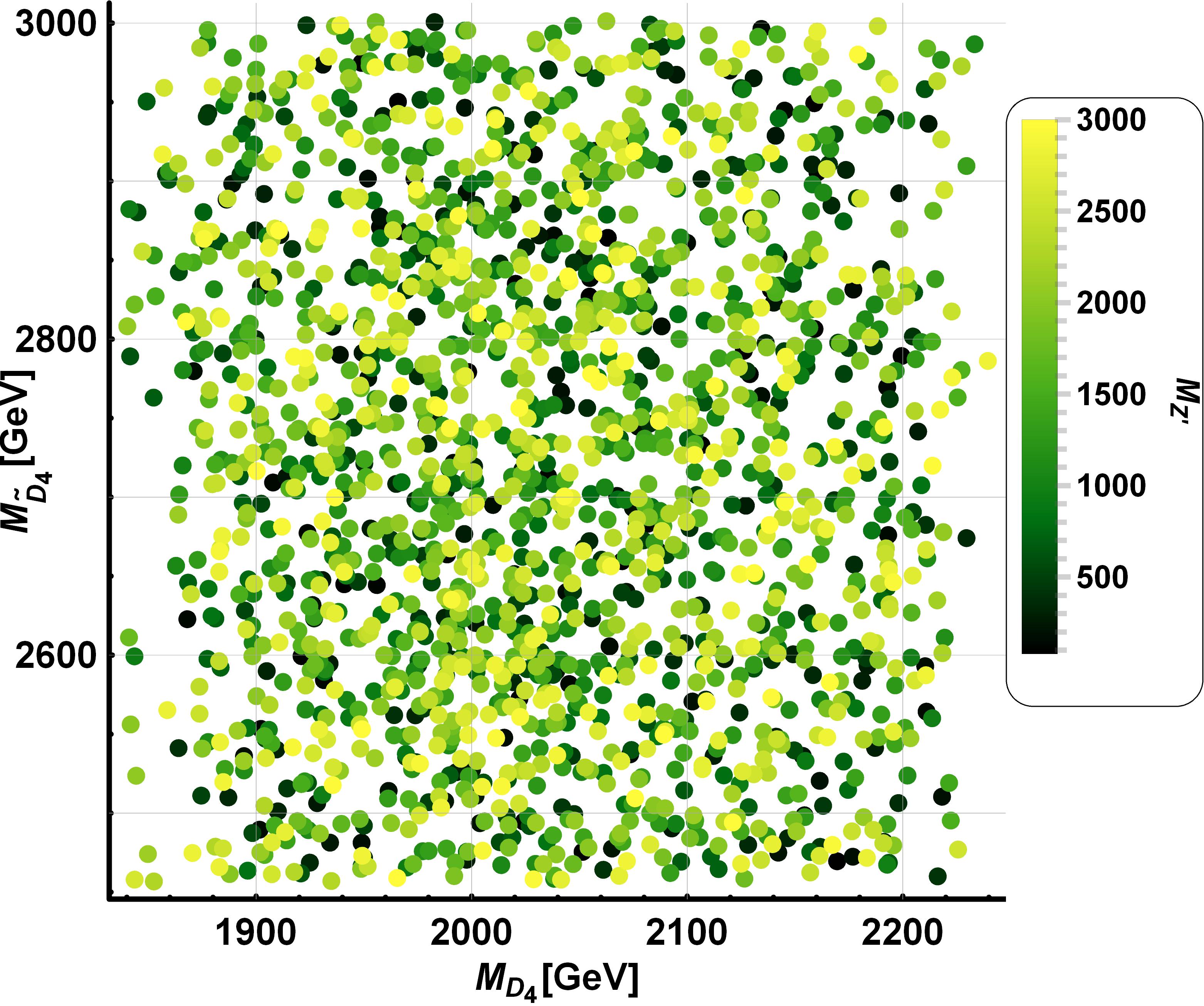}
	}
\end{subfigure} \par
\begin{subfigure}{0.49\textwidth}
	\scalebox{0.9}{
	\includegraphics[keepaspectratio,width=1.0\textwidth]{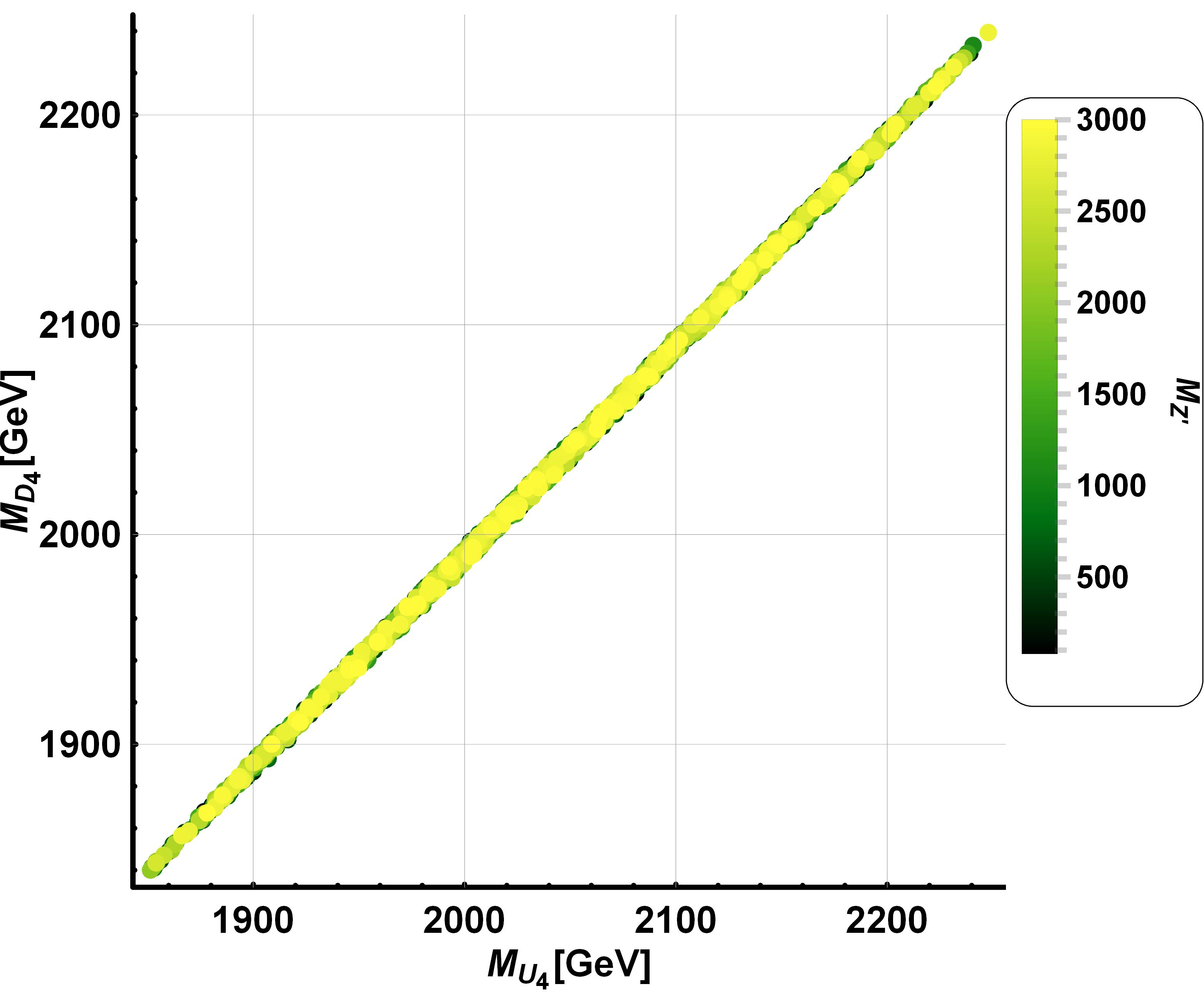}
	}
\end{subfigure} 
\begin{subfigure}{0.49\textwidth}
	\scalebox{0.9}{
	\includegraphics[keepaspectratio,width=1.0\textwidth]{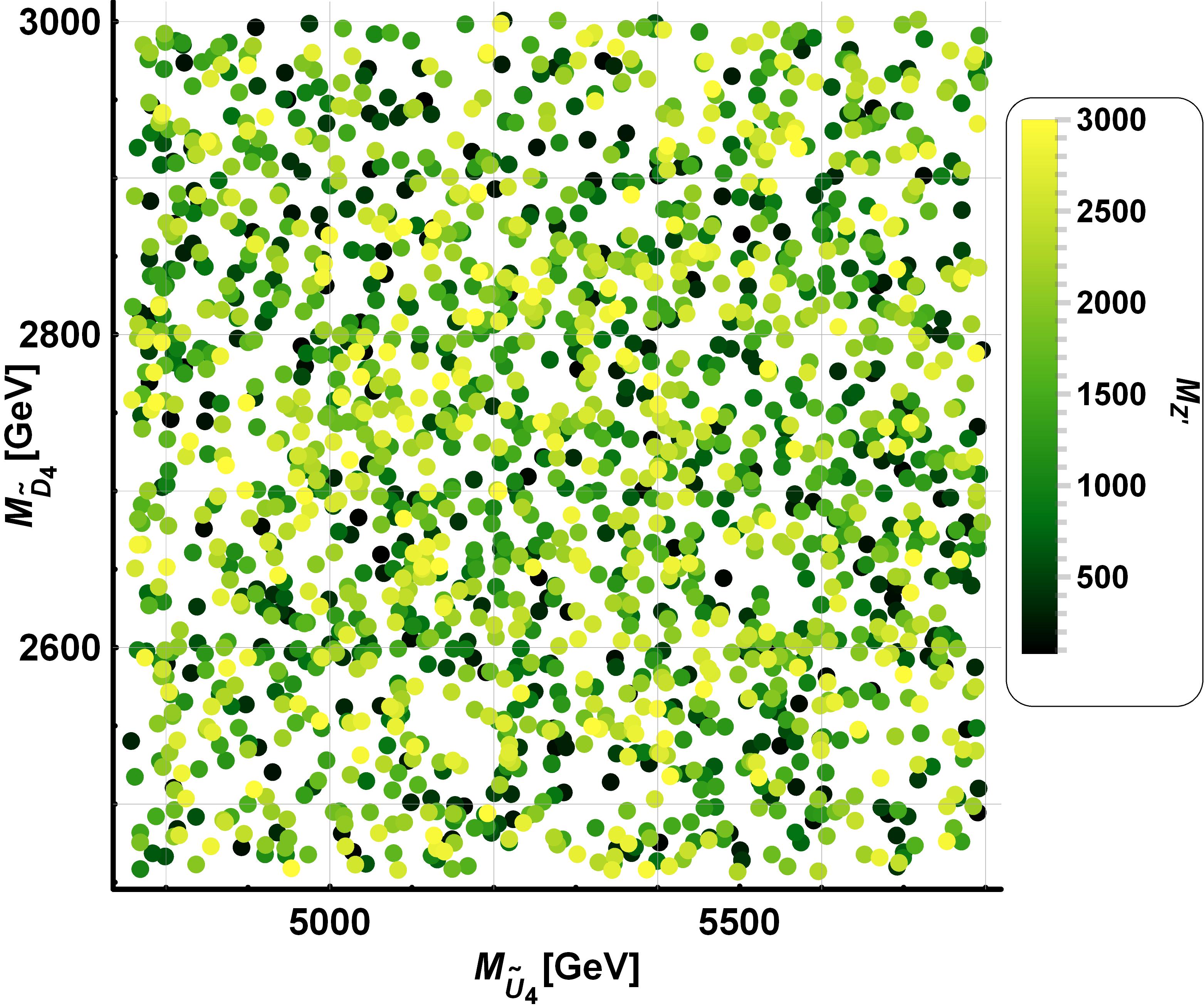}
	} 
\end{subfigure}
\caption{Scanned mass ranges of vectorlike quarks. $U_{4}$ and $D_{4}$ are vectorlike doublet up- and down-type quarks respectively and $\widetilde{U}_{4}$ and $\widetilde{D}_{4}$ are vectorlike singlet up- and down-type quarks respectively. The two above plots are exactly consistent with ones obtained in our previous work~\cite{CarcamoHernandez:2021yev}. The vectorlike doublet quarks $M_{U_{4}}$ and $M_{D_{4}}$ shows strong correlation between their masses, which can be directly seen from the fifth column of mass matrices of Equation~\ref{eqn:diff_up_down}, and this correlation plays an important role in suppressing fermion contributions to the oblique parameters $T,S,U$ as we will see soon. Here, the $Z^{\prime}$ mass is still free parameter.}
\label{fig:scanned_vectorlike_quarks}
\end{figure}
\endgroup
The upper-left plot in Figure~\ref{fig:scanned_vectorlike_quarks} shows mass ranges for the vectorlike doublet up-type quark $M_{U_{4}}$ against the vectorlike singlet up-type quark $M_{\widetilde{U}_{4}}$. The upper-right plot shows mass ranges for the vectorlike doublet down-type quark $M_{D_{4}}$ against the vectorlike singlet down-type quark $M_{\widetilde{D}_{4}}$. The reported vectorlike doublet and singlet quark masses in our quark sector simulation are free from the most recent experimental vectorlike mass constraints at $95\%$ C.L. given by the ATLAS experiments~\cite{ATLAS:2021ibc}:
\begingroup
\begin{equation}
\begin{split}
M_{U_{4}} &> 1.46 \func{TeV}, \\
M_{\widetilde{U}_{4}} &> 1.27 \func{TeV}, \\
M_{D_{4}} &> 1.32 \func{TeV}, \\
M_{\widetilde{D}_{4}} &> 1.20 \func{TeV}.
\end{split}
\end{equation}
\endgroup 
Considering 100$\%$ branching ratio for $U \rightarrow Zt$ and for $D \rightarrow Zb$, the upper-limits for the vectorlike up- and down-type quark masses increase up to $1.60\func{TeV}$ and $1.42\func{TeV}$ at $95\%$ C.L. respectively~\cite{ATLAS:2021ibc}:
\begingroup
\begin{equation}
\begin{split}
M_{U_{4}}, M_{\widetilde{U}_{4}} &> 1.60 \func{TeV}, \\
M_{D_{4}}, M_{\widetilde{D}_{4}} &> 1.42 \func{TeV}. \\
\end{split}
\end{equation}
\endgroup 
All plots in Figure~\ref{fig:scanned_vectorlike_quarks} show no correlation, displaying only the allowed range of the vectorlike quark masses, excepting for the lower-left plot. The lower-left plot shows that the vectorlike doublet up-type and down type quarks are degenerate and this property can be directly seen from the quark mass matrices of Equation~\ref{eqn:diff_up_down}. The equal masses for the vectorlike up-type and down type quarks doublets are quite important to fit the experimental oblique $S, T, U$ parameters. In other words, if their mass difference is large, it will yield maximal violation of the custodial symmetry thus implying unacceptably large contributions to these oblique parameters. Thus, a large up and down type vector like doublet quark mass splitting will not allow to successfully accommodate the experimental bounds for the oblique parameters. Therefore, the fact that the structure of the quark mass matrices can provide the degenerate masses for the vectorlike doublet quarks in an analytic way makes this BSM theory more convincible. Going back to scanning the $Z^{\prime}$ coupling constant to $b$ and $s$ quarks, they are given by (we also included the derived $Z^{\prime}$ coupling constant to muon pair together for convenience):
\begin{table}[H]
\centering
\resizebox{0.6\textwidth}{!}{
\centering\renewcommand{\arraystretch}{1.3} 
\begin{tabular}{*{2}{c}}
\toprule
\toprule
Coupling constant & Scanned range \\ 
\midrule
$g_{\mu\mu}^{L}$ & $4.888 \times 10^{-6} < g_{\mu\mu}^{L} < 2.113 \times 10^{-3}$ \\
$g_{\mu\mu}^{R}$ & $-1.195 \times 10^{-2} < g_{\mu\mu}^{R} < -3.756 \times 10^{-4}$ \\
$g_{bs}^{L}$ & $-1.439 \times 10^{-2} < g_{bs}^{L} < 1.404 \times 10^{-2}$ \\
$g_{bs}^{R}$ & $-9.773 \times 10^{-4} < g_{bs}^{R} < 9.723 \times 10^{-4}$ \\
$g_{\mu\mu}^{L} + g_{\mu\mu}^{R}$ & $-1.175 \times 10^{-2} < g_{\mu\mu}^{L} + g_{\mu\mu}^{R} < 1.179 \times 10^{-3}$ \\
\bottomrule
\bottomrule
\end{tabular}}%
\caption{Scanned ranges of the $Z^{\prime}$ coupling constant to $b$ and $s$ quarks as well as to muon pair, required to derive mass range of $Z^{\prime}$ gauge boson from the $R_{K^{*}}$ anomaly and from the $B_{s}$ meson oscillation.}
\label{tab:coupling_constants_gbs_gmumu}
\end{table}
\subsubsection{$M_{Z^{\prime}}$ mass ranges resulting from the $R_{K^{*}}$ anomaly and $B_{s}$ meson oscillation}
In the model under consideration, the most relevant mass ranges of $Z^{\prime}$ come from the $R_{K^{*}}$ anomaly and the $B_{s}$ meson oscillation. We start from the $R_{K^{*}}$ anomaly first.
\\~\\ 
\textbf{\underline{$R_{K^{*}}$ anomaly}}
\\~\\
Reminding the experimental bounds for the Wilson coefficients $G_{bs\mu}^{L,R}$ in both ``theoretically clean fit" (Equation~\ref{eqn:RK_const_tcfit}) and ``global fit" (Equation~\ref{eqn:RK_const_gfit}), they read respectively:
\begingroup
\begin{equation}
\begin{split}
\frac{1}{\left( 44.44\func{TeV} \right)^{2}} &< G_{bs\mu}^{L} = -\frac{g_{bs}^{L} g_{\mu\mu}^{L}}{M_{Z^{\prime}}^{2}} < \frac{1}{\left( 40\func{TeV} \right)^{2}}, \\
-\frac{1}{\left( 35.9\func{TeV} \right)^{2}} &< G_{bs\mu}^{R} = -\frac{g_{bs}^{L} g_{\mu\mu}^{R}}{M_{Z^{\prime}}^{2}} < \frac{1}{\left( 205\func{TeV} \right)^{2}}, \\
\label{eqn:RK_const_tcfit}
\end{split}
\end{equation}
\endgroup
\begingroup
\begin{equation}
\begin{split}
\frac{1}{\left( 39.75\func{TeV} \right)^{2}} &< G_{bs\mu}^{L} = -\frac{g_{bs}^{L} g_{\mu\mu}^{L}}{M_{Z^{\prime}}^{2}} < \frac{1}{\left( 36.67\func{TeV} \right)^{2}}, \\
-\frac{1}{\left( 83.8\func{TeV} \right)^{2}} &< G_{bs\mu}^{R} = -\frac{g_{bs}^{L} g_{\mu\mu}^{R}}{M_{Z^{\prime}}^{2}} < \frac{1}{\left( 44.44\func{TeV} \right)^{2}}. \\
\label{eqn:RK_const_gfit}
\end{split}
\end{equation}
\endgroup
In order to confirm $M_{Z^{\prime}}$ ranges, the first task is to determine numerical ranges of the combined coupling constants $g_{bs} g_{\mu\mu}$ using the varied range of each coupling constant in Table~\ref{tab:coupling_constants_gbs_gmumu} as given in Table~\ref{tab:combined_cc}:
\begingroup
\begin{table}[H]
\centering
\resizebox{0.8\textwidth}{!}{
\centering\renewcommand{\arraystretch}{1.3} 
\begin{tabular}{*{2}{c}}
\toprule
\toprule
Combination of the coupling constants & Range of the combination \\ 
\midrule
$g_{bs}^{L} g_{\mu\mu}^{L}$ & $-3.041 \times 10^{-5} < g_{bs}^{L} g_{\mu\mu}^{L} < 2.967 \times 10^{-5}$ \\[1ex]
$g_{bs}^{L} g_{\mu\mu}^{R}$ & $-1.678 \times 10^{-4} < g_{bs}^{L} g_{\mu\mu}^{R} < 1.720 \times 10^{-4}$ \\[1ex]
\bottomrule
\bottomrule
\end{tabular}}%
\caption{Combined coupling constants, required to confirm mass ranges of $Z^{\prime}$ from the $R_{K^{*}}$ anomaly.}
\label{tab:combined_cc}
\end{table}
\endgroup
Using each boundary value of the combined coupling constants $g_{bs} g_{\mu\mu}$, we can confirm mass ranges of $Z^{\prime}$ when both the CMS and $\tau \rightarrow \mu \gamma$ experimental constraints are considered together as given in Table~\ref{tab:Range_MZp_RK}:
\begingroup
\begin{table}[H]
\centering\renewcommand{\arraystretch}{1.5} 
\begin{tabular*}{500pt}{@{\extracolsep{\fill}}*{3}{c}}
\toprule
\toprule
Fit & Combination of the coupling constants & Mass range of $Z^{\prime}$ ($\func{GeV}$) \\ 
\midrule
\multirow{4}*{Theoretically clean fit} & $g_{bs}^{L} g_{\mu\mu}^{L} = -3.041 \times 10^{-5}$ & $220.581 < M_{Z^{\prime}} < 245.066$ \\[1ex]
 & $g_{bs}^{L} g_{\mu\mu}^{L} = 2.967 \times 10^{-5}$ & $\text{no result}$ \\[1ex]
(CMS + $\left( \tau \rightarrow \mu \gamma \right)$) & $g_{bs}^{L} g_{\mu\mu}^{R} = -1.678 \times 10^{-4}$ & $M_{Z^{\prime}} > 2655.522$ \\[1ex]
 & $g_{bs}^{L} g_{\mu\mu}^{R} = 1.720 \times 10^{-4}$ & $M_{Z^{\prime}} > 470.824$ \\[1ex]
\midrule
\multirow{4}*{Global fit} & $g_{bs}^{L} g_{\mu\mu}^{L} = -3.041 \times 10^{-5}$ & $202.218 < M_{Z^{\prime}} < 219.202$ \\[1ex]
 & $g_{bs}^{L} g_{\mu\mu}^{L} = 2.967 \times 10^{-5}$ & $\text{no result}$ \\[1ex]
(CMS + $\left( \tau \rightarrow \mu \gamma \right)$) & $g_{bs}^{L} g_{\mu\mu}^{R} = -1.678 \times 10^{-4}$ & $M_{Z^{\prime}} > 575.665$ \\[1ex]
 & $g_{bs}^{L} g_{\mu\mu}^{R} = 1.720 \times 10^{-4}$ & $M_{Z^{\prime}} > 1099.027$ \\[1ex]
\bottomrule
\bottomrule
\end{tabular*}%
\caption{Scanned ranges of $M_{Z^{\prime}}$, based on the combined coupling constants, from the theoretically clean and global fit in the $R_{K^{*}}$ anomaly. The parenthesis below each fit means the constraints considered. ``No result" means there is no possible range due to overall sign difference.}
\label{tab:Range_MZp_RK}
\end{table}
\endgroup
Simplifying the given results in Table~\ref{tab:Range_MZp_RK} by contracting each combined coupling constants, it can be rearranged as follows:
\begingroup
\begin{table}[H]
\centering\renewcommand{\arraystretch}{1.5} 
\begin{tabular*}{500pt}{@{\extracolsep{\fill}}*{3}{c}}
\toprule
\toprule
Fit & Combination of the coupling constants & Mass range of $Z^{\prime}$ ($\func{GeV}$) \\ 
\midrule
\multirow{2}*{Theoretically clean fit} & $-3.041 \times 10^{-5} < g_{bs}^{L} g_{\mu\mu}^{L} < 2.967 \times 10^{-5}$ & $220.581 < M_{Z^{\prime}} < 245.066$ \\[1ex]
(CMS + $\left( \tau \rightarrow \mu \gamma \right)$) & $-1.678 \times 10^{-4} < g_{bs}^{L} g_{\mu\mu}^{R} < 1.720 \times 10^{-4}$ & $M_{Z^{\prime}} > 2655.522$ \\[1ex]
\midrule
\multirow{2}*{Theoretically clean fit} & $-1.927 \times 10^{-4} < g_{bs}^{L} g_{\mu\mu}^{L} < 1.880 \times 10^{-4}$ & $555.266 < M_{Z^{\prime}} < 616.900$ \\[1ex]
$\left( \tau \rightarrow \mu \gamma \right)$ & $-1.726 \times 10^{-4} < g_{bs}^{L} g_{\mu\mu}^{R} < 1.769 \times 10^{-4}$ & $M_{Z^{\prime}} > 2693.235$ \\[1ex]
\midrule
\multirow{2}*{Global fit} & $-3.041 \times 10^{-5} < g_{bs}^{L} g_{\mu\mu}^{L} < 2.967 \times 10^{-5}$ & $202.218 < M_{Z^{\prime}} < 219.202$ \\[1ex]
(CMS + $\left( \tau \rightarrow \mu \gamma \right)$) & $-1.678 \times 10^{-4} < g_{bs}^{L} g_{\mu\mu}^{R} < 1.720 \times 10^{-4}$ & $M_{Z^{\prime}} > 1099.027$ \\[1ex]
\midrule
\multirow{2}*{Global fit} & $-1.927 \times 10^{-4} < g_{bs}^{L} g_{\mu\mu}^{L} < 1.880 \times 10^{-4}$ & $509.040 < M_{Z^{\prime}} < 551.795$ \\[1ex]
$\left( \tau \rightarrow \mu \gamma \right)$ & $-1.726 \times 10^{-4} < g_{bs}^{L} g_{\mu\mu}^{R} < 1.769 \times 10^{-4}$ & $M_{Z^{\prime}} > 1114.570$ \\[1ex]
\bottomrule
\bottomrule
\end{tabular*}%
\caption{Ranges of $M_{Z^{\prime}}$ simplified. Apart from the given results in Table~\ref{tab:Range_MZp_RK}, we also include $Z^{\prime}$ mass ranges derived from the $R_{K^{*}}$ anomaly when the CLFV $\tau \rightarrow \mu \gamma$ experimental constraint is considered alone to show that how the derived $Z^{\prime}$ mass ranges get affected and that the $Z^{\prime}$ and CP-odd scalar can be connected via the scalar potential in this case only.}
\label{tab:simplified_Range_MZp_RK}
\end{table}
\endgroup
What we found via this numerical study for the $R_{K^{*}}$ anomaly is three features: the first is there is no overlapped region between $M_{Z^{\prime}}$ ranges from $G_{bs\mu}^{L}$ and those from $G_{bs\mu}^{R}$ in both theoretically clean and global fit, which indicates that the $R_{K^{*}}$ anomaly is quite likely to be explained by only left-handed interactions, and the second is there is no overlapped region for $G_{bs\mu}^{L}$ even in theoretically clean fit and in global fit. Lastly, we include the derived $Z^{\prime}$ mass ranges from the $R_{K^{*}}$ anomaly when the CLFV $\tau \rightarrow \mu \gamma$ constraint is considered alone to show that how the $Z^{\prime}$ mass ranges get affected and that the $Z^{\prime}$ and the CP-odd scalar can be connected in this case only via the scalar potential. Next we discuss the $B_{s}$ meson oscillation.
\\~\\
\textbf{\underline{$B_{s}$ meson oscillation}}
\\~\\
Reminding the two theoretical constraints of Equation~\ref{eqn:Bs_const_theor_1}, \ref{eqn:Bs_const_theor_2} and the experimental constraint of Equation~\ref{eqn:Bs_const_exp}, they read respectively:
\begingroup
\begin{equation}
\begin{gathered}
G_{bs} = \frac{g_{bs}^{L2} + g_{bs}^{L} g_{bs}^{R*} + g_{bs}^{R2}}{M_{Z^{\prime}}^{2}} < \frac{1}{\left( 330\func{TeV} \right)^{2}}, \quad \text{from FLAG$^{\prime}$19}
\\
G_{bs} = \frac{g_{bs}^{L2} + g_{bs}^{L} g_{bs}^{R*} + g_{bs}^{R2}}{M_{Z^{\prime}}^{2}} < \frac{1}{\left( 220\func{TeV} \right)^{2}}, \quad \text{from Average$^{\prime}$19}
\\ 
-1.07237 \times 10^{-13} \func{GeV} < \frac{g_{bs}^{L2} + g_{bs}^{L} g_{bs}^{R*} + g_{bs}^{R2}}{M_{Z^{\prime}}^{2}} \times 0.188 \func{GeV}^{3} < 9.125 \times 10^{-13} \func{GeV}.
\end{gathered}
\end{equation}
\endgroup
A numerical range of summing over the flavor violating coupling constants $g_{bs}^{L,R}$, referring to the coupling constants in Table~\ref{tab:coupling_constants_gbs_gmumu}, reads:
\begingroup
\begin{equation}
1.833 \times 10^{-4} < g_{bs}^{2} = g_{bs}^{L2} + g_{bs}^{L} g_{bs}^{R*} + g_{bs}^{R2} < 2.106 \times 10^{-4}
\end{equation}
\endgroup
Then, $M_{Z^{\prime}}$ can be determined as in the $R_{K^{*}}$ anomaly and the final result for the $B_{s}$ meson oscillation is given in Table~\ref{tab:Range_MZp_Bs}:
\begingroup
\begin{table}[H]
\centering\renewcommand{\arraystretch}{1.5} 
\begin{tabular*}{500pt}{@{\extracolsep{\fill}}*{3}{c}}
\toprule
\toprule
Fit & Combination of the coupling constants & Mass range of $Z^{\prime}$ ($\func{GeV}$) \\ 
\midrule
Theoretical fit (FLAG) & \multirow{3}{*}{$1.833 \times 10^{-4} < g_{bs}^{2} = g_{bs}^{L2} + g_{bs}^{L} g_{bs}^{R*} + g_{bs}^{R2} < 2.106 \times 10^{-4}$} & $M_{Z^{\prime}} > 4788.980$ \\[1ex]
Theoretical fit (Average) &  & $M_{Z^{\prime}} > 3192.650$ \\[1ex]
Experimental result &  & $M_{Z^{\prime}} > 6591.740$ \\[1ex]
\bottomrule
\bottomrule
\end{tabular*}%
\caption{Ranges of $M_{Z^{\prime}}$ derived from the $B_{s}$ meson oscillation experiments. They are derived from two theoretical fits, FLAG and Average, and from one experimental result respectively.}
\label{tab:Range_MZp_Bs}
\end{table}
\endgroup
The confirmed $Z^{\prime}$ mass ranges from the $B_{s}$ meson oscillation fits are too much tightened when compared to those derived from the $R_{K^{*}}$ anomaly as well as from the neutrino trident production. For comparison, it is good to look at the derived results from the diverse constraints at one sight in Table~\ref{tab:Range_MZp_final}:
\begingroup
\begin{table}[H]
\centering\renewcommand{\arraystretch}{1.5} 
\begin{tabular*}{500pt}{@{\extracolsep{\fill}}*{4}{c}}
\toprule
\toprule
Observable & Fit & Coupling constants & Mass range of $Z^{\prime}$ ($\func{GeV}$) \\ 
\midrule
$\nu$ Trident &  &  & $M_{Z^{\prime}} > 0.920 \func{GeV}$ \\
\midrule
\multirow{8}{*}{$R_{K^{*}}$ anomaly}  & \multirow{2}{*}{Theoretically clean fit} & $-3.041 \times 10^{-5} < g_{bs}^{L} g_{\mu\mu}^{L} < 2.967 \times 10^{-5}$ & $220.581 < M_{Z^{\prime}} < 245.066$ \\[1ex] 
\multirow{8}{*}{} & (CMS + $\left( \tau \rightarrow \mu \gamma \right)$) & $-1.678 \times 10^{-4} < g_{bs}^{L} g_{\mu\mu}^{R} < 1.720 \times 10^{-4}$ & $M_{Z^{\prime}} > 2655.522$ \\[1ex]
\cmidrule{2-4}
\multirow{8}{*}{}  & \multirow{2}{*}{Theoretically clean fit} & $-1.927 \times 10^{-4} < g_{bs}^{L} g_{\mu\mu}^{L} < 1.880 \times 10^{-4}$ & $555.266 < M_{Z^{\prime}} < 616.900$ \\[1ex] 
\multirow{8}{*}{} & $\left( \tau \rightarrow \mu \gamma \right)$ & $-1.726 \times 10^{-4} < g_{bs}^{L} g_{\mu\mu}^{R} < 1.769 \times 10^{-4}$ & $M_{Z^{\prime}} > 2693.235$ \\[1ex] 
\cmidrule{2-4}
\multirow{8}{*}{} & \multirow{2}*{Global fit} & $-3.041 \times 10^{-5} < g_{bs}^{L} g_{\mu\mu}^{L} < 2.967 \times 10^{-5}$ & $202.218 < M_{Z^{\prime}} < 219.202$ \\[1ex]
\multirow{8}{*}{} & (CMS + $\left( \tau \rightarrow \mu \gamma \right)$) & $-1.678 \times 10^{-4} < g_{bs}^{L} g_{\mu\mu}^{R} < 1.720 \times 10^{-4}$ & $M_{Z^{\prime}} > 1099.027$ \\[1ex]
\cmidrule{2-4}
\multirow{8}{*}{} & \multirow{2}*{Global fit} & $-1.927 \times 10^{-4} < g_{bs}^{L} g_{\mu\mu}^{L} < 1.880 \times 10^{-4}$ & $509.040 < M_{Z^{\prime}} < 551.795$ \\[1ex]
\multirow{8}{*}{} & $\left( \tau \rightarrow \mu \gamma \right)$ & $-1.726 \times 10^{-4} < g_{bs}^{L} g_{\mu\mu}^{R} < 1.769 \times 10^{-4}$ & $M_{Z^{\prime}} > 1114.570$ \\[1ex]
\midrule
\multirow{3}{*}{$B_{s} - \overline{B}_{s}$} & Theoretical fit (FLAG) & \multirow{3}{*}{$1.833 \times 10^{-4} < g_{bs}^{2} < 2.106 \times 10^{-4}$} & $M_{Z^{\prime}} > 3192.650$ \\[1ex] \cmidrule{2-2} \cmidrule{4-4}
\multirow{3}{*}{} & Theoretical fit (Average) & \multirow{3}{*}{} & $M_{Z^{\prime}} > 4788.980$ \\[1ex] \cmidrule{2-2} \cmidrule{4-4}
\multirow{3}{*}{} & Experimental result & \multirow{3}{*}{} & $M_{Z^{\prime}} > 6591.740$ \\[1ex] 
\bottomrule
\bottomrule
\end{tabular*}%
\caption{Ranges of $M_{Z^{\prime}}$ derived from the neutrino trident production ($\nu$ Trident), the $R_{K^{*}}$ anomaly and finally the $B_{s}$ meson oscillation.}
\label{tab:Range_MZp_final}
\end{table}
\endgroup
The first constraint we discuss in Table~\ref{tab:Range_MZp_final} is the neutrino trident production and it can not constrain $Z^{\prime}$ mass compared to ones from the $R_{K^{*}}$ anomaly and $B_{s}$ meson oscillation. Next we discuss constraints resulting from the $R_{K^{*}}$ anomaly and we consider two cases where the first is to consider both the CMS and $\tau \rightarrow \mu \gamma$ experimental constraints and the other is to consider only the $\tau \rightarrow \mu \gamma$ experimental constraint. The reasons why we consider both the cases are as follows:
\begingroup
\begin{enumerate}
\item The first reason is to show that how the $Z^{\prime}$ mass range is affected when the left-handed muon coupling constant $g_{L}^{\mu\mu}$ is suppressed by the CMS experimental limit.
\item The second reason is the $R_{K^{*}}$ anomaly global fit with the experimental CLFV $\tau \rightarrow \mu \gamma$ decay alone is the only possible region where non-SM scalars for the muon $g-2$ and the $Z^{\prime}$ gauge boson for the $R_{K^{*}}$ anomaly can be connected via the scalar potential under consideration, while fitting the oblique parameters $T,S,U$ as well as the $W$ mass anomaly and not changing out conclusion that the muon $g-2$ and $R_{K^{*}}$ anomaly can not be explained by the same new physics. So we call the region ``theoretically interesting $Z^{\prime}$ mass range".
\end{enumerate}
\endgroup
Taking the reasons listed into account, we consider the $R_{K^{*}}$ anomaly constraint with both the CMS and $\tau \rightarrow \mu \gamma$ experiments first. The first important feature is there is no overlapped region between the two combination of coupling constants in both the theoretically clean fit and global fit and what this implements is the $R_{K^{*}}$ anomaly is likely to be explained by the only left-handed interactions. This tendency is not changed for the constraints with the $\tau \rightarrow \mu \gamma$ experimental limit alone. The second feature is the $Z^{\prime}$ mass range from the $R_{K^{*}}$ anomaly global fit in Table~\ref{tab:Range_MZp_final} with both the experimental bounds, CMS and $\tau \rightarrow \mu \gamma$, is not compatible to the lightest $Z^{\prime}$ mass, $261\func{GeV}$, derived in the squared gauge mass matrix of Equation~\ref{eqn:Mgaugesquared_partial} under this BSM model when we turn off the vev $v_{3}$ (even though we consider the $R_{K^{*}}$ anomaly constraint at $3\sigma$, giving rise to $186.942\func{GeV} < M_{Z^{\prime}} < 237.786\func{GeV}$, its tension is not relaxed enough to cover the lightest $Z^{\prime}$ mass). On top of that, the vev $v_{3}$ is somewhat constrained by both the CKM mixing matrix and the muon $g-2$. Looking at the up-type mass matrix of Equation~\ref{eqn:MuMd_diagMuMd}, giving rise to the best fitted CKM prediction of Equation~\ref{eqn:CKM_bestfit} in the model under consideration, one can know that the vev $v_{3}$ must be heavier than $340\func{GeV}$, which is given by diving the $m_{35} = 1206.340\func{GeV}$ by the purturbative limit of a Yukawa constant $\sqrt{4\pi} = 3.54$. Plus, the vev $v_{3}$ is also constrained by the muon $g-2$ prediction of Equation~\ref{eqn:muong2prediction_scalar}. Referring to our previous work~\cite{Hernandez:2021tii}, one of the Yukawa constant, $y_{24}^{e}$, in the muon prediction with scalar exchange remains as a free parameter whereas the other, $x_{42}^{e}$, is governed by the vev $v_{3}$ locating in the denominator. Therefore, the vev $v_{3}$ can not be heavy, for example, up to $\func{TeV}$ scale in the model under consideration, otherwise the muon $g-2$ prediction will have a smaller order compared to its experimental order, therefore not possible to fit its experimental bound at all. What this implements is the $R_{K^{*}}$ anomaly and muon $g-2$ can not be explained by the same new physics and their new physics sources, $Z^{\prime}$ for the $R_{K^{*}}$ anomaly and non-SM scalars for the muon $g-2$ can not be connected to each other in the scalar potential under consideration, therefore concluding the $R_{K^{*}}$ anomaly and muon $g-2$ should be considered as an independent new physics signal in this BSM model. Considering the ``theoretically interesting $Z^{\prime}$ mass range" is less interesting for the reason that the CMS experimental bound is not considered, however we will show that the $Z^{\prime}$ for the $R_{K^{*}}$ anomaly and the non-SM scalars for the muon $g-2$ can be connected via the scalar potential while fitting oblique parameters $T,S,U$ and $W$ mass anomaly at their $1$ or $2\sigma$ constraints in the range. Lastly, since the $Z^{\prime}$ mass ranges derived from the $B_{s}$ meson oscillations are too heavy when compared to ones from the neutrino trident and $R_{K^{*}}$ anomaly, we consider them to be explained by some other heavier new physics sources, not considered in this work.
\section{Muon anomalous magnetic moment $g-2$ revisited with scalar exchange}
We have seen that the muon $g-2$ mediated by $Z^{\prime}$ at one loop level is not an answer since it can not explain the experimental muon $g-2$ bound at all by yielding only negative contributions. In order to cure this problem, it requires another approach to muon $g-2$ and we have shown that scalar exchange at one loop level can be an answer for muon $g-2$~\cite{Hernandez:2021tii}. The corresponding muon $g-2$ diagram with scalar exchange at one loop level can be drawn by closing the scalars of the diagrams in Figure~\ref{fig:diagrams_charged_leptons_mass_insertion}:
\begingroup
\begin{figure}[H]
\centering
\begin{subfigure}{0.48\textwidth}
	\includegraphics[width=1.0\textwidth]{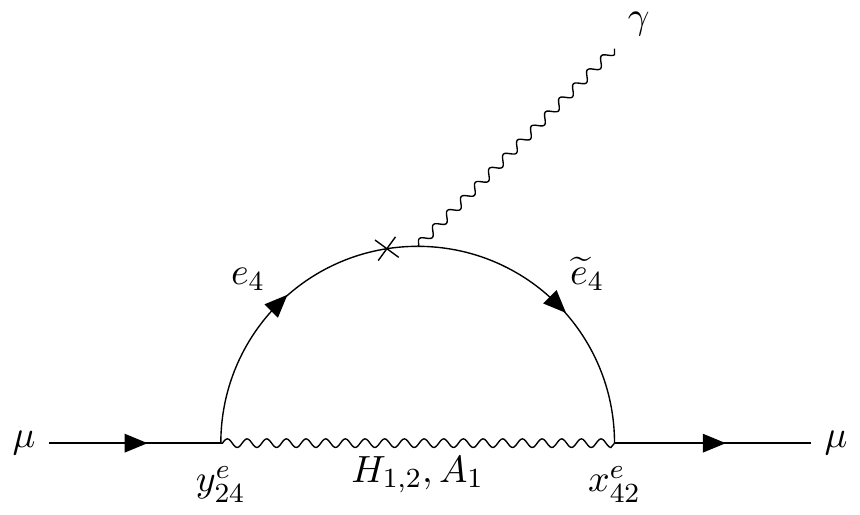}
\end{subfigure} 
\caption{Feynman diagram contributing to the muon $g-2$ mediated by the vectorlike singlet charged leptons and by the CP-even scalars $H_{1,2}$ and the CP-odd scalar $A_{1}$,}
\label{fig:muong2_scalar}
\end{figure}
\endgroup
where it is worth mentioning that the muon $g-2$ mediated by the vectorlike doublet charged leptons does not work in this BSM model as there is no seesaw operator mediated by the vectorlike doublet charged lepton in the fully rotated mass matrix of Equation~\ref{eqn:cl_2} for the purpose of imposing the SM hierarchy dynamically. In order to setup an analytic prediction for the muon $g-2$ by scalar exchange, it requires to discuss the scalar potential and its relevant features and this process is interconnected with the $M_{Z^{\prime}}$ as we will see soon.
\subsection{Scalar potential}
In this BSM model, the Higgs alignments in the interaction basis read~\cite{Hernandez:2021tii} ($v_{u} = v_{1}, v_{d} = v_{2}, v_{\phi} = v_{3}$):
\begingroup
\begin{equation}
\begin{split}
H_{u} &= 
\begin{pmatrix}
H_{u}^{+} \\
v_{u} + \frac{1}{\sqrt{2}} \left( \func{Re} H_{u}^{0} + i\func{Im} H_{u}^{0} \right)
\end{pmatrix},
\\
H_{d} &= 
\begin{pmatrix}
v_{d} + \frac{1}{\sqrt{2}} \left( \func{Re} H_{d}^{0} + i\func{Im} H_{d}^{0} \right) \\,
H_{d}^{-}
\end{pmatrix},
\\
\phi &= \frac{1}{\sqrt{2}} \left( v_{\phi} + \func{Re} \phi + i\func{Im} \phi \right).
\end{split}
\end{equation}
\endgroup
The scalar potential takes the form of~\cite{Hernandez:2021tii}:
\begingroup
\begin{equation}
\begin{split}
V &= \mu_{1}^{2} (H_{u}^{\dagger} H_{u}) + \mu_{2}^{2} (H_{d}^{\dagger} H_{d}) + \mu_{3}^{2} (\phi^{*} \phi)
\\
&+ \lambda_{1} (H_{u}^{\dagger} H_{u})^{2} + \lambda_{2} (H_{d}^{\dagger} H_{d})^{2} + \lambda_{3} (H_{u}^{\dagger} H_{u}) (H_{d}^{\dagger} H_{d}) + \lambda_{4} (H_{u}^{\dagger} H_{d}) (H_{d}^{\dagger} H_{u})
\\
&+ \lambda_{5} (\epsilon_{ij} H_{u}^{i} H_{d}^{j} \phi^{2} + \func{H.c.}) + \lambda_{6} (\phi^{*}\phi)^{2} + \lambda_{7} (\phi^{*}\phi)(H_{u}^{\dagger}H_{u}) + \lambda_{8} (\phi^{*}\phi)(H_{d}^{\dagger}H_{d}),
\end{split}
\end{equation}
\endgroup
where the $\mu_{1,2,3}^{2}$ are dimensionful mass parameters and the $\lambda_{1,2,\cdots,8}$ are dimensionless quartic coupling constants. The minimization conditions of the scalar potential allow the dimensionful mass parameters to be written in terms of the input parameters of the scalar potential as follows:
\begingroup
\begin{equation}
\begin{split}
\mu_{1}^{2} &= -2 \lambda_{1} v_{1}^{2} - \lambda_{3} v_{2}^{2} + \frac{\lambda_{5} v_{2} v_{3}^{2}}{2v_{1}} - \lambda_{7} v_{3}^{2},
\\
\mu_{2}^{2} &= -2 \lambda_{2} v_{2}^{2} - \lambda_{3} v_{1}^{2} + \frac{\lambda_{5} v_{1} v_{3}^{2}}{2v_{2}} - \lambda_{8} v_{3}^{2},
\\
\mu_{3}^{2} &= -2 \lambda_{6} v_{3}^{2} - \lambda_{7} v_{1}^{2} - \lambda_{8} v_{2}^{2} + 2\lambda_{5} v_{1} v_{2}.
\label{eqn:dimensionful_mass_parameters}
\end{split}
\end{equation}
\endgroup
The dimensionful mass parameters are important in exploring vacuum stability of this BSM model and will be discussed in the next subsection.
\subsection{Mass matrix for CP-even, CP-odd neutral and charged scalars}
The squared CP-even mass matrix in the interaction basis of $\left( \func{Re} H_{u}^{0}, \func{Re} H_{d}^{0}, \func{Re} \phi \right)$ reads:
\begingroup
\begin{equation}
\mathbf{M}_{\func{CP-even}}^{2} 
= 
\left(
\begin{array}{ccc}
 4 \lambda _1 v_1^2+\frac{v_2 v_3^2 \lambda _5}{2 v_1} & 2 v_1 v_2 \lambda _3-\frac{1}{2} v_3^2 \lambda _5 & \sqrt{2} v_3 \left(v_1 \lambda _7-v_2 \lambda _5\right) \\
 2 v_1 v_2 \lambda _3-\frac{1}{2} v_3^2 \lambda _5 & 4 \lambda _2 v_2^2+\frac{v_1 v_3^2 \lambda _5}{2 v_2} & \sqrt{2} v_3 \left(v_2 \lambda _8-v_1 \lambda _5\right) \\
 \sqrt{2} v_3 \left(v_1 \lambda _7-v_2 \lambda _5\right) & \sqrt{2} v_3 \left(v_2 \lambda _8-v_1 \lambda _5\right) & 2 v_3^2 \lambda _6 \\
\end{array}
\right).
\label{eqn:mass_matrix_for_CP_even}
\end{equation}
\endgroup
In order to make our analysis for the scalar sector simpler, we apply decoupling limit to the scalar sector in this BSM model. Under the decoupling limit, the SM Higgs $h$ has no mixing with CP-even scalars $H_{1}$ and $H_{2}$ and it can be achieved by turning off $(12)$, $(13)$, $(21)$ and $(31)$ elements of the CP-even mass matrix of Equation~\ref{eqn:mass_matrix_for_CP_even}. The decoupling limit can be implemented by setting a couple of relations between the relevant quartic coupling constants as follows:
\begingroup
\begin{equation}
\begin{split}
\lambda_{5} &= \frac{4v_{1} v_{2}}{v_{3}^{2}} \lambda_{3},
\\
\lambda_{7} &= \frac{v_{2}}{v_{1}} \lambda_{5} = \frac{4v_{2}^{2}}{v_{3}^{2}} \lambda_{3}.
\label{eqn:decoupling_limit}
\end{split}
\end{equation}
\endgroup
Then, the squared CP-even mass matrix after implementing the decoupling limit is reduced to:
\begingroup
\begin{equation}
\textbf{M}_{\func{CP-even}}^{2}
=
\left(
\begin{array}{ccc}
 4 \lambda _1 v_1^2+2 v_2^2 \lambda _3 & 0 & 0 \\
 0 & 2 \lambda _3 v_1^2+4 v_2^2 \lambda _2 & \sqrt{2} v_3 \left(v_2 \lambda _8-\frac{4 v_1^2 v_2 \lambda _3}{v_3^2}\right) \\
 0 & \sqrt{2} v_3 \left(v_2 \lambda _8-\frac{4 v_1^2 v_2 \lambda _3}{v_3^2}\right) & 2 v_3^2 \lambda _6 \\
\end{array}
\right).
\end{equation}
\endgroup
Diagonalizing the squared CP-even mass matrix, it reveals masses of the physical SM Higgs and non-SM scalars $H_{1,2}$ in the physical basis of $\left( h, H_{1}, H_{2} \right)$:
\begingroup
\begin{equation}
R_{\func{CP-even}} \mathbf{M}_{\func{CP-even}}^{2} R_{\func{CP-even}}^{\dagger} = \func{diag} \left( m_{h}^{2}, m_{H_{1}}^{2}, m_{H_{2}}^{2} \right).
\label{eqn:diag_CPeven}
\end{equation}
\endgroup
The squared CP-odd mass matrix in the interaction basis of $\left( \func{Im} H_{u}^{0}, \func{Im} H_{d}^{0}, \func{Im} \phi \right)$ features:
\begingroup
\begin{equation}
\mathbf{M}_{\func{CP-odd}}^{2}
=
\left(
\begin{array}{ccc}
 \frac{v_2 v_3^2 \lambda _5}{2 v_1} & \frac{1}{2} v_3^2 \lambda _5 & \sqrt{2} v_2 v_3 \lambda _5 \\
 \frac{1}{2} v_3^2 \lambda _5 & \frac{v_1 v_3^2 \lambda _5}{2 v_2} & \sqrt{2} v_1 v_3 \lambda _5 \\
 \sqrt{2} v_2 v_3 \lambda _5 & \sqrt{2} v_1 v_3 \lambda _5 & 4 v_1 v_2 \lambda _5 \\
\end{array}
\right)
\end{equation}
\endgroup
After diagonalizing the squared CP-odd mass matrix, it reveals masses of one CP-odd scalar $A_{1}$ and two Goldstone bosons $G_{Z}$ and $G_{Z^{\prime}}$ in the physical basis of $\left( G_{Z}, A_{1}, G_{Z^{\prime}} \right)$, which will be the SM $Z$ gauge boson and a massive neutral $Z^{\prime}$ gauge boson as follows:
\begingroup
\begin{equation}
R_{\func{CP-odd}} \mathbf{M}_{\func{CP-odd}}^{2} R_{\func{CP-odd}}^{\dagger} = \func{diag} \left( 0, m_{A_{1}}, 0 \right).
\label{eqn:diag_CPodd}
\end{equation}
\endgroup
The squared charged mass matrix in the interaction basis of $\left( H_{1}^{\pm}, H_{2}^{\pm} \right)$ has the form of:
\begingroup
\begin{equation}
\textbf{M}_{\func{charged}}^{2} 
=
\left(
\begin{array}{cc}
 \lambda _4 v_2^2+\frac{v_3^2 \lambda _5 v_2}{2 v_1} & \frac{1}{2} \lambda _5 v_3^2+v_1 v_2 \lambda _4 \\
 \frac{1}{2} \lambda _5 v_3^2+v_1 v_2 \lambda _4 & \lambda _4 v_1^2+\frac{v_3^2 \lambda _5 v_1}{2 v_2} \\
\end{array}
\right)
\end{equation}
\endgroup
After diagonalizing the squared charged mass matrix, it reveals masses of one non-SM charged scalar and one Goldstone boson for the SM $W^{\pm}$ gauge boson. The squared physical charged mass matrix can be written in the basis of $\left( G_{W^{\pm}}, H^{\pm} \right)$:
\begingroup
\begin{equation}
R_{\func{charged}} \mathbf{M}_{\func{charged}}^{2} R_{\func{charged}}^{\dagger} = \func{diag} \left( 0, m_{H^{\pm}} \right).
\end{equation}
\endgroup
\subsection{SM Higgs diphoton signal strength}
As the BSM model under consideration generates the scalars which are the SM Higgs and the CP-even and -odd scalars, the SM Higgs diphoton signal strength should be discussed and we will see that the Higgs diphoton signal strength is quite close to the SM prediction under the decoupling limit. The decay rate for the $h \rightarrow \gamma\gamma$ reads~\cite{Hernandez:2021tii}:
\begingroup
\begin{equation}
\Gamma \left( h \rightarrow \gamma \gamma \right) = \frac{\alpha_{\func{em}}^{2} m_{h}^{3}}{256\pi^{3} v^{2}} \lvert \sum_{f} a_{hff} N_{C} Q_{f}^{2} F_{1/2}(\rho_{f}) + a_{hWW} F_{1} (\rho_{W}) + \frac{C_{hH^{\pm}H^{\mp}} v}{2m_{H^{\pm}}^{2}} F_{0} (\rho_{H_{k}^{\pm}}) \rvert^{2},
\end{equation}
\endgroup
where $\alpha_{\func{em}}$ is the fine structure constant, $\rho_{i}$ is the mass ratio $\rho_{i} = m_{h}^{2}/(4M_{i}^{2}$ with $M_{i} = m_{f}, M_{W}$, $N_{C}$ is the number of colors, and $Q_{f}$ is the electric charge of the fermion running in the loop, $C_{hH^{\pm}H^{\mp}}$ is the coupling constant between the SM Higgs $h$ and a charged Higgs pair $H^{\pm}$, $a_{htt}$ and $a_{hWW}$ are the deviation factors
\begingroup
\begin{equation}
\begin{split}
a_{htt} &\simeq 1, 
\\
a_{hWW} &= \frac{1}{\sqrt{v_{1}^{2} + v_{2}^{2}}} \frac{\partial}{\partial h} \sum_{i,j=1,2,3} v_{i} \left( R_{\func{CP-even}}^{T} \right)_{ij} \left( h, H_{1}, H_{2} \right)_{j}^{T} = \frac{v_{1}}{\sqrt{ v_{1}^{2} + v_{2}^{2}}},
\end{split}
\end{equation}
\endgroup
which are very close to unity in the SM, and lastly $F_{0,1/2,1}$ are loop functions for scalar, fermion and boson respectively. The loop functions read~\cite{Hernandez:2021tii}:
\begingroup
\begin{equation}
\begin{split}
F_{0} (z) &= -\left( z-f(z) \right) z^{-2},
\\
F_{1/2} (z) &= 2\left( z + (z-1) f(z) \right) z^{-2},
\\
F_{1} (z) &= -2 \left( 2z^{2} + 3z + 3(2z-1) f(z) \right) z^{-2}.
\end{split}
\end{equation}
\endgroup
where
\begin{equation}
f(z)=\left\{ 
\begin{array}{lcc}
\arcsin ^{2}\sqrt{2} & \text{for} & z\leq 1 \\ 
&  &  \\ 
-\frac{1}{4}\left( \ln \left( \frac{1+\sqrt{1-z^{-1}}}{1-\sqrt{1-z^{-1}}%
-i\pi }\right) ^{2}\right) & \text{for} & z>1
\end{array}%
\right.
\end{equation}
The defined decay width for $h \rightarrow \gamma\gamma$ is appeared in the Higgs diphoton signal strength $R_{\gamma\gamma}$, which takes the form of:
\begingroup
\begin{equation}
R_{\gamma\gamma} = \frac{\sigma\left( pp \rightarrow h \right) \Gamma\left( h \rightarrow \gamma\gamma \right)}{\sigma\left( pp \rightarrow h \right)_{\func{SM}} \Gamma\left( h \rightarrow \gamma\gamma \right)_{\func{SM}}} \simeq a_{htt}^{2} \frac{\Gamma\left( h \rightarrow \gamma\gamma \right)}{\Gamma\left( h \rightarrow \gamma\gamma \right)_{\func{SM}}}.
\end{equation}
\endgroup
The experimental $R_{\gamma\gamma}$ is given by ATLAS and CMS collaborations~\cite{ATLAS:2019nkf,CMS:2018piu}:
\begingroup
\begin{equation}
R_{\gamma\gamma}^{\func{ATLAS}} = 0.96 \pm 0.14., \quad R_{\gamma\gamma}^{\func{CMS}} = 1.18_{-0.14}^{+0.17}
\end{equation}
\endgroup
\subsection{Vacuum stability}
In order for all the scalars appearing in this BSM model to be physical, it is important to secure vacuum stability of the scalar potential. We generally follow the argument for vacuum stability covered in our previous work~\cite{Hernandez:2021tii}, however there is a big difference between the previous work and the current work: the SM gauge symmetry is extended by global $U(1)^{\prime}$ symmetry in the previous work so there is one more CP-odd scalar, whereas the freedom for the CP-odd scalar is replaced by $Z^{\prime}$ gauge boson as the SM gauge symmetry is extended by local $U(1)^{\prime}$ symmetry in the current work. The first condition for vacuum stability is the mass parameters $\mu_{1,2,3}^{2}$ given by the minimization condition should be negative otherwise it would yield massless scalar particles
\begingroup
\begin{equation}
\begin{split}
\mu_{1}^{2} &= -2 \lambda_{1} v_{1}^{2} - \lambda_{3} v_{2}^{2} + \frac{\lambda_{5} v_{2} v_{3}^{2}}{2v_{1}} - \lambda_{7} v_{3}^{2} = -2\lambda_{1} v_{1}^{2} - \lambda_{3} v_{2}^{2} < 0,
\\
\mu_{2}^{2} &= -2 \lambda_{2} v_{2}^{2} - \lambda_{3} v_{1}^{2} + \frac{\lambda_{5} v_{1} v_{3}^{2}}{2v_{2}} - \lambda_{8} v_{3}^{2} = -2\lambda_{2} v_{2}^{2} + \lambda_{3} v_{1}^{2} - \lambda_{8} v_{3}^{2} < 0,
\\
\mu_{3}^{2} &= -2 \lambda_{6} v_{3}^{2} - \lambda_{7} v_{1}^{2} - \lambda_{8} v_{2}^{2} + 2\lambda_{5} v_{1} v_{2} = \frac{4\lambda_{3} v_{1}^{2} v_{2}^{2}}{v_{3}^{2}} - \lambda_{6} v_{3}^{2} - \lambda_{8} v_{2}^{2} < 0,
\label{eqn:d_masses}
\end{split}
\end{equation}
\endgroup
where at the second equality of each equation the decoupling limit of Equation~\ref{eqn:decoupling_limit} was used. From the squared mass parameter $\mu_{2}^{2}$, one can know that the quartic coupling constant $\lambda_{3}$ must be negative, taking into account hierarchy among vevs $(\mathcal{O}(v_{2}) \ll \mathcal{O}(v_{1}) \approx \mathcal{O}(v_{3}))$ as we will see soon in the following numerical discussion. The negative sign of $\lambda_{3}$ also leads to negative sign of $\lambda_{5}$ and $\lambda_{7}$ due to the decoupling limit. As order of the vev $v_{1}$ is compatible to that of $v_{3}$, it is difficult to derive analytically further conditions for vacuum stability at the current stage as the other parameters depend on their numerical values. The next condition we can use is to reduce degree of freedom of the scalar potential by developing the singlet flavon $\phi$ to its vev $v_{3}$ after spontaneously breaking the $U(1)^{\prime}$ local symmetry, and then to apply the vacuum stability conditions for 2HDM~\cite{Bhattacharyya:2015nca,Maniatis:2006fs} to the scalar potential under consideration. The scalar potential after developing the vev $v_{3}$ takes the form of~\cite{Hernandez:2021tii}:
\begingroup
\begin{equation}
\begin{split}
V &= \mu_{1}^{2} (H_{u}^{\dagger} H_{u}) + \mu_{2}^{2} (H_{d}^{\dagger} H_{d}) + \mu_{3}^{2} (\frac{v_{3}^{2}}{2})
\\
&+ \lambda_{1} (H_{u}^{\dagger} H_{u})^{2} + \lambda_{2} (H_{d}^{\dagger} H_{d})^{2} + \lambda_{3} (H_{u}^{\dagger} H_{u}) (H_{d}^{\dagger} H_{d}) + \lambda_{4} (H_{u}^{\dagger} H_{d}) (H_{d}^{\dagger} H_{u})
\\
&+ \lambda_{5} (\epsilon_{ij} H_{u}^{i} H_{d}^{j} \frac{v_{3}^{2}}{2} + \func{H.c.}) + \lambda_{6} (\frac{v_{3}^{2}}{2})^{2} + \lambda_{7} (\frac{v_{3}^{2}}{2})(H_{u}^{\dagger}H_{u}) + \lambda_{8} (\frac{v_{3}^{2}}{2})(H_{d}^{\dagger}H_{d}),
\end{split}
\label{eqn:scalar_potential_dev_v3}
\end{equation}
\endgroup
The reduced scalar potential of Equation~\ref{eqn:scalar_potential_dev_v3} can be further simplified to by dropping all number and then by rearranging same order terms (the $\lambda_{4}$ term can be safely ignored as it has nothing to do with the neutral scalar fields but with charged scalar fields)~\cite{Hernandez:2021tii}:
\begingroup
\begin{equation}
\begin{split}
V &= \left( \mu_{1}^{2} + \lambda_{7} \frac{v_{3}^{2}}{2} \right) (H_{u}^{\dagger} H_{u}) + \left( \mu_{2}^{2} + \lambda_{8} \frac{v_{3}^{2}}{2} \right) (H_{d}^{\dagger} H_{d}) + \lambda_{5} \frac{v_{3}^{2}}{2} (\epsilon_{ij} H_{u}^{i} H_{d}^{j} + \func{H.c.})
\\
&+ \lambda_{1} (H_{u}^{\dagger} H_{u})^{2} + \lambda_{2} (H_{d}^{\dagger} H_{d})^{2} + \lambda_{3} (H_{u}^{\dagger} H_{u}) (H_{d}^{\dagger} H_{d}),
\end{split}
\label{eqn:scalar_potential_simplified}
\end{equation}
\endgroup
where the renewed dimensionful mass parameters can give rise to the renewed vacuum stability conditions as follows:
\begingroup
\begin{equation}
\begin{split}
\mu_{1}^{2} + \lambda_{7} \frac{v_{3}^{2}}{2} &= \mu_{1}^{2} + 2\lambda_{3} v_{2}^{2} = -2 \lambda_{1} v_{1}^{2} + \lambda_{3} v_{2}^{2} < 0, \\
\mu_{2}^{2} + \lambda_{8} \frac{v_{3}^{2}}{2} &= -2\lambda_{2} v_{2}^{2} + \lambda_{3} v_{1}^{2} - \frac{\lambda_{8}}{2} v_{3}^{2} < 0,
\end{split}
\end{equation}
\endgroup
where it can confirm that the quartic coupling constant $\lambda_{3}$ must be negative as $\lambda_{1}$, fixed by the SM Higgs mass, is positive in the upper condition of Equation~\ref{eqn:scalar_potential_simplified} and this reidentifies what we discussed earlier. Then we are ready to apply the simplified scalar potential of Equation~\ref{eqn:scalar_potential_simplified} to vacuum stability condition for 2HDM and we rearrange the mass parameters and coupling constants as follows~\cite{Hernandez:2021tii}:
\begingroup
\begin{equation}
\begin{gathered}
m_{11}^{2} = \mu_{1}^{2} + \lambda_{7} \frac{v_{3}^{2}}{2}, \quad m_{22}^{2} = \mu_{2}^{2} + \lambda_{8} \frac{v_{3}^{2}}{2}, \quad m_{12}^{2} = \lambda_{5} \frac{v_{3}^{2}}{2},
\\
\beta_{1} = 2\lambda_{1}, \quad \beta_{2} = 2\lambda_{2}, \quad \beta_{3} = \lambda_{3}, \quad \beta_{4} = 0, \quad \beta_{5} = 0.
\end{gathered}
\end{equation}
\endgroup
Then, the vacuum stability conditions are given by~\cite{Bhattacharyya:2015nca,Maniatis:2006fs} (the rightarrow means the condition can be simplified since $\beta_{4,5}$ are zero and $\tan\beta$ is much larger than $2\beta_{1}/\beta_{2}$ in the model under consideration):
\begingroup
\begin{equation}
\begin{gathered}
\beta_{1} \geq 0, \quad \beta_{2} \geq 0, \quad \beta_{3} + \sqrt{\beta_{1} \beta_{2}} \geq 0,
\\[1ex]
\beta_{3} + \beta_{4} + \sqrt{\beta_{1} \beta_{2}} > \lvert \beta_{5} \rvert \rightarrow \beta_{3} + \sqrt{\beta_{1} \beta_{2}} >0,
\\[1ex]
m_{12}^{2} \left( m_{11}^{2} - m_{22}^{2} \sqrt{\frac{\beta_{1}}{\beta_{2}}} \right) \left( \tan\beta - 2\frac{\beta_{1}}{\beta_{2}} \right) > 0 \rightarrow m_{12}^{2} \left( m_{11}^{2} - m_{22}^{2} \sqrt{\frac{\beta_{1}}{\beta_{2}}} \right) > 0,
\end{gathered}
\end{equation}
\endgroup
where the conditions are implemented in our numerical study and it is worth mentioning that the most related quartic coupling constants $\lambda_{1,2,3,5,7}$ for the vacuum stability conditions are somewhat tightly constrained in the scalar potential, so providing less freedom to the parameters and we will confirm this feature in the following numerical study.

\subsection{Muon anomalous magnetic moment $g-2$ with scalar exchange}
As we have seen that the muon $g-2$ mediated by $Z^{\prime}$ at one loop gives only negative contributions with mostly order of $10^{-11}$, we require another approach to muon $g-2$ and scalar exchange can be an answer for the approach. We confirmed that the most sizeable mixing in the charged lepton sector is left-handed $34$ mixing, which can be identified by the charged lepton mass matrix of Equation~\ref{eqn:cl_2} and by the mass hierarchy shown in Table~\ref{tab:parameter_setup_chargedleptonsector} and is not very relevant for muon $g-2$, and the other left- or right-handed angles are generally quite suppressed. However, the $34$ mixing angle eventually gets to be suppressed due to the vectorlike doublet charged lepton mass constraint, $M_{E_{4}} > 790\func{GeV}$, given by the CMS experiment~\cite{Bhattiprolu:2019vdu,CMS:2019hsm}. Therefore, it is possible to follow the approximated muon $g-2$ prediction given in our previous work~\cite{Hernandez:2021tii} without loss of generality. The relevant Yukawa interactions for muon $g-2$ can be derived from the charged lepton renormalizable Yukawa interactions of Equation~\ref{eqn:general_charged_lepton_Yukawa_Mass_Lagrangian}:
\begingroup
\begin{equation}
\mathcal{L}_{\Delta a_{\mu}} = \frac{y_{24}^{e}}{\sqrt{2}} \overline{e}_{4} \left( \func{Re} H_{d}^{0} - i\gamma^{5} \func{Im} H_{d}^{0} \right) \mu + \frac{x_{42}^{e}}{\sqrt{2}} \overline{\widetilde{e}}_{4} \left( \func{Re} \phi - i\gamma^{5} \func{Im} \phi \right) \mu + M_{44}^{e} \overline{\widetilde{e}}_{4} e_{4} + \func{H.c.},
\end{equation}
\endgroup
which give rise to the one loop muon $g-2$ diagram:
\begingroup
\begin{figure}[H]
\centering
\begin{subfigure}{0.48\textwidth}
	\includegraphics[width=1.0\textwidth]{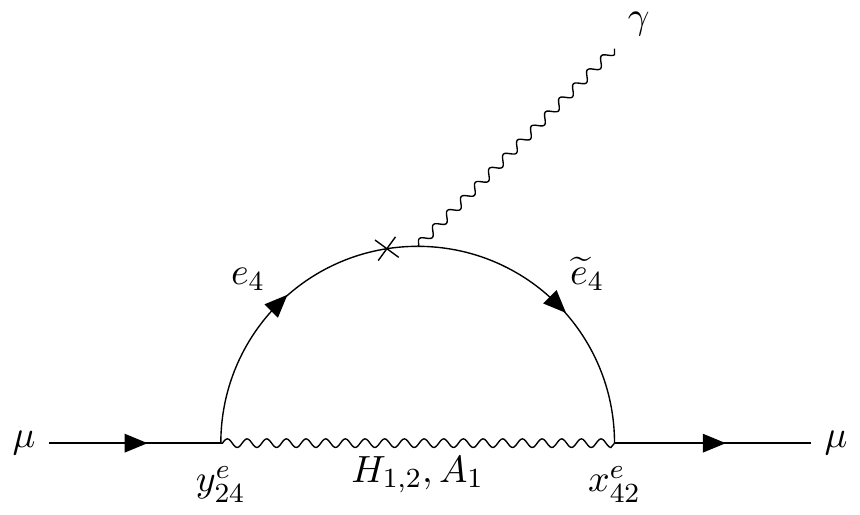}
\end{subfigure} 
\caption{One loop diagram for muon $g-2$ with scalar exchange in the model under consideration. $H_{1,2}$ are the CP-even scalars and $A_{1}$ is the CP-odd scalar.}
\end{figure}
\endgroup
The corresponding muon $g-2$ prediction with scalar exchange reads~\cite{Lindner:2016bgg,Diaz:2002uk,Jegerlehner:2009ry,Kelso:2014qka,Kowalska:2017iqv}:
\begingroup
\begin{equation}
\Delta a_{\mu} = \frac{y_{24}^{e}}{\sqrt{2}} \frac{x_{42}^{e}}{\sqrt{2}} \frac{m_{\mu}^{2}}{8\pi} \left[ (R_{e}^{T})_{22} (R_{e}^{T})_{32} I_{S}^{\mu} \left( M_{\widetilde{E}_{4}}, m_{H_{1}} \right) + (R_{e}^{T})_{23} (R_{e}^{T})_{33} I_{S}^{\mu} \left( M_{\widetilde{E}_{4}}, m_{H_{2}} \right) - (R_{o}^{T})_{22} (R_{o}^{T})_{32} I_{P}^{\mu} \left( M_{\widetilde{E}_{4}}, m_{A_{1}} \right) \right]
\label{eqn:muong2prediction_scalar}
\end{equation}
\endgroup
where $R_{e,o}$ mean the mixing matrices for the CP-even scalars of Equation~\ref{eqn:diag_CPeven} and -odd scalars of Equation~\ref{eqn:diag_CPodd} respectively and $I_{S,P}^{\mu}$ mean the loop functions for either scalar or pseudo-scalar defined by~\cite{Lindner:2016bgg,Diaz:2002uk,Jegerlehner:2009ry,Kelso:2014qka,Kowalska:2017iqv}:
\begingroup
\begin{equation}
I_{S,P}^{\mu} = \int_{0}^{1} dx \frac{x^{2} \left( 1 - x \pm \frac{M_{\widetilde{E}_{4}}}{m_{\mu}} \right)}{m_{\mu}^{2} x^{2} + \left( M_{\widetilde{E}_{4}} - m_{\mu}^{2} \right) x + m_{S,P}^{2} \left( 1 - x \right)},
\end{equation}
\endgroup
where $M_{\widetilde{E}_{4}}$ is the vectorlike singlet charged lepton mass.
\section{ELECTROWEAK PRECISION OBSERVABLES} \label{sec:VII}
The electroweak precision observables, represented by the oblique parameters $T,S$ and $U$, have been an important cornerstone to check whether theoretical observables from a BSM theory are consistent with their experimental fit. As the model under consideration extends the SM particle spectrum by a complete vectorlike family as well as one more SM-like Higgs plus a singlet flavon, it becomes important to confirm new physics contributions arising from this BSM to the oblique parameters are within their experimental bounds. On top of that, the CDF collaboration has recently reported the SM $W$ mass anomaly at $7\sigma$~\cite{CDF:2022hxs}. The interesting $W$ mass anomaly can be explained mainly by new physics contributions from the extended gauge sector and from the extended scalar sector if it generally assumes that fermion contributions led by vectorlike fermions are suppressed to the anomaly. The reason why we assume fermions contributions are suppressed is they give unacceptably large contributions to the oblique parameters when the vectorlike mass differences are sizeable, thus not possible to fit their experimental bounds at all and at the same time leading to maximal violation of the custodial symmetry. In order to avoid this catastrophe arising in the oblique parameters, many assume that vectorlike doublet fermion masses, which are main new physics contributions to the parameters in the fermion sector, are degenerate~\cite{Peskin:1991sw} and thus the vectorlike family safely does not contribute to the oblique parameters at all and the model under consideration can provide the degenerate vectorlike doublet masses as seen in Figure~\ref{fig:scanned_vectorlike_quarks} in an analytic way. We will see that fitting the oblique parameters up to $2\sigma$ at most can naturally explain not just the $W$ mass anomaly but also muon $g-2$ simultaneously in the following numerical discussion. The first task to investigate the oblique parameters and $W$ mass anomaly is to discuss the covariant derivatives of the extended scalar sector:
\begingroup
\begin{equation}
\begin{split}
D_{\mu} H_{u} &= \left( \partial_{\mu} - ig\tau^{a} A_{\mu}^{a} - ig^{\prime} Y B_{\mu} - ig_{X} q_{X} V_{\mu} \right) H_{u},
\\
D_{\mu} H_{d} &= \left( \partial_{\mu} - ig\tau^{a} A_{\mu}^{a} - ig^{\prime} Y B_{\mu} - ig_{X} q_{X} V_{\mu} \right) H_{d},
\\
D_{\mu} \phi &= \left( \partial_{\mu} - ig\tau^{a} A_{\mu}^{a} - ig^{\prime} Y B_{\mu} - ig_{X} q_{X} V_{\mu} \right) \phi,
\end{split}
\end{equation}
\endgroup
where $g,g^{\prime},g_{X}$ are coupling constants for the SM $SU(2)_{L}$ gauge bosons $A_{\mu}^{1,2,3}$, the $U(1)_{Y}$ gauge boson $B_{\mu}$ and the $U(1)^{\prime}$ gauge boson $V_{\mu}$, which will be promoted to $Z^{\prime}$ gauge boson by the Goldstone boson $G_{Z^{\prime}}$, respectively and $q_{X}$ is the $U(1)^{\prime}$ charge assigned in the particle content of Table~\ref{tab:BSM_model}. Then, the gauge sector in this BSM is described by the following Lagrangian:
\begingroup
\begin{equation}
\mathcal{L}_{\func{gauge}} = \left( D^{\mu} H_{u} \right)^{\dagger} \left( D_{\mu} H_{u} \right) + \left( D^{\mu} H_{d} \right)^{\dagger} \left( D_{\mu} H_{d} \right) + \left( D^{\mu} \phi \right)^{\dagger} \left( D_{\mu} \phi \right),
\label{eqn:Lagrangian_gauge}
\end{equation}
\endgroup
which gives rise to the squared mass matrix for the gauge bosons in the interaction basis:
\begingroup
\begin{equation}
\begin{split}
M_{\func{gauge}}^2 &= \frac{1}{4}\left( 
\begin{array}{c|ccc}
& B_{\mu} & A_{\mu}^{3} & V_{\mu} \\ \hline
B_{\mu} & g^{\prime 2} \left( v_{u}^2 + v_{d}^2 \right) & -g g^{\prime} \left( v_{u}^2 + v_{d}^2 \right) & -2g^{\prime} g_{X} \left( v_{u}^2 - v_{d}^2 \right) \\ 
A_{\mu}^{3} & -g g^{\prime} \left( v_{u}^2 + v_{d}^2 \right) & g^2 \left( v_{u}^2 + v_{d}^2 \right) & 2g g_{X} \left( v_{u}^2 - v_{d}^2 \right) \\ 
V_{\mu} & -2g^{\prime} g_{X} \left( v_{u}^2 - v_{d}^2 \right) & 2g g_{X} \left( v_{u}^2 - v_{d}^2 \right) & 4g_{X}^{2} \left( v_{u}^2 + v_{d}^2 + \frac{1}{2} v_{\phi}^2 \right)
\end{array}
\right),
\\
&= \frac{1}{4}\left( 
\begin{array}{c|ccc}
& B_{\mu} & A_{\mu}^{3} & V_{\mu} \\ \hline
B_{\mu} & g^{\prime 2} v^2 & -g g^{\prime} v^2 & -2g^{\prime} g_{X} \left( v_{u}^2 - v_{d}^2 \right) \\ 
A_{\mu}^{3} & -g g^{\prime} v^2 & g^2 v^2 & 2g g_{X} \left( v_{u}^2 - v_{d}^2 \right) \\ 
V_{\mu} & -2g^{\prime} g_{X} \left( v_{u}^2 - v_{d}^2 \right) & 2g g_{X} \left( v_{u}^2 - v_{d}^2 \right) & 4g_{X}^{2} \left( v^2 + \frac{1}{2} v_{\phi}^2 \right)
\end{array}
\right).
\end{split}
\end{equation}
\endgroup
After the weak mixing, all the entries in the first column and row of the squared mass matrix are vanished and then the familiar form for the SM $Z$ gauge boson appears in the partially rotated mass matrix:
\begingroup
\begin{equation}
M_{\func{gauge}}^2 = \frac{1}{4}\left( 
\begin{array}{c|ccc}
& B_{\mu}^{\prime} & A_{\mu}^{3 \prime} & V_{\mu} \\ \hline
B_{\mu}^{\prime} & 0 & 0 & 0 \\ 
A_{\mu}^{3 \prime} & 0 & \left( g^2 + g^{\prime 2} \right) v^2 & -2\sqrt{g^2 + g^{\prime 2}} g_{X} v^2 \\ 
V_{\mu} & 0 & -2\sqrt{g^2 + g^{\prime 2}} g_{X} v^2 & 4g_{X}^{2} \left( v^2 + \frac{1}{2}v_{\phi}^2 \right)
\end{array}
\right) 
\label{eqn:Mgaugesquared_partial}
\end{equation}
\endgroup
We arrive at the fully diagonalized squared mass matrix for the physical gauge bosons $A_{\mu}, Z_{\mu}, Z_{\mu}^{\prime}$ after carrying out $23$ mixing:
\begingroup
\begin{equation}
M_{\func{gauge}}^2 = 
\left( 
\begin{array}{c|ccc}
& A_{\mu} & Z_{\mu} & Z_{\mu}^{\prime} \\ \hline
A_{\mu} & 0 & 0 & 0 \\ 
Z_{\mu} & 0 & M_{Z}^{2} & 0 \\ 
Z_{\mu}^{\prime} & 0 & 0 & M_{Z^{\prime}}^{2}
\end{array}
\right) 
\end{equation}
\endgroup
where the $23$ mixing angle $\gamma$ is given by:
\begingroup
\begin{equation}
\sin\gamma = \frac{\sqrt{g^{2} + g^{\prime 2}} v^{2}}{\sqrt{\left( g^{2} + g^{\prime 2} \right) v^{4} + 4 g_{X}^{2} \left( v^{2} + \frac{1}{2} v_{\phi}^{2} \right)^{2}}},
\end{equation}
\endgroup
and the gauge bosons in the interactions basis are connected to those in the physical basis via the gauge boson mixing matrices:
\begingroup
\begin{equation}
\begin{split}
\begin{pmatrix}
A_{\mu} \\
Z_{\mu} \\
Z_{\mu}^{\prime}
\end{pmatrix}
&=
\begin{pmatrix}
1 & 0 & 0 \\
0 & \cos\gamma & \sin\gamma \\
0 & -\sin\gamma & \cos\gamma
\end{pmatrix}
\begin{pmatrix}
1 & 0 & 0 \\
0 & 1 & 0 \\
0 & 0 & 1
\end{pmatrix}
\begin{pmatrix}
\cos\theta_{W} & \sin\theta_{W} & 0 \\
-\sin\theta_{W} & \cos\theta_{W} & 0 \\
0 & 0 & 1
\end{pmatrix}
\begin{pmatrix}
B_{\mu} \\
A_{\mu}^{3} \\
V_{\mu}
\end{pmatrix}
\\
&=
\begin{pmatrix}
\cos\theta_{W} & \sin\theta_{W} & 0 \\
-\cos\gamma \sin\theta_{W} & \cos\theta_{W} \cos\gamma & \sin\gamma \\
\sin\theta_{W} \sin\gamma & -\cos\theta_{W} \sin\gamma & \cos\gamma
\end{pmatrix}
\begin{pmatrix}
B_{\mu} \\
A_{\mu}^{3} \\
V_{\mu}
\end{pmatrix}.
\end{split}
\end{equation}
\endgroup
The last preparation for discussing the oblique parameters is to express the charged Higgs $H_{u}^{+}, H_{d}^{-}$ in terms of neutral fields, as in the SM $W$ gauge fields $W_{\mu}^{\pm} = \frac{1}{\sqrt{2}} \left( A_{\mu}^{1} \mp iA_{\mu}^{2} \right)$, in order to relax the electric charge conservation and then to access $\Pi_{11}$ contributions of $T$ parameter in person. We rewrite the charged Higgs fields in terms of the neutral fields as follows:
\begingroup
\begin{equation}
\begin{split}
H_{u}^{\pm} &= \frac{1}{\sqrt{2}} \left( H_{u+}^{1} \mp i H_{u+}^{2} \right), 
\\
H_{d}^{\mp} &= \frac{1}{\sqrt{2}} \left( H_{d-}^{1} \pm i H_{d-}^{2} \right),
\end{split}
\end{equation}
\endgroup
where $H_{u+}^{1,2}$ and $H_{d-}^{1,2}$ are the neutral scalar fields. Then, the gauge Lagrangian of Equation~\ref{eqn:Lagrangian_gauge} can be expressed by only neutral fields and it is of great help as we explore the $\Pi_{11}$ contributions of $T$ parameter. The oblique parameters $T,S,U$ read~\cite{Peskin:1991sw,Altarelli:1990zd,Barbieri:2004qk,CarcamoHernandez:2017pei,CarcamoHernandez:2015smi}:
\begingroup
\begin{equation}
\begin{split}
T &= \left. \frac{\Pi_{33}\left( q^{2} \right) - \Pi_{11}\left( q^{2} \right)}{\alpha_{\func{EM}} \left( M_{Z} \right) M_{W}^{2}} \right|_{q^{2} = 0},
\\
S &= \left. \frac{2\sin2\theta_{W}}{\alpha_{\func{EM}} \left( M_{Z} \right)} \frac{d\Pi_{30}\left( q^{2} \right)}{dq^{2}} \right|_{q^{2}=0},
\\
U &= \left. \frac{4\sin^{2}\theta_{W}}{\alpha_{\func{EM}} \left( M_{Z} \right)} \left( \frac{d\Pi_{33}\left( q^{2} \right)}{dq^{2}} - \frac{d\Pi_{11}\left( q^{2} \right)}{dq^{2}} \right) \right|_{q^{2} = 0},
\end{split}
\end{equation}
\endgroup
where $\Pi_{ij}$ means the vacuum polarization amplitude consisting of the initial gauge boson $i$ and the final gauge boson $j$ in the interaction basis and $q$ is the external momentum for the initial and final gauge boson. In the model under consideration, the oblique parameters $T,S,U$ can be separated by their SM and NP contributions as follows:
\begingroup
\begin{equation}
\begin{split}
T &= T_{\func{SM}} + T_{\func{NP}}, \\
S &= S_{\func{SM}} + S_{\func{NP}}, \\
U &= U_{\func{SM}} + U_{\func{NP}},
\end{split}
\end{equation}
\endgroup
where we equate $T_{\func{NP}}, S_{\func{NP}}, U_{\func{NP}}$ by $\Delta T, \Delta S, \Delta U$, respectively.
\subsection{Oblique parameter $T$}
The oblique parameter $T$ is given by~\cite{Peskin:1991sw,Altarelli:1990zd,Barbieri:2004qk,CarcamoHernandez:2017pei,CarcamoHernandez:2015smi}:
\begingroup
\begin{equation}
T = \left. \frac{\Pi_{33}\left( q^{2} \right) - \Pi_{11}\left( q^{2} \right)}{\alpha_{\func{EM}} \left( M_{Z} \right) M_{W}^{2}} \right|_{q^{2} = 0},
\end{equation}
\endgroup
where $\alpha_{\func{EM}}\left( M_{Z} \right)$ is the running fine structure constant having an approximated value of $1/129$. In order to calculate the $T$ parameter, it requires to calculate the vacuum polarization amplitudes $\Pi_{33,11}$. We start from the $\Pi_{33}$ contributions first and the corresponding diagrams are given in Figure~\ref{fig:Pi33_contributions}:
\begingroup
\begin{figure}[H]
\centering
\begin{subfigure}{0.49\textwidth}
	\scalebox{0.9}{
	\includegraphics[keepaspectratio,width=1.0\textwidth]{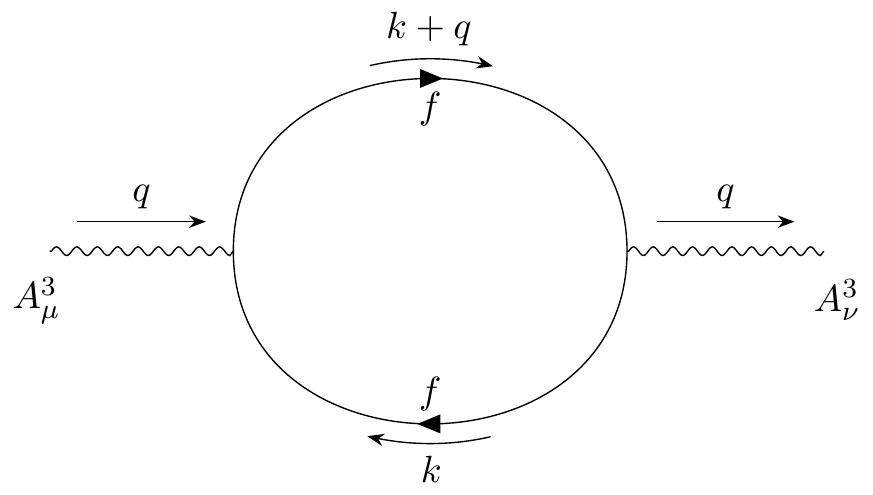}
	}
\end{subfigure} 
\begin{subfigure}{0.49\textwidth}
	\scalebox{0.9}{
	\includegraphics[keepaspectratio,width=1.0\textwidth]{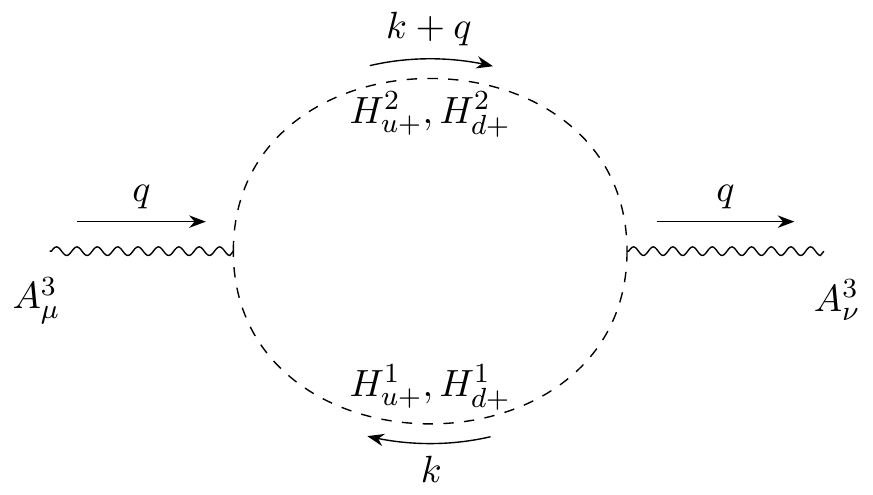}
	}
\end{subfigure} \par
\begin{subfigure}{0.49\textwidth}
	\scalebox{0.9}{
	\includegraphics[keepaspectratio,width=1.0\textwidth]{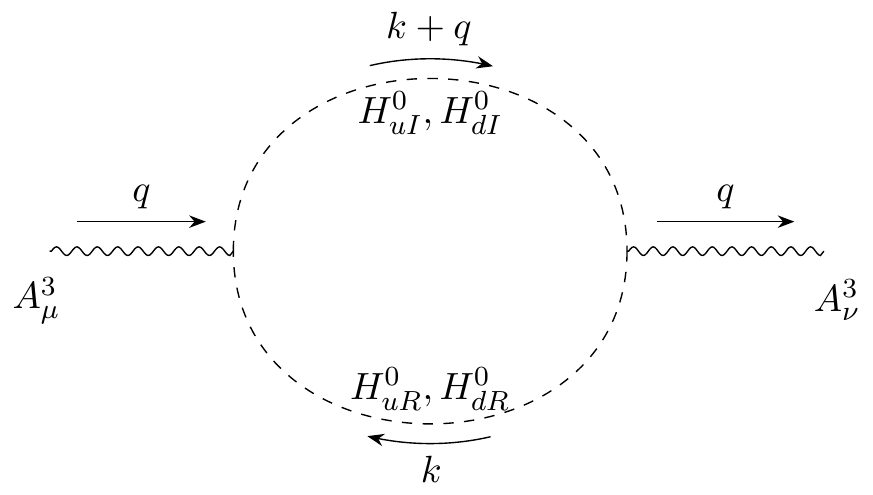}
	}
\end{subfigure} 
\begin{subfigure}{0.49\textwidth}
	\scalebox{0.9}{
	\includegraphics[keepaspectratio,width=1.0\textwidth]{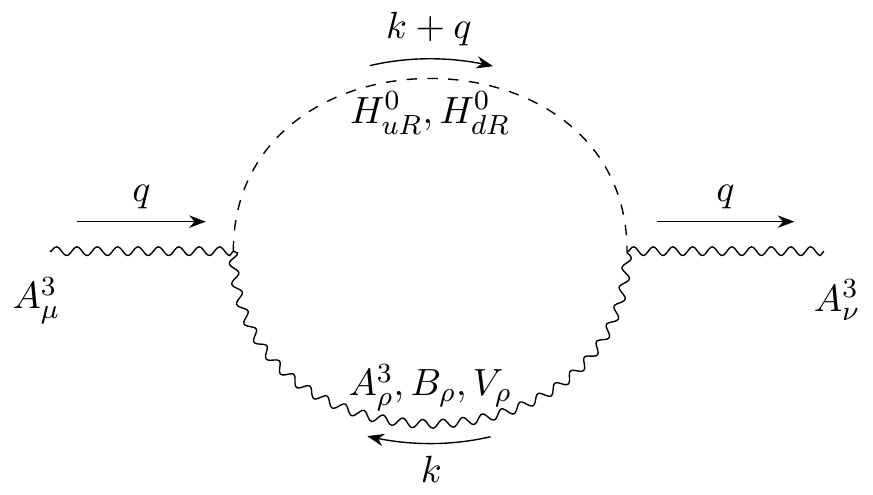}
	} 
\end{subfigure}
\caption{Feynman diagrams contributing to the vacuum polarization amplitude $\Pi_{33}$ in the model under consideration. We refer from the upper-left to lower-right diagram by $\Pi_{33}^{1,2,3,4}$ respectively for distinction. The $\Pi_{33}^{1}$ is mediated by the $SU(2)_{L}$ doublet fermions such as $t$ and $b$ quarks in the SM and vectorlike doublet fermions. The $\Pi_{33}^{2}$ is mediated by the neutral scalar fields of the charged scalar fields $H_{u,d}^{\pm}$. The $\Pi_{33}^{3}$ is mediated by the neutral scalar fields $H_{uR,uI}^{0},H_{dR,dI}^{0}$. The $\Pi_{33}^{4}$ is mediated by one gauge boson and one neutral scalar field.}
\label{fig:Pi33_contributions}
\end{figure}
\endgroup
Then, we evaluate the vacuum polarization amplitudes $\Pi_{33}^{1,2,3,4}$ and they are given in Appendix~\ref{app:A}. Next we discuss the vacuum polarization amplitude $\Pi_{11}$ and the relevant Feynman diagrams read in Figure~\ref{fig:Pi11_contributions}:
\begingroup
\begin{figure}[H]
\centering
\begin{subfigure}{0.49\textwidth}
	\scalebox{0.9}{
	\includegraphics[keepaspectratio,width=1.0\textwidth]{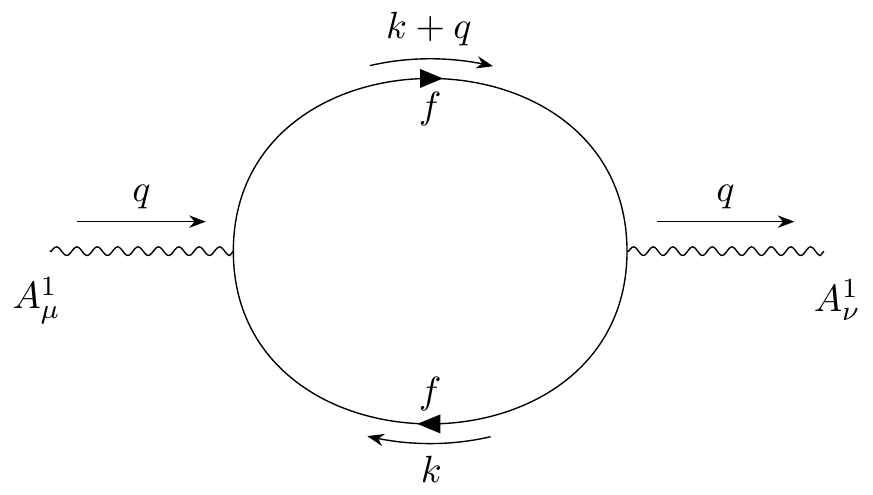}
	}
\end{subfigure} 
\begin{subfigure}{0.49\textwidth}
	\scalebox{0.9}{
	\includegraphics[keepaspectratio,width=1.0\textwidth]{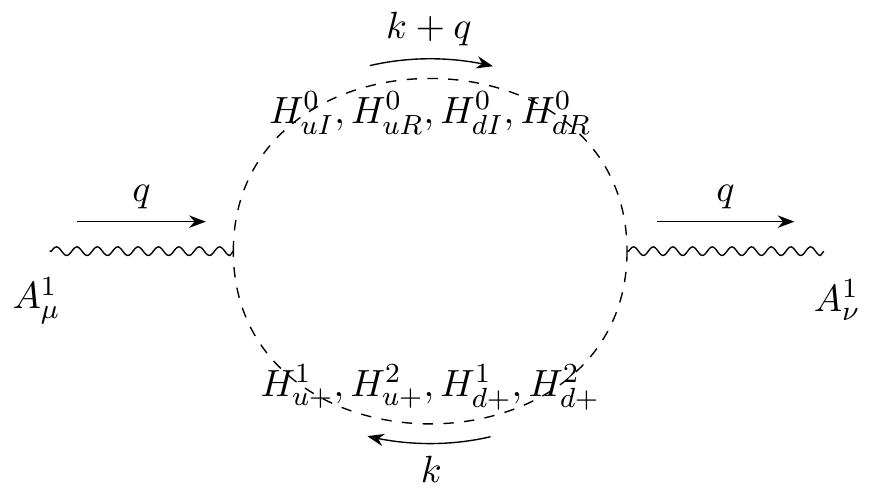}
	}
\end{subfigure} \par
\begin{subfigure}{0.49\textwidth}
	\scalebox{0.9}{
	\includegraphics[keepaspectratio,width=1.0\textwidth]{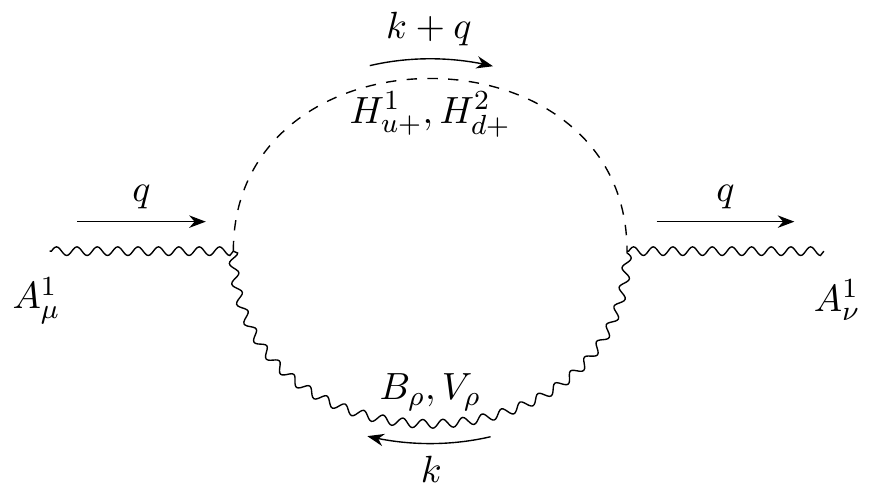}
	}
\end{subfigure} 
\begin{subfigure}{0.49\textwidth}
	\scalebox{0.9}{
	\includegraphics[keepaspectratio,width=1.0\textwidth]{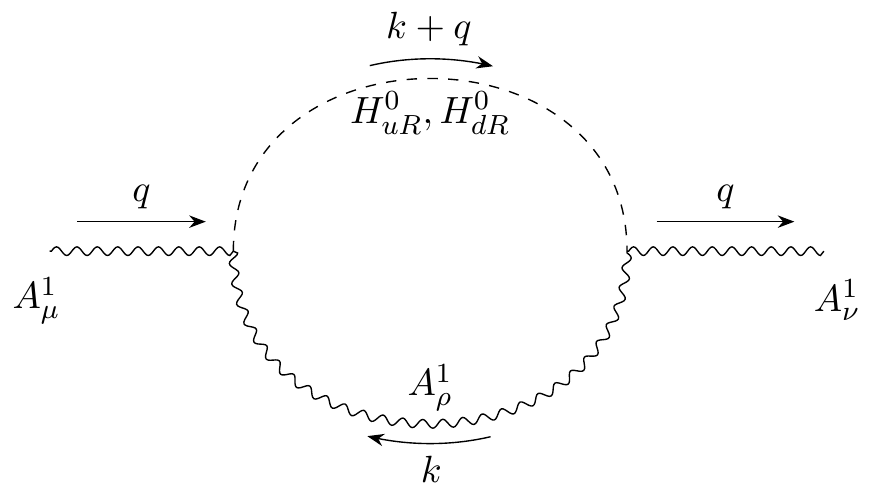}
	} 
\end{subfigure}
\caption{Feynman diagrams contributing to the vacuum polarization amplitude $\Pi_{11}$ in the model under consideration. We refer from the upper-left to lower-right diagram by $\Pi_{11}^{1,2,3,4}$ respectively for distinction. The $\Pi_{11}^{1}$ is mediated by the $SU(2)_{L}$ doublet fermions such as $t$ and $b$ quarks in the SM and vectorlike doublet fermions. The $\Pi_{11}^{2}$ is mediated by one neutral scalar field of the charged scalar fields $H_{u,d}^{\pm}$ and one neutral field. The $\Pi_{11}^{3}$ is mediated by one gauge boson and one neutral field of the charged scalars. The $\Pi_{11}^{4}$ is mediated by one gauge boson and one neutral scalar field.}
\label{fig:Pi11_contributions}
\end{figure}
\endgroup
We evaluate the vacuum polarization amplitudes $\Pi_{11}^{1,2,3,4}$ as in the $\Pi_{33}^{1,2,3,4}$ in Appendix~\ref{app:B}. The SM contribution to the $T$ parameter in this model under consideration reads:
\begingroup
\begin{equation}
T_{\func{SM}} = -\frac{3}{16\pi \cos^{2}\theta_{W}} \ln \frac{m_{h}^{2}}{M_{W}^{2}} - \frac{3}{16\pi^{2} \alpha_{\func{EM}} v_{u}^{2}} \left( \frac{m_{b}^{2}}{2} \ln \frac{m_{t}^{2}}{m_{b}^{2}} - \frac{m_{t}^{2} + m_{b}^{2}}{2} \right),
\end{equation}
\endgroup
which is well consistent with the previous result~\cite{CarcamoHernandez:2017pei}. It is worth referring that the new physics contribution $\Delta T$ of the $T$ parameter is considered for explaining the $T$ experimental bound and $W$ mass anomaly. The experimental bound for the new physics contribution of $T$ parameter at $1\sigma$ reads~\cite{Lu:2022bgw}:
\begingroup
\begin{equation}
\begin{split}
\Delta T = 0.11 \pm 0.12
\end{split}
\end{equation}
\endgroup
\subsection{Oblique parameter $S$}
The definition for the oblique parameter $S$ reads~\cite{Peskin:1991sw,Altarelli:1990zd,Barbieri:2004qk,CarcamoHernandez:2017pei,CarcamoHernandez:2015smi}:
\begingroup
\begin{equation}
\begin{split}
S = \left. \frac{2\sin2\theta_{W}}{\alpha_{\func{EM}} \left( M_{Z} \right)} \frac{d\Pi_{30}\left( q^{2} \right)}{dq^{2}} \right|_{q^{2}=0}.
\end{split}
\end{equation}
\endgroup
The relevant Feynamn diagrams for $\Pi_{30}$ contributions read in Figure~\ref{fig:Pi30_contributions}:
\begingroup
\begin{figure}[H]
\centering
\begin{subfigure}{0.49\textwidth}
	\scalebox{0.9}{
	\includegraphics[keepaspectratio,width=1.0\textwidth]{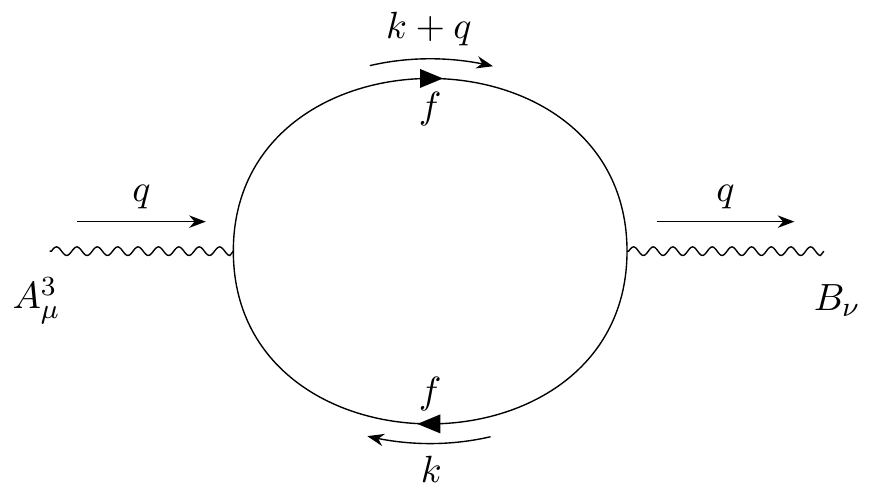}
	}
\end{subfigure} 
\begin{subfigure}{0.49\textwidth}
	\scalebox{0.9}{
	\includegraphics[keepaspectratio,width=1.0\textwidth]{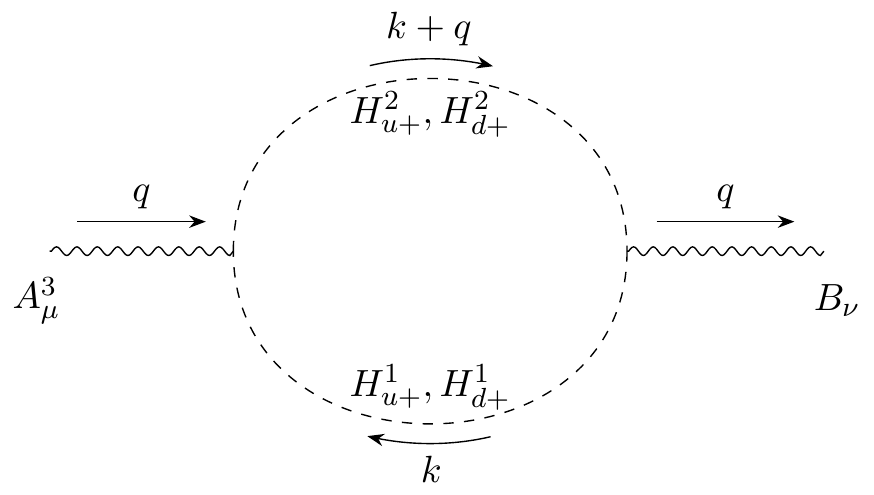}
	}
\end{subfigure} \par
\begin{subfigure}{0.49\textwidth}
	\scalebox{0.9}{
	\includegraphics[keepaspectratio,width=1.0\textwidth]{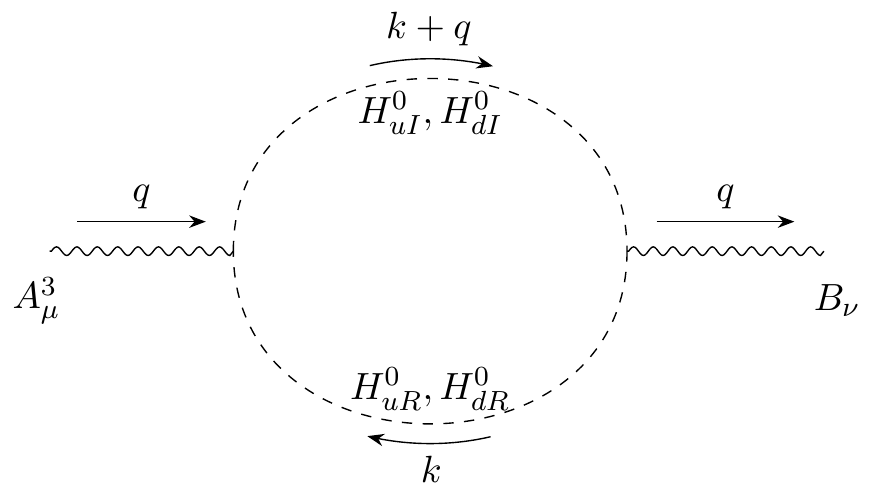}
	}
\end{subfigure} 
\begin{subfigure}{0.49\textwidth}
	\scalebox{0.9}{
	\includegraphics[keepaspectratio,width=1.0\textwidth]{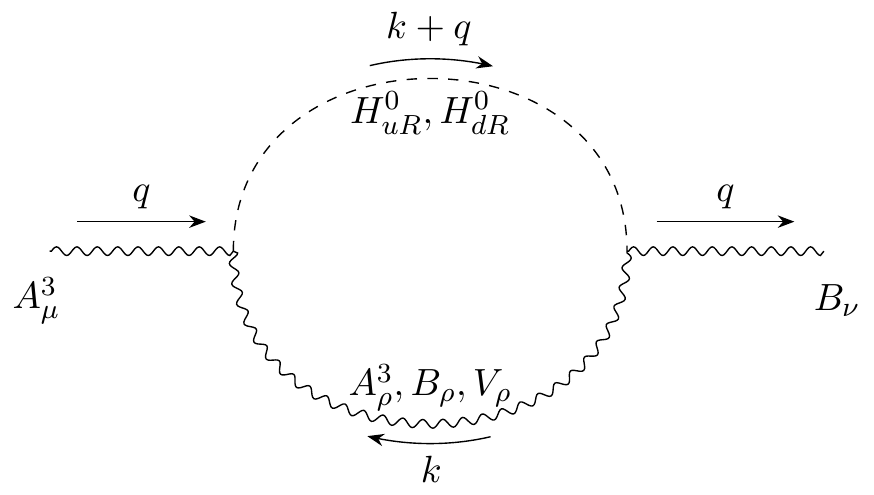}
	} 
\end{subfigure}
\caption{Feynman diagrams contributing to the vacuum polarization amplitude $\Pi_{30}$ in the model under consideration. We refer from the upper-left to lower-right diagram by $\Pi_{30}^{1,2,3,4}$ respectively for distinction. The $\Pi_{30}$ contributions are exactly same as the $\Pi_{33}$ contributions except that the outgoing gauge boson is replaced by the SM $B_{\nu}$ gauge boson in the interaction basis.}
\label{fig:Pi30_contributions}
\end{figure}
\endgroup
The analytic expressions for $\Pi_{30}$ contributions after differentiating the $\Pi_{30}$ contributions with respect to $q^{2}$ read in Appendix~\ref{app:C}. As for the SM contribution to the $S$ parameter, we found that the coefficients of our SM prediction for the $S$ parameter is not exactly consistent with the previous known result~\cite{CarcamoHernandez:2017pei}. The SM contribution to the $S$ parameter reads:
\begingroup
\begin{equation}
\begin{split}
S_{\func{SM}} = -\frac{1}{6\pi} \ln \frac{m_{t}^{2}}{m_{b}^{2}} + \left( \frac{C_{1}}{\pi} \ln \frac{m_{h}^{2}}{M_{W}^{2}} + \frac{C_{2}}{\pi} \right),
\end{split}
\end{equation}
\endgroup
where the coefficient $C_{1}$ reads
\begin{equation}
\begin{split}
C_{1} \ln \frac{m_{h}^{2}}{M_{W}^{2}} &= \frac{1}{12} \ln \frac{\Lambda^{2}}{M_{W}^{2}} - \frac{1}{12} \ln \frac{\Lambda^{2}}{m_{h}^{2}} - \left( \frac{M_{W}^{2} - m_{h}^{2}}{4} - \frac{M_{W}^{2}}{6} \right) \frac{M_{W}^{4}}{\left( M_{W}^{2} - m_{h}^{2} \right)^{3}} \ln \frac{m_{h}^{2}}{M_{W}^{2}}
\\
&- \left( \frac{1}{2} - \frac{M_{W}^{2}}{3\left( M_{W}^{2} - m_{h}^{2} \right)} \right) \frac{M_{W}^{4}}{4\left( M_{W}^{2} - m_{h}^{2} \right)^{2}} \ln \frac{m_{h}^{2}}{M_{W}^{2}} - \left( \frac{1}{2} - \frac{M_{B}^{2}}{3\left( M_{B}^{2} - m_{h}^{2} \right)} \right) \frac{M_{B}^{4}}{4\left( M_{B}^{2} - m_{h}^{2} \right)^{2}} \ln \frac{m_{h}^{2}}{M_{B}^{2}}
\\
&= \left[ \frac{1}{12} - \left( \frac{3M_{W}^{4}}{8\left( M_{W}^{2} - m_{h}^{2} \right)^{2}} - \frac{3M_{W}^{6}}{12\left( M_{W}^{2} - m_{h}^{2} \right)^{3}} \right) - \left( \frac{1}{2} - \frac{M_{B}^{2}}{3\left( M_{B}^{2} - m_{h}^{2} \right)} \right) \frac{M_{B}^{4}}{4\left( M_{B}^{2} - m_{h}^{2} \right)^{2}} \right] \ln \frac{m_{h}^{2}}{M_{W}^{2}}
\\
&\approx -0.193123 \ln \frac{m_{h}^{2}}{M_{W}^{2}}
\end{split}
\end{equation}
and the coefficient $C_{2}$ reads
\begin{equation}
\begin{split}
C_{2} &= \frac{1}{18} - \left( \frac{M_{W}^{2} - m_{h}^{2}}{4} - \frac{M_{W}^{2}}{6} \right) \left( \frac{1}{2\left( M_{W}^{2} - m_{h}^{2} \right)} + \frac{M_{W}^{2}}{\left( M_{W}^{2} - m_{h}^{2} \right)^{2}} \right)
\\
&+ \frac{M_{W}^{2}}{24\left( M_{W}^{2} - m_{h}^{2} \right)} - \left( \frac{1}{2} - \frac{M_{W}^{2}}{3\left( M_{W}^{2} - m_{h}^{2} \right)} \right) \left( \frac{M_{W}^{2}}{4\left( M_{W}^{2} - m_{h}^{2} \right)} \right)
\\
&+ \frac{M_{B}^{2}}{24\left( M_{B}^{2} - m_{h}^{2} \right)} - \left( \frac{1}{2} - \frac{M_{B}^{2}}{3\left( M_{B}^{2} - m_{h}^{2} \right)} \right) \left( \frac{M_{B}^{2}}{4\left( M_{B}^{2} - m_{h}^{2} \right)} + \frac{M_{B}^{4}}{4\left( M_{B}^{2} - m_{h}^{2} \right)^{2}} \ln \frac{1}{t_{W}^{2}} \right)
\\
&\approx 0.240712.
\end{split}
\end{equation}
Comparing our theoretical prediction for $S_{\func{SM}}$ with that in \cite{CarcamoHernandez:2017pei}, one can know that the fermion contribution to the $S$ parameter in each work is consistent, however the SM gauge sector contribution is not. On top of that, the pure coefficient without logarithmic dependence appearing in \cite{CarcamoHernandez:2017pei} $3/(2\pi)$ can not appear in our prediction since the same fermion pair (for example, $t\overline{t}$ and $b\overline{b}$) enters in the fermion one loop contributions and their analytic expression can not have the pure number. In other words, if the different SM fermions, such as $t$ and $b$ quarks in $\Pi_{11}$, enter in the loop, they can lead to a pure number by a ratio by their mass differences. Thus, we argue that the coefficient $C_{2}$ must arise in the SM scalar- and gauge-mediated one loops and the derived coefficients $C_{1}$ and $C_{2}$ are correct in the model under consideration. As in $T$ parameter, the new physics contribution to $S$ parameter is considered for fitting the $S$ parameter experimental bound and $W$ mass anomaly. The experimental bound for the new physics contribution of $S$ parameter at $1\sigma$ reads~\cite{Lu:2022bgw}:
\begingroup
\begin{equation}
\begin{split}
\Delta S = 0.06 \pm 0.10
\end{split}
\end{equation}
\endgroup
\subsection{Oblique parameter $U$}
The last oblique parameter $U$ reads~\cite{Peskin:1991sw,Altarelli:1990zd,Barbieri:2004qk,CarcamoHernandez:2017pei,CarcamoHernandez:2015smi}:
\begingroup
\begin{equation}
U = \left. \frac{4\sin^{2}\theta_{W}}{\alpha_{\func{EM}} \left( M_{Z} \right)} \left( \frac{d\Pi_{33}\left( q^{2} \right)}{dq^{2}} - \frac{d\Pi_{11}\left( q^{2} \right)}{dq^{2}} \right) \right|_{q^{2} = 0}.
\end{equation}
\endgroup
The diagrams contributing to the $U$ parameter share the exactly same structure shown in Figure~\ref{fig:Pi33_contributions} and \ref{fig:Pi11_contributions}, so we avoid putting the same diagrams for the $U$ parameter, however it requires for the vacuum polarization amplitudes $\Pi_{33}^{1,2,3,4}$ and $\Pi_{11}^{1,2,3,4}$ to be differentiated with respect to $q^{2}$. Analytical predictions for the differentiated $\Pi_{33}^{1,2,3,4}$ with respect to $q^{2}$ read in Appendix~\ref{app:D} and analytic predictions for the differentiated $\Pi_{11}^{1,2,3,4}$ with respect to $q^{2}$ read in Appendix~\ref{app:E}. The SM contribution to the $U$ parameter reads in the model under consideration:
\begingroup
\begin{equation}
\begin{split}
U_{\func{SM}} &\approx \frac{1}{2\pi} \ln \frac{m_{t}^{2}}{m_{b}^{2}} + \frac{C_{3}}{\pi} \ln \frac{m_{h}^{2}}{M_{W}^{2}} + \frac{C_{4}}{\pi}
\end{split}
\end{equation}
\endgroup
where $C_{3} = 0.421970$ and $C_{4} = -0.822971$. As in the $T$ and $S$ parameter, the only new physics contribution to $U$ parameter is considered to fit its experimental bound as well as $W$ mass anomaly. The experimental bound for the new physics contribution to $U$ parameter at $1\sigma$ reads~\cite{Lu:2022bgw}:
\begingroup
\begin{equation}
\begin{split}
\Delta U = 0.13 \pm 0.09
\end{split}
\end{equation}
\endgroup
\subsection{$W$ mass anomaly}
The $W$ mass anomaly, reporting nearly $7\sigma$ SM deviation by the CDF collaboration~\cite{CDF:2022hxs}, can be written in terms of new physics contributions to the oblique parameters $T,S,U$~\cite{Peskin:1991sw}:
\begingroup
\begin{equation}
\begin{split}
\left( M_{W}^{\func{exp}} \right)^{2} = \left( M_{W}^{\func{SM}} \right)^{2} + \frac{\alpha_{\func{EM}} \left( M_{Z} \right) \cos^{2}\theta_{W} M_{Z}^{2}}{\cos^{2}\theta_{W} - \sin^{2}\theta_{W}} \left[ -\frac{\Delta S}{2} + \cos^{2}\theta_{W} \Delta T + \frac{\cos^{2}\theta_{W} - \sin^{2}\theta_{W}}{4\sin^{2}\theta_{W}} \Delta U \right],
\label{eqn:dMW_TSU}
\end{split}
\end{equation}
\endgroup
where the SM prediction for the $W$ gauge boson reads~\cite{ATLAS:2017rzl,LHCb:2021abm,Lyons:1988rp,Valassi:2003mu}:
\begingroup
\begin{equation}
M_{W}^{\func{SM}} = 80.377 \pm 0.012 \func{GeV},
\end{equation}
\endgroup
and the experimental $W$ mass determined by the CDF collaboration reads~\cite{CDF:2022hxs}:
\begingroup
\begin{equation}
\begin{split}
M_{W}^{\func{exp}} = 80.433 \pm 0.0064 \func{GeV}
\end{split}
\end{equation}
\endgroup
The theoretical $W$ mass deviation of Equation~\ref{eqn:dMW_TSU} can be rewritten as follows:
\begingroup
\begin{equation}
\begin{split}
\Delta M_{W} = M_{W}^{\func{exp}} - M_{W}^{\func{SM}} \approx \frac{\alpha_{\func{EM}} \left( M_{Z} \right) \cos^{2}\theta_{W} M_{Z}^{2}}{2 M_{W}^{\func{SM}} \left( \cos^{2}\theta_{W} - \sin^{2}\theta_{W} \right)} \left[ -\frac{\Delta S}{2} + \cos^{2}\theta_{W} \Delta T + \frac{\cos^{2}\theta_{W} - \sin^{2}\theta_{W}}{4\sin^{2}\theta_{W}} \Delta U \right].
\end{split},
\end{equation}
\endgroup
where it is worth mentioning that the new physics contributions $\Delta T, \Delta S, \Delta U$ arise from the scalar- or gauge-mediated one loop interactions and contributions of fermion sector are forbidden for the reason that those gives unacceptably large contributions. Then, the experimental bound for the new physics contribution to the $W$ mass anomaly $\Delta M_{W}$ at $1\sigma$ reads:
\begingroup
\begin{equation}
\begin{split}
0.0496 < \Delta M_{W} < 0.0624
\end{split}
\end{equation}
\endgroup
\subsection{Numerical analysis for the scalar-mediated observables}
In this subsection, we carry out numerical scan for the scalar- and gauge-mediated observables, which are the muon $g-2$ with scalar exchange, the oblique parameters $T,S,U$, and lastly the $W$ mass anomaly $\Delta M_{W}$, and we also constrain masses of the CP-even scalars $H_{1,2}$, CP-odd scalar $A_{1}$, the non-SM charged scalars $H^{\pm}$, and lastly $Z^{\prime}$ based on the $Z^{\prime}$ mass range derived from the $R_{K^{*}}$ anomaly global fit with the CLFV $\tau \rightarrow \mu \gamma$ experimental bound alone (or equally ``theoretically interesting $Z^{\prime}$ mass range") and the muon $g-2$ experimental bound. The $Z^{\prime}$ gauge boson does not directly participate in the scalar-mediated observables, however its mass is determined by the scalar potential input parameters discussed and by the Goldstone boson $G_{Z^{\prime}}$ in the CP-odd scalar sector, therefore the $Z^{\prime}$ mass shown in this numerical study is not a free parameter any more but a physical mass and we already discussed the $M_{Z^{\prime}}$ should be put between $\left[ 510, 550 \func{GeV} \right]$ as constrained in the $R_{K^{*}}$ anomaly when the $Z^{\prime}$ coupling constant $g_{X}$ is unity in the model under consideration. We start by discussing the initial parameter setup for the numerical study.
\subsubsection{Parameter setup for the muon $g-2$ with scalar exchange and for the oblique parameters plus $W$ mass anomaly}
The initial parameter setup for the scalar- and gauge-mediated observables is given by:
\begin{table}[H]
\resizebox{\textwidth}{!}{
\centering\renewcommand{\arraystretch}{1.3} 
\begin{tabular}{cccccccc}
\toprule
\toprule
Parameter & Value & Parameter & Value & Parameter & Value & Parameter & Value \\
\midrule
$\tan\beta$ & $50$ & $y_{24}^{e}$ & $\pm [0.1,1.0]$ & $m_{24}$ & $y_{24}^{e} v_{2}$ & $\lambda_{1}$ & $\frac{\left( m_{h}^{2} - 2 v_{2}^{2} \lambda_{3} \right)}{4 v_{1}^{2}}$ \\
$v_{1}$ & $\frac{\tan\beta}{\sqrt{1 + \tan^2 \beta}} \times 246$ & $y_{34}^{e}$ & $\pm [0.1,2.0]$ & $m_{34}$ & $y_{34}^{e} v_{2}$ & $\lambda_{2}$ & $[0.5,12.0]$ \\
$v_{2}$ & $\frac{1}{\sqrt{1 + \tan^2 \beta}} \times 246$ & $x_{34}^{L}$ & $\pm [0.1,0.4]$ & $m_{35}$ & $x_{34}^{L} v_{3}$ & $\lambda_{3}$ & $-[0.1,3.0]$ \\
$v_{3}$ & $\pm \left[ 300, 500 \right]$ & $y_{43}^{e}$ & $\pm [1.0,2.0]$ & $m_{43}$ & $y_{43}^{e} v_{2}$ & $\lambda_{4}$ & $\pm [0.5,12.0]$ \\
 &  & $x_{42}^{e}$ & $\pm [0.1,0.4]$ & $m_{52}$ & $x_{42}^{e} v_{3}$ & $\lambda_{5}$ & $\frac{4 v_{1} v_{2}}{v_{3}^{2}} \lambda_{3}$ \\
 &  & $x_{43}^{e}$ & $\pm [0.1,0.4]$ & $m_{53}$ & $x_{43}^{e} v_{3}$ & $\lambda_{6}$ & $\left[ 10^{-5}, 10^{-2} \right] \times \pm [0.5,12.0]$ \\
 &  & $M_{45}^{L}$ & $\pm [150,2000]$ & $M_{45}^{L}$ & $M_{45}^{L}$ & $\lambda_{7}$ & $4 \frac{v_{2}^{2}}{v_{3}^{2}} \lambda_{3}$ \\
 &  & $M_{54}^{e}$ & $\pm [150,2000]$ & $M_{54}^{e}$ & $M_{54}^{e}$ & $\lambda_{8}$ & $[0.5,12.0]$ \\
\bottomrule
\bottomrule
\end{tabular}}%
\caption{Input parameters for numerical scan in the charged lepton and scalar sector in the model under consideration.}
\label{tab:parameter_setup_scalar_mediated_observables}
\end{table}
There are a few important conditions implemented in the parameter setup of Table~\ref{tab:parameter_setup_scalar_mediated_observables}:
\begingroup
\begin{enumerate}
\item The vev $v_{1}$ is very close to the SM Higgs vev $246\func{GeV}$ whereas $v_{2}$ runs from $1$ to $10\func{GeV}$, and they satisfy the relation $v_{1}^{2} + v_{2}^{2} = \left( 246\func{GeV} \right)^{2}$. The vev $v_{3}$ is a free parameter and is set to fit the $Z^{\prime}$ mass range, $510 < M_{Z^{\prime}} < 550$, constrained by the $R_{K^{*}}$ anomaly.
\item It requires to scan the charged lepton sector as the singlet vectorlike charged lepton $\widetilde{E}_{4}$ is relevant for the muon $g-2$ with scalar exchange at one loop level. We follow the numerical study done in our previous work~\cite{CarcamoHernandez:2021yev} and respect all the conditions for the charged lepton sector scan, which are $m_{\mu,\tau}^{\func{pred}} < \left[ 1 \pm 0.1 \right] m_{\mu,\tau}^{\func{exp}}$ and $\theta_{23} < 0.2$ and all the charged lepton Yukawa constants should be within their perturbative limit in the entire fitting processes.
\item The vacuum stability conditions are imposed in the quartic coupling constants $\lambda_{1,2,3}$ and the vacuum stability conditions provide less freedom to the relevant quartic coupling constants. On top of that, the decoupling limit is implemented in the quartic coupling constants $\lambda_{3,5,7}$, so consequently the SM Higgs $h$ does not mix with the other CP-even scalars $H_{1,2}$ in the model under consideration. As in the charged leptonic Yukawa constants, all the quartic coupling constants should be within their perturbative limit in the entire fitting processes.
\item The quartic coupling constant $\lambda_{6}$, mainly relevant for mass of the CP-even scalar $H_{2}$, has a pre-factor of range $\left[ 10^{-5}, 10^{-2} \right]$ and this is the reason that we want to find correlations between the scalar masses.
\item The vectorlike doublet charged lepton $E_{4}$ does not play a leading role for the observables considered in this numerical scan. However, its experimental constraint given by the CMS experiment~\cite{Bhattiprolu:2019vdu,CMS:2019hsm}, $M_{E_{4}} > 790\func{GeV}$, must be kept and this condition is imposed in the entire fitting processes as already reflected in Figure~\ref{fig:numerical_scan_chargedlepton1}.
\end{enumerate}
\endgroup
Next we set up a fitting function $\chi^{2}$ to find a better input parameter for each one as follows:
\begingroup
\begin{equation}
\begin{split}
\chi^{2} &= \frac{\left( m_{h}^{\func{pred}} - m_{h}^{\func{exp}} \right)^{2}}{\left( \delta m_{h} \right)^{2}} + \frac{\left( a_{hWW}^{\func{pred}} - a_{hWW}^{\func{exp}} \right)^{2}}{\left( \delta a_{hWW} \right)^{2}} + \frac{\left( R_{\gamma\gamma}^{\func{pred}} - R_{\gamma\gamma}^{\func{exp}} \right)^{2}}{\left( \delta R_{\gamma\gamma} \right)^{2}} + \frac{\left( \Delta a_{\mu}^{\func{pred}} - \Delta a_{\mu}^{\func{exp}} \right)^{2}}{\left( \delta \Delta a_{\mu} \right)^{2}}
\\
&+ \frac{\left( \Delta T^{\func{pred}} - \Delta T^{\func{exp}} \right)^{2}}{\left( \delta \Delta T \right)^{2}} + \frac{\left( \Delta S^{\func{pred}} - \Delta S^{\func{exp}} \right)^{2}}{\left( \delta \Delta S \right)^{2}} + \frac{\left( \Delta U^{\func{pred}} - \Delta U^{\func{exp}} \right)^{2}}{\left( \delta \Delta U \right)^{2}} + \frac{\left( \Delta M_{W}^{\func{pred}} - \Delta M_{W}^{\func{exp}} \right)^{2}}{\left( \delta \Delta M_{W} \right)^{2}},
\end{split}
\end{equation}
\endgroup
where the superscript ``pred" means our prediction to the physical quantity and $\delta$ means $1\sigma$ deviation from the central value of the physical quantity. The predicted SM Higgs mass $m_{h}$, the deviation factor $a_{hWW}$ and disphoton signal strength $R_{\gamma\gamma}$ are protected by the decoupling limit, so they are very close to the experimental Higgs mass and unity in the model under consideration and thus we focus further on the rest of the observables such as $\Delta a_{\mu}, \Delta T, \Delta S, \Delta U$, and $\Delta M_{W}$. Using the fitting function, the varied input parameters are given in Table~\ref{tab:parameter_setup_scalar_observables_varied}:
\begin{table}[H]
\resizebox{\textwidth}{!}{
\centering\renewcommand{\arraystretch}{1.3} 
\begin{tabular}{cccccccc}
\toprule
\toprule
Parameter & Value & Parameter & Value & Parameter & Value & Parameter & Value \\
\midrule
$\tan\beta$ & $50$ & $y_{24}^{e}$ & $\left[ 1 \pm \kappa \right] \times y_{24r}^{e}$ & $m_{24}$ & $y_{24}^{e} v_{2}$ & $\lambda_{1}$ & $\frac{\left( m_{h}^{2} - 2 v_{2}^{2} \lambda_{3} \right)}{4 v_{1}^{2}}$ \\
$v_{1}$ & $\frac{\tan\beta}{\sqrt{1 + \tan^2 \beta}} \times 246$ & $y_{34}^{e}$ & $\left[ 1 \pm \kappa \right] \times y_{34r}^{e}$ & $m_{34}$ & $y_{34}^{e} v_{2}$ & $\lambda_{2}$ & $\left[ 1 \pm \kappa \right] \times \lambda_{2r}$ \\
$v_{2}$ & $\frac{1}{\sqrt{1 + \tan^2 \beta}} \times 246$ & $x_{34}^{L}$ & $\left[ 1 \pm \kappa \right] \times x_{34r}^{L}$ & $m_{35}$ & $x_{34}^{L} v_{3}$ & $\lambda_{3}$ & $\left[ 1 \pm \kappa \right] \times \lambda_{3r}$ \\
$v_{3}$ & $\left[ 1 \pm \kappa \right] \times v_{3r}$ & $y_{43}^{e}$ & $\left[ 1 \pm \kappa \right] \times y_{43r}^{e}$ & $m_{43}$ & $y_{43}^{e} v_{2}$ & $\lambda_{4}$ & $\left[ 1 \pm \kappa \right] \times \lambda_{4r}$ \\
  &  & $x_{42}^{e}$ & $\left[ 1 \pm \kappa \right] \times x_{42r}^{e}$ & $m_{52}$ & $x_{42}^{e} v_{3}$ & $\lambda_{5}$ & $\frac{4 v_{1} v_{2}}{v_{3}^{2}} \lambda_{3}$ \\
  &  & $x_{43}^{e}$ & $\left[ 1 \pm \kappa \right] \times x_{43r}^{e}$ & $m_{53}$ & $x_{43}^{e} v_{3}$ & $\lambda_{6}$ & $\left[ 1 \pm \kappa \right] \times \lambda_{6r}$ \\
  &  & $M_{45}^{L}$ & $\left[ 1 \pm \kappa \right] \times M_{45r}^{L}$ & $M_{45}^{L}$ & $M_{45}^{L}$ & $\lambda_{7}$ & $4 \frac{v_{2}^{2}}{v_{3}^{2}} \lambda_{3}$ \\ \cmidrule{1-2}
\multicolumn{2}{c}{$\kappa = 0.1$} & $M_{54}^{e}$ & $\left[ 1 \pm \kappa \right] \times M_{54r}^{e}$ & $M_{54}^{e}$ & $M_{54}^{e}$ & $\lambda_{8}$ & $\left[ 1 \pm \kappa \right] \times \lambda_{8r}$ \\ 
\bottomrule
\bottomrule
\end{tabular}}%
\caption{Varied input parameters by a factor of $\left[ 1 \pm \kappa \right]$ where $\kappa = 0.1$ The parameters having an index $r$ in the subscript mean ones obtained in the previous numerical scan.}
\label{tab:parameter_setup_scalar_observables_varied}
\end{table}
\subsubsection{Numerical scan result for the scalar-mediated observables}
After repeating the varying process many times, we find a most fitted benchmark point with $\chi^{2} = 5.64216$ and collect benchmark points whose $\chi^{2}$ is less than $8.0$. The numerical results with the collected benchmark points read in both Figure~\ref{fig:scalar-mediated_observables1} and \ref{fig:scalar-mediated_observables2}. We start by analyzing the numerical results shown in Figure~\ref{fig:scalar-mediated_observables1} first:
\newpage

\begingroup
\begin{figure}[H]
\centering
\begin{subfigure}{0.49\textwidth}
	\scalebox{0.90}{
	\includegraphics[keepaspectratio,width=1.0\textwidth]{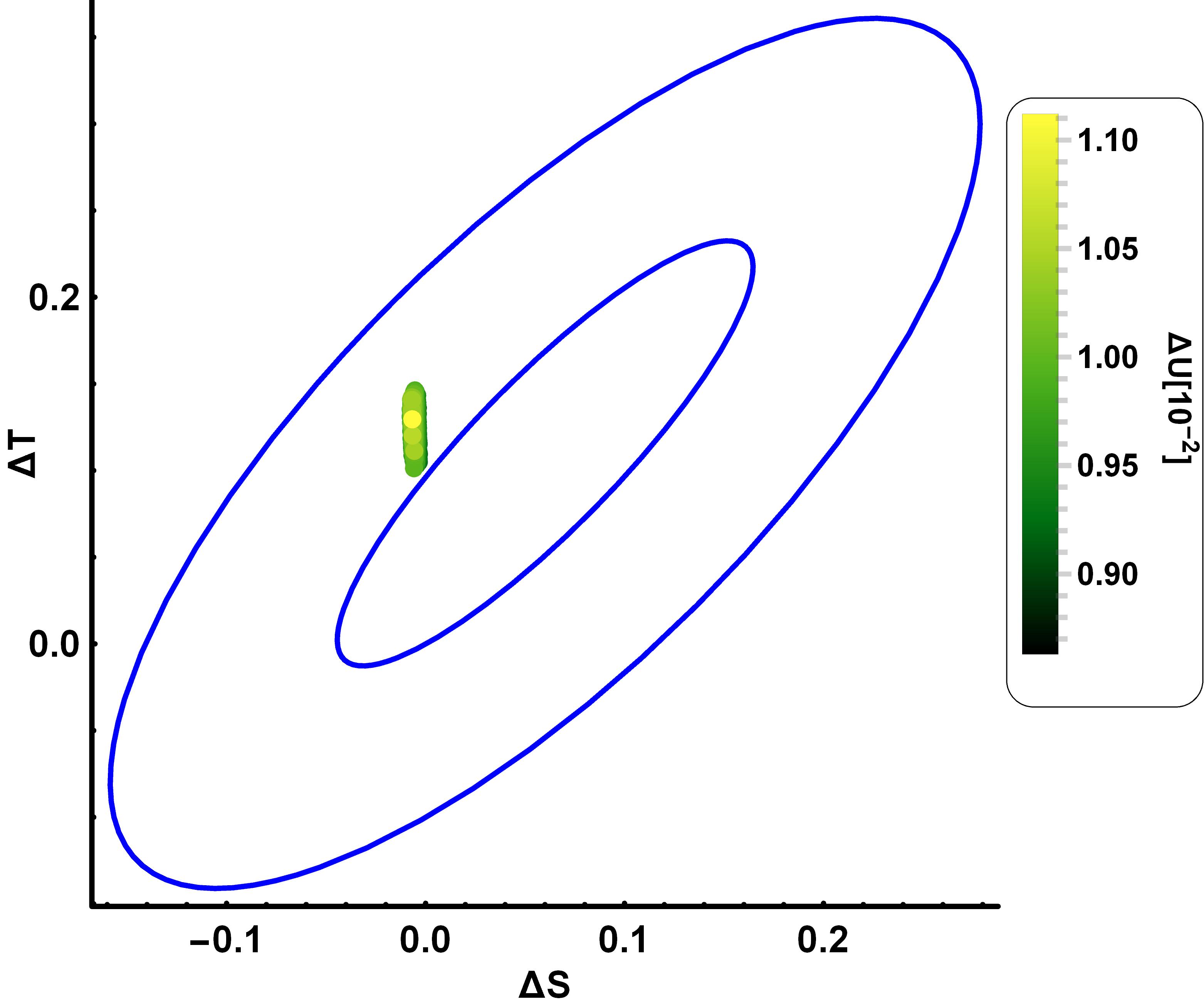} 
	}
\end{subfigure} 
\begin{subfigure}{0.49\textwidth}
	\scalebox{0.90}{
	\includegraphics[keepaspectratio,width=1.0\textwidth]{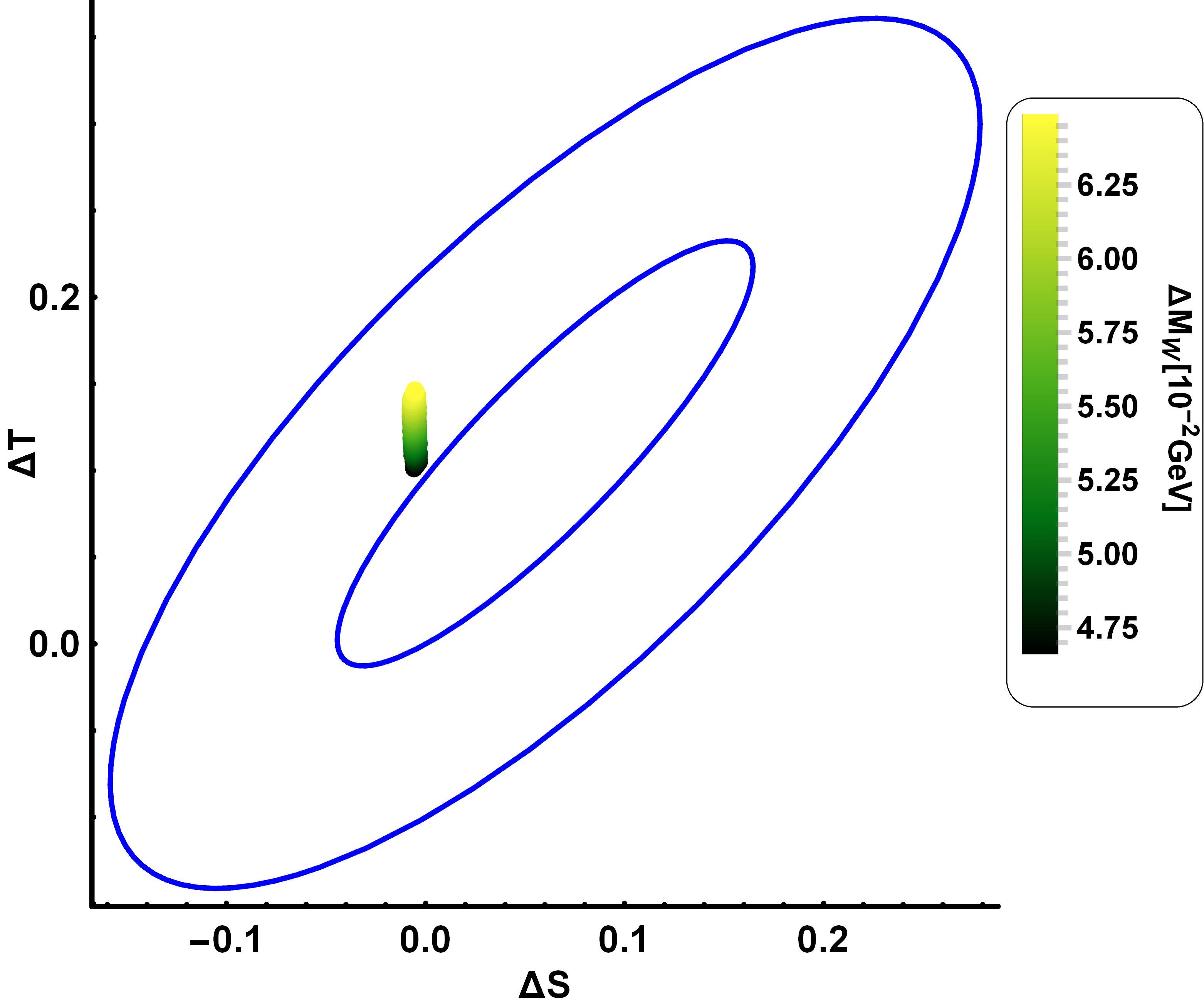}
	} 
\end{subfigure} \par
\begin{subfigure}{0.49\textwidth}
	\scalebox{0.90}{
	\includegraphics[keepaspectratio,width=1.0\textwidth]{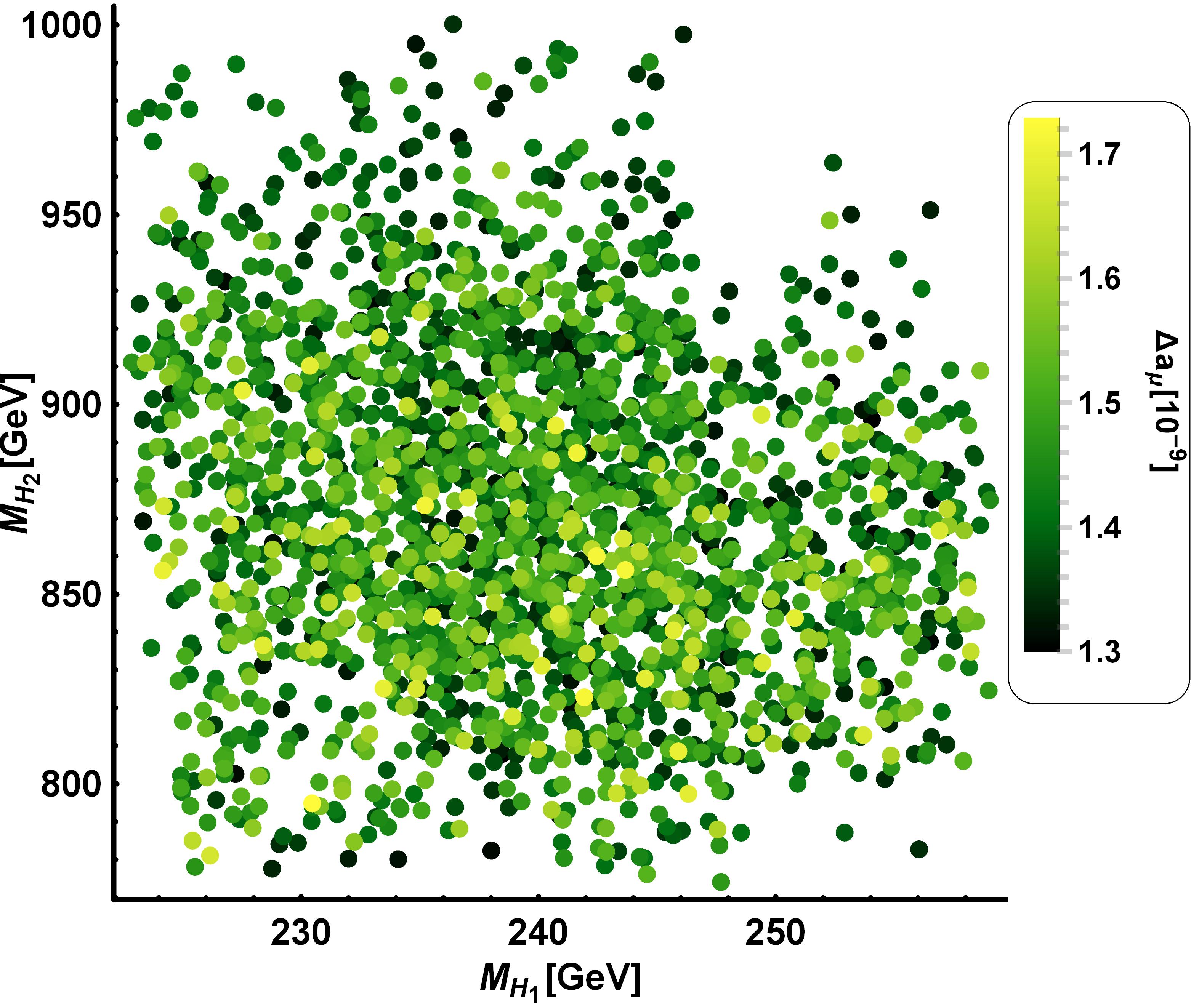}
	}
\end{subfigure} 
\begin{subfigure}{0.49\textwidth}
	\scalebox{0.90}{
	\includegraphics[keepaspectratio,width=1.0\textwidth]{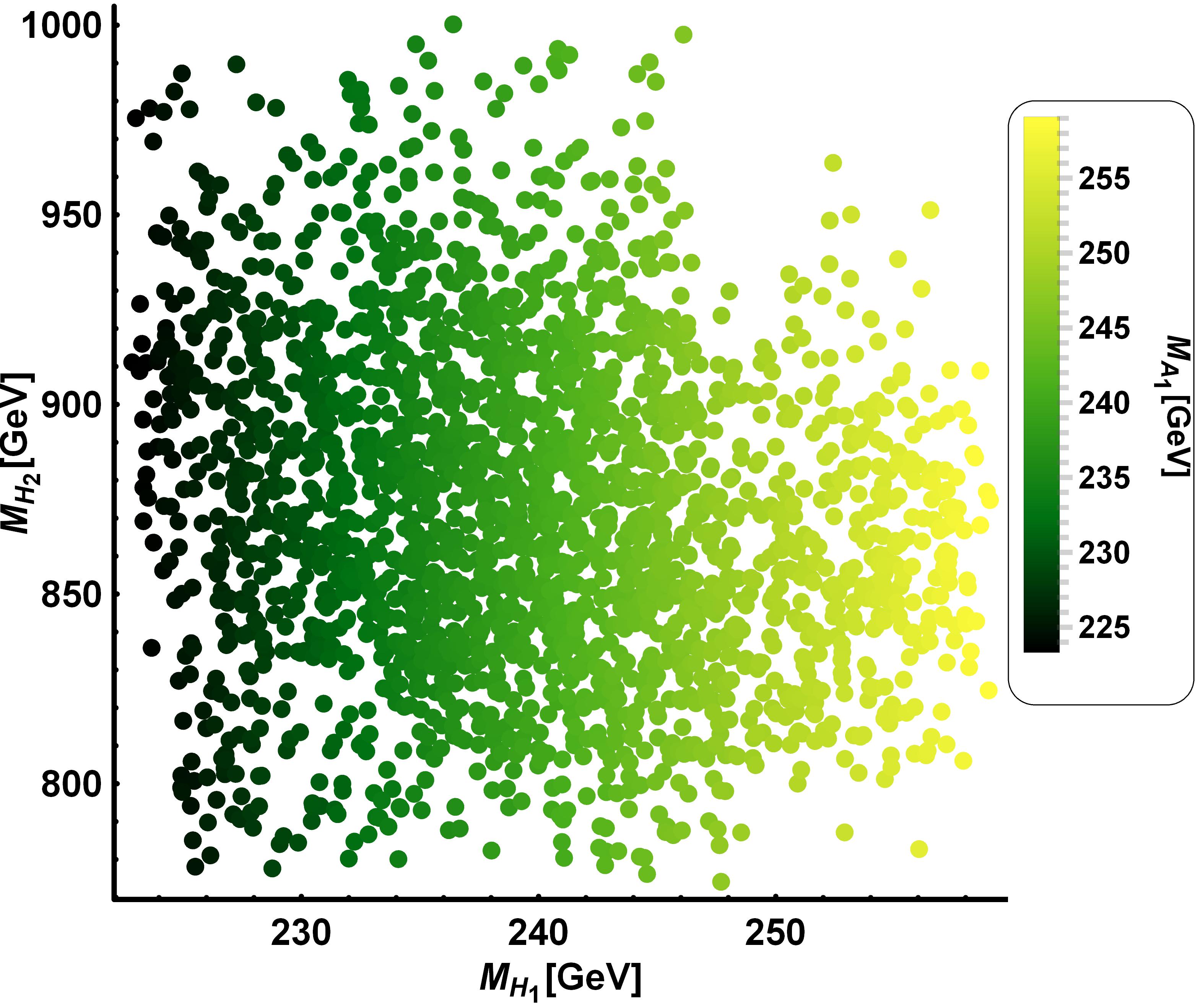}
	}
\end{subfigure} \par
\begin{subfigure}{0.49\textwidth}
	\scalebox{0.9}{
	\includegraphics[keepaspectratio,width=1.0\textwidth]{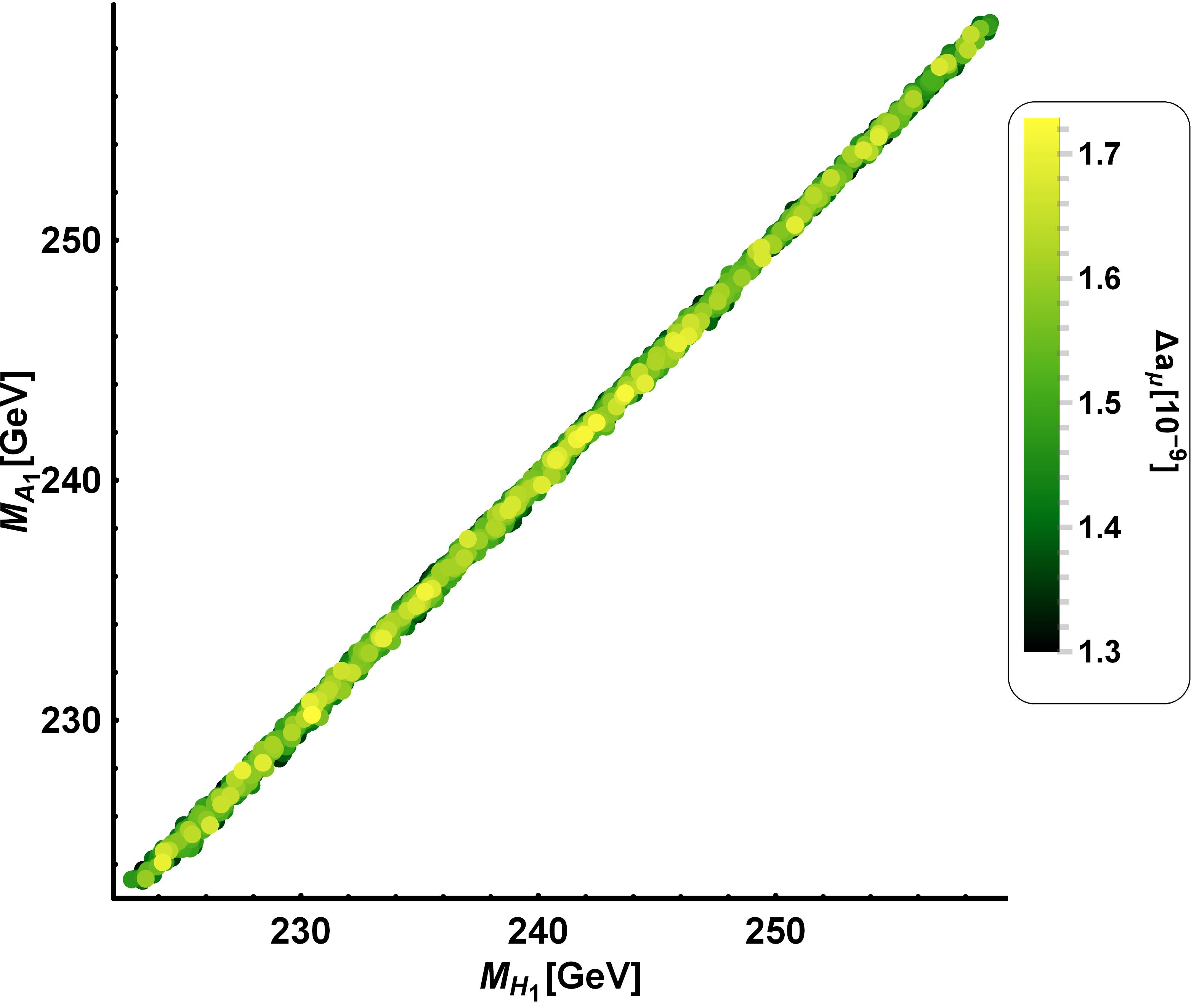}
	}
\end{subfigure} 
\begin{subfigure}{0.49\textwidth}
	\scalebox{0.9}{
	\includegraphics[keepaspectratio,width=1.0\textwidth]{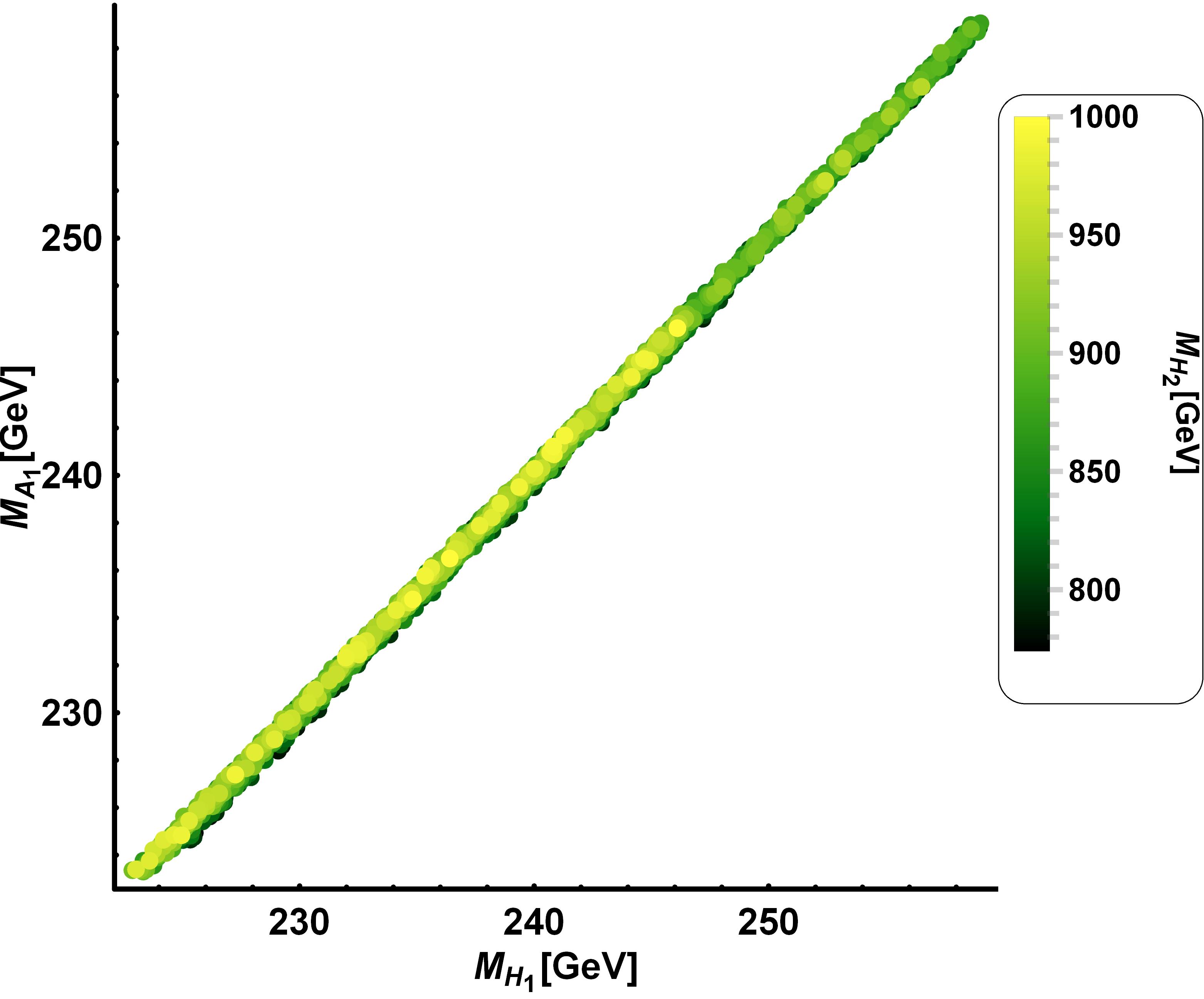}
	} 
\end{subfigure}
\caption{First numerical scan result for the scalar-mediated observables. The upper-left plot shows that all the oblique parameters $T,S,U$ can be explained up to their $1, 2\sigma$ constraints at most. The upper-right plot shows the $W$ mass anomaly can be mainly explained at its $1\sigma$ constraint. The middle two plots show relations between masses of the CP-even scalar $H_{1,2}$ against the muon $g-2$ (left) with scalar exchange and against the CP-odd scalar $A_{1}$ (right). The muon $g-2$ in the middle left plot tells that it can be explained by its $2\sigma$ constraint at most. The lower two plots show a strong correlation between masses of the CP-even scalar $H_{1}$ and -odd scalar $A_{1}$ against the muon $g-2$ (left) and against mass of $H_{2}$ (right).}
\label{fig:scalar-mediated_observables1}
\end{figure}
\endgroup
The upper-left plot of Figure~\ref{fig:scalar-mediated_observables1} tells that the oblique parameters $T,S,U$ can be fitted up to their $1$ or $2$ sigma constraints. The upper-right plot has the same $x$ and $y$ axis of the upper-left plot, however the $U$ parameter in the legend of the left is replaced by the $W$ mass anomaly and it can be mainly fitted up to its $1\sigma$ constraint. The middle two plots display how masses of the CP-even scalar $H_{1}$ are distributed against those of the CP-even scalar $H_{2}$ with the muon $g-2$ reported (left) and with mass range of the CP-odd scalar $A_{1}$ (right). For the lower two plots, we investigated relations between masses of $A_{1}$ and $H_{1}$ with the muon $g-2$ (left) and with masses of $H_{2}$ (right) and they show a very strong correlation between the mass parameters. Next we discuss other new physics parameters such as mass of $Z^{\prime}$, of vectorlike singlet charged lepton $\widetilde{E}_{4}$, and of non-SM charged scalars $H^{\pm}$ in the second numerical results in Figure~\ref{fig:scalar-mediated_observables2}:
\begingroup
\begin{figure}[H]
\centering
\begin{subfigure}{0.49\textwidth}
	\scalebox{0.9}{
	\includegraphics[keepaspectratio,width=1.0\textwidth]{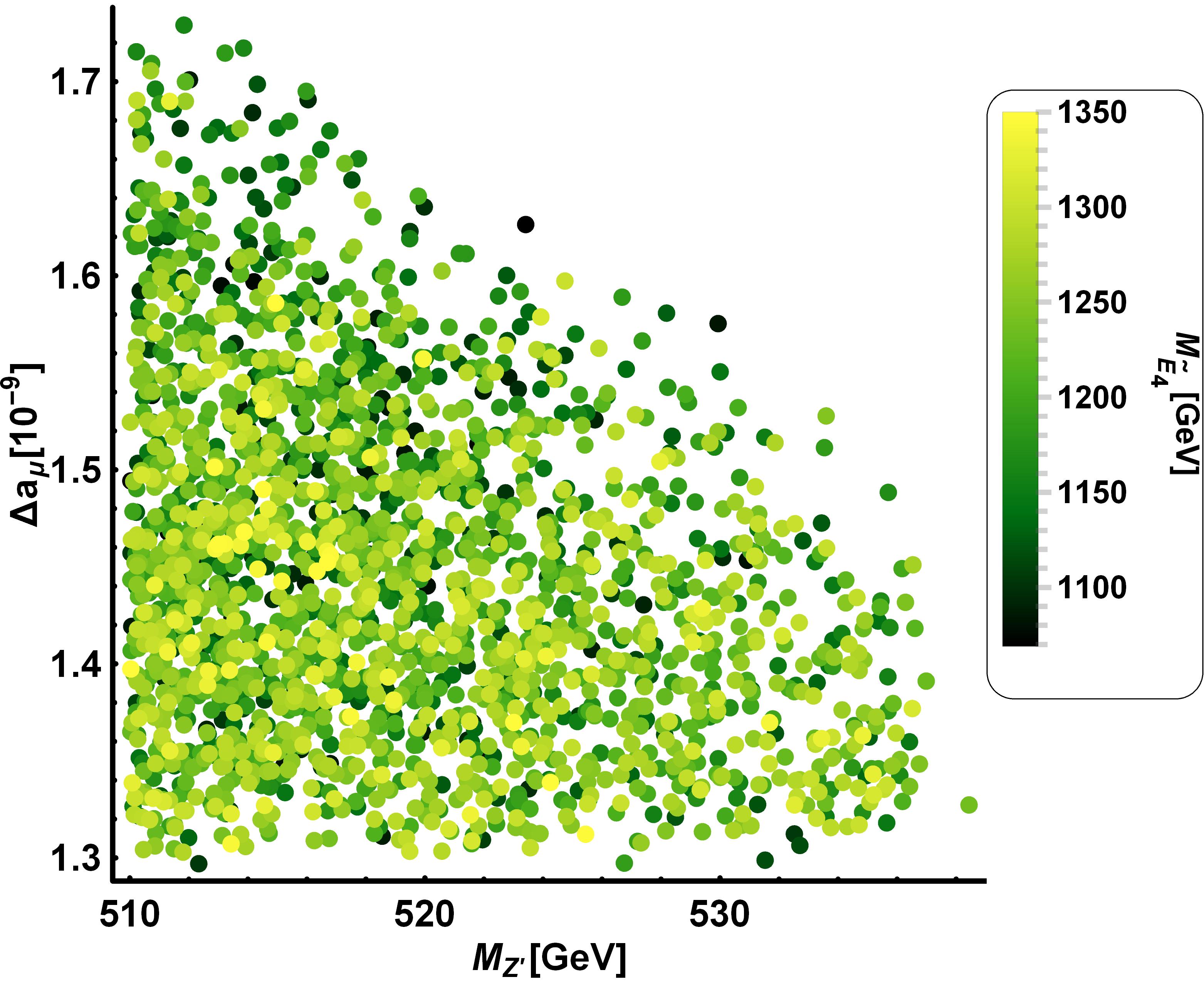}
	}
\end{subfigure} 
\begin{subfigure}{0.49\textwidth}
	\scalebox{0.9}{
	\includegraphics[keepaspectratio,width=1.0\textwidth]{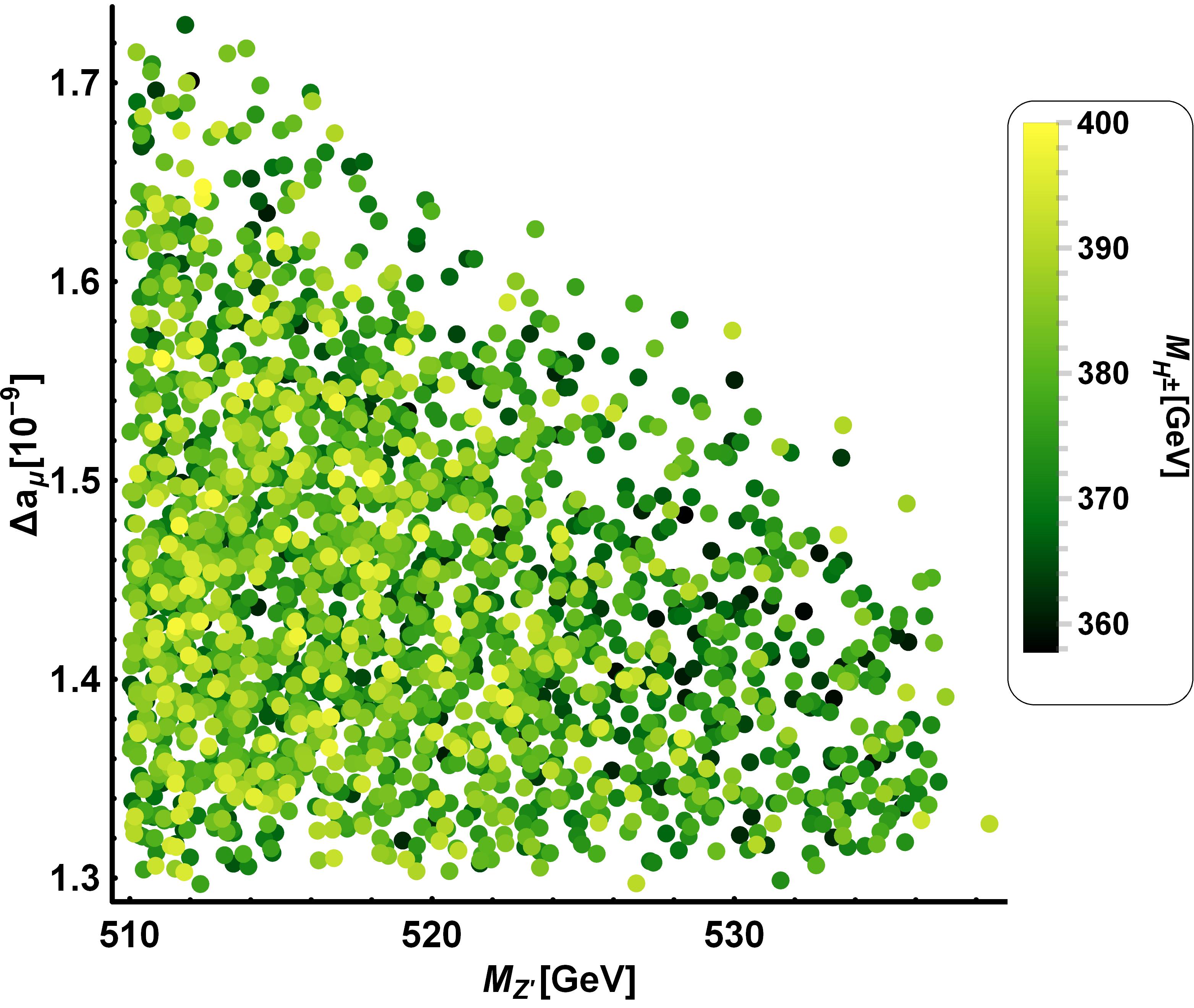}
	}
\end{subfigure} \par
\begin{subfigure}{0.49\textwidth}
	\scalebox{0.9}{
	\includegraphics[keepaspectratio,width=1.0\textwidth]{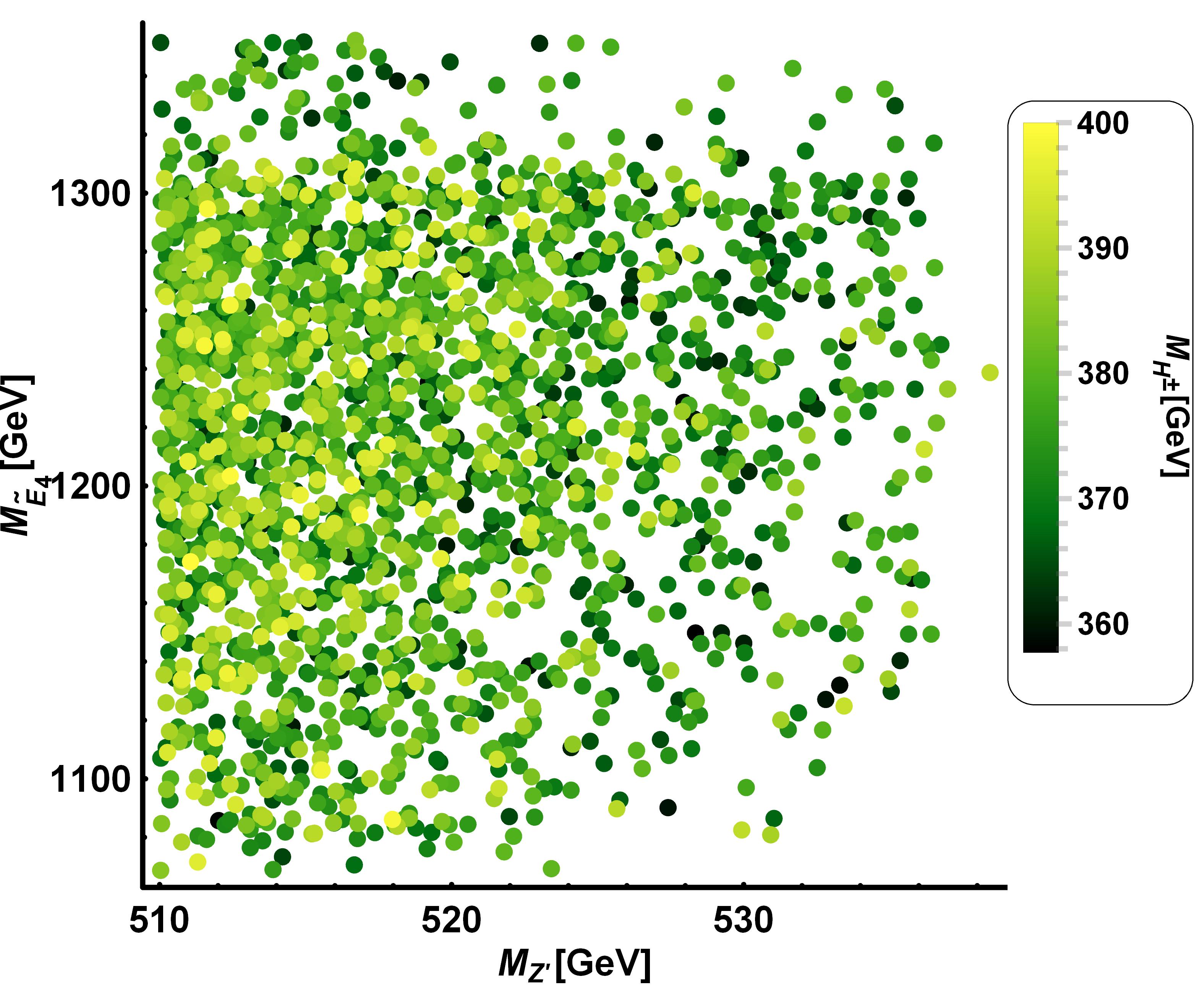}
	}
\end{subfigure} 
\begin{subfigure}{0.49\textwidth}
	\scalebox{0.9}{
	\includegraphics[keepaspectratio,width=1.0\textwidth]{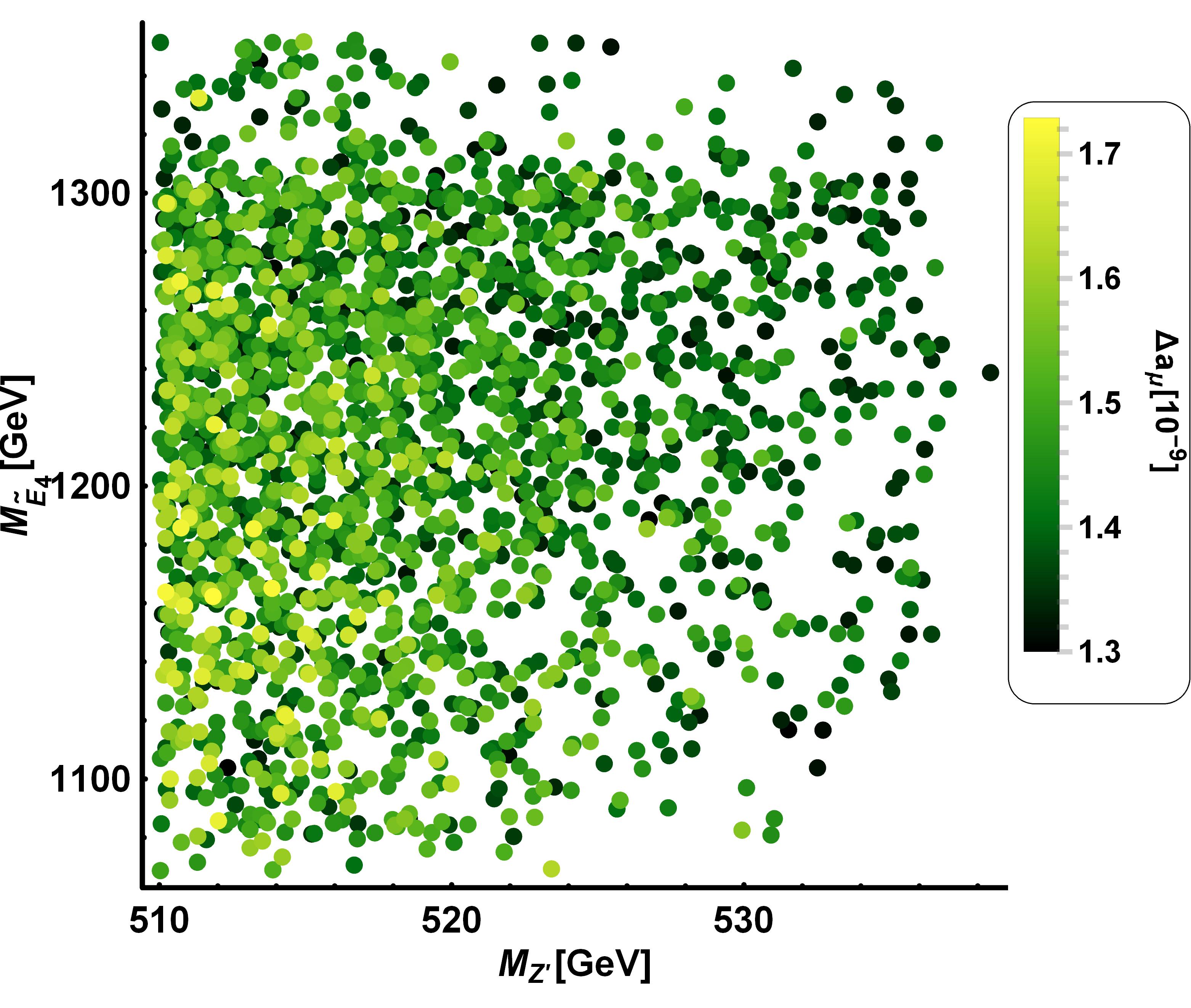}
	}
\end{subfigure} \par
\begin{subfigure}{0.49\textwidth}
	\scalebox{0.9}{
	\includegraphics[keepaspectratio,width=1.0\textwidth]{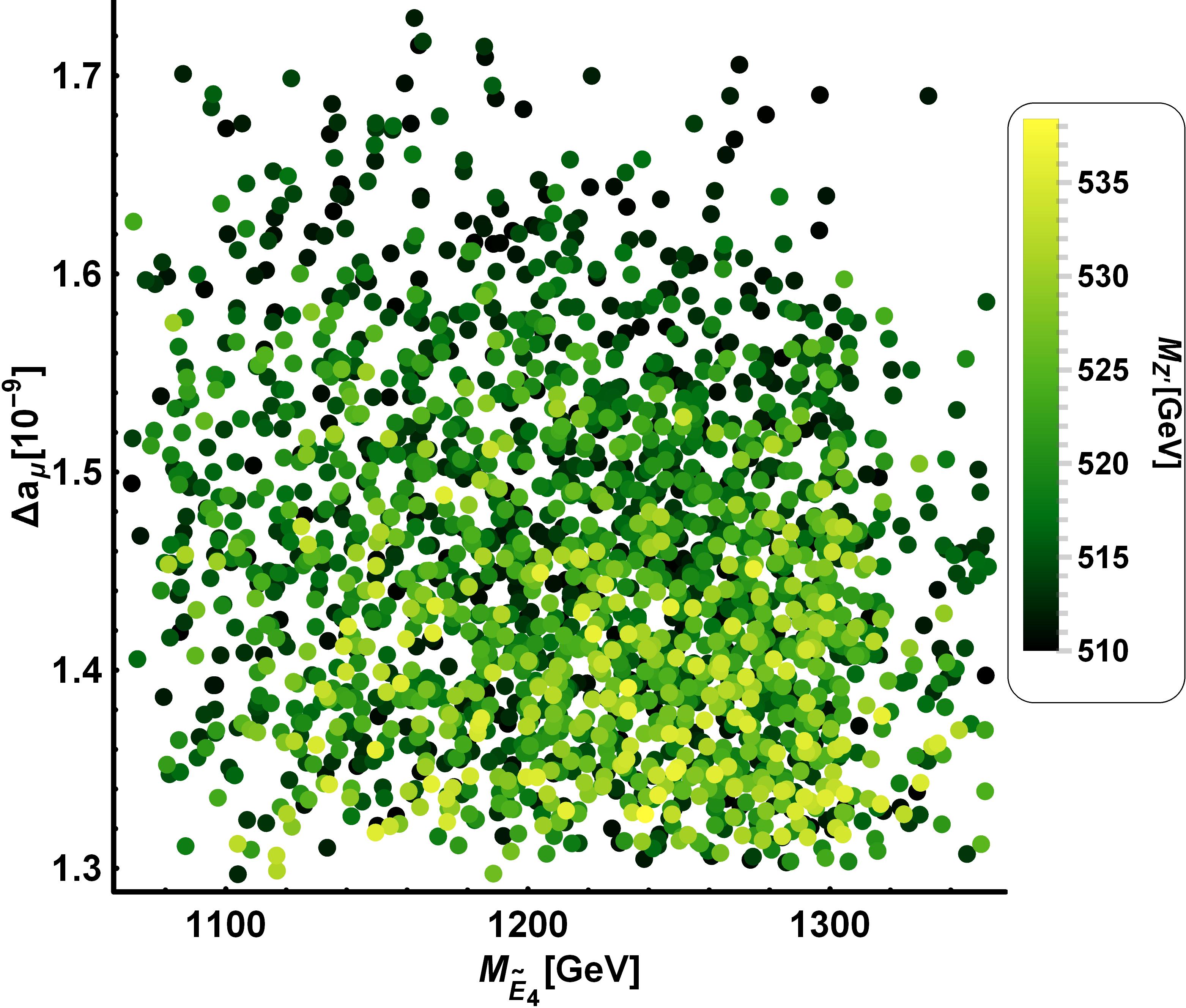}
	}
\end{subfigure} 
\begin{subfigure}{0.49\textwidth}
	\scalebox{0.9}{
	\includegraphics[keepaspectratio,width=1.0\textwidth]{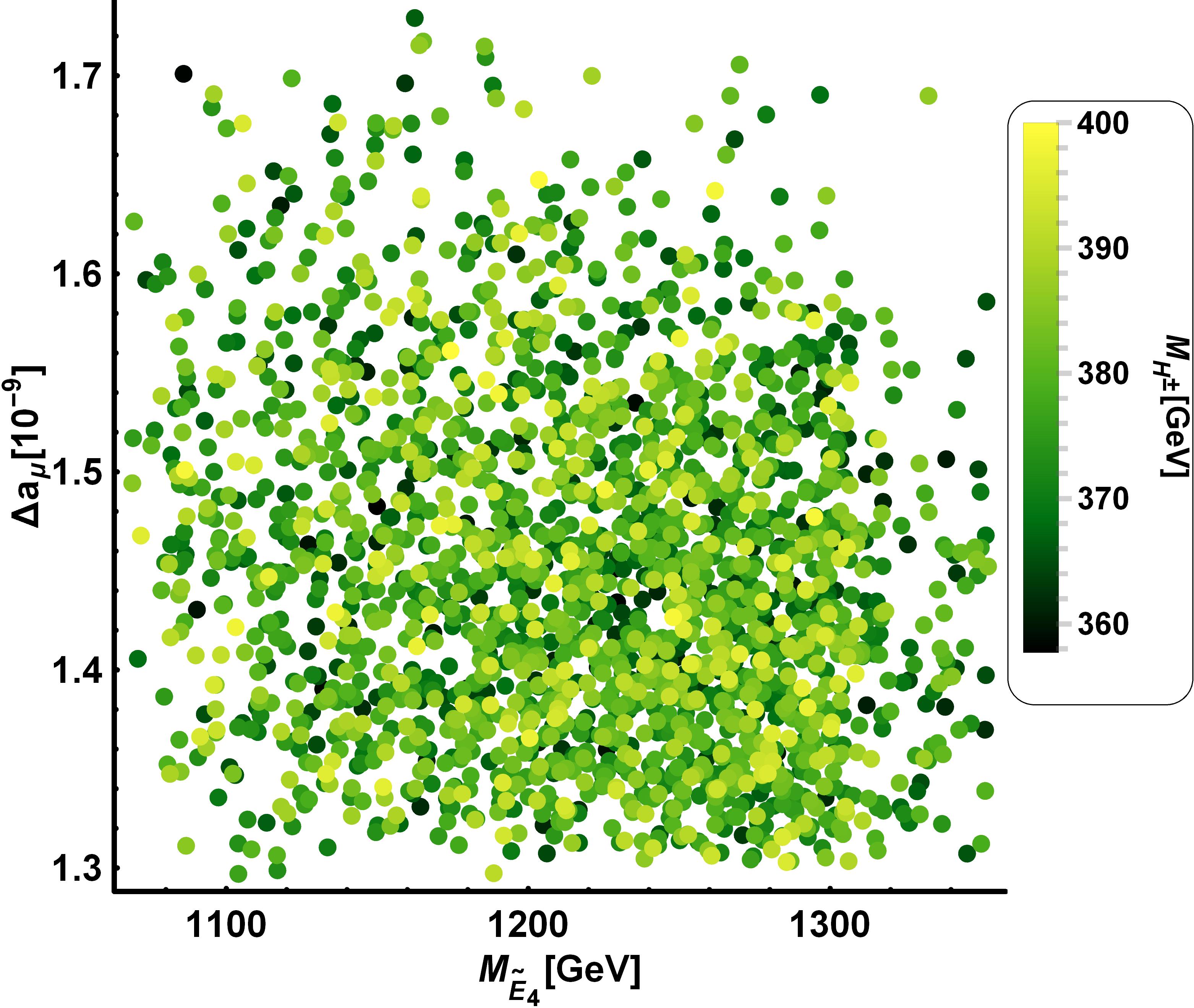}
	} 
\end{subfigure}
\caption{Second numerical scan result for the scalar-mediated observables. The upper two plots show a correlation between $M_{Z^{\prime}}$ and muon $g-2$ with scalar exchange with vectorlike singlet charged lepton mass $M_{\widetilde{E}_{4}}$ in legend (left) and non-SM charged Higgs mass $M_{H^{\pm}}$ in legend (right). The middle two plots show a loose correlation between $M_{Z^{\prime}}$ and $M_{\widetilde{E}_{4}}$ with $M_{H^{\pm}}$ (left) and with muon $g-2$ (right) than the ones showed in the upper two plots. The lower two plots show a loose correlation between $M_{\widetilde{E}_{4}}$ and muon $g-2$ with $M_{Z^{\prime}}$ (left) and with $M_{H^{\pm}}$ (right).}
\label{fig:scalar-mediated_observables2}
\end{figure}
\endgroup
In Figure~\ref{fig:scalar-mediated_observables2}, we discuss interplays among muon $g-2$, vectorlike singlet charged lepton mass $M_{\widetilde{E}_{4}}$, $Z^{\prime}$ mass, and non-SM charged scalar mass $M_{H^{\pm}}$. The upper-left plot in Figure~\ref{fig:scalar-mediated_observables2} shows an interesting correlation between $Z^{\prime}$ mass and muon $g-2$ with vectorlike singlet charged lepton mass $M_{\widetilde{E}_{4}}$. Here, the $Z^{\prime}$ does not participate in the muon $g-2$ with scalar exchange, however the $Z^{\prime}$ mass is determined by the scalar potential as well as the Goldstone boson $G_{Z^{\prime}}$ arising in the CP-odd scalar sector. What the upper-left plot of Figure~\ref{fig:scalar-mediated_observables2} tells is the muon $g-2$ and $R_{K^{*}}$ anomaly can not be explained by the same new physics, as the muon $g-2$ is explained by the non-SM scalars $H_{1,2}, A_{1}$ whereas the $R_{K^{*}}$ anomaly is explained by the $Z^{\prime}$ gauge boson whose mass runs from $510$ to $550\func{GeV}$, however their new physics sources $Z^{\prime}$ and non-SM scalars $H_{1,2},A_{1}$ are quite closely interconnected via the scalar potential. On top of that, the upper-left plot tells that the fact we are able to reach up to unity for the $Z^{\prime}$ coupling constant $g_{X}$ is a pleasing result, which implies that if the $Z^{\prime}$ coupling constant has order of $10^{-1}$ instead of unity, the vev $v_{3}$ must have order of $\func{TeV}$ scale in order to fit the $M_{Z^{\prime}}$ range constrained by $R_{K^{*}}$ anomaly and this makes explaining the muon $g-2$ much more challenging. The upper-right plot shares the same $x$ and $y$ axis as in the upper-left plot but $M_{\widetilde{E}_{4}}$ in the legend replaced by the charged non-SM scalar mass $M_{H^{\pm}}$. The middle two plots investigate correlations between $M_{Z^{\prime}}$ and $M_{\widetilde{E}_{4}}$ with $M_{H^{\pm}}$ (left) and with muon $g-2$ (right) and it is less correlated than the ones shown in the upper two plots. The lower two plots show a loose correlation between $M_{\widetilde{E}_{4}}$ and muon $g-2$ with $M_{Z^{\prime}}$ (left) and with $M_{H^{\pm}}$ (right). Concluding this numerical study, the muon $g-2$ and $R_{K^{*}}$ anomalies can not be simultaneously explained via the same new physics, however they can be explained by the different new physics, $Z^{\prime}$ for the $R_{K^{*}}$ anomaly and non-SM scalars $H_{1,2},A_{1}$ for the muon $g-2$, and the new physics sources are closely interconnected to each other via the scalar potential while fitting all the oblique parameters $T,S,U$ and $W$ mass anomaly at their $1,2\sigma$ constraints.
\section{CONCLUSION} \label{sec:VIII}
Searching for a simultaneous explanation for the muon and electron anomalous magnetic moment $g-2$ as a proof of new physics has been an interesting study. This topic has received a lot of attention by the particle physics community due to their overall sign difference in the $g-2$ electron and muon anomalies and their relative magnitude difference until the renewed results for the muon and electron $g-2$ were revealed in 2020. The renewed results for the muon and electron $g-2$ has significantly changed the situation as the muon $g-2$ anomaly gets more important by increasing its $3.3\sigma$ SM deviation up to $4.2\sigma$ whereas the electron $g-2$ gets less important since it has been reduced from $-2.4\sigma$ SM deviation up to $1.6\sigma$. This experimental situation suggests that studying the muon $g-2$ and the $B$ anomalies is more important than addressing the muon and the electron $g-2$.  
 The muon $g-2$ and $B$ anomalies can be connected via a massive neutral $Z^{\prime}$ gauge boson or via a leptoquark, and here in this work we use the $Z^{\prime}$ gauge boson in order to address these anomalies. 
\\~\\
We start from this question: is it possible to simultaneously explain the muon $g-2$ and $B$ anomalies in an extended $Z^{\prime}$ model? Generally, answering to this question is not easy since there are many different $Z^{\prime}$ models. The work~\cite{Navarro:2021sfb}, which has been a good motivation for this paper and a guideline, showed that the muon $g-2$ and $B$ anomalies can be simultaneously explained by the $Z^{\prime}$ exchange with a fewer number of free parameters. However, the Fermionic $Z^{\prime}$ model considered in~\cite{Navarro:2021sfb} can not gain access to the CLFV $\tau \rightarrow \mu \gamma$ decay, as the BSM model does not have mixings between the SM fermions, which appears to be one of the most stringent constraints in this BSM theory. The BSM theory under consideration can feature all the mixings between all fermions, thus allowing to accommodate the CLFV decays and then our analysis reveals a very different direction, which is that the muon $g-2$ and $R_{K^{*}}$ anomalies can not be simultaneously explained by the same new physics under all the current constraints as the predicted lightest $Z^{\prime}$ mass, $261\func{GeV}$, given by the squared gauge mass matrix can not be accommodated by the $Z^{\prime}$ mass range derived from the $R_{K^{*}}$ anomaly.
\\~\\
For this investigation, we consider an extended 2HDM theory where the SM gauge symmetry is augmented by the inclusion of a $U(1)^{\prime}$ local symmetry and the SM fermionic spectrum is extended by adding a complete vectorlike family, whereas the scalar sector is enlarged by a singlet flavon, whose inclusion is necessary to trigger the spontaneous breaking of the $U(1)^{\prime}$ gauge symmetry. In the model under consideration, the SM fermions acquire their masses via the low energy scale seesaw mechanism mediated by heavy fermions belonging to the vectorlike family. Based on the low energy scale seesaw mechanism, we build the mass matrices for the SM fermions and then explicitly diagonalize them. Using the mixing matrices arising from the diagonalization of the fermion mass matrices, 
we determine the $Z^{\prime}$ coupling constants in the physical basis. Unlike the other charged fermions, we only require a $Z^{\prime}$ coupling constant to neutrino pair for the neutrino trident constraint, so we derive it via an analytical approach instead of setting an explicit neutrino mass matrix.
\\~\\
After determining the $Z^{\prime}$ coupling constants in the physical basis, we explore the charged lepton sector observables such as the muon $g-2$, the CLFV $\tau \rightarrow \mu \gamma, \tau \rightarrow 3\mu$ decays  
for the purpose of constraining the 
$Z^{\prime}$ coupling constant $g_{X}$ and to find that its upper bound  
can reach up to unity. At the same time, the muon $g-2$ mediated by the $Z^{\prime}$ gauge boson only gives negative contributions whose corresponding magnitudes are of the order of $10^{-11}$ thus implying that the $Z^{\prime}$ gauge boson exchange does not successfully explain the muon anomalous magnetic moment.  
 Next we move to the analysis of the quark phenomenology and then we proceed 
 to derive mass ranges from the experimental constraints on the $R_{K^{*}}$ anomaly and $B_{s}$ meson oscillation. 
\\~\\
As for the $R_{K}$ anomaly, we discussed both the theoretically clean fit, which can explain both the $R_{K^{*}}$ anomaly and the $B_{s} \rightarrow \overline{\mu}\mu$ data without theoretical uncertainty, and the global fit, which is an experimental fit with sizeable uncertainty. We further separated each fit by two cases where one conserves all the constraints, the CMS and $\tau \rightarrow \mu \gamma$ experimental bound, and the other conserves only the $\tau \rightarrow \mu \gamma$ constraint and the reason why we separated each fit is it is enough to conclude that $R_{K^{*}}$ anomaly and muon $g-2$ can not be explained by the same new physics simultaneously as there is no overlapped $Z^{\prime}$ mass range, however the other fit can accommodate both the $Z^{\prime}$ for the $R_{K^{*}}$ anomaly and non-SM scalars for the muon $g-2$ via the scalar potential under consideration while fitting the oblique parameters $T,S,U$ as well as the $W$ mass anomaly at their $1$ or $2\sigma$ constraints, so we call $Z^{\prime}$ mass range derived from the other fit ``theoretically interesting $Z^{\prime}$ mass range". And then we investigated the left- and right-handed Wilson coefficients and found that the right-handed Wilson coefficient has a very limited overlapped region compared to the left-handed one. 
 We also confirmed that our numerical prediction shows that there is no overlapped region between the left- and right-handed Wilson coefficients in both theoretically clean fit and global fit, so confirming that the $R_{K^{*}}$ anomaly is likely to be explained by only left-handed fermion interactions and constraining $M_{Z^{\prime}}$ range to be from $202.218\func{GeV}$ up to $219.202\func{GeV}$ under all the constraints in the BSM model under consideration. We also analyzed the $B_{s}$ meson oscillation with three fits, which are FLAG, Average, and experimental fit, and found that the experimental fit gives the most stringent result: $M_{Z^{\prime}} > 6.5\func{TeV}$ and
that there is no overlapped region between the ones obtained from the $R_{K^{*}}$ anomaly and from the $B_{s}$ meson oscillation, so concluding that the $B_{s}$ meson oscillation is likely to be explained by another much heavier new physics source. The collider constraints, compared to the $R_{K^{*}}$ anomaly and $B_{s}$ meson oscillation, can not constrain the parameter space at all and lastly the CKM mixing matrix is considered as one of the most stringent constraints. The CKM mixing matrix in this BSM theory can be fitted up to a good approximation and we constrain the flavor changing coupling constants such as $g_{bs}$ using the CKM mixing matrix and then the derived ranges of $g_{bs}$ are used in the analysis of the $R_{K^{*}}$ anomaly and $B_{s}$ meson oscillation in order to derive $M_{Z^{\prime}}$ ranges as previously mentioned.  
The derived vectorlike quark mass ranges fullfill the ATLAS constraints and, especially, we obtain that the vectorlike doublet quark masses $M_{U_{4}}$ and $M_{D_{4}}$ are degenerate each other and it is greatly helpful to explain the oblique $T,S,U$ parameters since if the vectorlike doublet quarks have a significant mass difference between them, it gives unacceptably large contributions, which yield values for the oblique parameters outside their experimentally allowed range.
\\~\\
We have determined the highest allowed value of the 
 $Z^{\prime}$ coupling constant $g_{X}$ in the charged lepton sector and then computed $M_{Z^{\prime}}$ ranges from the experimental data on the $R_{K^{*}}$ anomaly and $B_{s}$ meson oscillation. As the contribution to the muon $g-2$ anomaly arising from the one loop virtual $Z^{\prime}$ exchange is negative, the one loop level scalar exchange is required to successfully accommodate the muon anomalous magnetic moment.  
 As a first step to consider scalar exchange, we construct a scalar potential and then define the squared CP-even and -odd and charged mass matrices and also determine the degrees of freedom for the Goldstone bosons $G_{Z}, G_{Z^{\prime}}, G_{W^{\pm}}$. Next we discuss the diphoton signal strength $R_{\gamma\gamma}$, the vacuum stability conditions, the muon $g-2$ anomaly induced by the scalar exchange and we determine analytic expressions for the oblique $T,S,U$ parameters and we determine the implications of the model in the $W$ mass anomaly. Through the numerical study, we confirm that the one loop induced muon anomalous magnetic moment via the scalar exchange can be successfully accommodated within the $2\sigma$ experimentally allowed range whereas the obtained values of the oblique parameters and the $W$ mass anomalies are consistent with their $1$ or $2\sigma$ constraints in the ``theoretically interesting $Z^{\prime}$ mass range". We have further obtained some correlations between these observables.
  Answering to the question in the title of this work, the muon $g-2$ and $R_{K^{*}}$ anomalies cannot be explained by the same new physics within the framework of the extended 2HDM theory considered in this work, since the lightest $Z^{\prime}$ mass, $261\func{GeV}$, resulting from the squared gauge mass matrix can not be fitted up to the $Z^{\prime}$ mass range derived from the $R_{K^{*}}$ anomaly global fit with all the constraints and we conclude the $R_{K^{*}}$ anomaly and muon $g-2$ are an independent new physics signal in this BSM theory.

\section*{Acknowledgements}
This research is supported by the National Science Centre (Poland) under the research Grant No. 2017/26/E/ST2/00470.
This research has received funding from Fondecyt (Chile), Grant No.~1210378, ANID PIA/APOYO AFB180002 and ANID- Programa Milenio - code ICN2019\_044.

\appendix

\section{Vacuum polarization amplitude $\Pi_{33}$} \label{app:A}
The correct analytic expressions for the vacuum polarization amplitudes $\Pi_{33}^{1,2,3,4}$ contributing to the $T$ parameter read in Equation~\ref{eqn:Pi33_T}:
\begingroup
\begin{equation}
\begin{split}
\Pi_{33}^{1} &= -2\left( \frac{g^{2}}{4} \right) \frac{3}{\left( 4\pi \right)^{2}} \left[ m_{t}^{2} \ln \frac{\Lambda^{2}}{m_{t}^{2}} + m_{b}^{2} \ln \frac{\Lambda^{2}}{m_{b}^{2}} + 2 M_{U_{4}}^{2} \ln \frac{\Lambda^{2}}{M_{U_{4}}^{2}} + 2 M_{D_{4}}^{2} \ln \frac{\Lambda^{2}}{M_{D_{4}}^{2}} \right] 
\\
&-2\left( \frac{g^{2}}{4} \right) \frac{1}{\left( 4\pi \right)^{2}} \left[ 2M_{\nu_{4}}^{2} \ln \frac{\Lambda^{2}}{M_{\nu_{4}}^{2}} + 2M_{E_{4}}^{2} \ln \frac{\Lambda^{2}}{M_{E_{4}}^{2}} \right],
\\
\Pi_{33}^{2} &= -\left( \frac{g^{2}}{2} \right) \frac{1}{\left( 4\pi \right)^{2}} \left[ M_{W}^{2} \ln \frac{\Lambda^{2}}{M_{W}^{2}} + M_{H_{d}^{-}}^{2} \ln \frac{\Lambda^{2}}{M_{H_{d}^{-}}^{2}} \right],
\\
\Pi_{33}^{3} &= -\left( \frac{g^{2}}{2} \right) \frac{1}{\left( 4\pi \right)^{2}} \left[ \frac{m_{h}^{2} + M_{W}^{2}}{2} \left[ \ln \frac{\Lambda^{2}}{M_{W}^{2}} + \frac{1}{2} \right] + \frac{m_{h}^{4}}{2\left( m_{h}^{2} - M_{W}^{2} \right)} \ln \frac{M_{W}^{2}}{m_{h}^{2}} \right]
\\
&-\left( \frac{g^{2}}{2} \right) \frac{1}{\left( 4\pi \right)^{2}} \left[ \frac{M_{H_{dR}^{0}}^{2} + M_{H_{dI}^{0}}^{2}}{2} \left[ \ln \frac{\Lambda^{2}}{M_{H_{dI}^{0}}^{2}} + \frac{1}{2} \right] + \frac{M_{H_{dR}^{0}}^{4}}{2\left( M_{H_{dR}^{0}}^{2} - M_{H_{dI}^{0}}^{2} \right)} \ln \frac{M_{H_{dI}^{0}}^{2}}{M_{H_{dR}^{0}}^{2}} \right]
\\
\Pi_{33}^{4} &= \left( \frac{g^{4}}{16} \right) \frac{2v_{u}^{2}}{\left( 4\pi \right)^{2}} \left[ \frac{3}{4} \ln \frac{\Lambda^{2}}{m_{h}^{2}} + \frac{3}{4} \frac{M_{A}^{2}}{M_{A}^{2} - m_{h}^{2}} \ln \frac{m_{h}^{2}}{M_{A}^{2}} - \frac{7}{8} \right] + \left( \frac{g^{4}}{16} \right) \frac{2v_{d}^{2}}{\left( 4\pi \right)^{2}} \left[ \frac{3}{4} \ln \frac{\Lambda^{2}}{M_{H_{dR}^{0}}^{2}} + \frac{3}{4} \frac{M_{A}^{2}}{M_{A}^{2} - M_{H_{dR}^{0}}^{2}} \ln \frac{M_{H_{dR}^{0}}^{2}}{M_{A}^{2}} - \frac{7}{8} \right],
\\
&+ \left( \frac{g^{2} g^{\prime 2}}{4} \right) \frac{2v_{u}^{2}}{\left( 4\pi \right)^{2}} \left[ \frac{3}{4} \ln \frac{\Lambda^{2}}{m_{h}^{2}} + \frac{3}{4} \frac{M_{B}^{2}}{M_{B}^{2} - m_{h}^{2}} \ln \frac{m_{h}^{2}}{M_{B}^{2}} - \frac{7}{8} \right] + \left( \frac{g^{2} g^{\prime 2}}{4} \right) \frac{2v_{d}^{2}}{\left( 4\pi \right)^{2}} \left[ \frac{3}{4} \ln \frac{\Lambda^{2}}{M_{H_{dR}^{0}}^{2}} + \frac{3}{4} \frac{M_{B}^{2}}{M_{B}^{2} - M_{H_{dR}^{0}}^{2}} \ln \frac{M_{H_{dR}^{0}}^{2}}{M_{B}^{2}} - \frac{7}{8} \right]
\\
&+ \left( g^{2} g_{X}^{2} \right) \frac{2v_{u}^{2}}{\left( 4\pi \right)^{2}} \left[ \frac{3}{4} \ln \frac{\Lambda^{2}}{m_{h}^{2}} + \frac{3}{4} \frac{M_{V}^{2}}{M_{V}^{2} - m_{h}^{2}} \ln \frac{m_{h}^{2}}{M_{V}^{2}} - \frac{7}{8} \right] + \left( g^{2} g_{X}^{2} \right) \frac{2v_{d}^{2}}{\left( 4\pi \right)^{2}} \left[ \frac{3}{4} \ln \frac{\Lambda^{2}}{M_{H_{dR}^{0}}^{2}} + \frac{3}{4} \frac{M_{V}^{2}}{M_{V}^{2} - M_{H_{dR}^{0}}^{2}} \ln \frac{M_{H_{dR}^{0}}^{2}}{M_{V}^{2}} - \frac{7}{8} \right],
\label{eqn:Pi33_T}
\end{split}
\end{equation}
\endgroup
where $\Lambda$ is an arbitrary high energy scale and cancels out when we consider the whole oblique parameters $T,S,U$, therefore not affecting the oblique parameters, and it should be clarify that all the theoretical predictions $\Pi_{33}^{1,2,3,4}$ for the oblique parameter $T$ are derived without any approximation or assumption and this correct derivation will be pursued while derive the rest of theoretical predictions for the oblique parameters. 
\section{Vacuum polarization amplitude $\Pi_{11}$} \label{app:B}
The correct analytic expressions for the vacuum polarization amplitudes $\Pi_{11}^{1,2,3,4}$ contributing to the $T$ parameter read in Equation~\ref{eqn:Pi11_T}:
\begingroup
\begin{equation}
\begin{split}
\Pi_{11}^{1} &= -2 \left( \frac{g^{2}}{4} \right) \frac{3}{\left( 4\pi \right)^{2}} \left[ \frac{m_{t}^{2} + m_{b}^{2}}{2} \left[ \frac{\Lambda^{2}}{m_{t}^{2}} + \frac{1}{2} \right] + \frac{m_{b}^{4}}{2\left( m_{b}^{2} - m_{t}^{2} \right)} \ln \frac{m_{t}^{2}}{m_{b}^{2}} \right]
\\
&-2 \left( \frac{g^{2}}{4} \right) \frac{3}{\left( 4\pi \right)^{2}} \left[ \frac{m_{t}^{2} + m_{b}^{2}}{2} \left[ \frac{\Lambda^{2}}{m_{b}^{2}} + \frac{1}{2} \right] + \frac{m_{t}^{4}}{2\left( m_{t}^{2} - m_{b}^{2} \right)} \ln \frac{m_{b}^{2}}{m_{t}^{2}} \right]
\\
&-2 \left( \frac{g^{2}}{4} \right) \frac{3}{\left( 4\pi \right)^{2}} \left[ 2 M_{U_{4}}^{2} \ln \frac{\Lambda^{2}}{M_{U_{4}}^{2}} + 2 M_{D_{4}}^{2} \ln \frac{\Lambda^{2}}{M_{D_{4}}^{2}} \right] -2 \left( \frac{g^{2}}{4} \right) \frac{1}{\left( 4\pi \right)^{2}} \left[ 2 M_{\nu_{4}}^{2} \ln \frac{\Lambda^{2}}{M_{\nu_{4}}^{2}} + 2 M_{E_{4}}^{2} \ln \frac{\Lambda^{2}}{M_{E_{4}}^{2}} \right],
\\
\Pi_{11}^{2} &= -\left( \frac{g^{2}}{2} \right) \frac{1}{\left( 4\pi \right)^{2}} \left[ M_{W}^{2} \ln \frac{\Lambda^{2}}{M_{W}^{2}} \right] -\left( \frac{g^{2}}{2} \right) \frac{1}{\left( 4\pi \right)^{2}} \left[ \frac{M_{W}^{2} + m_{h}^{2}}{2} \ln \frac{\Lambda^{2}}{m_{h}^{2}} + \frac{1}{2} \right] + \frac{M_{W}^{4}}{2\left( M_{W}^{2} - m_{h}^{2} \right)} \ln \frac{m_{h}^{2}}{M_{W}^{2}}
\\
&-\left( \frac{g^{2}}{2} \right) \frac{1}{\left( 4\pi \right)^{2}} \left[ \frac{M_{H_{dI}^{0}}^{2} + M_{H_{d}^{-}}^{2}}{2} \left[ \ln \frac{\Lambda^{2}}{M_{H_{dI}^{0}}^{2}} + \frac{1}{2} \right] + \frac{M_{H_{d}^{-}}^{4}}{2\left( M_{H_{d}^{-}}^{2} - M_{H_{dI}^{0}}^{2} \right)} \ln \frac{M_{H_{dI}^{0}}^{2}}{M_{H_{d}^{-}}^{2}} \right]
\\
&-\left( \frac{g^{2}}{2} \right) \frac{1}{\left( 4\pi \right)^{2}} \left[ \frac{M_{H_{dR}^{0}}^{2} + M_{H_{d}^{-}}^{2}}{2} \left[ \ln \frac{\Lambda^{2}}{M_{H_{dR}^{0}}^{2}} + \frac{1}{2} \right] + \frac{M_{H_{d}^{-}}^{4}}{2\left( M_{H_{d}^{-}}^{2} - M_{H_{dR}^{0}}^{2} \right)} \ln \frac{M_{H_{dR}^{0}}^{2}}{M_{H_{d}^{-}}^{2}} \right],
\\
\Pi_{11}^{3} &= \left( \frac{g^{2} g^{\prime 2}}{4} \right) \frac{2v_{u}^{2}}{\left( 4\pi \right)^{2}} \left[ \frac{3}{4} \ln \frac{\Lambda^{2}}{M_{W}^{2}} + \frac{3}{4} \frac{M_{B}^{2}}{M_{B}^{2} - M_{W}^{2}} \ln \frac{M_{W}^{2}}{M_{B}^{2}} - \frac{7}{8} \right] 
\\
&+ \left( g^{2} g_{X}^{2} \right) \frac{2v_{u}^{2}}{\left( 4\pi \right)^{2}} \left[ \frac{3}{4} \ln \frac{\Lambda^{2}}{M_{W}^{2}} + \frac{3}{4} \frac{M_{V}^{2}}{M_{V}^{2} - M_{W}^{2}} \ln \frac{M_{V}^{2}}{M_{B}^{2}} - \frac{7}{8} \right]
\\
&+ \left( \frac{g^{2} g^{\prime 2}}{4} \right) \frac{2v_{u}^{2}}{\left( 4\pi \right)^{2}} \left[ \frac{3}{4} \ln \frac{\Lambda^{2}}{M_{H_{d}^{-}}^{2}} + \frac{3}{4} \frac{M_{B}^{2}}{M_{B}^{2} - M_{H_{d}^{-}}^{2}} \ln \frac{M_{H_{d}^{-}}^{2}}{M_{B}^{2}} - \frac{7}{8} \right] 
\\
&+ \left( g^{2} g_{X}^{2} \right) \frac{2v_{u}^{2}}{\left( 4\pi \right)^{2}} \left[ \frac{3}{4} \ln \frac{\Lambda^{2}}{M_{H_{d}^{-}}^{2}} + \frac{3}{4} \frac{M_{V}^{2}}{M_{V}^{2} - M_{H_{d}^{-}}^{2}} \ln \frac{M_{H_{d}^{-}}^{2}}{M_{V}^{2}} - \frac{7}{8} \right]
\\
\Pi_{11}^{4} &= \left( \frac{g^{4}}{16} \right) \frac{2v_{u}^{2}}{\left( 4\pi \right)^{2}} \left[ \frac{3}{4} \ln \frac{\Lambda^{2}}{m_{h}^{2}} + \frac{3}{4} \frac{M_{A}^{2}}{M_{A}^{2} - m_{h}^{2}} \ln \frac{m_{h}^{2}}{M_{A}^{2}} - \frac{7}{8} \right]
\\
&+ \left( \frac{g^{4}}{16} \right) \frac{2v_{d}^{2}}{\left( 4\pi \right)^{2}} \left[ \frac{3}{4} \ln \frac{\Lambda^{2}}{M_{H_{dR}^{0}}^{2}} + \frac{3}{4} \frac{M_{A}^{2}}{M_{A}^{2} - M_{H_{dR}^{0}}^{2}} \ln \frac{M_{H_{dR}^{0}}^{2}}{M_{A}^{2}} - \frac{7}{8} \right],
\label{eqn:Pi11_T}
\end{split}
\end{equation}
\endgroup
where it should be clarified that the new physics contributions led by vectorlike doublet fermions in $\Pi_{11}^{1}$ are derived when masses of the vectorlike doublet fermions are degenerate, therefore not affecting the oblique parameters at all.
\section{Vacuum polarization amplitude $\frac{d\Pi_{30}}{dq^{2}}$} \label{app:C}
The correct analytic expressions for the vacuum polarization amplitudes $\Pi_{30}^{1,2,3,4}$ differentiated with respect to $q^{2}$ read in Equation~\ref{eqn:dPi30_S}:
\begingroup
\begin{equation}
\begin{split}
\frac{d\Pi_{30}^{1}}{dq^{2}} &= \left( \frac{g g^{\prime}}{12} \right) \frac{3}{(4\pi)^2} \frac{2}{3} \Bigg[ \ln \frac{\Lambda^2}{m_{t}^{2}} - \ln \frac{\Lambda^2}{m_{b}^{2}} + 2\left( \ln \frac{\Lambda^2}{M_{U_4}^{2}} - \ln \frac{\Lambda^2}{M_{D_4}^{2}} \right) + \left( -\frac{g g^{\prime}}{4} \right) \frac{1}{(4\pi)^2} \frac{2}{3} \Bigg[ 2\left( \ln \frac{\Lambda^2}{M_{\nu_4}^{2}} - \ln \frac{\Lambda^2}{M_{E_4}^{2}} \right) \Bigg],
\\
\frac{d\Pi_{30}^{2}}{dq^{2}} &= \frac{g g^{\prime}}{\left( 4\pi \right)^2} \frac{1}{12}  \left[ \ln \frac{\Lambda^2}{M_{W}^{2}} + \ln \frac{\Lambda^2}{M_{H_{d}^{-}}^{2}} \right],
\\
\frac{d\Pi_{30}^{3}}{dq^2} &= -\frac{g g^{\prime}}{\left( 4\pi \right)^{2}} \Bigg[ \frac{1}{12} \ln \frac{\Lambda^{2}}{m_{h}^{2}} - \frac{1}{18} + \left( \frac{M_{W}^{2}-m_{h}^{2}}{4} - \frac{M_{W}^{2}}{6} \right) \left[ \frac{1}{2\left( M_{W}^{2} - m_{h}^{2} \right)} + \frac{M_{W}^{2}}{\left( M_{W}^{2} - m_{h}^{2} \right)^{2}} + \frac{M_{W}^{4}}{\left( M_{W}^{2} - m_{h}^{2} \right)^{3}} \ln \frac{m_{h}^{2}}{M_{W}^{2}} \right] 
\\
&+ \left( m_{h}^{2} \leftrightarrow M_{H_{dR}^{0}}^{2}, M_{W}^{2} \leftrightarrow M_{H_{dI}^{0}}^{2} \right) \Bigg],
\\
\frac{d\Pi_{30}^{4}}{dq^{2}} &= - \frac{g g^{\prime}}{\left( 4\pi \right)^{2}} \left[ - \frac{M_{W}^{2}}{24\left( M_{W}^{2} - m_{h}^{2} \right)} + \left( \frac{1}{2} - \frac{M_{W}^{2}}{3\left( M_{W}^{2} - m_{h}^{2} \right)} \right) \left( \frac{M_{W}^{2}}{4\left( M_{W}^{2} - m_{h}^{2} \right)} + \frac{M_{W}^{4}}{4\left( M_{W}^{2} - m_{h}^{2} \right)^{2}} \ln \frac{m_{h}^{2}}{M_{W}^{2}} \right) \right]
\\
&- \frac{g g^{\prime}}{\left( 4\pi \right)^{2}} \left[ -\frac{M_{B}^{2}}{24\left( M_{B}^{2} - m_{h}^{2} \right)} + \left( \frac{1}{2} - \frac{M_{B}^{2}}{3\left( M_{B}^{2} - m_{h}^{2} \right)} \right) \left( \frac{M_{B}^{2}}{4\left( M_{B}^{2} - m_{h}^{2} \right)} + \frac{M_{B}^{4}}{4\left( M_{B}^{2} - m_{h}^{2} \right)^{2}} \ln \frac{m_{h}^{2}}{M_{B}^{2}} \right) \right]
\\
&- \frac{g g^{\prime} g_{X}^{2}}{\left( 4\pi \right)^{2}} \frac{2M_{W}^{2}}{g^{2}} \left[ - \frac{1}{6\left( M_{V}^{2} - m_{h}^{2} \right)} + \left( \frac{1}{2} - \frac{M_{V}^{2}}{3\left( M_{V}^{2} - m_{h}^{2} \right)} \right) \left( \frac{1}{\left( M_{V}^{2} - m_{h}^{2} \right)} + \frac{M_{V}^{2}}{\left( M_{V}^{2} - m_{h}^{2} \right)^{2}} \ln \frac{m_{h}^{2}}{M_{V}^{2}} \right) \right]
\\
&+ \left( m_{h}^{2} \leftrightarrow M_{H_{dR}^{0}}^{2} \right) \Bigg],
\label{eqn:dPi30_S}
\end{split}
\end{equation}
\endgroup
where it is worth emphasizing that all the analytical expressions for the vacuum polarization amplitudes $d\Pi_{30}^{1,2,3,4}/dq^{2}$ are derived without any approximation and assumption, as in the $\Pi_{33,11}$ in the $T$ parameter, 
\section{Vacuum polarization amplitude $\frac{d\Pi_{33}}{dq^{2}}$} \label{app:D}
The correct analytic expressions for the vacuum polarization amplitudes $\Pi_{33}^{1,2,3,4}$ differentiated with respect to $q^{2}$ read in Equation~\ref{eqn:dPi33_U}:
\begingroup
\begin{equation}
\begin{split}
\frac{d\Pi_{33}^{1}}{dq^{2}} &= \left( \frac{g^{2}}{4} \right) \frac{3}{(4\pi)^2} \frac{2}{3} \Bigg[ \ln \frac{\Lambda^2}{m_{t}^{2}} + \ln \frac{\Lambda^2}{m_{b}^{2}} + 2\left( \ln \frac{\Lambda^2}{M_{U_4}^{2}} + \ln \frac{\Lambda^2}{M_{D_4}^{2}} \right) \Bigg] + \left( \frac{g^{2}}{4} \right) \frac{1}{(4\pi)^2} \frac{2}{3} \Bigg[ 2\left( \ln \frac{\Lambda^2}{M_{\nu_4}^{2}} + \ln \frac{\Lambda^2}{M_{E_4}^{2}} \right) \Bigg]
\\
\frac{d\Pi_{33}^{2}}{dq^{2}} &= \frac{g^{2}}{\left( 4\pi \right)^2} \frac{1}{12}  \left[ \ln \frac{\Lambda^2}{M_{W}^{2}} + \ln \frac{\Lambda^2}{M_{H_{d}^{-}}^{2}} \right]
\\
\frac{d\Pi_{33}^{3}}{dq^{2}} &= \frac{g^{2}}{\left( 4\pi \right)^{2}} \Bigg[ \frac{1}{12} \ln \frac{\Lambda^{2}}{M_{W}^{2}} - \frac{1}{18} + \left( \frac{m_{h}^{2}-M_{W}^{2}}{4} - \frac{m_{h}^{2}}{6} \right) \left[ \frac{1}{2\left( m_{h}^{2} - M_{W}^{2} \right)} + \frac{m_{h}^{2}}{\left( m_{h}^{2} - M_{W}^{2} \right)^{2}} + \frac{m_{h}^{4}}{\left( m_{h}^{2} - M_{W}^{2} \right)^{3}} \ln \frac{M_{W}^{2}}{m_{h}^{2}} \right] 
\\
&+ \left( m_{h}^{2} \leftrightarrow M_{H_{dR}^{0}}^{2}, M_{W}^{2} \leftrightarrow M_{H_{dI}^{0}}^{2} \right) \Bigg]
\\
\frac{d\Pi_{33}^{4}}{dq^{2}} &= \frac{g^{2}}{\left( 4\pi \right)^{2}} \left[ - \frac{M_{W}^{2}}{48\left( M_{W}^{2} - m_{h}^{2} \right)} + \left( \frac{1}{2} - \frac{M_{W}^{2}}{3\left( M_{W}^{2} - m_{h}^{2} \right)} \right) \left( \frac{M_{W}^{2}}{8\left( M_{W}^{2} - m_{h}^{2} \right)} + \frac{M_{W}^{4}}{8\left( M_{W}^{2} - m_{h}^{2} \right)^{2}} \ln \frac{m_{h}^{2}}{M_{W}^{2}} \right) \right]
\\
&+ \frac{g^{2}}{\left( 4\pi \right)^{2}} \left[ -\frac{M_{B}^{2}}{12\left( M_{B}^{2} - m_{h}^{2} \right)} + \left( \frac{1}{2} - \frac{M_{B}^{2}}{3\left( M_{B}^{2} - m_{h}^{2} \right)} \right) \left( \frac{M_{B}^{2}}{2\left( M_{B}^{2} - m_{h}^{2} \right)} + \frac{M_{B}^{4}}{2\left( M_{B}^{2} - m_{h}^{2} \right)^{2}} \ln \frac{m_{h}^{2}}{M_{B}^{2}} \right) \right]
\\
&+ \frac{g_{X}^{2} 2M_{W}^{2}}{\left( 4\pi \right)^{2}} \left[ - \frac{1}{6\left( M_{V}^{2} - m_{h}^{2} \right)} + \left( \frac{1}{2} - \frac{M_{V}^{2}}{3\left( M_{V}^{2} - m_{h}^{2} \right)} \right) \left( \frac{1}{\left( M_{V}^{2} - m_{h}^{2} \right)} + \frac{M_{V}^{2}}{\left( M_{V}^{2} - m_{h}^{2} \right)^{2}} \ln \frac{m_{h}^{2}}{M_{V}^{2}} \right) \right]
\\
&+ \left( m_{h}^{2} \leftrightarrow M_{H_{dR}^{0}}^{2} \right) \Bigg].
\label{eqn:dPi33_U}
\end{split}
\end{equation}
\endgroup
\section{Vacuum polarization amplitude $\frac{d\Pi_{11}}{dq^{2}}$} \label{app:E}
The correct analytic expressions for the vacuum polarization amplitudes $\Pi_{11}^{1,2,3,4}$ differentiated with respect to $q^{2}$ read in Equation~\ref{eqn:dPi11_U}:
\begingroup
\begin{equation}
\begin{split}
\frac{d\Pi_{11}^{1}}{dq^{2}} &= \left( g^{2} \right) \frac{3}{(4\pi)^2} \Bigg[ \frac{1}{6} \ln \frac{\Lambda^2}{m_{t}^{2}} - \frac{1}{9} + \left( \frac{m_{b}^{2}-m_{t}^{2}}{2} - \frac{m_{b}^{2}}{3} \right) \left( \frac{1}{2\left( m_{b}^{2} - m_{t}^{2} \right)} + \frac{m_{b}^{2}}{\left( m_{b}^{2} - m_{t}^{2} \right)^{2}} + \frac{m_{b}^{4}}{\left( m_{b}^{2} - m_{t}^{2} \right)^{3}} \ln \frac{m_{t}^{2}}{m_{b}^{2}} \right) \Bigg]
\\
&+ \left( m_{b}^{2} \leftrightarrow m_{t}^{2} \right)
\\
&+ \left( g^{2} \right) \frac{3}{\left( 4\pi \right)^{2}} \left[ \frac{1}{3} \left( \ln \frac{\Lambda^{2}}{M_{U_{4}}^{2}} + \ln \frac{\Lambda^{2}}{M_{D_{4}}^{2}} \right) \right] + \left( g^{2} \right) \frac{1}{\left( 4\pi \right)^{2}} \left[ \frac{1}{3} \left( \ln \frac{\Lambda^{2}}{M_{\nu_{4}}^{2}} + \ln \frac{\Lambda^{2}}{M_{E_{4}}^{2}} \right) \right]
\\
\frac{d\Pi_{11}^{2}}{dq^{2}} &= \frac{g^{2}}{\left( 4\pi \right)^2} \frac{1}{2}  \Bigg[ \frac{1}{6} \ln \frac{\Lambda^2}{M_{W}^{2}} 
\\
&+ \left[ \frac{1}{6} \ln \frac{\Lambda^{2}}{m_{h}^{2}} - \frac{1}{9} + \left( \frac{M_{W}^{2} - m_{h}^{2}}{2} - \frac{M_{W}^{2}}{3} \right) \left( \frac{1}{2\left( M_{W}^{2} - m_{h}^{2} \right)} + \frac{M_{W}^{2}}{\left( M_{W}^{2} - m_{h}^{2} \right)^{2}} + \frac{M_{W}^{4}}{\left( M_{W}^{2} - m_{h}^{2} \right)^{3}} \ln \frac{m_{h}^{2}}{M_{W}^{2}} \right) \right] \Bigg] 
\\
&+ \left( m_{h}^{2} \leftrightarrow M_{H_{dI}^{0}}^{2}, M_{W}^{2} \leftrightarrow M_{H_{d}^{-}}^{2} \right)
\\
&+ \left( m_{h}^{2} \leftrightarrow M_{H_{dR}^{0}}^{2}, M_{W}^{2} \leftrightarrow M_{H_{d}^{-}}^{2} \right)
\\
\frac{d\Pi_{11}^{3}}{dq^{2}} &=  -\frac{g^{2} g^{\prime 2} 2v_{u}^{2}}{4\left( 4\pi \right)^{2}} \frac{2}{4M_{B}^{2}} \Bigg[ \frac{M_{B}^{2}}{6\left( M_{B}^{2} - M_{W}^{2} \right)} - \left( \frac{M_{B}^{2}}{2} - \frac{M_{B}^{4}}{3\left( M_{B}^{2} - M_{W}^{2} \right)} \right) \left( \frac{1}{\left( M_{B}^{2} - M_{W}^{2} \right)} + \frac{M_{B}^{2}}{\left( M_{B}^{2} - M_{W}^{2} \right)^{2}} \ln \frac{M_{W}^{2}}{M_{B}^{2}} \right) 
\\
&+ \left( M_{B}^{2} \leftrightarrow M_{V}^{2}, \quad \frac{g^{2} g^{\prime 2}}{4} \leftrightarrow g^{2} g_{X}^{2} \right) 
\\
&+ \left( M_{W}^{2} \leftrightarrow M_{H_{d}^{-}}^{2}, \quad v_{u}^{2} \leftrightarrow v_{d}^{2} \right) 
\\
&+ \left( M_{W}^{2} \leftrightarrow M_{H_{d}^{-}}^{2}, \quad v_{u}^{2} \leftrightarrow v_{d}^{2}, \quad M_{B}^{2} \leftrightarrow M_{V}^{2}, \quad \frac{g^{2} g^{\prime 2}}{4} \leftrightarrow g^{2} g_{X}^{2} \right) \Bigg]
\\
\frac{d\Pi_{11}^{4}}{dq^{2}} &= \left( \frac{g^{4}}{16} \right) \frac{v_{u}^{2}}{\left( 4\pi \right)^{2}} \left[ - \frac{1}{6\left( M_{W}^{2} - m_{h}^{2} \right)} + \left( \frac{1}{2} - \frac{M_{W}^{2}}{3\left( M_{W}^{2} - m_{h}^{2} \right)} \right) \left( \frac{1}{\left( M_{W}^{2} - m_{h}^{2} \right)} + \frac{M_{W}^{2}}{\left( M_{W}^{2} - m_{h}^{2} \right)^{2}} \ln \frac{m_{h}^{2}}{M_{W}^{2}} \right) \right]
\\
&+ \left( m_{h}^{2} \leftrightarrow M_{H_{dR}^{0}}^{2}, \quad v_{u}^{2} \leftrightarrow v_{d}^{2} \right) \Bigg]
\label{eqn:dPi11_U}
\end{split}
\end{equation}
\endgroup

\end{document}